\documentclass[12pt]{article}
\usepackage[final]{graphicx}
\usepackage{amsmath}
\usepackage{amssymb}
\usepackage{xcolor}
\usepackage{tensor}
\usepackage{braket}
\usepackage{units}
\usepackage[hypertex,colorlinks=true,linkcolor=red,citecolor=blue]{hyperref}

\usepackage{amssymb}
\newcounter{comment}

{\refstepcounter{comment}%
\begin{quote}
\ttfamily\small$\blacksquare$ \textbf{\underline{Comment} $\sharp$\thecomment:}}%
{\end{quote}}

{
\begin{quote}
\ttfamily\small$\blacktriangleright$ \textbf{\underline{Reply} $\sharp$\thecomment:}}%
{\end{quote}}

\renewcommand{\theequation}{\arabic{section}.\arabic{equation}}%

\newcommand{\tffH}{{\mathcal H}}
\newcommand{\tffE}{{\mathcal E}}
\newcommand{\tffF}{{\mathcal F}}
\newcommand{\tfftH}{\widetilde{\mathcal H}}
\newcommand{\tfftE}{\widetilde{\mathcal E}}
\newcommand{\tffbE}{\overline{\mathcal E}}

\newcommand{\SigG}{{\text{S}}}
\newcommand{\pS}{{\mathrm{pS}}}

\newcommand{\qT}{T}

\newcommand{\req}[1]{(\ref{#1})}

\newcommand{\cQ}{{\cal Q}}
\newcommand{\cE}{{\cal E}}

\newcommand{\xB}{x_{\rm B}}
\newcommand{\muR}{\mu_{\rm R}}
\newcommand{\muF}{\mu_{\rm F}}

\newcommand{\muphi}{\mu_{\varphi}}

\newcommand{\qplmi}{{q^{(\pm)}}}
\newcommand{\qpl}{{q^{(+)}}}
\newcommand{\upl}{{u^{(+)}}}
\newcommand{\dpl}{{d^{(+)}}}
\newcommand{\spl}{{s^{(+)}}}
\newcommand{\cpl}{{c^{(+)}}}
\newcommand{\qmi}{{q^{(-)}}}
\newcommand{\umi}{u^{(-)}}
\newcommand{\dmi}{d^{(-)}}
\newcommand{\smi}{s^{(-)}}

\newcommand{\NS}{{\rm NS}}
\newcommand{\NSpl}{{{\rm NS}^{(+)}}}
\newcommand{\NSplmi}{{{\rm NS}^{(\pm)}}}
\newcommand{\g}{{\text{G}}}

\newcommand{\GeV}{{\rm GeV}}

\newcommand{\eps}{\epsilon}

\def\muFgpd{\relax\ifmmode\mu_\text{F,GPD}^2\else{$\mu_\text{F,GPD}^2${ }}\fi}
\def\muFda{\relax\ifmmode\mu_\text{F,DA}^2\else{$\mu_\text{F,DA}^2${ }}\fi}
\def\muO{\relax\ifmmode{\mu_{0}^{2}}\else{$\mu_{0}^{2}${ }}\fi}
\def\Mev{\relax\ifmmode{\text{MeV}}\else{MeV{ }}\fi}

\def\by{\overline{y} \,}
\def\bz{\overline{z} \,}
\def\bu{\overline{u} \,}
\def\bv{\overline{v} \,}

\def\CF{C_\text{F} }
\def\CG{C_\text{G} }
\def\CA{C_\text{A} }
\def\NC{N_\text{C} }

\def\Li{\relax\ifmmode{\text{Li}_{2}}\else{Li$_2${ }}\fi}
\def\im{\Im{\rm m}}
\def\re{\Re{\rm e}}

\newcommand{\nn}{\nonumber}

\font\cmss=cmss12 
\def\1{\hbox{{1}\kern-.25em\hbox{l}}}
\def\bfZ{\relax{\hbox{\cmss Z\kern-.4em Z}}}
\begin{document}

\begin{titlepage}

\centerline{\large \bf Towards a fitting procedure to deeply virtual meson production}
\centerline{\large \bf  -- the next-to-leading order case --}

\vspace{10mm}

\centerline{D.~M\"{u}ller$^a$,
            T.~Lautenschlager$^b$,
            K.~Passek-Kumeri\v{c}ki$^d$,
            A.~Sch\"{a}fer$^b$}

\vspace{8mm}

\centerline{\it $^a$Institut f\"ur Theoretische
Physik II, Ruhr-Universit\"at Bochum}
\centerline{\it  D-44780 Bochum, Germany}

\vspace{4mm}

\centerline{\it $^b$ Institut f\"ur Theoretische Physik, University Regensburg}
\centerline{\it D-93040 Regensburg, Germany}

\vspace{4mm}
\centerline{\it $^d$Theoretical Physics Division, Rudjer Bo{\v s}kovi{\'c} Institute}
\centerline{\it HR-10002 Zagreb, Croatia}

\vspace{10mm}

\centerline{\bf Abstract}
{\ }

\noindent
Based on the collinear factorization approach, we present a comprehensive
perturbative next-to-leading (NLO) analysis of  deeply virtual meson production (DVMP). Our representation
in conformal Mellin space can serve as basis for a global fitting procedure to access generalized parton distributions from experimental
measurements of DVMP and deeply virtual Compton scattering (DVCS).
We introduce a rather general formalism for the evaluation of conformal moments  that can be developed further
beyond the considered order. We also confirm previous diagrammatical findings in the pure  singlet quark
channel. Finally, we use the analytic properties of the hard scattering amplitudes to estimate qualitatively  the size of radiative corrections
and  illustrate these considerations with  some numerical examples. The results suggest that global NLO GPD fits,
including both DVMP and DVCS data, could be more stable than often feared.

\vspace{0.5cm}

\noindent

\vspace*{12mm}



\end{titlepage}

\newpage


\tableofcontents

\newpage

\section{Introduction}

Besides DVCS \cite{Mueller:1998fv,Radyushkin:1996nd,Ji:1996nm},
exclusive electroproduction of mesons in the deeply virtual regime (DVMP), belongs to
the class of hard exclusive processes that allows us to access GPDs  from experimental measurements
\cite{Collins:1996fb}.
One of the main goals of such fits is to resolve the transverse distribution of
partons inside the nucleon
\cite{Burkardt:2000za,Ralston:2001xs,Diehl:2002he}.
Triggered by the link of GPDs to the partonic spin decomposition of the
nucleon \cite{Ji:1996ek},
GPDs have been intensively studied for some time in theory and  a whole
framework is now build up around
them, see the reviews \cite{Diehl:2003ny,Belitsky:2005qn}.  The heart of this
framework is the phenomenological access to GPDs, based on factorization
theorems which ensure that unobservable transverse degrees of freedom can be integrated out
if the exchanged photon in DVMP  (DVCS) is longitudinally
(transversally) polarized. This factorization property of DVMP amplitudes
has been
shown by diagrammatical considerations for light (pseudo)scalar and
longitudinal vector mesons \cite{Collins:1996fb}.
Thereby, it has been stated that in leading order of $1/\cQ$ the DVMP
amplitude factorizes into a hard scattering part
and two non-perturbative and process independent distributions. The
formation of the meson is described by the corresponding leading twist-two
meson distribution amplitude (DA) while the transition from the initial
nucleon to the final hadronic state is encoded in twist-two
GPDs.  Various DVMP channels have been considered
to leading order (LO) accuracy of perturbation theory in numerous papers
\cite{Frankfurt:1995jw,Radyushkin:1996ru,Frankfurt:1997fj,Mankiewicz:1997uy,Mankiewicz:1997aa,Mankiewicz:1998kg,Frankfurt:1999xe,Frankfurt:1999fp}.
Knowing that these hard scattering amplitudes are only classified by a flavor non-singlet or singlet label and a signature factor, one can easily
extend the processes of phenomenological interest to the level of next-to-leading order (NLO)
perturbation theory \cite{Belitsky:2001nq,Ivanov:2004zv} (for DVCS related processes see\cite{Belitsky:1997rh,Mankiewicz:1997bk,Ji:1997nk,Ji:1998xh,Pire:2011st}).
Note that the naive calculation of so-called `power-corrections'
\cite{Vanderhaeghen:1999xj} is maybe not consistent with the idea that one
integrates out transverse degrees of freedom, yielding both perturbative and
power-suppressed contributions.
Thus, such a simple minded treatment can not be used if one likes to stay
with a systematic field theoretical framework. We add that a
calculation of kinematical power-corrections to DVMP, as it is
feasible in DVCS \cite{Braun:2011zr,Braun:2011dg,Braun:2012hq,Braun:2012bg},
is a challenging task which has not been studied so far.

Furthermore,
much effort has been spend during the last decade to measure the exclusive
processes in question in the collider experiments H1 and ZEUS  \cite{Aid:1996ee,Adloff:1997jd,Aaron:2009xp,Breitweg:1998nh,Breitweg:2000mu,Chekanov:2005cqa,Chekanov:2007zr}, fixed target
experiments HERMES \cite{Airapetian:2000ni,Airapetian:2009ad,Airapetian:2007aa,Airapetian:2009ac} and at JLAB \cite{Hadjidakis:2004zm,Morand:2005ex,Morrow:2008ek,Santoro:2008ai,DeMasi:2007id,Bedlinskiy:2012be,Blok:2008jy}.
Unfortunately,  on the phenomenological side -- apart from some earlier model dependent estimates as well as
more recent data  descriptions for $\pi^+$ \cite{Bechler:2009me} and light
vector mesons \cite{Meskauskas:2011aa} at leading order accuracy -- the
collinear framework has still not been confronted to the increasing amount
of experimental DVMP data. However, we like to emphasize here that a GPD inspired  hand-bag model approach
has been used to link GPD models to DVMP measurements
\cite{Goloskokov:2005sd,Goloskokov:2007nt,Goloskokov:2009ia}.
On the other hand some effort has been spent to analyze DVCS data with
flexible GPD models
\cite{Kumericki:2007sa,Kumericki:2009uq,Kumericki:2013br}, while the idea
to describe present DVCS data with some given class of models might be not
considered as an appropriate approach \cite{Kumericki:2008di},  see the review
\cite{Guidal:2013rya}.
Furthermore, it has been shown that utilizing the model for the dominant $H$
GPD, based on the popular Radyushkin ansatz \cite{Radyushkin:1997ki}, from
the hand-bag approach provides predictions for DVCS on unpolarized protons
that reproduce collider DVCS data and are roughly compatible with
fixed target DVCS data \cite{Kumericki:2011zc,Meskauskas:2011aa}. Very
similar results are obtained if one utilizes the complete GPD content of this
model for polarized proton DVCS data \cite{Kroll:2012sm}. This together with
the above mentioned DVMP LO description provides a hint that a global
analysis of DVMP and DVCS data might be possible.

In particular in the small-$\xB$ region flexible GPD models are needed  and
used to control both the size and the evolution flow of Compton form factors
(CFFs).
This was realized when GPDs were directly parameterized in terms of (conformal)
Mellin moments \cite{Kumericki:2007sa}
rather than in momentum fraction representation.
Apart from providing an easy possibility to parameterize GPDs, this technique allows
also to set up robust and fast numerics
\cite{Kumericki:2007sa,Kumericki:2009uq}. To apply this technique
for a global DVCS and DVMP analysis, the NLO corrections to DVMP are needed,
which we will provide in this paper. We will also present explicit formulae
for the evaluation of the imaginary part of DVMP amplitudes to NLO accuracy
in the momentum fraction representation. Combined with dispersion relation
technique,
this may offer an alternative possibility for an efficient numerical
treatment at least for the purpose to confront some given GPD models
in momentum fraction representation with experimental measurements.

In this article we systematize the perturbative framework for DVMP at
NLO in such a manner that it can be utilized in a straightforward manner
in existing fitting routines for a global
analysis of DVCS and  DVMP processes.
To do so, we will first define transition form factors (TFFs), which allow
for a clear separation of  observables
and the perturbative evaluation procedure on amplitude level. We also
complete the set of observables for a  DVMP
process from two to four. This  allow at least in principle  for an
disentanglement of the imaginary and real parts of  TFFs in longitudinal
photoproduction
if in future the polarization of the final state proton is experimentally
measurable, which  would provide an additional handle for the access of twist-two GPDs. We also
give for the first time  a generic
discussion of radiative corrections for TFFs and compare them with those of
CFFs.
The detailed outline of our presentation is as follows.

In Sec.~\ref{sec:Pre} we introduce our nomenclature for TFFs.  We
parameterize then the
longitudinal photon helicity amplitudes in terms of intrinsic parity even
and odd  TFFs and calculate the longitudinal
photon cross section for all possible target polarizations, as well as for the longitudinal polarization
of the outgoing nucleon. Furthermore, we perform the charge and flavor
decomposition of these TFFs for some important DVMP channels. This allows us
in return to present the perturbative corrections in the flavor
non-singlet and singlet channel in a compact manner. In Sec.~\ref{sec:factorization} we recall the
collinear framework for DVMP in momentum fraction representation, point out
the general analytic properties of hard scattering amplitudes, and introduce
our conventions.
We then explain the evaluation of TFFs from GPDs by means of both the
dispersion relation integral and the Mellin--Barnes integral, and shortly discuss
mixed representations. Moreover, we develop a method that allows to evaluate the conformal moments
by means of a standard Mellin transform.
In Sec.~\ref{sec:NLO} we introduce first building blocks for the NLO hard scattering amplitudes in momentum fraction
representation, calculate their imaginary parts and their conformal moments. We confirm the result for the pure singlet part
at NLO in momentum fraction space \cite{Ivanov:2004zv},
present the whole NLO corrections in a more economical manner
in this space. From these results we derive compact expressions, so far not listed in the literature, for the
imaginary parts of the hard scattering amplitudes and their conformal moments.
In Sec.~\ref{sect-numerics} we set up  GPD models in Mellin space, discuss the size of radiative
NLO corrections from the generic point of view, and  provide some numerical examples for the size of radiative corrections.
Finally, we give our conclusions and an outlook for the application of this
work. Appendix \ref{app:def} contains our GPD conventions as well as  the conventions for evolution kernels and anomalous dimensions.
In App.~\ref{app:ImRe} we list the expressions for the real part of NLO
building blocks and in App.~\ref{app:H0} we discuss
some properties of the non-separable building block for the hard scattering amplitude.

\section{Preliminaries}
\label{sec:Pre}

Although  we are primary interested to use DVMP to access GPDs,
we prefer  to distinguish  clearly between observables and their
partonic description, which are conventionally defined
w.r.t.~a light-cone direction (since momentum is transferred in the $t$-channel in DVMP, one has great liberty
to define the light-cone direction in which partons travel).
In the following we define first a form factor decomposition of the  $\gamma_L^\ast N \to M N$ amplitude,
where for the goal of accessing twist-two GPDs it is sufficient to restrict ourselves to longitudinal polarized photons and scalar components,
e.g., longitudinally polarized vector mesons. Note that due to helicity conservation the
contributions of transversally polarized mesons connected to quark transversity GPDs vanish to all orders in
the strong coupling constant \cite{Collins:1999un,Hoodbhoy:2001da,Diehl:2003ny}.
For the two  TFFs of each channel%
\footnote{Alternatively, (light-cone) helicity amplitudes are adopted to describe the nucleon states in DVCS/DVMP \cite{Diehl:2001pm}.}
the same nomenclature will be adopted that is used for twist-two GPDs.
Hence, one can immediately read off from cross section expressions which information
can be accessed in an experiment. To our best knowledge polarization measurements of
the recoiled nucleon have not been much discussed with respect
to GPD phenomenology, except for $J/\Psi$ electroproduction in \cite{Koempel:2011rc}. We will fill this gap and show that in a complete
measurement, the number of observables matches twice the number
of complex valued TFFs. If one can measure these transition
form factors, one has the most
complete experimental information to access twist-two GPDs. One may, however, employ other frameworks to facilitate their interpretation. Moreover,
we will classify the TFFs with respect to  parity and $t$-channel charge conjugation parity and decompose them according to $t$-channel flavor flow.
Such decomposition can be also used in (GPD) phenomenology as a flavor filter.

\renewcommand{\arraystretch}{2}
\begin{table}[t]
\begin{center}
\begin{tabular}{||c|cl||c|c||}
\hline \hline
TFF $\tffF_M^\text{A}$ & TFF $\tffF_M$ & in Eq.  &(MF,DR,MB) &  $\sigma(F^\text{A})$
\\
\hline \hline
$\tffH^{\rm S}_{{\rm V}^0_L}, \tffE^{\rm S}_{{\rm V}^0_L}$
 & $\tffH_{{\rm V}^0_L}, \tffE_{{\rm V}^0_L}$
 & \req{eqs:tffV0-SU}
 & (\ref{eqs:tffF^S},\ref{eqs:DRF^S_V},\ref{eq:tffSF_M-MBI})
 &$+1$
\\
$\tffH^{\NSpl}_{{\rm V}_L},  \tffE^{\NSpl}_{{\rm V}_L}$
 & $\tffH_{{\rm V}_L}, \tffE_{{\rm V}_L}$
 & (\ref{eqs:tffV0-SU},\ref{eq:rho+MTFFs})
 & (\ref{eq:tffF^A},\ref{eqs:DRtffFqC},\ref{eq:tffqF_M-MBI})
 &$+1$
\\  \hline
$\tffH^{\qmi}_{{\rm V}^\pm_L}, \tffE^{\qmi}_{{\rm V}^\pm_L}$
 & $\tffH_{{\rm V}^\pm_L}$, $\tffE_{{\rm V}^\pm_L}$
 & \req{eq:rho+MTFFs}
 & (\ref{eq:tffF^A},\ref{eqs:DRtffFqC},\ref{eq:tffqF_M-MBI})
 &$-1$
\\ \hline \hline
 $\tfftH^{\qmi}_{\rm PS}, \tfftE^{\qmi}_{\rm PS}$
 & $\tfftH_{{\rm PS}}$, $\tfftE_{{\rm PS}}$
 & (\ref{eq:PSMTFFs},\ref{eq:pi+MTFFs})
 & (\ref{eq:tffF^A},\ref{eqs:DRtffFqC},\ref{eq:tffqF_M-MBI})
 &$+1$
\\  \hline
$\tfftH^{\NSpl}_{{\rm PS}^\pm}, \tfftE^{\NSpl}_{{\rm PS}^\pm}$
 & $\tfftH_{{\rm PS}^\pm}$, $\tfftE_{{\rm PS}^\pm}$
 & \req{eq:pi+MTFFs}
 & (\ref{eq:tffF^A},\ref{eqs:DRtffFqC},\ref{eq:tffqF_M-MBI})
 &$-1$
\\
  \hline
  \hline
\end{tabular}
\end{center}
\caption{
\label{tab:definitions}\small
Nomenclature of flavor dependent TFFs (first column) appearing
in the parametrization of the
$\gamma_L^\ast N \to M N$ amplitude for longitudinal vector ($\text{V}_L$)  and pseudo scalar (PS) mesons $M$
with $J^{PC}$ quantum  numbers $1^{--}$  and $0^{-+}$, respectively. The references to the decomposition in which particular TFFs appear are given next.
In the third column we refer to formulae which allow to evaluate  TFFs from the corresponding GPD $F^\text{A}$
in momentum fraction (MF), dispersion relation (DR),
and Mellin-Barnes integral (MB) representations, depending on the signature factor $\sigma(F^\text{A})$ that is given next.
The label $\text{A}\in\{\NSpl, \text{S}\equiv \text{S}^{(+)}, \qmi\}$ encodes information about flavor decomposition (non-singlet, singlet, quark species)
with definite ($t$-channel) charge parity $C=\pm 1$, given by a superscript $(C)$.
}
\end{table}

In Sec.~\ref{sec:Pre-TFFs} we introduce the aforementioned TFFs, e.g., usable for longitudinal photoproduction of (pseudo)scalar
and longitudinal polarized (axial)vector mesons. Furthermore,
we calculate  the longitudinal photoproduction cross section in terms of these TFFs exactly, including the polarization state of the
outgoing nucleon. In Sec.~\ref{sec:pre-SU} we give our conventions for the flavor decomposition of TFFs including their  parity and
charge conjugation parity assignments. This is exemplified for longitudinal vector and pseudoscalar  meson production, which are the
phenomenologically most important DVMP processes.
The reader, who is only interested in our conventions and defining equations, can find them in Tab.~\ref{tab:definitions},
which lists our TFF nomenclature, DVMP processes of interest, and GPD
factorization formulae (given in the next section for three different representations).
We add that in the twist-two approximation the name of the quark TFFs matches the name of the GPDs.

\subsection{Longitudinal photoproduction cross section}
\label{sec:Pre-TFFs}

Let us first introduce our reference frame for exclusive electroproduction,
which is the same as in  \cite{Belitsky:2001ns}.
The incoming electron momentum has a positive $x$-component,
the longitudinal photon with momentum $q_1$ travels in the direction
of the negative $z$-axis and the nucleon with momentum $p_1=(M_N,0,0,0)$
and polarization vector $s_1$, is at rest.
The outgoing nucleon  has momentum $p_2$ and may be polarized
along the direction $s_2$.
Finally, the momentum of the produced meson is called $q_2$.
The longitudinal polarization vector of the photon can be expressed in terms
of the incoming nucleon $p_1$ and photon $q_1$ momenta
\begin{eqnarray}
\label{eps_1(0)}
\varepsilon_1^\mu(0)&\!\!\! =\!\!\!&
-\frac{1}{\cQ\sqrt{1 + \epsilon^2}}\, q_1^\mu - \frac{2 \xB}{\cQ\sqrt{1 + \epsilon^2}}\, p_1^\mu\,,\quad \epsilon= \frac{2\xB M_N}{\cQ}\,.
\end{eqnarray}
We parameterize the photon helicity amplitude for longitudinal
photoproduction of a (pseudo)scalar meson $M$ in terms of
transition form factors.
We are left with four TFFs  or, alternatively, nucleon helicity amplitudes,
however, by parity conservation these are reduced to two independent ones.
We adopt the parametrization for helicity dependent Compton form factors from \cite{Belitsky:2012ch}.
By means of the free Dirac equation (Gordon identity) it is easy to see
that for the case of even or odd intrinsic parity the form factor basis
can be chosen to be:
\begin{eqnarray}
\label{tffF-def}
\epsilon_1^\mu(0) \langle M N| j_\mu |N\rangle \!\!\!&=&\!\!\!  \left\{{
\overline{u}(p_2,s_2) \bigg[
 {\not\!\! m}\, \tffH_{M} + i \sigma_{\alpha \beta} \frac{m^\alpha \Delta^\beta}{2 M_N}\, \tffE_{M} \bigg]u(p_1,s_1)
\quad \mbox{parity even}
 \atop
 \overline{u}(p_2,s_2) \bigg[
 {\not\!\! m} \gamma_5 \,  \tfftH_{M} + \gamma_5 \frac{m\cdot\Delta}{2 M_N}\, \tfftE_{M} \bigg]u(p_1,s_1)
 \quad \mbox{parity odd}
 }\right. ,
\end{eqnarray}
where $\Delta^\mu= p_2^\mu-p_1^\mu = q_1^\mu-q_2^\mu$ is the momentum transfer in the $t$-channel ($t\equiv \Delta^2$).
The choice of the vector $m^\mu$ is not unique. To stay close to the conventions,
used by us for DVCS, as well as to have  a $s\leftrightarrow u$ symmetric energy variable, a favored choice for dispersion relation analysis,
we choose the following vector \cite{Belitsky:2012ch}
\begin{eqnarray}
m^\mu= \frac{q^\mu}{P\cdot q},\quad \mbox{where}\quad q^\mu=\frac{1}{2}\left(q_1^\mu+q_2^\mu\right)\,,
\quad P^\mu = p_1^\mu + p_2^\mu\,.
\end{eqnarray}

The photoproduction cross sections in terms of these TFFs (\ref{tffF-def}) is straightforwardly calculated. In fact,
if the meson mass is neglected, the formulae for an unpolarized outgoing nucleon can be read off from the expressions for
DVCS \cite{Belitsky:2012ch}. For the conversion of electroproduction to photoproduction cross section we adopt the
Hand convention \cite{Hand:1963bb}, which fixes the photon flux and yields
\begin{eqnarray}
\label{dX^V}
\frac{d\sigma^{\gamma_{\rm L}^*\,N\to M\,N}}{dt d\varphi} & \!\!\!= & \!\!\!
  \frac{2\pi \alpha_{\rm em}}{\cQ^4 \sqrt{1+ \epsilon^2}}\,\frac{\xB^2}{1 - \xB} \frac{1}{2}
  \Big\{ {\cal C}_{\rm unp}(\tffF_M,\tffF_M^\ast|s_2)  +  \Lambda \cos(\theta) {\cal C}_{\rm LP}(\tffF_M,\tffF_M^\ast|s_2)
\\
 &&\phantom{\frac{2\pi \alpha_{\rm em}}{\cQ^4 }}+ \Lambda \cos(\varphi) \sin(\theta)\, {\cal C}_{{\rm TP}+}(\tffF_M,\tffF_M^\ast|s_2) + \Lambda \sin(\varphi) \sin(\theta)\, {\cal C}_{{\rm TP}-}(\tffF_M,\tffF_M^\ast|s_2) \Big\}\,.
\nonumber
\end{eqnarray}
Here, $\alpha_{\rm em}\approx 1/137$ is the electromagnetic fine structure constant,  $q_1^2=-\cQ^2$ is the photon virtuality, $\xB= \cQ^2/p_1\cdot q_1$ is the Bjorken variable and $\varphi=\Phi-\phi$, where $\Phi$ appears in the transverse part of the polarization vector
$$s_1 =(0,\cos\Phi \cos\theta,\sin\Phi \cos\theta,\sin\theta)$$ and $\phi$ is the azimuthal angle between the electron plane and the recoiled proton.
Furthermore, the squared scattering amplitudes $\cal C$ are labeled  by the polarization of the incoming nucleon.
Note that when summed over the final state proton polarization,
the conventional factor $1/2$ on the r.h.s.~will disappear.

The bilinear $\cal C$-coefficients depend on the polarization direction of the outgoing nucleon, which provides the possibility to measure
various combinations of TFFs. In experiments where the outgoing  protons are unpolarized
one can only access the cross section for a transversally polarized nucleon, which contains
for scalar or longitudinally polarized vector meson production the terms
\begin{eqnarray}
\label{Cunp}
{\cal C}_{\rm unp}(\tffF,\tffF^\ast)&\!\!\!=\!\!\!&
\frac{4 (1-\xB) \left(1-\xB \frac{m^2-t}{\cQ^2}\right)-\frac{m^2}{M^2}\epsilon^2}{\left(2-\xB-\xB\frac{ m^2-t}{\cQ^2}\right)^2}
\Bigg|
\tffH-\frac{\xB^2\left(1+\frac{m^2-t}{\cQ^2}\right)^2+4\xB^2 \frac{t}{\cQ^2}}{4 (1-\xB) \left(1-\xB \frac{m^2-t}{\cQ^2}\right)-\frac{m^2}{M^2}\epsilon^2}\tffE \Bigg|^2
\nonumber\\
&&+ \frac{1}{(1-\xB)\left(1-\xB \frac{m^2-t}{\cQ^2}\right)-\frac{m^2}{4 M^2}\epsilon^2}\, \frac{\widetilde{K}^2 }{4 M^2}\left|\tffE\right|^2\,,
\\
\label{CTP-}
{\cal C}_{{\rm TP}-}(\tffF,\tffF^\ast)&\!\!\!=\!\!\!&   -\frac{2}{2-\xB -\xB\frac{m^2-t}{\cQ^2}}\, \frac{\widetilde{K}}{M} \im \tffH \tffE^\ast\,,
\end{eqnarray}
and for pseudo scalar or longitudinal polarized axial-vector   meson production the terms
\begin{eqnarray}
\label{tCunp}
\widetilde{\cal C}_{\rm unp}(\tffF,\tffF^\ast)&\!\!\!=\!\!\!&
\frac{4(1-\xB) \left(1- \xB\frac{m^2-t}{\cQ^2}\right)+\epsilon ^2\left(2-\frac{m^2 }{M^2}-\frac{2 m^2-t }{\cQ^2}\right)}{
\left(2-\xB -\xB\frac{m^2-t}{\cQ^2}\right)^2}
\\
&&
\times \Bigg| \tfftH -\frac{\xB \left(1+\frac{m^2}{\cQ^2}\right)^2 \left(2-\xB -\xB\frac{m^2-t}{\cQ^2}\right)}{4(1-\xB) \left(1- \xB\frac{m^2-t}{\cQ^2}\right)+\epsilon^2\left(2-\frac{m^2 }{M^2}-\frac{2 m^2-t }{\cQ^2}\right)} \tffbE \Bigg|^2
\nonumber\\
&&
+ \frac{4 \left(1+\frac{m^2}{\cQ^2}\right)^2}{
4(1-\xB) \left(1- \xB\frac{m^2-t}{\cQ^2}\right)+\epsilon ^2\left(2-\frac{m^2 }{M^2}-\frac{2 m^2-t }{\cQ^2}\right)}\,
\frac{\widetilde{K}^2}{4 M^2} \left|\tffbE \right|^2\,,
\nonumber
\\
\label{tCTP-}
\widetilde{\cal C}_{{\rm TP}-}(\tffF,\tffF^\ast)&\!\!\!=\!\!\!&
\frac{2\left(1+\frac{m^2}{\cQ^2}\right)}{2-\xB -\xB\frac{m^2-t}{\cQ^2}}\, \frac{\widetilde{K}}{M} \im \tfftH \tffbE^\ast\,.
\end{eqnarray}
Here, the kinematical factor
\begin{eqnarray}
\label{tK}
\widetilde{K} =\sqrt{
-M^2 \xB^2 \left[\!\left(1-\frac{m^2-t}{\cQ^2}\right)^2+\frac{4 m^2}{\cQ^2}\left(1-\frac{t}{4 M^2}\right)\! \right]-
(1-\xB) t \left(1- \xB \frac{m^2-t}{\cQ^2}\right)
}
\end{eqnarray}
vanishes at the minimal and maximal allowed value of $-t$,
\begin{eqnarray}
\label{tmin}
t_{\rm min/max} = -\cQ^2\frac{ 2 \left(1-\xB- \xB\frac{m^2}{\cQ^2}\right)+\epsilon^2 \left(1-\frac{m^2}{\cQ^2}\right)\mp 2 \sqrt{1+\epsilon ^2} \sqrt{\left(1-\xB- \xB\frac{m^2}{\cQ^2}\right)^2-\frac{m^2}{\cQ^2}\epsilon ^2}}{4 (1-\xB) \xB+\epsilon ^2},\quad
\end{eqnarray}
where the lower and upper sign applies for $t_{\rm min}$ and  $t_{\rm max}$, respectively.
The unpolarized ${\cal C}$-coefficients (\ref{Cunp},\ref{tCunp})  are build from two squared terms while the square of target spin flip TFFs is naturally accompanied by a $\widetilde{K}^2/4 M^2$ suppression factor. Relying on this kinematical suppression, one can essentially  extract from unpolarized cross section measurements at smaller values of $\xB$ the modulus of the TFFs $\tffH$ or $\tfftH$. Having a transversally polarized proton at hand,
the single target spin  asymmetry offers an access to the combinations (\ref{CTP-}) and (\ref{tCTP-}), see, e.g., phenomenologically discussions in
\cite{Goloskokov:2008ib,Goloskokov:2009ia,Bechler:2009me}.
Note that the quantity
$\overline{\cal E} = \xB\, \tfftE/(2-\xB -\xB\frac{m^2-t}{\cQ^2})$ absorbs one additional power of $\xB$, and thus has the same Regge counting
as the other TFFs.

In experiments where the polarization of the outgoing  proton can be measured, one can  access further TFF combinations.
However, it turns out that the transverse-to-transverse proton spin contribution does not contain new information rather it offers access to
the unpolarized TFF combinations (\ref{Cunp},\ref{tCunp}), while the final state transverse single spin asymmetry provides again the imaginary parts (\ref{CTP-}) and (\ref{tCTP-}). The remaining terms will project on the longitudinal component of the final state polarization vector.
Hence, choosing the longitudinal magnetization direction
$$
s^{\parallel}_2 = \frac{1}{\sqrt{\left(p_2^0\right)^2- \left( p_2^3\right)^2}}\left(p_2^3,0,0,p_2^0\right)
$$
provides the most appropriate handle to access two new TFF combinations.
We find the following combinations
\begin{eqnarray}
\label{CLP}
{\cal C}_{{\rm LP}}(\tffF,\tffF^\ast|s_2^\parallel)&\!\!\!=\!\!\!&
\frac{\sqrt{1+\epsilon ^2}}{\sqrt{1+\epsilon ^2+\frac{\widetilde{K}^2}{M^2}}}
\Bigg\{{\cal C}_{\rm unp}(\tffF,\tffF^\ast) - \frac{2 \widetilde{K}^2}{M^2\left(2-\xB-\xB\frac{m^2-t}{\cQ^2}\right)^2}
\\
&&\phantom{\frac{\sqrt{1+\epsilon ^2}}{\sqrt{1+\epsilon ^2+\frac{\widetilde{K}^2}{M^2}}}}\times
\Bigg( \left|\tffH+\tffE\right|^2
-\frac{1}{1+\epsilon ^2}\left|\tffH +\frac{\xB}{2}\Big(1+\frac{m^2-t}{\cQ^2}\Big)\tffE \right|^2\Bigg)\Bigg\},
\nonumber\\
\label{CTP+}
{\cal C}_{{\rm TP}+}(\tffF,\tffF^\ast|s_2^\parallel)&\!\!\!=\!\!\!&
\frac{-\widetilde{K}}{M\sqrt{1+\epsilon ^2}\sqrt{1+\epsilon ^2+\frac{\widetilde{K}^2}{M^2}}}
\Bigg\{
{\cal C}_{\rm unp}(\tffF,\tffF^\ast)  + \frac{2 \epsilon^2}{2-\xB-\xB\frac{m^2-t }{\cQ^2}} \Big|\tffH+\frac{t}{4 M^2}\tffE\Big|^2
\nonumber\\
&&+\frac{2(1+\epsilon ^2)}{2-\xB-\xB\frac{m^2-t}{\cQ^2}}\,\Re{\rm e}\tffH \tffE^\ast -\frac{\widetilde{K}^2-t \left(1+\epsilon ^2\right) }{2 M^2 \left(2-\xB-\xB\frac{m^2-t }{\cQ^2}\right)} \left|\tffE \right|^2
\Bigg\},
\end{eqnarray}
for longitudinally polarized vector or scalar meson production and
\begin{eqnarray}
\label{tCLP}
\widetilde{\cal C}_{{\rm LP}}(\tffF,\tffF^\ast|s_2^\parallel)&\!\!\!=\!\!\!&
\frac{\sqrt{1+\epsilon ^2}}{\sqrt{1+\epsilon ^2+\frac{\widetilde{K}^2}{M^2}}}
\Bigg\{\widetilde{\cal C}_{\rm unp}(\tffF,\tffF^\ast) -\frac{8\xB \widetilde{K}^2}{\cQ^2 \left(1+\epsilon ^2\right)}\frac{1-\xB+\xB\frac{M^2}{\cQ^2}}{ \left(2-\xB-\xB \frac{ m^2-t}{\cQ^2}\right)^2} \left|\tfftH\right|^2
\qquad\\
&&- \frac{2\left(1+\frac{m^2}{\cQ^2}\right) \left(1+\xB\frac{2 M^2 }{\cQ^2}\right)}{\left(1+\epsilon ^2\right) \left(2-\xB-\xB \frac{ m^2-t}{\cQ^2}\right)}\frac{\widetilde{K}^2}{M^2} \Re{\rm e} \tfftH \tffbE^\ast
-\frac{\left(1+\frac{m^2}{\cQ^2}\right)^2 }{1+\epsilon^2}\frac{\widetilde{K}^2}{2 M^2} \left|\tffbE\right|^2
\Bigg\},
\nonumber\\
\label{tCTP+}
\widetilde{\cal C}_{{\rm TP}+}(\tffF,\tffF^\ast|s_2^\parallel)&\!\!\!=\!\!\!&
\frac{\widetilde{K}}{M\sqrt{1+\epsilon ^2}\sqrt{1+\epsilon ^2+\frac{\widetilde{K}^2}{M^2}}}
\Bigg\{\!
\left(\!1+\frac{2 \xB M^2}{\cQ^2}\! \right)\widetilde{\cal C}_{\rm unp}(\tffF,\tffF^\ast)
\\
&& +
\frac{2 \epsilon ^2 \left(1+\frac{m^2}{\cQ^2}\right) \left(1-\xB+\frac{\xB M^2}{\cQ^2}\right)}{
\left(2-\xB-\xB \frac{\left(m^2-t\right)}{\cQ^2}\right)^2} \Big|\widetilde{\tffH}\Big|^2
+\frac{2+\frac{4(1-\xB) \xB t }{\cQ^2}+\left(2+\frac{t}{\cQ^2}\right) \epsilon^2}{2-\xB-\xB \frac{m^2-t}{\cQ^2}}
\nonumber\\
&&
\times \left(1+\frac{m^2}{\cQ^2}\right) \Re{\rm e}\tfftH \tffbE^\ast + \frac{\left(1+\frac{m^2}{\cQ^2}\right)^2
\left(t- \xB  M^2 \left(1+\frac{m^2-2 t}{\cQ^2}\right)\right)}{2M^2} \left|\tffbE \right|^2
\Bigg\},
\nonumber
\end{eqnarray}
for longitudinally polarized axial-vector or pseudoscalar meson production.

In particular in the smaller-$\xB$ region we have the following combinations, e.g., for longitudinally polarized vector meson production
\begin{eqnarray}
\label{C-smalxB}
{\cal C}_{\rm unp}(\tffF,\tffF^\ast)&\!\!\!\simeq\!\!\!& |\tffH|^2 -\frac{t}{4 M^2} |\tffE|^2\,,\quad\;\;
{\cal C}_{\rm LP}(\tffF,\tffF^\ast|s_2^\parallel) \simeq \sqrt{1-\frac{t}{M^2}} \left[|\tffH|^2   +\frac{t\, |2\tffH+\tffE|^2}{4
\left(M^2 -t\right)}\right],
\\
{\cal C}_{{\rm TP}-}(\tffF,\tffF^\ast) &\!\!\!\simeq\!\!\!&  -\frac{\sqrt{-t} }{M} \im \tffH \tffE^\ast\,, \qquad
{\cal C}_{{\rm TP}+}(\tffF,\tffF^\ast|s_2^\parallel) \simeq \frac{\sqrt{-t}\sqrt{1-\frac{t}{M^2}} }{4M } \left[
 |\tffE|^2 - \frac{|2\tffH+\tffE|^2}{1-\frac{t}{M^2}} \right].
 \nonumber
\end{eqnarray}
An analog formula set is also valid for pseudo scalar meson production, obtained by substituting
$$\tffH \to \tfftH\,,  \quad \tffE \to (1+ m^2/\cQ^2)\tffbE\,, \quad\mbox{and}\quad  {\cal C}_{{\rm TP}\pm}\to -\widetilde{\cal C}_{{\rm TP}\pm}\,. $$
Since in the ${\cal C}_{{\rm TP}+}$ expression both kinds of TFFs enter on the same kinematical level,  one clearly realizes  that transverse-to-longitudinal target spin flip measurements yield a handle on $\tffE$ (and $\tffbE$) for $-t \gg 0$. One the other hand, longitudinal-to-longitudinal spin flip
cross sections are expected to be dominated for $-t \ll 4 M^2$ by the modulus $|\tffH|$. However, since $\tffbE$ may contain a pion pole contribution, e.g.,
in $\pi^+$ production, the $t/4 M^2$ suppression factor can be overcompensated. Hence, such a measurement would be helpful for a complete disentanglement of parity odd TFFs, where of course one should bear in mind that in contrast to $\tffE$ the TFF $\tffbE$ does  not contain a `pomeron' exchange.

\subsection{Flavor decomposition of transition form factors}
\label{sec:pre-SU}

\begin{figure}[t]
\begin{center}
\includegraphics[width=10cm]{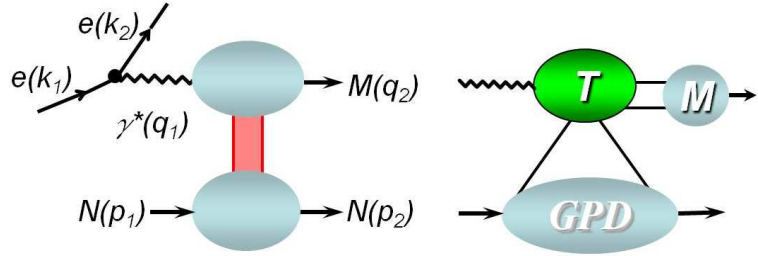}
\end{center}
\vspace{-5mm}
\caption{\small DVMP process from the $t$-channel view (left) and partonic view (right).}
\label{Fig-TFF}
\end{figure}

To perform a flavor decomposition of the TFFs (\ref{tffF-def}) we rely on the quark picture and consider the processes of interest from the
$t$-channel point of view as an exchange of colorless degrees of freedom that can be associated with a quark-antiquark  or gluon pair, see Fig.~\ref{Fig-TFF}.
We note that $t$-channel contributions
are the dominant ones for both large $\cQ^2$, i.e., when partonic
$t$-channel exchanges are justified by (diagrammatical) power counting,
and high-energy limit, e.g., in Regge phenomenology where mesonic degrees
of freedom are utilized.
Based on the $t$-channel exchange of quark-antiquark pairs we start by performing a flavor decomposition of the TFFs (\ref{tffF-def}). This more general classification scheme matches with the nomenclature in the partonic description  of DVMP amplitudes in terms of twist-two GPDs and meson DAs.

First, however, we introduce discrete $t$-channel quantum numbers
which are used to label GPDs and TFFs.
It is instructive to see that from a partonic point of view
in the $t$-channel reaction
$\gamma^\ast q \bar{q} (gg) \to M^0$ in which the
photon scatters on a $q \bar{q}$ ($g g$) pair, picked up from
the proton and described by GPD,
and then forms a meson.
Note that corresponding crossed processes where analyzed on
similar basis in \cite{Baier:1982,Baier:1985wv}
and \cite{Ji:2000id}. From $C_{\gamma}=-1$ and charge parity conservation
follows that the $q \overline{q}$ ($g g$) state
has to satisfy
\begin{eqnarray}
\label{C-t-channel}
C=C_\gamma C_{M^0}=-C_{M^0}
\end{eqnarray}
or, in other words, that the $t$-channel charge parity
is given by $(-C_{M^0})$.
The charge parity of the meson can be read off from the $J^{PC}$
nomenclature with angular momentum $J$, parity $P$, and charge parity $C$,
and for vector (${\text V}^0$) mesons  of interest is $1^{--}$ while for
pseudoscalar ($\text{PS}^0$) mesons $0^{-+}$.
Due to the fact that $C_{gg}=1$ it follows trivially that
there is no $gg$ contribution for $\text{PS}^0$ production
(as well as, there are no $gg$ Fock states in ${\text V}^0$ mesons).
Furthermore, since $P_{\gamma}=-1$,
the $C P$ quantum number is given by
$C P=C_{M^0} P_{M^0}$
and it corresponds to intrinsic GPD parity%
\footnote{When considering $q \bar{q}$
states, this terminology seems quite obvious
since $C_{q \bar{q}}=(-1)^{l+s}$ and $P_{q \bar{q}}=-(-1)^{l}$,
and thus $C_{q \bar{q}}P_{q \bar{q}}=-(-1)^{s}$, so this quantum number
depending just on spin, i.e., intrinsic angular momentum, is
the same for different ${q \bar{q}}$ excitations.
}
or, to say it in \req{tffF-def} nomenclature,
the production of vector mesons is described by TFFs
$\tffF\in\{\tffH, \tffE\}$ with even intrinsic parity,
while the production of pseudoscalar mesons is
described by odd intrinsic parity TFFs $\tffF\in\{\tfftH, \tfftE\}$.
The TFFs derived by using GPDs with well defined charge
parity will then be denoted by $\tffF^{(C)}$.

Next we  define the flavor content of the considered final meson state in terms of quark-antiquark degrees of freedom. Normalizing all states
to one, we expand  the charged meson states in terms of the leading quark-antiquark Fock states
\begin{eqnarray}
|M^{0}\rangle = \sum_{q}  c_{M^0}^{q}\,  |q \overline{q}\rangle \quad\mbox{and}\quad
|M^\pm\rangle = \sum_{q\overline{q}^\prime}  c_{M^\pm}^{q\overline{q}^\prime}\,  |q \overline{q}^\prime\rangle\,,
\end{eqnarray}
respectively.
For neutral mesons the Clebsch-Gordon coefficients are flavor diagonal, while they are flavor off-diagonal for charged meson.
The interaction of the longitudinal photon with a quark or an antiquark gives us then a fractional quark charge factor
\begin{equation}
\label{e_q}
e_q \in \left\{e_u=\frac{2}{3}, e_d=-\frac{1}{3}, e_c = \frac{2}{3}, e_s=-\frac{1}{3},\cdots \right\}\,,
\end{equation}
which is together with the Clebsch-Gordon coefficients factorized out.
This defines us the quark decomposition of TFFs
\begin{subequations}
\label{tff-dec}
\begin{equation}
\label{tff-dec1}
\tffF_{M^0} = \sum_q  e_{q}\, c_{M^0}^{q}\, \tffF_{M^0}^{q}
\quad \mbox{and}\quad
\tffF_{M^\pm} =   e_{q}\, \tffF_{M^\pm}^{q\overline{q}^\prime} + e_{q^\prime}\, \tffF_{M^\pm}^{q^\prime\overline{q}} \,.
\end{equation}

Assuming that isospin symmetry holds true, we may express the flavor off-diagonal nucleon TFFs in terms of flavor diagonal TFFs ones, see, e.g., \cite{Mankiewicz:1997aa}, yielding
\begin{eqnarray}
\label{tff-dec2}
\tffF_{M^\pm} =  \frac{e_{q}-e_{q^\prime}}{2} \left[ \tffF_{M^0}^{q^{(-C)}} - \tffF_{M^0}^{q^{\prime(-C)}} \right]
+ \frac{e_{q}+e_{q^\prime}}{2} \left[ \tffF_{M^0}^{q^{(C)}} - \tffF_{M^0}^{q^{\prime(C)}} \right].
\end{eqnarray}
\end{subequations}
The information about $t$-channel charge parity is encoded in the quark superscript,
e.g., $\tffH^{\upl}$ ($\tffE^{\upl}$) stands for a $t$-channel exchange
of a $u\overline{u}$ pair with even charge parity  and even intrinsic
parity, where loosely spoken proton helicity is (non)conserved.

Finally, in the charge even sector  also gluons can be exchanged in the $t$-channel which may have a  flavor quark singlet admixture.
Such a contribution will be denoted as $\tffF^\g + \tffF^\pS$, where subscript $\pS$ stands for a pure singlet quark. To separate quark degrees and
gluonic ones in the most clean manner it is necessary to change from a quark/gluon basis to group theoretical irreducible SU($n_f$) multiplets,
consisting out of the flavor non-singlet (NS) multiplets ($\tffF^{3}$, $\ldots$, $\tffF^{n_f^2-1}$) and the flavor singlet one ($\tffF^{0}$). Such
decomposition  solves also the quark-gluon mixing problem that appears in the perturbatively predicted evolution.
For $n_f=3$ ($n_f=4$) this group theoretical decomposition follows
from the multiplets (\ref{eq:SU(n)}) and reads
\begin{subequations}
\label{eqs:SU(n)-decomposition}
\begin{eqnarray}
\tffF^\upl &\!\!\!=&\!\!\!   \frac{1}{2} \tffF^{3^{(+)}} + \frac{1}{6} \tffF^{8^{(+)}} +
\frac{1}{3} \tffF^{0^{(+)}} +\frac{1}{12} \left(\tffF^{15^{(+)}} -  \tffF^{0^{(+)}}\right), \\
\tffF^\dpl &\!\!\!=&\!\!\! - \frac{1}{2} \tffF^{3^{(+)}} + \frac{1}{6} \tffF^{8^{(+)}} +
\frac{1}{3} \tffF^{0^{(+)}} +\frac{1}{12} \left(\tffF^{15^{(+)}} -  \tffF^{0^{(+)}}\right), \\
\tffF^\spl &\!\!\!=&\!\!\! - \frac{1}{3} \tffF^{8^{(+)}} +
\frac{1}{3} \tffF^{0^{(+)}} +\frac{1}{12} \left(\tffF^{15^{(+)}} -  \tffF^{0^{(+)}}\right), \\
\tffF^\cpl &\!\!\!=&\!\!\! -\frac{1}{4} \left(\tffF^{15^{(+)}} - \tffF^{0^{(+)}}\right).
\end{eqnarray}
\end{subequations}
Obviously, for $n_f=4$ this decomposition reduces smoothly
for $\tffF^\cpl=0$, i.e., $\tffF^{15^{(+)}}=\tffF^{0^{(+)}}$
to the well known SU(3) one.
We add that one may perform also such decomposition in the charge odd sector, however, this is not a necessity, since a gluon pair has
charge parity even and so no quark-gluon mixing problem appears.

In the following the flavor decomposition of the longitudinal photoproduction TFFs is listed  for the phenomenologically most important processes
as they appear in the exclusive light meson  electroproduction off proton. Namely, of vector mesons extensively measured in both collider and fixed target kinematics at H1 \cite{Aid:1996ee,Adloff:1997jd,Aaron:2009xp}, ZEUS \cite{Breitweg:1998nh,Breitweg:2000mu,Chekanov:2005cqa,Chekanov:2007zr},
HERMES \cite{Airapetian:2000ni,Airapetian:2009ad}, E665 \cite{Adams:1997bh},  NMC \cite{Arneodo:1994id}, COMPASS \cite{Adolph:2012ht}, CLAS \cite{Hadjidakis:2004zm,Morand:2005ex,Morrow:2008ek,Santoro:2008ai}, and CORNELL \cite{Cassel:1981sx}  as well as pseudoscalar mesons in fixed target kinematics at HERMES \cite{Airapetian:2007aa,Airapetian:2009ac}, CLAS \cite{DeMasi:2007id,Bedlinskiy:2012be}, and HALL-C \cite{Blok:2008jy}.

\begin{itemize}
\item DV{$\!V_L$}P: {\em longitudinal vector meson TFFs} $\tffH_{{\rm V}_L}^{\text{A}}$ {\em and} $\tffE_{{\rm V}_L}^{\text{A}}$ {\em for}
$\gamma_L^\ast\, p\to V_L^0\, p$ {\em and}  $\gamma_L^\ast\, p\to V_L^+\, n$.
\end{itemize}
\vspace{-2mm}
For longitudinal vector meson (V$=1^{--}$) photoproduction the TFFs are $\tffH_{\rm V},\tffE_{\rm V}$ and
for neutral ones we have definite $t$-channel charge parity $C=+1$, see (\ref{C-t-channel}).
The light neutral vector mesons have according to the (constituent) quark model the  Fock state expansion
\begin{eqnarray}
|\rho^0\rangle = \frac{1}{\sqrt{2}}
\left(|u\bu\rangle - |d\bar{d}\rangle\right)\,,\quad
|\omega\rangle = \frac{1}{\sqrt{2}}
\left(|u\bu\rangle + |d\bar{d}\rangle\right)\,,\quad
|\phi\rangle =  |s\bar{s}\rangle
\,.
\label{eq:VMexp}
\end{eqnarray}
As already noted, a two gluon component, which has charge parity even,
can not appear in these meson states.
We decompose the TFFs $\tffF \in \{\tffH,\tffE\}$ with respect to quark
and gluonic $t$-channel exchanges according to (\ref{tff-dec1}),
\begin{subequations}
\label{eqs:tffV0}
\begin{eqnarray}
\tffF_{\rho^0} &
\!\!\!=\!\!\!&
\frac{2}{3\sqrt{2}} \tffF_{\rho^0}^{\upl}
+ \frac{1}{3\sqrt{2}}\tffF_{\rho^0}^{\dpl}
+\frac{1}{\sqrt{2}}\,
  \left( \tffF_{\rho^0}^{\g}
         + \tffF_{\rho^0}^{\pS}
  \right)\,,
\label{eq:rho0TFF}
\\
\tffF_{\omega} &\!\!\!=\!\!\!&
\frac{2}{3\sqrt{2}}\tffF^{\upl}_\omega
- \frac{1}{3\sqrt{2}}\tffF^{\dpl}_\omega
+\frac{1}{3\sqrt{2}}\,
 \left( \tffF^{\g}_\omega
        +  \tffF^{\pS}_\omega
 \right)  \,,
\\
\tffF_{\phi} &\!\!\!=\!\!\!&
-\frac{1}{3}\tffF^{\spl}_\phi
-\frac{1}{3}\,
  \left( \tffF^{\g}_\phi
         +  \tffF^{\pS}_\phi
  \right)\,.
\end{eqnarray}
\end{subequations}
Since the $t$-channel exchanges of two gluons or pure singlet quark-antiquark pairs is flavor blind,
the factors in front of $\left(\tffF^{\g}_{{\rm V}^0}+\tffF^{\pS}_{{\rm V}^0}\right)$
are simply given as the sum over all quark coefficients in the corresponding formulae (\ref{eqs:tffV0}).
Here, the quark TFFs $\tffF^{\qpl}_{{\rm V}^0}$, the pure singlet TFFs $\tffF^{\pS}_{{\rm V}^0}$,
and the gluon TFFs $\tffF^{\g}_{{\rm V}^0}$ correspond to the underlying partonic subprocesses shown
on Figs.~\ref{f:contributions}a, \ref{f:contributions}b, and \ref{f:contributions}c, respectively.

To overcome the quark-gluon mixing, we plug the SU$(n_f)$ representation (\ref{eqs:SU(n)-decomposition}) for quarks into (\ref{eqs:tffV0}),
the TFFs $\tffF_{\text{V}^0}$ for neutral vector mesons $\text{V}^0\in \{\rho^0,\omega,\phi\}$  are then decomposed
into flavor non-singlet  multiplets and a singlet (S) one,
\begin{eqnarray}
\label{eq:rhoTFFsing}
\tffF_{{\rm V}^0}^{\SigG} = \tffF_{{\rm V}^0}^{\g}+  \tffF_{{\rm V}^0}^{\Sigma}
\quad\mbox{with}\quad  \tffF_{{\rm V}^0}^{\Sigma}  =   \frac{1}{n_f} \tffF_{{\rm V}^0}^{0^{(+)}} + \tffF_{{\rm V}^0}^{\pS}\,,
\end{eqnarray}
which is given as sum of gluon  and  flavor singlet quark ($\Sigma$) contributions.  The latter is build from the
group theoretical part ($0^{(+)}$), weighted with the Clebsch--Gordon coefficient $1/n_f$, and the pure singlet piece.
Note that
$\tffF_{{\rm V}^0}^{\SigG}$ and $\tffF_{{\rm V}^0}^{\Sigma}$
are charge even by definition and so an additional superscript $(+)$
is omitted. Using \req{eqs:SU(n)-decomposition}
the  TFFs \req{eqs:tffV0} assume in their group theoretical representation the following form
\begin{subequations}
\label{eqs:tffV0-SU}
\begin{eqnarray}
\label{eq:tffFrho-1}
\tffF_{\rho^0} &
\!\!\!=\!\!\!&
\tffF_{\rho^0}^{\text{NS}(+)}
+\frac{1}{\sqrt{2}} \tffF_{\rho^0}^{\SigG}
\,,
\\
\label{eq:tffFomega-1}
\tffF_{\omega} &
\!\!\!=\!\!\!&
\tffF_{\omega}^{\text{NS}(+)}
+
\frac{1}{3 \sqrt{2}}\tffF_{\omega}^{\SigG}\,,
\\
\label{eq:tffFphi-1}
\tffF_{\phi}  &
\!\!\!=\!\!\!&
\tffF_{\phi}^{\text{NS}(+)}
- \frac{1}{3} \tffF_{\phi}^{\SigG}
\,,
\end{eqnarray}
\end{subequations}
where the flavor non-singlet (NS) combinations for three (four) active quarks read as following
\begin{subequations}
\label{eqs:tffV0-NS}
\begin{eqnarray}
\label{eq:tffFrho-1-NS}
\tffF_{\rho^0}^{\text{NS}(+)} &
=&
\frac{1}{6\sqrt{2}} \tffF_{\rho^0}^{3^{(+)}}
+ \frac{1}{6\sqrt{2}}\tffF_{\rho^0}^{8^{(+)}}
(+ \frac{1}{12\sqrt{2}}\tffF_{\rho^0}^{15^{(+)}})
\,,
\\
\label{eq:tffFomega-1-NS}
\tffF_{\omega}^{\text{NS}(+)} &
=&
\frac{1}{2 \sqrt{2}}\tffF_{\omega}^{3^{(+)}} +
\frac{1}{18 \sqrt{2}}\tffF_{\omega}^{8^{(+)}}
(+
\frac{1}{36 \sqrt{2}}  \tffF_{\omega}^{15^{(+)}})
\,,
\\
\label{eq:tffFphi-1-NS}
\tffF_{\phi}^{\text{NS}(+)}
&=&
\frac{1}{9} \tffF_{\phi}^{8^{(+)}}
(- \frac{1}{36}  \tffF_{\phi}^{15^{(+)}})
\,.
\end{eqnarray}
\end{subequations}
Note that using (\ref{eq:SU(n)}) these non-singlet combinations could be directly expressed in terms of
$\tffF_{\text{V}_0}^{q(+)}$.

Charged vector meson $|\rho^+ \rangle =  |u\overline{d}\rangle$ production in $\gamma_L^\ast p\to \rho_L^+ n$ is given in
terms of flavor off-diagonal TFF $\tffF^{u\overline{d}}$. We  rely on isospin symmetry,  and from (\ref{tff-dec})
we find
\begin{eqnarray}
\label{eq:rho+MTFFs}
\tffF_{\rho^+} &
\!\!\!=\!\!\!&
\frac{1}{2}\tffF^{\umi}_{\rho^0}
-\frac{1}{2}\tffF^{\dmi}_{\rho^0}
+ \frac{1}{6}\tffF^{\upl}_{\rho^0}
-\frac{1}{6}\tffF^{\dpl}_{\rho^0} = \frac{1}{2}\tffF^{3^{(-)}}_{\rho^0} +  \frac{1}{6}\tffF^{3^{(+)}}_{\rho^0}\,,
\quad \tffF \in \{\tffH,\tffE\}\,.
\end{eqnarray}
Note that the flavor off-diagonal quark TFFs splits then in a diagonal flavor isotriplet  with charge even ($\qpl$) and  charge odd ($\qmi$), resulting the
prefactors $(e_u-e_d)/2 = 1/2$ and $(e_u+e_d)/2 = 1/6$.

\pagebreak[3]
\begin{itemize}
\item DV$\!PS$P: {\em pseudoscalar meson TFFs} $\tfftH_{{\rm PS}}$ {\em and} $\tfftE_{{\rm PS}}$ {\em for} $\gamma_L^\ast\, p\to {\rm PS}^0\, p$ {\em and} $\gamma_L^\ast\, p\to {\rm PS}^+\, n$.
\end{itemize}
\vspace{-2mm}
The TFFs (\ref{tffF-def}) for pseudoscalar mesons (PS$^{-+}$) longitudinal photoproduction
are assigned with even parity, i.e., they are called $\tfftH_{\rm PS}$ and $\tfftE_{\rm PS}$.
The neutral pseudoscalar mesons have even charge parity. Hence, we have odd ($t$-channel) charge parity and,
consequently, a two-gluon exchange in the $t$-channel can not occur. The normalized meson states read
\begin{eqnarray}
|\pi^0\rangle &\!\!\! = \!\!\! &\frac{1}{\sqrt{2}}
\left(|u\bu\rangle - |d\bar{d}\rangle\right)\,,
\quad
|\eta^{(8)}\rangle  = \frac{1}{\sqrt{6}}
\left(|u\bu\rangle + |d\bar{d}\rangle
-2 |s\bar{s}\rangle\right)\,,
\nn \\
|\eta^{(0)}\rangle &\!\!\! = \!\!\! & \frac{1}{\sqrt{3}}
\left(|u\bu\rangle + |d\bar{d}\rangle  + |s\bar{s}\rangle \right)
\,.
\end{eqnarray}
Note that we here do not discuss the $\eta$/$\eta^\prime$ mixing problem and rather provide only the formulae
for the pure octet and singlet states.
Furthermore, in the flavor singlet state $|\eta^{(0)}\rangle$
a two gluon component contributes, which is also
beyond  the scope of our considerations here%
\footnote{Strictly speaking the factorization proof from
\cite{Collins:1996fb} did not encompass mesons with quantum numbers which
allow the decay into two gluons, e.g., $\eta^0$. We believe that such a proof
should be straightforward.}.
Reading off the Clebsch-Gordon coefficients, we find then from (\ref{tff-dec1})
\begin{subequations}
\label{eq:PSMTFFs}
\begin{eqnarray}
\tffF_{\pi^0} &
\!\!\!=\!\!\!&
\frac{2}{3\sqrt{2}} \tffF^{\umi}_{\pi^0}
+ \frac{1}{3\sqrt{2}}  \tffF^{\dmi}_{\pi^0}\,,
\\
\tffF_{\eta^{(8)}} &
\!\!\!=\!\!\!&
\frac{2}{3 \sqrt{6}} \tffF^{\umi}_{\eta^{(8)}}
- \frac{1}{3 \sqrt{6}}  \tffF^{\dmi}_{\eta^{(8)}}
+ \frac{2}{3 \sqrt{6}}  \tffF^{\smi}_{\eta^{(8)}}\,,
\\
\tffF_{\eta^{(0)}} &
\!\!\!=\!\!\!&
\frac{2}{3 \sqrt{3}} \tffF^{\umi}_{\eta^{(0)}}
- \frac{1}{3 \sqrt{3}}  \tffF^{\dmi}_{\eta^{(0)}}
- \frac{1}{3 \sqrt{3}}  \tffF^{\smi}_{\eta^{(0)}}\,.
\label{TffF_{eta^{(0)}}}
\end{eqnarray}
\end{subequations}

Analogously to $\rho^+$ case discussed above, for DV$\!\pi^+$P the quark content is flavor off-diagonal,
however, employing isospin symmetry, it can be expressed by diagonal flavor non-singlet ones
\begin{eqnarray}
\label{eq:pi+MTFFs}
\tffF_{\pi^+} &\!\!\!=\!\!\!&
\frac{1}{2}\tffF^{\upl}_{\pi^0}- \frac{1}{2}\tffF^{\dpl}_{\pi^0} + \frac{1}{6}\tffF^{\umi}_{\pi^0} -\frac{1}{6}\tffF^{\dmi}_{\pi^0}
= \frac{1}{2}\tffF^{3^{(+)}}_{\pi^0} + \frac{1}{6}\tffF^{3^{(-)}}_{\pi^0} \,,
\quad \tffF\in\{\widetilde\tffH,\widetilde\tffE\}\,,
\end{eqnarray}
implying that both charge even ($\qpl$) and odd ($\qmi$) contributions enter.

\begin{itemize}
\item {\em Exclusive longitudinal photoproduction  of other mesons.}
\end{itemize}
\vspace{-2mm}
Supposing that the dominant mechanism is a quark-antiquark (or gluon pair)
$t$-channel exchange, the meson quantum numbers  that allow to access
the intrinsic parity even or odd TFFs (\ref{tffF-def})
in longitudinal photoproduction
of neutral (pseudo)scalar and longitudinal (axial-)vector mesons are:
\begin{equation}
\label{eq:mesontable}
\begin{tabular}{|c|c||c|c|}
\hline
 $\tffH^\qpl_M,\tffE^\qpl_M $,
$\tffH^{\text{G}}_M,\tffE^{\text{G}}_M $
& $1_L^{--}$
&  $\tffH^\qmi_M,\tffE^\qmi_M $
& $0^{++}$   \\
 $\tfftH^\qmi_M,\tfftE^\qmi_M $
& $0^{-+}$
& $\tfftH^\qpl_M,\tfftE^\qpl_M $,
$\tfftH^{\text{G}}_M,\tfftE^{\text{G}}_M $
& $1^{+-}_L $
\\ \hline
\end{tabular}
\end{equation}
The quantum number assignments given
on the basis of parity and charge parity conservation%
\footnote{In a nutshell, since $C_\gamma=P_\gamma=-1$, and
for $q \bar{q}$ states $P=(-1)^{l+1}$, $C=(-1)^{l+s}$,
while for $gg$ states $P=(-1)^l$ and $C=1$,
only the transitions corresponding to
$\gamma (J \text{odd})^{--} \to (J \text{even})^{++}$
and
$\gamma (J \text{even})^{-+} \to (J \text{odd})^{+-}$
and reversed are allowed in the quark model.}
are in more detail explained in \cite{Baier:1982,Baier:1985wv}
where relevant hard processes have been discussed
in the crossed channel, as well as in \cite{Ji:2000id}
(see also Tabs.~in \cite{Diehl:2003ny,Goeke:2001tz}).

The longitudinal vector mesons ($1_L^{--}$)
and pseudo scalar mesons ($0^{-+}$) we have discussed.
One may  also include neutral or charged kaon production,
where an initial proton state transforms to a hyperon.
The flavor decomposition is straightforwardly done,
however, if one likes to reduce off-diagonal flavor TFFs
to flavor diagonal ones one must rely on SU(3) flavor symmetry.
Various of these DVMP channels have been already considered and were given in terms
of LO GPD factorization formulae, see reviews
\cite{Diehl:2003ny,Belitsky:2005qn} for references therein and
explicit expressions.

On the same footing as pseudoscalar and longitudinal vector mesons,
one can also consider  $\gamma_L^\ast\, p \to 0^{++}\, p$ process for a scalar meson
(e.g., $0^{++}=f_0$)
where  $\tffH^{\qmi}$, $\tffE^{\qmi}$ contribute.
Remember that under scalar meson one usually understands
$q \bar{q}$ state with $l=1$, $s=1$ while, of course, $J=0$ %
\footnote{Scalar states generally satisfy $l+s$ even and $l$ odd so that
$P=(-1)^{l+1}=1$ and $C=(-1)^{l+s}=1$ and higher ones, i.e., with $l>1$ are
$2^{++}, \ldots$.}. As in the case of pseudoscalar mesons, there also exists
a scalar two-gluon component which mixes with the quark
flavor singlet component. Note that $0^{++}$ carries the same quantum numbers as
$t$-channel $q\bar{q}$ and $gg$ pairs described by
$H^{\qpl}$, $E^{\qpl}$, $H^{\text{G}}$, $E^{\text{G}}$
in the case of production of longitudinal vector mesons, and similarly for
$H^{\qmi}$, $E^{\qmi}$ and $1^{--}_L$. This is to be expected since these two processes
are, in a sense, reversed, as well as, production of $0^{-+}$ and $1^{+-}_L$
(see \cite{Baier:1982,Baier:1985wv} for crossed channel examples).
In the production of axial-vector meson whose $l+s$ and $l$ are odd
$\gamma_L p \to 1^{+-}_L\, p$
(e.g.,  $1^{+-}=h_0$) both
$\widetilde H^{\qpl}$, $\widetilde  E^{\qpl}$ and $\widetilde  H^\g$, $\widetilde  E^\g$ can, in principle,
be accessed. We add that in the literature there  are also suggestions to analyze
in the perturbative DVMP formalism  the production of exotic  meson states
\cite{Anikin:2004vc,Anikin:2004ja}, for example, hybrid mesons $1^{-+}$.

\section{Factorization of transition form factors}
\label{sec:factorization}
\setcounter{equation}{0}

Employing power counting, it has been shown that the dominant production
mechanism for longitudinal DVMP is the $t$-channel exchange
of a quark-antiquark pair or, if it is allowed,  a color singlet gluon pair
\cite{Collins:1996fb}.
Furthermore, it has been perturbatively proved
to all orders that the hard scattering  amplitude,
describing the interaction of the photon with collinear partons,
can be systematically calculated as an expansion w.r.t.~the strong
coupling constant $\alpha_s$.
Thereby, the collinear singularities which appear in
such a diagrammatical calculation can be factorized out and dress the bare DAs and GPDs. This procedure provides then also
a prediction how the DVMP amplitude changes w.r.t.~the variation of the photon virtuality, which is given in terms
of linear evolution equations.  Beyond the leading $1/\cQ$ order, i.e., in which only twist-two GPDs and DAs enters,
the authors state that factorization is maybe broken by final state interaction. In other words it remains questionable if one
can utilize for DVMP  a factorizable $t$-channel picture to access GPDs
in the twist-three sector%
\footnote{
Nevertheless, it was shown in \cite{Kivel:2001qw}
that factorization may not be violated for
twist-3 helicity-flip amplitudes for the hard meson electroproduction
in the Wandzura-Wilczek approximation on a scalar pion target.
}.

Based on this factorization proof, we can say that a flavor decomposed  TFF $\tffF^p_M$ with $\tffF \in \{\tffH,\tffE,\widetilde\tffH,\widetilde\tffE\}$, introduced in Sec.~\ref{sec:Pre},
factorizes in a elementary scattering amplitude,
depicted in Fig.~\ref{f:contributions},
twist-two GPDs $F^p(x,\xi,t)$, $F \in \{H,E,{\widetilde H},\widetilde{E}\}$
describing the transition of the
initial to the final nucleon state by emitting and reabsorbing a parton $p\in \{u,d,s,\cdots G\}$,
and a twist-two meson distribution amplitude (DA) $\varphi_M(v)$,
describing the transition of a quark-antiquark pair $q \bar{q}$ to the meson state $M$.
Thereby, one sums over the partonic exchanges $j$ and integrates over the momentum fractions $x$ and $v$.

Relying on SU$(n_f)$ symmetry and measurements in various channels, DVMP
can serve  to access GPDs with definite partonic content. However,
quark and gluon GPDs  will mix in the charge even sector.
Thus, it is more appropriate to employ in this sector
a group theoretical SU$(n_f)$ decomposition  in flavor non-singlet and singlet contributions,  which allow to solve
the quark-gluon mixing problem. In the charge odd sector it is just a question of taste if we use partonic or group theoretical labeling.
 From this perspective the hard scattering  amplitude has for the considered class of processes some universal features. Namely,
we have only one scattering amplitude in all flavor non-singlet
and charge odd channels.
However, as we will see in the following, different parts of this
hard scattering amplitude will be projected out in the charge even and odd sector.
In principle we have two charge even sectors in which quark and gluons mix, namely for GPD  $H(E)$  and $\widetilde{H}(\widetilde{E})$.
We consider here only the former one since it is relevant for phenomenology ($1^{--}$) and, fortunately, next-to-leading order results were calculated.
We should also mention here that for DVMP of pseudo scalar mesons
in the $\eta^{0}$ case (\ref{TffF_{eta^{(0)}}})
a mixing of quark and gluon DAs appears.
At LO accuracy the two gluon component in the singlet DA vanishes
in the collinear factorization approach
\cite{Baier:1982,Baier:1985wv,Huang:2003jd}.
So far this amplitude has not be calculated at NLO, however,
the mixing of quark and gluon DA and
factorization scale independence indicate that a contribution
from the gluonic meson DA enters the hard scattering  amplitude at NLO.
In the following we will not consider this case. Note that charge conjugation conservation tells us that a charge even meson DA can never appear
together with a charge even GPD and so a quark-gluon mixing can not simultaneously occur  for DAs and GPDs.

In the next section we give our definitions for the hard scattering amplitudes in the common momentum fraction representation,
explain the role of symmetries, show that together with the conventions from Sec.~\ref{sec:pre-SU} we recover the known LO results,
and predict then the NLO factorization and renormalization logarithms. For the phenomenological application we consider two other representations
as more appropriate. In Sec.~\ref{sec:prel-DR} we give simple convolution formulae for the imaginary parts, while the real parts can be obtained from  dispersion relations.
We also show how the  hard scattering amplitude can be decomposed into two parts which have only discontinuities on the negative
or positive $x$-axis in the complex plane. In particular for the purpose of global fitting, we give in Sec.~\ref{sec:prel-MB} a short introduction into the Mellin-Barnes integral representation, a discussion about the resummation of  evolution effects, and spell out our conventions
for conformal partial wave amplitudes.  Based on the aforementioned decomposition of the hard scattering amplitude,
we provide also a method for both the analytic and  numerical evaluation of complex valued conformal partial wave amplitudes.
Finally, in Sec.~\ref{sec:prel-mixed} we show how mixed representations are build with our conventions.

\subsection{Momentum fraction representation}
\label{sec:prel-MF}

For the sake of a compact presentation,
we employ
in the factorization formulae of  TFF $\tffF^{\text{A}}_{\rm M}$
only the quark GPDs with definite charge parity
\begin{subequations}
\label{gpdF^{q(pm)}}
\begin{eqnarray}
F^{q(C)}(x, \eta, t)=
F^{q}(x, \eta, t) -\sigma F^{q} (-x, \eta, t)\;,
\label{gpdF^{q(pm)a}}
\end{eqnarray}
which by construction have definite symmetry $(-\sigma)$
under $x\to-x$ reflection
\begin{eqnarray}
F^{q(C)}(-x, \eta, t) = -\sigma F^{q(C)}(x, \eta, t)
\,.
\label{gpdF^{q(pm)b}}
\end{eqnarray}
\end{subequations}
The GPDs $F \in \{H^q,E^q,\widetilde{H}^q,\widetilde{E}^q\}$
are defined in (\ref{eq:FtoHE}),
and $\sigma=+1(-1)$ for $C=+1(-1)$ and intrinsic parity even GPDs
$H^q,E^q$ and for $C=-1(+1)$ and intrinsic parity odd GPDs
$\widetilde{H}^q,\widetilde{E}^q$.
Hence, from the table in \req{eq:mesontable} it is clear that
for both neutral vector meson and pseudoscalar electroproduction
the signature assignment is $\sigma=+1$.
We notify that we adopt here PDF terminology, allowing us to solve
the quark-gluon and also quark-antiquark mixing problem,
see e.g.~\cite{Furmanski:1981cw}.
The GPD choice (\ref{gpdF^{q(pm)}}) will assign
a charge parity $\tffF^{(C)}_{\text{M}}$ or, equivalent, a signature label
$\tffF^{\sigma}_{\text{M}}$ to our TFFs. Note that sometimes in the literature such a superscript is
used to label the symmetry of the GPD rather than the signature.
Moreover, it allows us to work with an unsymmetrized elementary
hard scattering  amplitude $T$ %
\footnote{The full $T$ obtained directly from Feynman diagrams is
naturally symmetrized and it mirrors the symmetry properties
of the process at hand.},
which is perturbatively given as expansion in the QCD coupling
constant $\alpha_s$.

For charge odd  or flavor non-singlet quark GPDs it arises only from the class of Feynman diagrams,
where the flavor content of the initial quark pair can not be changed, see  Fig.~\ref{f:contributions}a.
Thus, stripping off the electrical quark charges, as already done in the preceding section, we have in the non-singlet channel
and/or charge odd sector the same quark amplitude $T$.
Taking a convenient prefactor, we write the factorization formula for  $A\in \{\NSplmi,\qmi\}$ as
\begin{eqnarray}
\tffF^{\text{A}}_{\rm  M}(\xB,t,\cQ^2)
\stackrel{\rm tw-2}{=}
\frac{C_F f_{\rm M}}{N_c \cQ}
 F^{\text{A}}(x,\xi,t,\muF^2)
\stackrel{x}{\otimes}
\qT\!\left(\!\frac{\xi+x-i \epsilon}{2(\xi-i \epsilon)},v\Big|\alpha_s(\muR),
\frac{\cQ^2}{\muF^2},\frac{\cQ^2}{\muphi^2},\frac{\cQ^2}{\muR^2}\right)
 \stackrel{v}{\otimes}
\varphi_{\rm M}(v,\muphi^2)
\,,
\label{eq:tffF^A}
\nonumber\\
\end{eqnarray}
where
$C_F=4/3$ and $N_C=3$ are the common color factors,   and the convolution
symbols
$$f(x)\stackrel{x}{\otimes}g(x)\equiv \int_{-1}^1\!\frac{dx}{2\xi}\,f(x)g(x)
\quad\mbox{and}\quad f(v)\stackrel{v}{\otimes}g(v) \equiv \int_{0}^1\!dv\, f(v) g(v)
$$
stay for the integration over the momentum fraction $x\in [-1,1]$ and $v\in [0,1]$, respectively.
To obtain the imaginary part according to  Feynman`s causality prescription, the scaling variable $\xi$ is
decorated with an imaginary part $-i\epsilon$.
This partonic scaling variable $\xi \sim \xB/(2-\xB)$
is conventionally defined (see also App. \ref{app:def-GPD})
and here and in the following we set
\begin{eqnarray}
\label{xi2xB}
\xi = \frac{\xB}{2-\xB} \quad\mbox{or inversely}\quad  \xB= \frac{2\xi}{1+\xi}.
\end{eqnarray}
Since the meson decay constants $f_{\rm M}$ is included in the prefactor,
we can normalize the  meson  DAs,
\begin{eqnarray}
\int_0^1\! dv\; \varphi_{\rm M} (v,\mu^2) = 1\,.
\end{eqnarray}
Our TFFs are dimensionless, however, they are proportional to $f_{\rm M}/\cQ$,
where the meson decay constants $f_{\rm M}$ has mass dimension.    We notify that
this canonical $1/{\cQ}$ scaling originates from the contraction
$(p_M+\xi P)^\mu \epsilon_\mu(\lambda_\gamma=0)/{\cQ}^2$ with the photon polarization vector.
The hard scattering amplitude possesses besides a logarithmical $\cQ^2$ dependence also a renormalization ($\muR$), GPD factorization scale ($\muF$),
DA factorization scale ($\muphi$) dependence, while the TFFs possess
only a residual renormalization and factorization  scale dependence.

The SU$(n_f)$ group theoretical decomposition for DV$\!V^0_L$P from the preceding section,
provided us the form of the flavor singlet TFF (\ref{eq:rhoTFFsing}), consisting of charge even quark (\ref{gpdF^{q(pm)}}) and gluon (\ref{eq:FtoHE}) entries.
Therefore, we may generally introduce the vector valued GPDs
\begin{equation}
\label{eq:gpdFvec}
\mbox{\boldmath $F$}(\cdots)
= \left({F^\Sigma } \atop {F^{\rm G} }\right)(\cdots)\quad\mbox{with}\quad F^{\Sigma}(\cdots) = \sum_{q=u,d,s,\cdots} F^{\qpl}(\cdots)
\quad \mbox{and}\quad F \in\{H,E,\widetilde H, \widetilde E\}\,.
\end{equation}
Note that due to Bose symmetry the charge even gluon GPDs have definite
symmetry under $x\to-x$ reflection, which contrarily to quarks is $\sigma$
rather $-\sigma$.  Furthermore, in the DV$\!V^0_L$P case, on which we will us concentrate here, we have $F\in\{H,E\}$,  $\sigma=+1$ and the gluon GPDs are symmetric%
\footnote{For longitudinally neutral axial-vector meson production,  see table in \req{eq:mesontable},
 $\sigma=-1$ and the corresponding gluon GPDs $\widetilde{H}^\g$ and $\widetilde{E}^\g$  are antisymmetric.}.
In analogy to the factorization formula (\ref{eq:tffF^A}), we write a flavor singlet TFF  as
\begin{subequations}
\label{eqs:tffF^S}
\begin{eqnarray}
\label{eq:F^S_V-a}
\tffF_{{\rm  V}^0}^{\SigG}(\xB,t,\cQ^2)
&\!\!\! \stackrel{\rm Tw-2}{=}\!\!\!  &
\frac{C_F f_{{\rm  V}^0}}{N_c \cQ}\times
\\
&&
\varphi_{{\rm  V}^0}(v,\muphi^2)
     \stackrel{v}{\otimes}
   \mbox{\boldmath $T$}\!\left(\!\frac{\xi+x-i\epsilon}{2(\xi-i\epsilon)},v;\xi\Big|\alpha_s(\muR),
\frac{\cQ^2}{\muF^2},\frac{\cQ^2}{\muphi^2},\frac{\cQ^2}{\muR^2}\right)
\stackrel{x}{\otimes}
\mbox{\boldmath $F$}(x,\xi,t,\muF^2)
\,,
\nonumber
\end{eqnarray}
where the convolution $\stackrel{x}{\otimes}$ includes now the forming
of a scalar product, built from the GPD $\mbox{\boldmath $F$}$ in (\ref{eq:gpdFvec})
and the vector valued hard scattering amplitude
\label{eqs:T^S-all}
\begin{equation}
\label{eq:T^S}
\mbox{\boldmath $T$}(u,v;\xi|\cdots) =
\left({^\Sigma}T (u,v|\cdots), \; \;
(C_F \, \xi)^{-1}\,{^{\rm G}T }(u,v|\cdots) \right)\,,
\end{equation}
which contains according to the decomposition (\ref{eq:rhoTFFsing}) the charge even quark entry
\begin{eqnarray}
{^\Sigma}T (u,v|\cdots) =
\frac{1}{n_f} \qT (u,v|\cdots) + \, {^\pS}T (u,v|\cdots)\,.
\label{eq:T^Sigma}
\end{eqnarray}
In the gluon entry the color factor $1/\CF$  compensates $\CF$ from
the overall factor in (\ref{eq:F^S_V-a}),
while the factor $1/\xi$ stems from the peculiarity of the common gluon GPD definition, see also the
forward limits (\ref{eq:Fq2pdf}) and  (\ref{eq:Fg2pdf}).
\end{subequations}

We note that TFFs (\ref{eq:tffF^A},\ref{eqs:tffF^S}) have definite symmetry properties w.r.t.~$\xi\to -\xi$ reflection. For
(\ref{eq:tffF^A}) and the quark entry of (\ref{eqs:tffF^S})
we find under  simultaneous   $\xi\to-\xi$ and $x\to -x$ reflections:
\begin{equation}
\frac{1}{\xi}T\left(\frac{\xi+x-i \epsilon}{2(\xi-i \epsilon)},\cdots \right) F^\sigma(x, \xi, t)
\quad \Longrightarrow \quad
\frac{\sigma}{\xi} T\left(\frac{\xi+x+i \epsilon}{2(\xi+i \epsilon)},\cdots \right) F^\sigma(x, \xi, t)\,,
\label{eq:TFsym}
\end{equation}
where we used that GPDs are even functions in $\xi$ and have symmetry $-\sigma$ under $x$-reflection.
Hence, the real part of quark TFFs with definite signature is an even and odd function for $\sigma=+1$ and $\sigma=-1$, respectively.
Since under simultaneous reflection $-i\epsilon$ goes into $+i\epsilon$, see (\ref{eq:TFsym}),
the symmetry of the imaginary part is reversed compared to the real part.
The gluonic part in the singlet flavor TFF (\ref{eqs:tffF^S}) has the  same symmetry as the quark entry, since the sign change of the
additional factor $1/\xi$ in (\ref{eq:T^S}) is compensated by the different symmetry behavior of the gluon GPD under $x$-reflection. Furthermore, we
can restrict the integration region  in (\ref{eq:tffF^A},\ref{eqs:tffF^S})  to  positive $x$,
where now  hard scattering amplitudes with definite symmetry properties have to be taken, i.e., we replace
\begin{subequations}
\label{eq:Tsignum}
\begin{eqnarray}
\stackrel{x}{\otimes}\equiv \int_{-1}^1\!\frac{dx}{2\xi}
&\Rightarrow&
\int_0^1\!\frac{dx}{2\xi}\,,
\label{eq:Tsignum-int}\\
\qT(u,v|\cdots)
&\Rightarrow&
{^\sigma}T(u,v|\cdots) =
\qT(u,v|\cdots) -\sigma\, \qT(\bu,v|\cdots) \,,
\label{eq:Tsignum-q}\\
{^\Sigma}T(u,v|\cdots)
&\Rightarrow&
{^\Sigma}T(u,v|\cdots) -  {^\Sigma}T(\bu,v|\cdots)\,,
\label{eq:Tsignum-Sigma}
\\
{^\g}T(u,v|\cdots)
&\Rightarrow&
{^\g}T(u,v|\cdots) +  {^\g}T(\bu,v|\cdots)\,.
\label{eq:Tsignum-G}
\end{eqnarray}
\end{subequations}
Here, in the flavor singlet channel we only refer to the phenomenological important DV$\!V_L^0$P process, i.e., we explicitly use $\sigma=+1$.
We add that the hard scattering amplitudes for $M\leftrightarrow \gamma_L$ crossed exclusive (time-like) processes can be obtained from those of the
corresponding DVMP ones \cite{Muller:2012yq}.

\subsubsection{Symmetry properties and leading order result}
\label{sec:prel-MF-definitions}

As said above, we employ in the factorization formulae (\ref{eq:tffF^A},\ref{eqs:tffF^S}) only
GPDs with definite charge parity and symmetry behavior
under $x\to -x$ reflection.
The symmetry property is characterized by the signature factor,
where quark GPDs and gluon GPD have the same signature, however,  different symmetry%
\footnote{We note that this is analogous to the symmetry properties of
meson DAs, i.e.,
$\phi^{q\bar{q}}(u)=-P\phi^{q\bar{q}}(1-u)$
and
$\phi^{gg}(u)=P\phi^{gg}(1-u)$, e.g.,  a $q\bar{q}$ pair has  intrinsic parity $(-1)$
and $P=(-1)^{l+1}$ with $l$ being its angular momentum.
}
\begin{equation}
F^{q(C)}(-x,\cdots) = -\sigma\, F^{q(C)}(x,\cdots) \quad\mbox{and}\quad
F^{\g}(-x,\cdots) = \sigma\, F^{\g}(x,\cdots)\,.
\label{gpdF^{p(C)}}
\end{equation}
The signature $\sigma(F^{\text{A}})$ which can be considered as function
of the GPD type [see discussion below \req{gpdF^{q(pm)}}]
reads explicitly in our nomenclature as
\begin{eqnarray}
\sigma(F)=\left\{{+1 \atop -1}\right\}\quad\mbox{for}\quad
F\in \left\{
H^\NSpl,  H^\Sigma,  H^{\g},  E^\NSpl, E^\Sigma,E^{\g},\widetilde{H}^{\qmi}, \widetilde{E}^{\qmi} \atop
\widetilde{H}^\NSpl, \widetilde{H}^\Sigma, \widetilde{H}^{\g},  \widetilde{E}^\NSpl, \widetilde{E}^\Sigma, \widetilde{E}^{\g}, H^{\qmi}, E^{\qmi}  \right\}.
\label{eq:sigma4F}
\end{eqnarray}
If  SU(3) breaking effects are ignored,  meson DAs for both vector
($1^{--}$) and pseudo scalar ($0^{-+}$) mesons
are symmetric in $v \to \bv = 1-v$,
except for the antisymmetric two-gluon DA
that contributes to the pseudoscalar state $\eta^{0}$.
One may include SU(3) breaking effects, which induce then an antisymmetric component in the DA amplitude, e.g., for $K$ mesons
where according to \cite{Ball:2006wn} only a small admixture appears (at leading power of $1/\cQ^2$).

\begin{figure}[t]
\begin{tabular}{ccccc}
LO\hspace*{0.5cm}
&\includegraphics[scale=0.4]{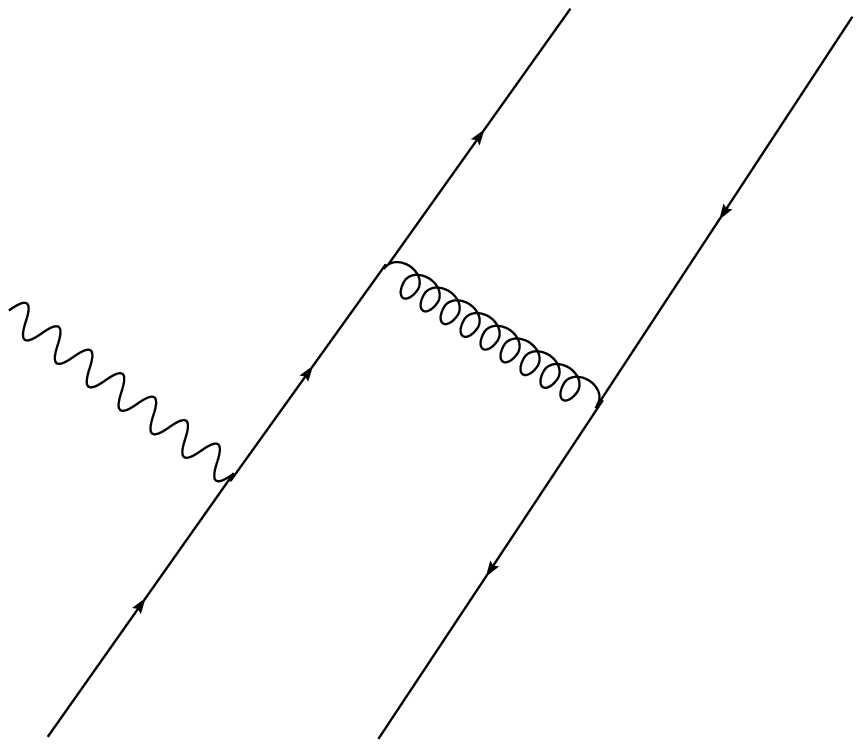}
&
&
&\includegraphics[scale=0.4]{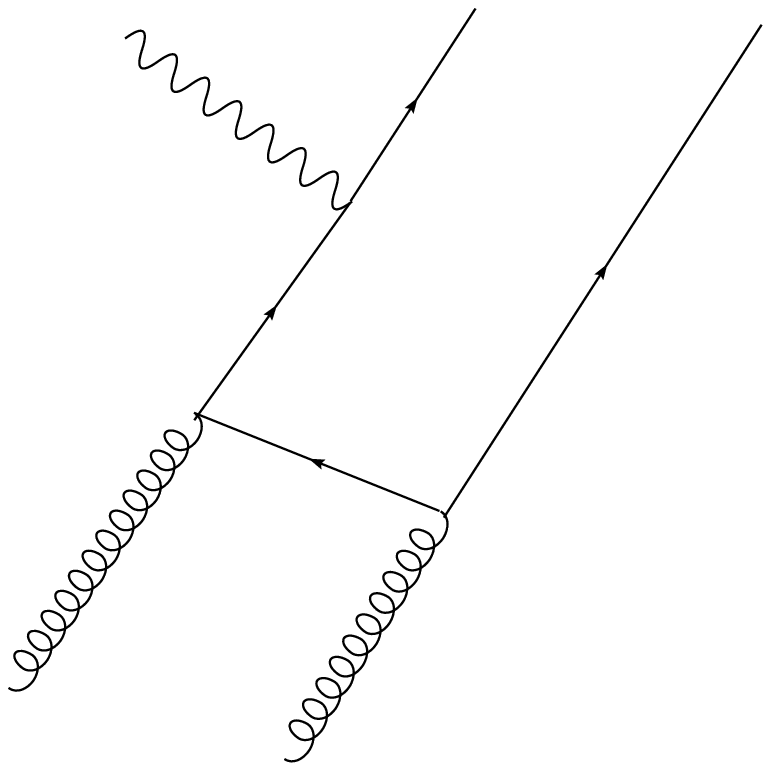} \\[0.5cm]
NLO\hspace*{0.5cm}
& \includegraphics[scale=0.4]{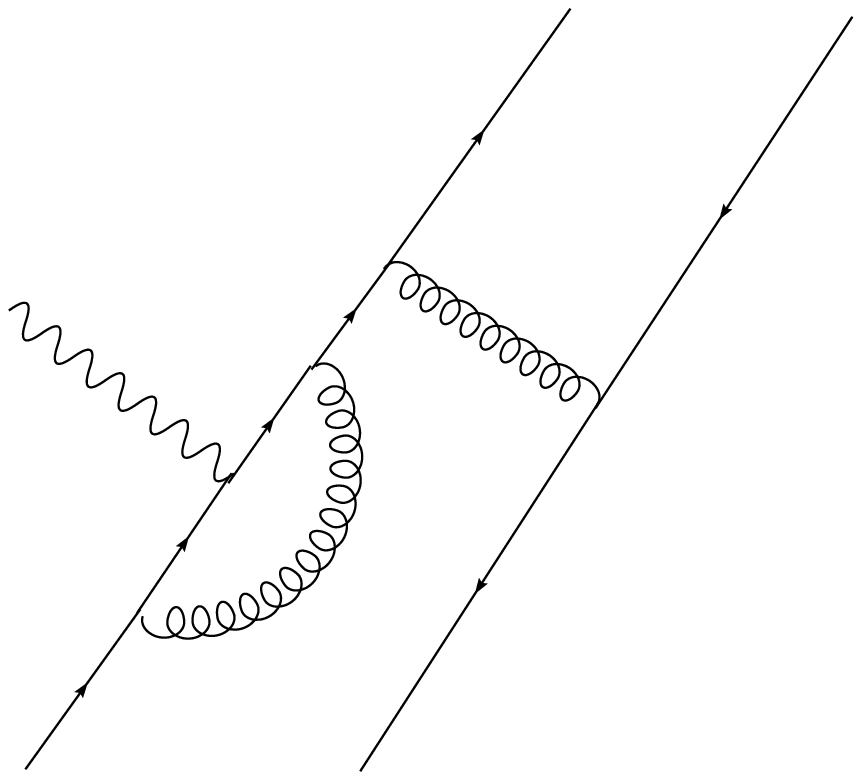}
&\includegraphics[scale=0.4]{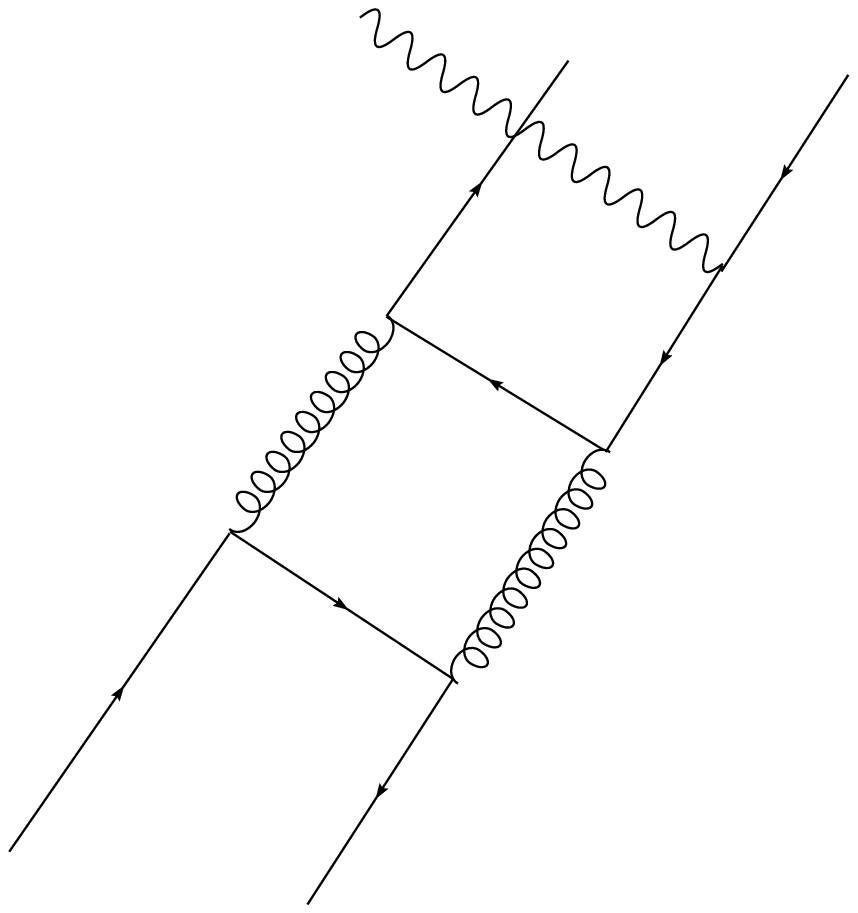}
&
&\includegraphics[scale=0.4]{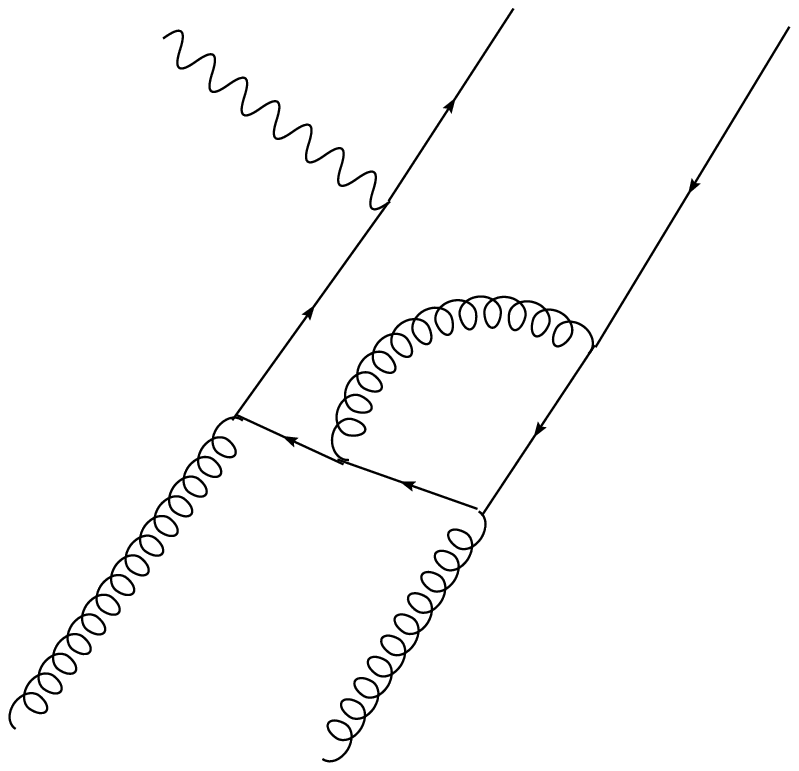} \\
& a) & b) & \phantom{space}  & c)
\end{tabular}
\caption{Representative partonic subdiagrams that contribute
to DVMP of neutral vector mesons.
In context of the GPD part, i.e., partons coming from
the proton, (lower ''legs'', while the upper ones denote
the meson) the contributions are organized as:
a) quark contribution b) pure singlet quark contribution
c) gluon contribution.}
\label{f:contributions}
\end{figure}
Let us now discuss the symmetry properties
of the hard scattering  amplitudes used above, which we will consider as functions $T(u,v)$.
To LO accuracy these amplitudes arise in the flavor non-singlet channel
from four Feynman diagrams, where
a representative one is depicted in Fig.~\ref{f:contributions}a).
Here  the initial quark and antiquark  $q_1(u) \bar{q}_2(\bu)$,
knocked out from the nucleon, have momentum fractions
\begin{eqnarray}
u=\frac{\xi+x}{2\xi} \, , &\mbox{and}&
\bu\equiv 1-u = \displaystyle \frac{\xi-x}{2\xi }
\label{eq:uxix}
\end{eqnarray}
of light-cone momentum $P_1^+ - P_2^+$  and the
quark and antiquark  $q_1(v) \bar{q}_2(\bv)$, forming the meson,  have momentum fraction $v \ge 0$ and $\bv \ge 0$ of the
meson light-cone momentum $P_{\rm M}^+$.
For $x \ge \xi$ the representative LO diagram can be  interpreted as a  partonic $s$-channel scattering subprocess
$$\gamma_L^\ast\, q_1(u)\to  [q_1(v) \bar{q}_2(\bv)]\, q_2(u-1),$$
where  a quark $q_1$ is knocked out from the nucleon with momentum fraction $(\xi+x)/2\xi$
and a quark $q_2$ with momentum fraction $(x-\xi)/2\xi$ is reabsorbed.
Exploiting the symmetry under $u\to v$ (photon couples to the in/outgoing $q_1$ quark line) and $u \to \bu$, $v \to \bv$ (photon couples to the $q_2$ quark lines) symmetries,
we can  obtain all four LO Feynman diagrams from the representative one in Fig.~\ref{f:contributions} a).
Generally,  according to the coupling of the photon to either the $q_1$ or $q_2$ quark we divide all Feynman diagrams in the quark channels in two classes:
\begin{subequations}
\begin{eqnarray}
+e_{q_1} \qT(u,v|\cdots)  &&
\mbox{if photon couples to $q_1$-quark line or}\;[\gamma_L^\ast q_1(u)] \overline{q}_2(\bu)\to q_1(v) \overline{q}_2(\bv)\,,
\qquad\quad
\label{eq:qT-a}
\\
-e_{q_2} \qT(\bu,\bv|\cdots)  &&
\mbox{if photon couples to $\overline{q}_2$-quark line or}\;  q_1(u)[\gamma_L^\ast \overline{q}_2(\bu)]\to q_1(v)q_2(\bv)\,,
\label{eq:qT-b}
\end{eqnarray}
\end{subequations}
where the (fractional)  quark charges $e_{q_i}$ are not included in the hard scattering amplitude $\qT(u,v|\cdots) $.
It is obvious that if the
quarks $q_1$ and $q_2$ have different flavors,
the $q_1 \leftrightarrow q_2$ exchange,
i.e., $(u,v) \leftrightarrow (\bu,\bv)$, goes hand in hand
with an exchange of quark charge factors $e_{q_1} \leftrightarrow e_{q_2}$.

To obtain the net contribution in a quark channel $\gamma_L^\ast\, q_1 \bar{q}_2 \to q_1 \bar{q}_2$, we obviously have to add to
the hard scattering amplitude (\ref{eq:qT-a})
the contributions from the second class (\ref{eq:qT-b}) and multiply them with the quark charges:
\begin{subequations}
\label{eq:qT-sym-asym-0}
\begin{equation}
e_{q_1 }\, \qT(u,v|\cdots)  - e_{q_2} \, \qT(\bu,\bv|\cdots)\,,
\label{eq:qT-sym}
\end{equation}
where the struck quark $q_1$ is exchanged with $q_2$. We may decompose the net amplitude (\ref{eq:qT-sym}) in
a charge even and odd part
\begin{equation}
\frac{e_{q_1 } + e_{q_2 }}{2} \left[\qT(u,v|\cdots)  - \qT(\bu,\bv|\cdots)\right]+
\frac{e_{q_1 } - e_{q_2 }}{2} \left[\qT(u,v|\cdots)  + \qT(\bu,\bv|\cdots)\right].
\label{eq:qT-sym-asym}
\end{equation}
For neutral meson production  $e_{q_2}=e_{q_1}$.
Hence, the second term drops out  and the net amplitudes are
antisymmetric under simultaneous $u \to \bu$ (or $x\to -x$)
and $v \to \bv$ exchange. Moreover,  DAs for neutral mesons are even
under $v\to \bv$ exchange and so  the convolution with a quark GPD projects
out their positive signature, i.e., antisymmetric parts.
For charged isotriplet meson production  the DAs are also symmetric
under  $v\to \bv$ and both positive and negative signature GPDs contribute. Employing symmetric meson DAs $\varphi(v)=\varphi(\bv)$, we can
replace in a convolution formula  the hard scattering amplitude (\ref{eq:qT-sym-asym}) by
\begin{equation}
\frac{e_{q_1 } + e_{q_2 }}{2} \left[\qT(u,v|\cdots)  - \qT(\bu,v|\cdots)\right]+
\frac{e_{q_1 } - e_{q_2 }}{2} \left[\qT(u,v|\cdots)  + \qT(\bu,v|\cdots)\right].
\label{eq:qT-sym-asym-1}
\end{equation}
\end{subequations}
Hence, after decomposition into contributions of definite signature
and pulling out of charge factors in the partonic decomposition of TFFs, as already done in Sec.~\ref{sec:pre-SU},
we can write down for all these cases the convolution formula (\ref{eq:tffF^A})  in terms of quark GPDs
with definite signature. Thus, the    definition (\ref{gpdF^{q(pm)}}) for GPDs with definite charge parity ensures that the $u\to\bu$ counterparts of $\qT(u,v)$ in
(\ref{eq:qT-sym-asym-0}) are taken into account.  We add that if one likes to include SU(3) symmetry breaking effects, an anti-symmetric
meson DA appears, too. Hence, the  relative signs in (\ref{eq:qT-sym-asym-1}) will change which implies that GPDs with reversed signature must
be taken into account.

For DV$\!V^0_L$P and DV$\!{PS}^0$P processes we have according to the table in (\ref{eq:mesontable}) to take the GPDs
$$
F^{q^{(+)}} \in \big\{ H^{\qpl}, E^{\qpl} \big\}
\quad\mbox{and}\quad
F^{q^{(-)}} \in \big\{ \widetilde{H}^{\qmi}, \widetilde{E}^{\qmi} \big\},
$$
respectively, which have different charge parity, inherited from the charge parity in the $t$-channel.
Obviously, only  in the flavor singlet channel with even charge parity,
i.e., for DV$\!V^0_L$P, both a pure singlet quark and gluon contribution can appear,
taken into account by the hard scattering amplitudes
(\ref{eq:T^S},\ref{eq:T^Sigma}) and depicted in Figs.~\ref{f:contributions}b) and \ref{f:contributions}c).
Note that a diagrammatical evaluation of the graphs in Figs.~\ref{f:contributions}b), \ref{f:contributions}c), and  other contributing ones provides
scattering amplitudes with explicit symmetry properties. Namely, the pure singlet quark contribution, which is absent at LO, is
antisymmetric under $u \to \bu$ and symmetric under $v \to \bv$,
\begin{equation}
\frac{1}{2}\Big[
{^\pS}T (u,v|\cdots)
-{^\pS}T (\bu,v|\cdots)
+{^\pS}T (u,\bv|\cdots)
-{^\pS}T (\bu,\bv|\cdots)\Big].
\label{eq:SigmaT-sym}
\end{equation}
The gluon contribution being
symmetric in both $u \to \bu$
and  $v \to \bv$,
\begin{equation}
\frac{1}{4}\Big[{^\text{G}}T (u,v|\cdots)
+{^\text{G}}T (\bu,v|\cdots)
+{^\text{G}} T (u,\bv|\cdots)
+{^\text{G}}T (\bu,\bv|\cdots)\Big].
\label{eq:GT-sym}
\end{equation}
The averaging factors $1/2$ and $1/4$ guarantee consistency with our normalization.
In defining \req{eqs:tffF^S} we have made use of the symmetry properties (\ref{eq:SigmaT-sym},\ref{eq:GT-sym})
of the contributing quark (antisymmetric) and gluon (symmetric) GPDs as well as meson DA (symmetric).

We note that due to symmetry the representation of the building block ${^\text{A}}T(u,v)$ in a definite signature sector is not unique. For instance,
we may add to such a building block a function $f(u,v)$ that is (anti-)symmetrized under $u\to \bu$ reflection,
$$
{^\text{A}}T(u,v)\pm {^\text{A}}T(\bu,v)\;\; \Rightarrow\;\; \left[{^\text{A}}T(u,v)+ f(u,v)\right] \pm \left[{^\text{A}}T(\bu,v) + f(\bu,v)\right]
\quad\mbox{with} \quad  f(\bu,v) =\mp f(u,v),
$$
which  cancels in the convolution with a GPD, having the proper signature.  As it will become obvious in Sec.~\ref{sec:prel-DR},
the ambiguity in choosing the building block ${^\text{A}}T(u,v)$ can be removed if we require that it possesses for real $v$ with
$0\le v\le 1$ only a discontinuity on the positive $u$-axis $[1,\infty]$.
This allows us in the following to deal with functions that are holomorphic in the second and third quadrant of the complex $u$-plane.

Let us add that our diagrammatical calculation yields
\begin{subequations}
\label{eqs:tff-LO}
\begin{equation}
\label{eqs:T-LO}
T(u,v)=  \frac{\alpha_s(\muR)}{\bu\bv} + O(\alpha_s^2) \quad\mbox{and}\quad  {^\g}T(u,v)=   \frac{\alpha_s(\muR)}{\bu\bv} + O(\alpha_s^2)\,.
\end{equation}
Plugging these results
into (\ref{eq:tffF^A}) and (\ref{eqs:tffF^S}),
we find the LO approximation of the quark TFFs with definite charge parity and the gluonic TFF,  respectively,
\begin{eqnarray}
\tffF_{\rm  M}^{q^{(\pm)}}(\xB,t,\cQ^2)
&\!\!\!\stackrel{\rm LO}{=}\!\!\!&
 \frac{C_F  f_{{\rm M}}\,\alpha_s(\muR)}{N_c \cQ}
  \int_{-1}^{1}\!dx\,\frac{F^{q^{(\pm)}}(x,\xi,t,\muF^2)}{\xi-x-i \epsilon} \int_{0}^1\!dv\, \frac{\varphi_{\rm M}(v,\muphi^2)}{\bv}\,,
\label{eq:TffF+--LO}
\\
\tffF_{{\rm  V}^0_L}^{\g}(\xB,t,\cQ^2)
&\!\!\!\stackrel{\rm LO}{=}\!\!\!&
\frac{f_{{\rm  V}^0_L}\,\alpha_s(\muR)}{N_c \cQ} \int_{-1}^{1}\!dx\, \frac{F^\g(x,\xi,t,\muF^2)}{\xi(\xi-x-i \epsilon)}
 \int_{0}^1\!dv\, \frac{\varphi_{{\rm  V}^0}(v,\muphi^2)}{\bv}
\,.
\end{eqnarray}
\end{subequations}
By means of the partonic decompositions, given in Sec.~\ref{sec:pre-SU}, we obtain then the well known LO expressions for the DVMP amplitudes,
see reviews \cite{Diehl:2003ny,Belitsky:2005qn} and references to original work therein.

\subsubsection{Perturbative expansion and  scale dependencies}
\label{sec:prel-MF-exp}

As alluded above, let us first shortly comment on the scale dependencies in the convolution formulae (\ref{eq:tffF^A},\ref{eqs:tffF^S}), where
one should bear in mind that the LO hard scattering amplitude starts with $\alpha_s(\muR)$. Afterwards, we present the
renormalization and factorization logarithms at NLO accuracy.
\begin{itemize}
\item {\em Renormalization scale independence.}
\end{itemize}
\vspace{-2mm}
The requirement that the hard scattering amplitude is independent of the renormalization scale is nothing but the famous renormalization group equation
\begin{eqnarray}
\label{T-RGE}
\left[\muR \frac{\partial}{\partial\muR}  + \beta(\alpha_s) \frac{\partial}{\partial \alpha_s}
\right] T\!\left(\cdots \Big|\alpha_s, \frac{\cQ^2}{\muF^2},\frac{\cQ^2}{\muphi^2},\frac{\cQ^2}{\muR^2}\right) =0\,.
\end{eqnarray}
We remind  that the running of $\alpha_s(\mu)$ is perturbatively controlled by the  equation%
\footnote{Note that the value of $\beta_0$ is here negative, contrary to common definitions in the literature.}
\begin{eqnarray}
\label{alpha_s}
\mu \frac{d}{d\mu} \alpha_s(\mu) = \beta_0\, \frac{\alpha_s^2(\mu)}{2\pi} + O(\alpha_s^3)\quad \mbox{with}\quad  \beta_0 = \frac{2 n_f}{3}  -11
\end{eqnarray}
and its solution is given as a function of $\ln(\mu^2/\Lambda^2_{\rm QCD})$, where the QCD scale  $\Lambda_{\rm QCD} \simeq 0.2\, \GeV$.
However, the perturbative expansion of the hard scattering amplitude induces a residual renormalization
scale dependence that is caused by the truncation
of the perturbative series.
This dependence appears in the QCD running coupling constant
$\alpha_s(\muR)$
and in $\ln(\cQ^2/\muR^2)$ terms,
and they partially compensate each other at any given order, see (\ref{T-RGE}).
At LO the residual dependence is of order $\alpha_s^2$
while the appearance of $\ln(\cQ^2/\muR^2)\alpha_s^2(\muR)$
terms at NLO weakens the $\muR$ dependence, leaving
us with an uncertainty of order $\alpha_s^3$.
In general, at order $n$ in perturbation theory one
is left with a renormalization scale uncertainty
of order $\alpha_s^{(n+1)}$.

\begin{itemize}
\item {\em Factorization scale dependencies.}
\end{itemize}
\vspace{-2mm}
The hard scattering amplitude explicitly
depends on factorization logarithms
$\ln(\cQ^2/\muF^2)$ and $\ln(\cQ^2/\muphi^2)$ ($\muF$ for GPDs and $\muphi$ for DAs).
In the convolution with the GPD and DA these scale dependencies of the
hard scattering amplitude are partially cancelled by those of GPDs and DAs, which
are perturbatively controlled by evolution equations.
The factorization scale dependencies of TFFs is of order $\alpha^2_s$ at LO,
entirely arising from the scale dependencies of GPD and DA, where one power of $\alpha_s$
stems from the LO hard scattering amplitude. Going to  NLO will push the residual factorization dependence
to order $\alpha_s^3$. At order $n$  in perturbation theory one
is left with the factorization scale uncertainties which is of order $\alpha_s^{(n+1)}$.
The independence of TFFs of the factorization scale can be easily formulated in terms of evolution equations for the hard scattering amplitudes
w.r.t.~both the factorization scale of the DA,
\begin{subequations}
\label{dT}
\begin{eqnarray}
\label{dT-muphi}
\muphi^2 \frac{d}{d\muphi^2}
\mbox{\boldmath $T$}\!\left(\frac{\xi+x}{2\xi},v\Big|\alpha_s,\frac{\cQ^2}{\muF^2},\frac{\cQ^2}{\muphi^2},\frac{\cQ^2}{\muR^2}\right)=-
\mbox{\boldmath $T$}\!\left(\frac{\xi+x}{2\xi},v^\prime\Big|\alpha_s,\frac{\cQ^2}{\muF^2},\frac{\cQ^2}{\muphi^2},\frac{\cQ^2}{\muR^2}\right) \! \stackrel{v^\prime}{\otimes}\!
V(v^\prime,v|\alpha_s),
\end{eqnarray}
and the factorization scale of the GPD
\begin{eqnarray}
\label{dT-muF}
\muF^2 \frac{d}{d\muF^2}
\mbox{\boldmath  $T$}\!\left(\frac{\xi+x}{2\xi},v\Big|\alpha_s,\frac{\cQ^2}{\muF^2},\cdots\!\right)
=-
\mbox{\boldmath $T$}\!\left(\frac{\xi+y}{2\xi},v\Big|\alpha_s,\frac{\cQ^2}{\muF^2},\cdots\!\right)\! \stackrel{y}{\otimes}\!
\mbox{\boldmath $V$} \!\left(\!\frac{\xi+y}{2 \xi}, \frac{\xi+x}{2 \xi};\xi \Big| \alpha_s\!\right),
\end{eqnarray}
\end{subequations}
where the evolution kernels $V$ and $\mbox{\boldmath $V$}$ are introduced in the App.~\ref{app:def-evo}.  Analogous equations hold for
the non-singlet hard scattering amplitude.

We add that the factorization scale dependencies are exploited to resum $\ln(\cQ^2/\Lambda^2_{\rm QCD})/\ln(\cQ_0^2/\Lambda^2_{\rm QCD})$ contributions
by means of the evolution equations, where one usually equates all scales with $\cQ^2$.  Note that in the general case
the evolution kernels are expanded w.r.t.~$\alpha_s(\muR)$ and their renormalization scale independency implies then that they also logarithmically
depend on the ratio of renormalization and factorization scales.

Depending on the mathematical representation one is using,
one may prefer one or the other method/philosophy to resum renormalization and/or factorization logarithms.
Results, which are obtained in one or the other way, will
formally differ by contributions that are beyond the order one takes into account. Various proposals, e.g.,  called `optimal' scale setting prescriptions and scheme independent evolution,  have been suggested to minimize the uncertainties due to the unknown higher radiative order  (or even power) corrections. Let us stress that the absorption of large radiative corrections may induce very low scales, i.e.,  one goes beyond the perturbative framework and, hence, additional assumptions and/or modeling is needed, e.g., by means of analytic perturbation theory.

\begin{itemize}
\item {\em NLO contributions.}
\end{itemize}
\vspace{-2mm}
To apply consistently the perturbative framework for DVMP at NLO accuracy in $\alpha_s$,  one
needs the one-loop corrections to the hard scattering amplitudes, entering in the partonic TFFs (\ref{eq:tffF^A},\ref{eqs:tffF^S}), and the two-loop corrections to the evolution effects.
Hence,  both the hard scattering amplitudes ($T$) and the evolution kernels ($V$) are expanded up  to $\alpha_s^2$ accuracy, where we use as
expansion parameter $\alpha_s/2\pi$,
\begin{eqnarray}
T\!\left(\!\cdots\Big|\alpha_s(\muR),
\frac{\cQ^2}{\muF^2},\frac{\cQ^2}{\muphi^2},\frac{\cQ^2}{\muR^2}
\right)
&\!\!\!=\!\!\! &
\alpha_s(\muR) \: T^{(0)}(\cdots)
+ \frac{\alpha^2_s(\muR)}{2\pi}\:
T^{(1)}\!\left(\!\cdots\Big|\frac{\cQ^2}{\muF^2},\frac{\cQ^2}{\muphi^2},\frac{\cQ^2}{\muR^2}\!\right) + O(\alpha_s^3)\,,
\nonumber \\ & &
\label{eq:T-pQCD} \\[0.3cm]
V(\cdots|\alpha_s(\mu)) &\!\!\!=\!\!\! &
\frac{\alpha_s(\mu)}{2\pi}\, V^{(0)}(\cdots) + \frac{\alpha_s^2(\mu)}{(2\pi)^2}\, V^{(1)}(\cdots)  + O(\alpha_s^3)\,.
\label{eq:V-pQCD}
\end{eqnarray}
The NLO corrections to both of these quantities are available from the literature.
To be sure, that no confusion is
left w.r.t.~the underlying conventions, we derive now the explicit renormalization and factorization scale dependencies of the
NLO hard scattering amplitude from the requirement that the TFFs are scale independent in the considered order.
Let us first shortly recall the form of the LO expressions, needed for the evaluation of (\ref{dT}), where we obviously can work without loss of
generality with the $\xi=1$ case, i.e., $u=(1+x)/2$.

The  LO expressions of the hard scattering amplitudes
${^\Sigma}T= \qT/n_f + {^\pS}T$ and ${^\g}T$
can be cast in the form
\begin{eqnarray}
\label{eq:T^{(0)}}
 \qT^{(0)}(u,v) = {^\g}T^{(0)}(u,v) = \frac{1}{\bu\bv}\quad\mbox{and} \quad {^\pS} T^{(0)}(u,v) =0\,.
\end{eqnarray}
The LO term of the flavor non-singlet evolution  kernel (\ref{V}) is well known and is written  as%
\footnote{Note that the terms with $\delta$-function are understood
in the following way $\delta(x - y)\left[ \theta (y - x) + \theta (x - y)
\right] = \delta(x - y)$ and the $+$-prescription as $\frac{u}{v} \frac{1}{(v-u)_+}  \tau(v) = \frac{1}{v-u} \left[\frac{u}{v}\, \tau(v)- \tau(u)\right]$.}
\begin{eqnarray}
\label{V^{(0)}}
 V^{(0)}(u, v)= \CF\, \theta(v - u ) \frac{u}{v}\left[ 1 + \frac{1}{(v-u)_+} + \frac{3}{2} \delta(u-v)\right]+
 \left\{ {u \to \bu \atop  v \to \bv } \right\}\,.
\end{eqnarray}
The matrix valued LO expression of the flavor singlet kernel \req{bV}, taken with $\eta=1$, reads
\begin{subequations}
\label{Vb^{(0)}}
\begin{eqnarray}
\label{bV^{(0)}-matrix}
\mbox{\boldmath $V$}^{(0)}(u, v;1)
&=&
\left(\!\!\!
\begin{array}{cc}
\phantom{2\,} {^{\Sigma\Sigma}}V^{(0)}   &   {^{\Sigma\g}}V^{(0)}/2 \\
        2\, {^{\g\Sigma}}V^{(0)}        &   {^{\g\g} V}^{(0)}\phantom{/2}
\end{array}
\!\!\!\right)\!\!
(u, v)\,,
\\
{^\text{AB} V}^{(0)}(u, v)
&=&
\theta(v - u )
\; {^\text{AB}}v^{(0)} (u,v) \pm
\left\{ {u \to \bar u \atop  v \to \bar v }\right\}
\mbox{\ for\ }
\left\{ {\text{A}=\text{B} \atop \text{A} \not= \text{B}}\right. ,
\nonumber
\end{eqnarray}
where the quark-quark entry is given by the non-singlet kernel (\ref{V^{(0)}})
since, as in the hard scattering amplitude, the pure singlet (pS) addenda is zero at LO.
We take the remaining three  entries from Ref.~\cite{Belitsky:1998gc},
\begin{eqnarray}
\label{V-QG}
{^{\Sigma\g}}v^{(0)} (u,v) &\!\!\! =\!\!\! &   n_f\, \frac{u}{v^2 \bv} (2 u - v - 1)\,,
\\
\label{V-GQ}
{^{\g\Sigma}}v^{(0)}(u,v)  &\!\!\! =\!\!\! & \CF\, \frac{u}{v} (2 v - u) \,,
\\
\label{V-GG}
{^{\g\g}}v^{(0)}(u,v)
&\!\!\! =\!\!\! & \CA\, \frac{u^2}{v^2} \left\{ \frac{1}{(v-u)_+} + 2\left[\bu + v ( 1 + 2 \bu )\right]\right\}  - \frac{\beta_0}{2} \delta(u-v)\,. \quad
\end{eqnarray}
Further details on evolution equations and kernels are summarized in  App.~\ref{app:def-evo}.
\end{subequations}

The scale dependencies in the NLO expression for the hard scattering amplitude of the quark TFF (\ref{eq:tffF^A}) follows from the NLO expansion of
(\ref{T-RGE}) and (\ref{dT}) [replace there $\mbox{\boldmath $T$}\to T$ and $\mbox{\boldmath $V$}\to V$],
\begin{subequations}
\label{qT^(1)}
\begin{eqnarray}
\label{T^(1)}
\qT^{(1)}(u,v|\cdots
) =
\left[\ln\frac{\cQ^2}{\muF^2}\, T^{(0)}\stackrel{u'}{\otimes}  V^{(0)}
+  \ln\frac{\cQ^2}{\muphi^2}\, T^{(0)}\stackrel{v'}{\otimes}  V^{(0)} +
\frac{\beta_0}{2}  \ln\frac{\cQ^2}{\muR^2} T^{(0)}+ \ldots \right]\!(u,v)\,.
\end{eqnarray}
The convolution of the LO evolution kernel with the LO hard scattering amplitudes yields
\begin{eqnarray}
\label{T^q-TcoV}
\left[ T^{(0)}\stackrel{u'}{\otimes}  V^{(0)} \right](u,v)=
\frac{\CF}{\bu \bv}  \left(\frac{3}{2} +  \ln \bu \!\right),
\quad
\left[T^{(0)}\stackrel{v'}{\otimes} V^{(0)}\right](u,v)=
\frac{\CF}{\bu \bv} \left(\frac{3}{2} +  \ln \bv \!\right),
\label{eq:cTqmuF}
\end{eqnarray}
\end{subequations}
which is known to be  consistent with diagrammatical findings.
Analogously, the scale dependencies of the NLO corrections
in the flavor singlet channel \req{eq:F^S_V-a} read in matrix notation
\begin{eqnarray}
\label{eq:sing-mudep}
\mbox{\boldmath $T$}^{(1)}\!\left(\! u,v\Big|\!\frac{\cQ^2}{\muF^2},\frac{\cQ^2}{\muphi^2},\frac{\cQ^2}{\muR^2}\!\right)  =
\left[\!
\ln\frac{\cQ^2}{\muF^2}
\mbox{\boldmath $T$}^{(0)}\stackrel{u'}{\otimes} \mbox{\boldmath $V$}^{(0)}
+
\ln\frac{\cQ^2}{\muphi^2}
\mbox{\boldmath $T$}^{(0)}\stackrel{v'}{\otimes} V^{(0)}
+ \frac{\beta_0}{2} \ln\frac{\cQ^2}{\muR^2}\mbox{\boldmath $T$}^{(0)} + \cdots\!\right]\!(u,v)\,,
\nonumber\\
\end{eqnarray}
where the $\mbox{\boldmath $T$}^{(i)}$ are row vectors
\req{eq:T^S} with the quark entry \req{eq:T^Sigma}.  Since (\ref{qT^(1)}) can be taken for granted,
in the quark entry a constraint appears only for the pure singlet quark  part and, of course, we have constraints
for the  gluon entry. To shorten the explicit expressions, we take in the following advantage of the
symmetry properties (\ref{eq:SigmaT-sym},\ref{eq:GT-sym}).

The  pure singlet quark contribution appears at NLO due to gluon-quark mixing,
which induces a $\muF$ factorization scale dependence
\begin{subequations}
\label{T^PS}
\begin{eqnarray}
{^\pS}T^{(1)}\!\left(\! u,v\Big|\frac{\cQ^2}{\muF^2}\!\right) &\!\!\!=\!\!\! &
 \frac{2}{C_F} \, \ln\frac{\cQ^2}{\muF^2}  \left[T^{(0)}
\stackrel{u'}{\otimes}
{^{{\rm G}\Sigma} V}^{(0)}
\right](u,v)
+ \ldots \,.
\label{eq:logSigma}
\end{eqnarray}
The  factor $2/C_F$ follows from
\req{eq:sing-mudep}
and the definitions
of $\mbox{\boldmath $V$}$ in \req{Vb^{(0)}}
and $\mbox{\boldmath $T$}$ in \req{eqs:T^S-all}.
The convolution of the gluon-quark entry (\ref{V-GQ})
with the gluonic hard scattering amplitude (\ref{eq:T^{(0)}}) then gives
\begin{eqnarray}
\frac{2}{C_F} \left[T^{(0)} \stackrel{u'}{\otimes} {^{{\rm G}\Sigma} V}^{(0)} \right]\left(u,v\right) =
2 \frac{\bu-u}{u \bv} \ln \bu
+[\ldots]
\, .
\label{eq:cTqIamuF}
\end{eqnarray}
\end{subequations}
The additional terms
that, due to symmetry properties, cancel in the expression
for the full scattering amplitude, are denoted by $[\ldots]$.
In our representation they vanish due to the antisymmetric properties
of $F^{q(+)}$.

The NLO corrections to the hard scattering amplitude of the gluonic entry in \req{eq:T^S} read
\begin{subequations}
\label{^GT^(1)}
\begin{eqnarray}
\label{^GcalT^(1)}
{^{\rm G} T}^{(1)}\!\left(\! u,v\Big|\!\frac{\cQ^2}{\muF^2},\frac{\cQ^2}{\muphi^2},\frac{\cQ^2}{\muR^2}\!\right) &\!\!\!=\!\!\!&
\ln\frac{\cQ^2}{\muF^2}
\left[ T^{(0)} \stackrel{u'}{\otimes} {^{\rm GG} V}^{(0)}
+ \frac{C_F}{2n_f}\, T^{(0)} \stackrel{u'}{\otimes}
  {^{\Sigma {\rm G}} V}^{(0)} \right](u,v)
\\
&&\!\!\!\!\!
+ \ln\frac{\cQ^2}{\muphi^2}
   \left[T^{(0)} \stackrel{v'}{\otimes} V^{(0)} \right](u,v)
+\frac{\beta_0}{2}
   \ln\frac{\cQ^2}{\muR^2} T^{(0)}(u,v) + \ldots \,.
\nonumber
\end{eqnarray}
Note that, as in the case of the pure singlet  quark  contribution,
the factor $C_F/(2 n_f)$ can be derived from
(\ref{eq:sing-mudep})
as a consequence of our definitions
\req{eqs:T^S-all}, \req{Vb^{(0)}}, and \req{eq:T^{(0)}}.
The convolution with the LO evolution kernels then yields
\req{T^q-TcoV} and
\begin{eqnarray}
\label{T^G-pQCD_TcoV}
\left[{T}^{(0)} \stackrel{u'}{\otimes} {^{\rm GG} V}^{(0)}
+
\frac{C_F}{2n_f}\,  T^{(0)} \stackrel{u'}{\otimes}
 {^{\Sigma {\rm G}} V}^{(0)}\right](u,v)&&
\\
&&\hspace{-3cm}=\frac{\CA}{\bu \bv}
\left(1+\frac{\bu^2}{u^2}\right)\ln \bu -
\frac{\beta_0}{2}  T^{(0)}(u,v)
-\frac{\CF}{2 \bv} \frac{\ln \bu}{u^2}
+[\ldots]
\, ,
\nonumber
\end{eqnarray}
\end{subequations}
where again $[\ldots]$ denotes  terms that vanish due to symmetry.
We note that the combination of terms  proportional to $\beta_0$
in \req{^GcalT^(1)}
results in a $\cQ^2$ independent $(\beta_0/2) \ln(\muF^2/\muR^2)$ term.

Let us stress that the $\ln(\cQ^2/\muF^2)$ terms
from  Eqs.~\req{T^(1)}, \req{eq:logSigma}, and \req{^GcalT^(1)}
enable us to check the relative normalization between the
separate contributions.
The corresponding forms are determined by demanding cancelation of
collinear singularities or, in other words, factorization scale
independence of the full expression for the TFFs
(where the use of evolution equations is made).

\subsection{Dispersion relations in the collinear framework}
\label{sec:prel-DR}

Instead of calculating the TFFs from the convolution formulae (\ref{eq:tffF^A},\ref{eqs:tffF^S}) one might equivalently use `dispersion relations' (DRs),  where the standard variable, i.e., the energy variable
$$\nu+i\epsilon \propto (s-u)+i\epsilon \propto  1/(\xi-i\epsilon) \quad\mbox{($\xi$ is here  meant as a physical variable)} $$
is replaced by $\xi-i\epsilon$. The quotation marks are meant to stress that the physical fixed $t$ dispersion relation is taken here
in leading power approximation, which also changes the integration region in the DR integral, see, e.g., the discussion for the DVCS case in Sec.~2.2 of \cite{Kumericki:2007sa}.
That such a DR is equivalent to the convolution formulae
has been shown in \cite{Teryaev:2005uj,Kumericki:2007sa,Diehl:2007jb}. Here, the polynomiality conditions of GPDs, implemented in the spectral or double distribution representation, are needed to establish the one-to-one correspondences.  By means of the DR we can evaluate the real part of a TFF from its imaginary part. This  has some advantages, e.g., one essentially
needs only to discuss the scale setting for the imaginary part%
\footnote{Although, as we will see below, this is not enough in the presence of a subtraction constant.}
\cite{Brodsky:2005vt} or one may drastically simplify the numerical treatment  in momentum
fraction representation. This DR framework is introduced in the next section. Further discussion on this subject can be found
in \cite{Kumericki:2008di}.  In a second section we discuss the dispersion relations for the hard scattering amplitudes and define their perturbative
expansion.

\subsubsection{Evaluation of TFFs from GPDs by means of dispersion relations}
\label{sec:prel-DR-TFF}
For TFFs with definite signature $\sigma$
we can utilize symmetrized  DR and restrict ourselves
to DR integrals over the positive region $x>0$. As in (\ref{eq:TFsym}), for $\sigma=+1$ ($\sigma=-1$) the real part of such TFFs
is a (anti)symmetric under $\xi\to -\xi$. It can be evaluated by means of the principal value integral
\begin{subequations}
\label{DR}
\begin{eqnarray}
\label{DR-Re}
\re\, \tffF(\xB,t,\cQ^2)
&\!\!\! \stackrel{\rm Tw-2}{=}\!\!\! &
{\cal P}\!\!\int_0^1\!\!\frac{dx}{\xi^2-x^2} \left\{ 2x\atop2\xi \right\}
\frac{1}{\pi} \im\, \tffF\!\left(\!\frac{2x}{1+x},t,\cQ^2\!\right)+\,{\cal C}_{\tffF}(t,\cQ^2)
\;\;\mbox{for}\;\; \sigma(\tffF) = \left\{+1 \atop -1\right.,
\nonumber\\
\end{eqnarray}
where again $\xB=2\xi/(1+\xi)$ and  the upper and lower expression applies for signature even and odd TFFs,
respectively. The signature of a TFF ${^\text{A}}\tffF$ is the same as for the associated GPD and it is
explicitly specified in
(\ref{eq:sigma4F}) (replace $F\to \tffF$).
Based on the common Regge arguments one may expect that for flavor non-singlet and charge odd TFFs, i.e., $A \in \{\NSplmi, \qmi\}$ ,
unsubtracted DR can be used,
while subtraction constants ${\cal C}_{\tffF}$ might be needed for the flavor singlet TFFs  $\tffH^{\text{S}}$ and $\tffE^{\text{S}}$, which
include both  quark  and gluon  contributions, see (\ref{eqs:tffF^S}).  This constant can be evaluated in various manners \cite{Kumericki:2008di},
one may simply take the unphysical limit $\xi\to\infty$ in (\ref{DR-Re}),
\begin{eqnarray}
\label{DR-C}
{\cal C}_{\tffF}(t,\cQ^2) \stackrel{\rm Tw-2}{=} \lim_{\xi\to\infty} \re\, \tffF\!\!\left(\!\frac{2\xi}{1+\xi},t,\cQ^2\!\right)\,.
\end{eqnarray}
\end{subequations}
Note that for the signature odd case a unsubtracted DR holds true \cite{Kumericki:2007sa}. However, an oversubtraction can be performed, which yields a new DR  with a subtraction constant \cite{Bechler:2009me},
\begin{subequations}
\label{DR^-}
\begin{eqnarray}
\label{DR^-_1}
\re\, \tffF^{-}\!(\xB,t,\cQ^2)
\stackrel{\rm Tw-2}{=}
{\cal P}\!\int_{0}^1 dx \frac{2x}{\xi^2-x^2} \; \frac{x}{\xi} \frac{1}{\pi}
\im\, \tffF^-\!\!\left(\!\frac{2x}{1+x},t,\cQ^2\!\right)
+ \frac{1}{\xi}{\cal C}_{\tffF^-}(t,\cQ^2)\,.
\end{eqnarray}
This subtraction constant can be again calculated from the unphysical limit $\xi\to \infty$,  which provides the constant in terms of the imaginary part,
cf.~(\ref{DR-Re}),
\begin{eqnarray}
\label{DR-C^-}
{\cal C}_{\tffF^-}(t,\cQ^2)
\stackrel{\rm Tw-2}{=}
\lim_{\xi\to\infty} \re\, \xi\,\tffF^-\!\!\left(\!\frac{2\xi}{1+\xi},t,\cQ^2\!\right)
\stackrel{\rm Tw-2}{=}
2 \int_0^1\!dx\,  \frac{1}{\pi}\im\, \tffF^-\!\!\left(\!\frac{2x}{1+x},t,\cQ^2\!\right).
\end{eqnarray}
\end{subequations}

\pagebreak[2]
\begin{itemize}
\item{\em Convolution integrals for the imaginary parts in the flavor non-singlet channel.}
\end{itemize}
\vspace{-2mm}
Since GPDs and DAs are real valued functions, the imaginary parts of  TFFs entirely
arise in the convolution formulae (\ref{eq:tffF^A},\ref{eqs:tffF^S}) from the hard scattering amplitude,
e.g.,  we find from (\ref{eq:tffF^A}),
\begin{eqnarray}
 \im \tffF_{\rm M}^{\text{A}}(\xB,t,\cQ^2)
 \stackrel{\rm Tw-2}{=}
  \frac{C_F f_{\rm M}}{N_c\cQ} F^{\text{A}}(x,\xi,t,\muF^2) \stackrel{x}{\otimes}
\im \, \qT\!\left(\!\frac{\xi+x-i\epsilon}{2(\xi-i\epsilon)},v\Big|\!\cdots\!\right)\!\stackrel{v}{\otimes}\varphi_{\rm M}(v,\muphi^2)\,.
\label{eq:imtffFq}
\end{eqnarray}
{F}rom the analytic properties of the hard scattering amplitudes,
which are real valued for
$$0\le u= \frac{\xi+x}{2\xi} \le 1,$$
it follows that only the outer GPD regions $ \xi\le x \le 1$
and  $ -1 \le x \le -\xi$ contribute in this convolution integral \req{eq:imtffFq}.
Note that in the partonic interpretation the GPD  can be viewed for $\xi\le x \le 1$  ($-1 \le x \le -\xi$) as probability amplitude that an
$s$-channel exchange of a quark (an antiquark) occurs, which is the analog of the familiar probability interpretation for a PDF, see (\ref{eq:Fq2pdf},\ref{eq:Fg2pdf}).
Thus, as done in \req{eq:Tsignum}, it is more appropriate to decompose the convolution integral in positive $x$ and negative $x$ regions.
Furthermore, motivated by the PDF convolution formulae, well known from deep inelastic structure functions, e.g.,
$$
F_1(\xB,\cQ^2) = \sum_{p} \int_{\xB}^{1}\!\frac{dx}{x}\, C_p\!\left(\!\frac{x}{\xB},\frac{\cQ^2}{\mu^2}\!\right) p(x,\mu^2)\quad
(\mbox{sum over all partons $p \in\{u,\overline{u},\cdots,g\}$})\,,
$$
we will write the imaginary part of TFFs in this fashion, too.
However, in our GPD case we consider it as more appropriate to use on the partonic side the scaling variable $\xi$ rather than
$\xB$. In contrast to PDF convolution integrals the GPD depends then on the scaling variable $\xi$, too.

The convolution integral (\ref{eq:imtffFq}) in the non-singlet channel has then the following form%
\footnote{With the transformation $x\to \xi/x$ of the integration variable, the convolution integral can be equivalently written
in two  different forms, namely,
$\int_\xi^1\!\frac{dx}{x}\, t(\xi/x) F(x,\cdots) =\int_\xi^1\!\frac{dx}{x}\, t(x) F(\xi/x,\cdots)$.
}
\begin{subequations}
\label{eqs:DRtffFqC}
\begin{eqnarray}
\im\, \tffF^{\text{A}}_{\rm M}\!(\xB,t,\cQ^2)
&\!\!\! \stackrel{\rm Tw-2}{=} \!\!\! &
\frac{\pi\, C_F  f_{\rm M}}{N_c\cQ}
\label{eq:imtffFqC}
\\
&&\times
\int_\xi^1\! \frac{dx}{x}\,
\varphi_{\rm M}(v,\muphi^2) \stackrel{v}{\otimes}  {^\sigma}t\!\left(\!\frac{\xi}{x},v\Big|\alpha_s(\muR),\frac{\cQ^2}{\muF^2},\frac{\cQ^2}{\muphi^2},\frac{\cQ^2}{\muR^2}\right) F^\text{A}(x,\xi,t,\muF^2)\,.
\nonumber
\end{eqnarray}
The new hard scattering amplitude ${^\sigma}t$ is calculated from the
imaginary part of ${^\sigma}T$, given in (\ref{eq:Tsignum-q}), with a signature $\sigma(F^\text{A})=\pm 1$ that
can be  read off from (\ref{eq:sigma4F}).
It is a function of the ratio $r= \xi/x $, obviously, restricted%
\footnote{The continuation of ${^\sigma}t(r)$ to negative $r$  is done by reflection $r\to -r$, where
its  symmetry is governed by the signature $\sigma(F^{\text{A}})$, entirely analogous as for the
GPD $F^{\text{A}}$, see (\ref{gpdF^{p(C)}}).} to $\xi\le r \le 1$, i.e., $0\le r \le 1$.
It is convenient to decompose ${^\sigma}t$  in a signature independent and dependent part,
\begin{eqnarray}
\label{^sigmat}
 {^\sigma}t(r,v|\cdots)
&\!\!\! =\!\!\! &  t(r,v|\cdots) - \sigma(F^{\text{A}})\; \overline{t}(r,v|\cdots)\quad\quad\;\;\mbox{for}\quad 0\le r= \xi/x\le 1\,,
\\
t(\xi/x,v|\cdots) &\!\!\! =\!\!\! &
\frac{x}{2\xi\pi} \im\,
\qT\!\left(\! u=\frac{\xi-i\epsilon+x}{2(\xi-i\epsilon)},v \Big|\cdots\!\right) \quad\mbox{for}\quad
 \Re{\rm e}\, u\ge 1\,,
\nonumber\\
\overline{t}(\xi/x,v|\cdots)
&\!\!\! =\!\!\! &
\frac{x}{2\xi\pi} \im\,
\qT\!\left(\! \bu=\frac{\xi-i\epsilon-x}{2(\xi-i\epsilon)},v \Big|\cdots\!\right)
\quad\mbox{for}\quad \Re{\rm e}\, \bu\le 0\,,
\nonumber
\end{eqnarray}
where the condition $\Re{\rm e}\, u\ge 1$  ($\Re{\rm e}\, \bu\le 0$)  ensure that only r.h.s.~(l.h.s.) discontinuities of $T(u,v)$ are picked up.
For instance, for $1/\bu$, appearing in the LO expressions (\ref{eq:T^{(0)}}), we find $\delta(1-r)$.
Note that the $\overline{t}$-contribution stems from a quark-antiquark mixing as it appears, e.g., in crossed ladder diagrams. Hence, it vanishes at LO.

\begin{itemize}
\item{\em Subtraction constant in the flavor non-singlet channel.}
\end{itemize}
\vspace{-2mm}
As argued above from analyticity and Regge arguments,  the real part of ${^\text{A}}\tffF$ for $A\in\{\NSplmi,\qmi\}$ can be calculated
from an unsubtracted DR (\ref{DR})
with signature $\sigma(F^\text{A})$, i.e., ${\cal C}_{\tffF^\text{A}} =0$. On the other hand, if one derives the DR (\ref{DR}) from the convolution formulae (\ref{eq:tffF^A}),
a subtraction constant for  $\tffH^\NSpl$ but not for the combination
$\tffH^\NSpl+\tffE^\NSpl$ is allowed. This subtraction constant, called  ${\cal D}^{\text{A}}={\cal C}_{\tffH^\text{A}} = -{\cal C}_{\tffE^{\text{A}}}$, can be calculated from the  convolution formula (\ref{eq:tffF^A}) by means of the limit (\ref{DR-C}). This procedure  yields
\begin{equation}
{\cal D}^\NSpl_{\rm M}(t,\cQ^2) \stackrel{\rm Tw-2}{=} \frac{C_F  f_{\rm M}}{N_c\cQ}\; \varphi_{\rm M}(v,\muphi^2)
\stackrel{v}{\otimes} T\left(u,v|\cdots\right) \stackrel{u}{\otimes} d^{\text{NS}}(u-\bu,t,\muF^2)\,,
\label{eq:tffD}
\end{equation}
where the function $d^{\text{NS}}$ is given by the following limit of the  GPD $H^{\NSpl}$ (or $-E^{\NSpl}$)
\begin{equation}
d^\NS(x,t,\muF^2) =   \lim_{\xi\to\infty} H^{\NSpl}(x\xi,\xi,t,\muF^2)\,.
\label{eq:gpdD}
\end{equation}
\end{subequations}
This function is antisymmetric in $x$ and vanishes for $|x|>1$. Essentially, it is the so-called $D$-term,
introduced in \cite{Polyakov:1999gs} to complete polynomiality%
\footnote{The term ${\rm sign}(\eta)\theta(|x| \le |\eta|)\,  d^q(x/\eta,\cdots)$ gives for odd $x$-moments of $H^\qpl$ the highest possible order in $\eta$,
$$
\int_{-1}^1\! dx\, x^{2n+1}\, {\rm sign}(\eta)\theta(|x| \le |\eta|)\,  d^q(x/\eta,\cdots) =  \eta^{2n+2} \int_{-1}^1\! dx\, x^{2n+1} d^q(x,\cdots) \quad\mbox{for}\quad n \in \{0,1,2,\cdots\}\,.
$$
In the gluonic sector $|\eta|\theta(|x| \le |\eta|) d^{\rm G}(x/\eta)$ completes polynomiality for even $x$-moments ($q\to {\rm G}, x^{2n+1}\to x^{2n}$).
\label{foo:Dterm}}, e.g., in the popular Radyushkin's double distribution ansatz for GPD $H^\qpl$ (or $E^\qpl$).
Note that alternative GPD representations in terms of double distributions exist, see \cite{Belitsky:2000vk,Teryaev:2001qm,Tiburzi:2004mh,Hwang:2007tb}, and that the limit (\ref{eq:gpdD}) projects onto the $D$-term, used in  the popular double distribution representation.
Rather analogously, one can view the subtraction constant of
$\tfftE^{3^{(+)}}_{\pi^+}$  in the oversubtracted DR (\ref{DR^-}) as a pion pole contribution. On GPD level one finds then the parametrization
which was suggested in \cite{Mankiewicz:1998kg,Frankfurt:1999fp}, further details and an alternative GPD representation of the pion pole contribution are given in \cite{Bechler:2009me}.

\begin{itemize}
\item{\em Convolution integrals for the imaginary parts in the flavor singlet channel.}
\end{itemize}
\vspace{-2mm}
In analogy to (\ref{eqs:DRtffFqC}), we evaluate now the flavor singlet TFF (\ref{eq:rhoTFFsing}) for DV$\!V_L^0$P, which possess signature $\sigma=+1$. Its imaginary part is taken from the convolution (\ref{eqs:tffF^S}),
\begin{subequations}
\label{eqs:DRF^S_V}
\begin{eqnarray}
\label{eq:imtffF^S_V}
\im\,\tffF_{{\rm  V}^0}^{\SigG}(\xB,t,\cQ^2)
&\!\!\! \stackrel{\rm Tw-2}{=} \!\!\! &
\frac{\pi C_F  f_{{\rm  V}^0}}{N_c\cQ}
\\
&&\times \int_\xi^1\! \frac{dx}{x}\,
   \varphi_{{\rm  V}^0}(v,\muphi^2)\stackrel{v}{\otimes} \mbox{\boldmath $t$}\!\left(\!\frac{\xi}{x},v;\xi\Big|\alpha_s(\muR),
\frac{\cQ^2}{\muF^2},\frac{\cQ^2}{\muphi^2},\frac{\cQ^2}{\muR^2}\right)\cdot \mbox{\boldmath $F$}(x,\xi,t,\muF^2)\,,
\nonumber
\end{eqnarray}
where the entries of the new vector valued  hard scattering amplitude,
\begin{eqnarray}
\label{Im^St}
\mbox{\boldmath $t$}(r,v;\xi|\cdots) &\!\!\! =\!\!\! &
\left(  {^\Sigma}t (r,v|\cdots), \CF^{-1}\,\xi^{-1}\,{^{\rm G}t }(r,v|\cdots) \right)\quad\mbox{for}\quad 0\le r= \xi/x\le 1\,,
\phantom{\Big|}
\\
\label{Im^SigmaT}
{^\Sigma}t (r,v|\cdots) &\!\!\! =\!\!\! &
\frac{1}{n_f}\; {^+}t(r,v|\cdots) + \,  {^\pS}t (r,v|\cdots)\;\;\;
\mbox{with}\;\;\;
{^+}t(r,v|\cdots) =t(r,v|\cdots) - \overline{t}(r,v|\cdots)\,,
\nonumber
\end{eqnarray}
follow from (\ref{eq:Tsignum-Sigma}) and (\ref{eq:Tsignum-G}):
\begin{eqnarray}
\label{Im^PST}
{^\pS}t(\xi/x,v|\cdots) &\!\!\! =\!\!\! &
\frac{x}{2\xi\pi} \im\left[
{^\pS}T\!\left(u=\frac{\xi-i\epsilon+x}{2(\xi-i\epsilon)},v\Big|\!\cdots\!\right)-
{^\pS}T\!\left(\bu=\frac{\xi-i\epsilon-x}{2(\xi-i\epsilon)},v\Big|\!\cdots\!\right)
\right],
\\
\label{Im^GT}
{^{\rm G} t}(\xi/x,v|\cdots) &\!\!\! =\!\!\! &
\frac{x}{2\xi\pi} \im\left[
{^{\rm G} T}\!\left(u=\frac{\xi-i\epsilon+x}{2(\xi-i\epsilon)},v\Big|\!\cdots\!\right)+
{^{\rm G} T}\!\left(\bu=\frac{\xi-i\epsilon-x}{2(\xi-i\epsilon)},v\Big|\!\cdots\!\right)
\right]\qquad\qquad
\end{eqnarray}
with the conditions $\Re{\rm e}\, u\ge 1$ and $\Re{\rm e}\, \bu\le 0$. Note that the second term in the square brackets on the r.h.s.~of (\ref{Im^PST}) and (\ref{Im^GT}) ensures that we can relax on the representation of the hard scattering amplitude {\boldmath \mbox{$T$}},
see (\ref{eq:SigmaT-sym}, \ref{eq:GT-sym}) and discussion below there%
\footnote{A fully (anti)symmetrized hard scattering amplitude  provides the same result as its minimal version that only contains a $[1,\infty]$ discontinuity on the real $u$-axis. For instance, in the convolution with a gluon GPD $H^\g$ both of
the expressions $(\bu +u)/2$ and $1/\bu$ can be taken, where in both cases (\ref{Im^GT}) provides $\delta(1-r)$.
\label{foot:t-example}}.

\begin{itemize}
\item{\em Subtraction constant in the flavor singlet channel.}
\end{itemize}
\vspace{-2mm}
The real part of the TFF $\tffF_{{\rm  V}^0}^{\SigG}(\xB,t,\cQ^2)$ is then calculated from the DR (\ref{DR}) with signature $\sigma=+1$.
The subtraction constant in terms of GPDs is analogously calculated as in (\ref{eq:tffD}) and reads
\begin{equation}
{\cal D}_{{\rm  V}^0}^{\SigG}(t,\cQ^2)
\stackrel{\rm Tw-2}{=}
 \frac{C_F  f_{\rm M}}{N_c\cQ}\;
\varphi_{{\rm  V}^0}(v,\muphi^2)\stackrel{v}{\otimes}\mbox{\boldmath $ T$}\!\left(\!u,v;1\Big|\alpha_s(\muR),
\frac{\cQ^2}{\muF^2},\frac{\cQ^2}{\muphi^2},\frac{\cQ^2}{\muR^2}
\right)\! \stackrel{u}{\otimes}  \mbox{\boldmath $d$}(u-\bu,t,\muF^2) \,,
\label{eq:tffD^S}
\end{equation}
where the limit of the vector valued GPD $\mbox{\boldmath $H$}$ (or $-\mbox{\boldmath $E$}$) yields
\begin{equation}
\mbox{\boldmath $d$}(x,t,\muF^2) \equiv \left( { d^\Sigma(x,t,\muF^2)\atop d^{\rm G}(x,t,\muF^2)} \right)=  \lim_{\xi\to\infty} \left( {H^\Sigma(x\xi,\xi,t,\muF^2) \atop \frac{1}{\xi} H^{\rm G}(x\xi,\xi,t,\muF^2)}\right).
\label{eq:gpdD^S}
\end{equation}
\end{subequations}
Note that the gluonic entry is symmetric in $x$ and as the antisymmetric quark entry it vanishes for $|x|>1$, see also footnote~\ref{foo:Dterm}.

\subsubsection{Properties and conventions of hard scattering amplitudes}
As advocated in Sec.~\ref{sec:prel-MF-definitions} and as the reader has maybe already realized, we can now represent the hard scattering amplitudes with
definite signature in such a manner that they possess only discontinuities on the positive real $u$-axis. Thus, their imaginary
parts on the $[1,\infty]$-cut are given for real $v$, restricted to $0\le v \le 1$,
in (\ref{eqs:DRtffFqC},\ref{eqs:DRF^S_V}). In standard manner we employ Cauchy theorem to
derive an unsubtracted single variable DR that provides the hard scattering amplitudes in the complex $u$-plane.
Adopting the notation of (\ref{^sigmat},\ref{Im^PST},\ref{Im^GT}) and  Feynman's causality prescription, the desired DR reads
for the hard scattering amplitudes  of interest as
\begin{eqnarray}
{^\text{A}}T(u,v|\cdots)=
\int_0^1\!dr\,  \frac{2\,{^\text{A}}t(r,v|\cdots)}{1+r- 2u r  -i\epsilon} \quad \mbox{for} \quad \text{A} \in \{+,-,\pS,\g\}\,.
\label{^AT(u,v)-DR-r}
\end{eqnarray}
We did not seek for a proof that a subtraction is not needed in this DR  to all orders of perturbation theory. However, it can be verified from the explicit expressions that the {\it unsubtracted} DR (\ref{^AT(u,v)-DR-r}) holds to  NLO accuracy.

Let us quote the general structure of the perturbative expansion of the
new hard coefficients,
\begin{equation}
{^\text{A}}t(r,v,|\cdots) = \alpha_s(\muR)\, {^\text{A}}t^{(0)}(r,v) + \frac{\alpha^2_s(\muR)}{2\pi}\, {^\text{A}}t^{(1)}(r,v|\cdots) + O(\alpha_s^3),
\label{eq:tT-pQCD}
\end{equation}
which is inherited from those of  hard scattering amplitudes $T$,
given in \req{eq:T-pQCD}.
The imaginary parts of the LO coefficients (\ref{eq:T^{(0)}}) are trivially calculated  by means of
(\ref{^sigmat},\ref{Im^PST},\ref{Im^GT}),
\begin{eqnarray}
\label{t^{(0)}}
{^\pm}t^{(0)}(r,v) = {^\g}t^{(0)}(r,v) = \frac{\delta(1-r)}{\bv} \quad \mbox{and} \quad
 {^\pS}t^{(0)}(r,v) =0\,,
\end{eqnarray}
where the $1/\bu$ pole in (\ref{eq:T^{(0)}}) yields $\delta(1-r)$  for quarks with even and odd signature as well as for gluons.

Consequently, formulae (\ref{eq:imtffFqC},\ref{eq:imtffF^S_V}) provide the imaginary parts of quark and gluon TFFs in
agreement with the LO approximations (\ref{eqs:tff-LO}),
\begin{eqnarray}
\im\, \tffF^{q^{(\pm)}}_{\rm M}\!(\xB,t,\cQ^2) &\!\!\! \stackrel{\rm LO}{=} \!\!\! &
\frac{\pi\, C_F  f_{\rm M}\, \alpha_s(\muR)}{N_c\cQ} F^{q^{(\pm)}}(\xi,\xi,t,\muF^2) \int_0^1\!dv\,\frac{\varphi_{\rm M}(v,\muphi^2)}{\bv}\,,
\label{eq:imtffFqC-LO}
\\
\im\, \tffF_{{\rm  V}^0}^{\rm G}(\xB,t,\cQ^2)
&\!\!\! \stackrel{\rm LO}{=} \!\!\! &
\frac{\pi\, f_{{\rm  V}^0}\, \alpha_s(\muR)}{N_c \cQ}\;  \frac{1}{\xi}F^{\text{G}}(\xi,\xi,t,\muF^2)
\int_0^1\!dv\,\frac{\varphi_{{\rm  V}^0}(v,\muphi^2)}{\bv}\,.
\end{eqnarray}
The corresponding real parts are evaluated from DR (\ref{DR-Re}) with the signature $\sigma(F^\text{A})$.
The possible subtraction constants can be easily evaluated from (\ref{eq:tffD},\ref{eq:tffD^S}), too,
\begin{eqnarray}
{\cal D}^q_{\rm M}(t,\cQ^2) &\!\!\! \stackrel{\rm LO}{=} \!\!\! & \frac{C_F  f_{\rm M}\, \alpha_s(\muR)}{N_c\cQ}\; \int_0^1\!du\, \frac{d^q(u-\bu,t,\muF^2)}{\bu}
\int_0^1\!dv\,\frac{\varphi_{\rm M}(v,\muphi^2)}{\bv}\,,
\\
{\cal D}_{{\rm  V}^0}^{\rm G}(t,\cQ^2)
&\!\!\! \stackrel{\rm LO}{=} \!\!\! &  \frac{f_{{\rm  V}^0}\, \alpha_s(\muR)}{N_c \cQ}
\int_0^1\!du\, \frac{d^{\text{G}}(u-\bu,t,\muF^2)}{\bu}
\int_0^1\!dv\,\frac{\varphi_{{\rm  V}^0}(v,\muphi^2)}{\bv}\,,
\end{eqnarray}
where the $d$-functions follow from the limiting procedures (\ref{eq:gpdD},\ref{eq:gpdD^S}).
 As it is now well realized, up to these subtraction constants,  the TFFs at LO arise  only from
GPDs on the cross-over line (antiquarks are included in GPDs with definite charge parity).
Neglecting evolution effects, these facts drastically simplify GPD phenomenology at LO accuracy. Furthermore,
if one  likes (or has) to implement evolution in momentum fraction representation, one needs only to evolve the GPD in the outer region. This may drastically
simplify the numerical treatment of the evolution operator in the momentum fraction representation.

\subsection{Mellin-Barnes integral representation}
\label{sec:prel-MB}

Instead of the momentum fraction representation, presented above, we may employ the conformal partial wave expansion (CPWE) for DAs and GPDs.
Before we adopt this expansion to TFFs, let us shortly remind of the well-known case of meson form factors in which the GPD is
replaced by a DA. The reader may find an introduction to conformal symmetry, as it is used here, in \cite{Braun:2003rp}.

The CPWE for (normalized) flavor non-singlet DAs  reads as
\begin{eqnarray}
\label{varphi-DA_CPWE}
\varphi(u,\mu^2) =  \sum_{{k=0\atop {\rm even}}}^\infty  6 u\bu  C_k^{3/2}(u-\bu)\,  \varphi_k(\mu^2)\,, \quad \varphi_0=1\,, \quad \bu=1-u\,,
\end{eqnarray}
where  $6 u\bu\, C^{3/2}_k(u-\bu)$ is a conformal partial wave (CPW),  expressed by the
Gegenbauer polynomial $ C^{3/2}_k$ with index $3/2$
and order $k$, and where $\varphi_k(\mu^2)$ are the CPW amplitudes.  Furthermore, for symmetric DAs  we can restrict ourselves to even $k$.
Utilizing the orthogonality relation for Gegenbauer polynomials,
the amplitudes in the CPWE (\ref{varphi-DA_CPWE}) are evaluated from the DA by forming integral conformal moments ($k\ge 0$):
\begin{eqnarray}
\label{eq:varphi_k}
\varphi_k(\mu^2)= \frac{2(2k+3)}{3(k+1)(k+2)}
\int_0^1\!dv\, C^{3/2}_k(v-\bv)\, \varphi(v,\mu^2)\,.
\end{eqnarray}
Plugging the CPWE (\ref{varphi-DA_CPWE}) into the factorization formulae for some meson form factor $F(\cQ^2)$, given as
convolution of a hard scattering amplitude $T$ with two DAs, yields its CPWE
\begin{eqnarray}
F \propto \varphi(u,\mu^2) \stackrel{u}{\otimes} T(u,v|\cdots)\stackrel{v}{\otimes} \varphi(v,\mu^2)
\quad \Leftrightarrow \quad
F \propto  \varphi_n(\mu^2) \stackrel{n}{\otimes}
T_{nk}(\cQ^2,\mu^2)\stackrel{k}{\otimes} \varphi_k(\mu^2)
\,.
\label{eq:FFrep}
\end{eqnarray}
Here, we find it convenient to write the series over $n$ and $k$ symbolically, such that the transition from the momentum fraction representation to the CPWE
(or reverse) is done by the replacement
$$
A(u)\stackrel{u}{\otimes} B(u)\equiv \int_{0}^1\!du A(u)B(u)\quad \Leftrightarrow \quad A_{k}\stackrel{k}{\otimes} B_k\equiv \sum_{{k=0\atop {\rm even}}}^\infty  A_k B_k\,.
$$
The new hard coefficients $T_{nk}(\cdots)$ in the CPWE (\ref{eq:FFrep}) are evaluated from convoluting the momentum fraction ones with
the CPWs, which we write as
\begin{equation}
\label{c_{nk}}
T_{nk}(\cdots) = 3^2\, c_{nk}(\cdots) \quad \mbox{with}\quad c_{nk}(\cdots) =2 u\bu C_n^{3/2}(u-\bu) \stackrel{u}{\otimes} T(u,v|\cdots) \stackrel{v}{\otimes} 2 v\bv C_k^{3/2}(v-\bv)\,.
\end{equation}
For the LO hard scattering amplitude $T^{(0)}(u,v)$ in (\ref{eq:T^{(0)}})
we have the simple correspondence
$$T^{(0)}(u,v)=\frac{1}{\bu}\frac{1}{\bv} \quad \Leftrightarrow \quad  c_{nk}^{(0)} =1\,.$$

One advantage of the CPWE is that the evolution operator to LO accuracy is diagonal
and so the conformal moments (\ref{eq:varphi_k}) evolve autonomously,
\begin{eqnarray}
\label{E_k-LO}
\varphi_k(\mu^2) \stackrel{\rm LO}{=} E_{k}(\mu,\mu_0) \varphi_k(\mu^2_0)\,,\quad
E_{k}(\mu,\mu_0) \stackrel{\rm LO}{=}
\left( \frac{\alpha_s(\mu)}{\alpha_s(\mu_0)}\right)^{{\gamma}^{(0)}_{k}/(11-2n_f/3)}\,,
\end{eqnarray}
where the anomalous dimensions,
\begin{eqnarray}
\label{eq:gamma0}
\gamma_{k}^{{\rm }(0)}
= \CF \left( 4 S_{1}(k + 1) - 3 - \frac{2}{(k + 1 )( k + 2 )} \right)
\,,
\end{eqnarray}
are (up to a factor $-1/2$) the eigenvalues of the LO evolution kernel (\ref{V^{(0)}}), which coincide with those known from deep inelastic scattering.
To LO accuracy the evolution operator (\ref{E_k-LO}) can be directly inserted into the CPWE (\ref{eq:FFrep}) of the form factor.
This implies advantages for the numerical treatment, namely, instead of solving numerically the evolution equation (\ref{ER-BLequation})
and performing then a two dimensional momentum fraction integral, one needs only to perform two summations. Practically, models for DAs
are specified by a finite number of conformal moments, which can be also viewed as an {\em effective} parameterization of DAs.
Consequently, for such popular models the numerical evaluation of the form factor at LO gets trivial in this representation.

Beyond LO conformal moments will mix under evolution, however,
the evolution operator, given now in terms of a triangular  matrix $\{E_{mn}\}$ with $n \le m$,  can be perturbatively diagonalized \cite{Mikhailov:1984ii,Mueller:1993hg}.
Moreover, instead of evolving the DA conformal moments, as in (\ref{E_k-LO}),
from an input scale $\mu_0$ to the factorization
scale $\mu$, we can convolute the evolution operator with the hard coefficients.
Consequently, we write here the convolution formula (\ref{eq:FFrep})
in the form
\begin{eqnarray}
\label{F(Q^2)}
F(\cQ^2)\propto
\varphi_n(\cQ_0^2) \stackrel{n}{\otimes}  {\rm T}_{nk}(\cQ^2,\cQ_0^2)
\stackrel{k}{\otimes} \varphi_k(\cQ_0^2)\,,
\end{eqnarray}
where the new hard coefficients
\begin{eqnarray}
{\rm T}_{nk}(\cQ^2,\cQ_0^2)&=&
 T_{n+m,k+l}(\cQ^2,\mu^2)
  \stackrel{l,m}{\otimes}
 E_{n+m,n}(\mu,\cQ_0)\,\,
 E_{k+l,k}(\mu,\cQ_0)\,,
\label{eq:Tnk(Q,mu0)}
\quad
\stackrel{l,m}{\otimes} \equiv \sum_{l=0\atop {\rm even}}^\infty \sum_{m=0\atop {\rm even}}^\infty
\end{eqnarray}
are `evolved backwards' from $\mu^2$ to the squared scale $\mu^2_0=\cQ^2_0$, which is taken to be of a few $\GeV^2$,
justifying our perturbative treatment.  Consequently, the new hard coefficients
possess only a residual $\mu$ dependence, which is not indicated on the l.h.s.~of (\ref{eq:Tnk(Q,mu0)}).
For a truncated DA model, given at the input scale $\cQ_0$, the factorization scale independent coefficients (\ref{eq:Tnk(Q,mu0)})
 are given as a finite dimensional matrix.
The two infinite sums which remain in (\ref{eq:Tnk(Q,mu0)}) can be numerically precalculated for some given experimental $\cQ^2$ values.
Hence, the CPWE allows to have fast fitting procedures with a limited set of conformal DA moments (\ref{eq:varphi_k}) as fitting parameters.

Adopting this popular form factor treatment to  TFFs will provide a powerful tool, as it does already for the analysis of DVCS data \cite{Kumericki:2009uq}.
To do so,  GPDs and  CFFs are expanded in terms of complex  CPWs by means of  Mellin-Barnes integrals \cite{Mueller:2005ed,Kumericki:2007sa}.
An introduction to this representation, where we spell out our conventions, and its adoption to TFFs is given in the next section.
In Sec.~\ref{sec:prel-MB-analytic} we introduce an efficient method for the evaluation of complex  CPW amplitudes.

\subsubsection{Conformal partial wave expansion of GPDs and TFFs}
\label{sec:prel-MB-TFF}

For a quark GPD we can use the same CPWs as for the DA, however, for integer $n \in \{0,1,2,\cdots\}$ their support is restricted to the inner GPD region $|x| \le \eta$ and, moreover, for convenience the normalization is changed. We define these integral CPWs for a quark GPD  as in \cite{Mueller:2005ed}:
\begin{equation}
\label{eq:p_n}
p_n(x,\eta) = \eta^{-n-1} p_n\!\left(\!\frac{x}{\eta}\!\right)\,,\quad
p_n(x) = \theta(1-x^2)\,\frac{2^{n} \Gamma\big(n+\frac{5}{2}\big)} {\Gamma\big(\frac{3}{2}\big)\Gamma(n+3)} (1-x^2)  C_n^{3/2}(-x).
\end{equation}
This normalization ensures that conformal  moments of a quark GPD (\ref{gpdF^{q(pm)}}) with  definite charge parity
\begin{eqnarray}
\label{Fj} F_{n}^{q^{(\pm)}}(\eta,t,\mu^2) = \frac{
\Gamma\big(\frac{3}{2}\big)\Gamma(n+1)}{2^{n}  \Gamma\big(n+\frac{3}{2}\big)}
\;\frac{1}{2}\int_{-1}^1\! dx\; \eta^{n} \, C_n^{3/2}\!\left(\!\frac{x}{\eta}\!\right)
F^{q^{(\pm)}}(x,\eta,t,\mu^2)\,,
\end{eqnarray}
coincide for  $F=H$ in forward kinematics ($\eta=0,t=0$) with the common integral Mellin moments, taken for positive $x$,
of a unpolarized quark PDF with  definite charge parity,
$$
H_{n}^{q^{(\pm)}}(\eta=0,t=0,\mu^2) = \int_0^1\!dx\, x^n \left[q(x,\mu^2) \pm  \overline{q}(x,\mu^2)\right]\,.
$$
Thus, it is ensured that signature $\sigma=+1$ and $\sigma=-1$ GPDs \req{gpdF^{p(C)}} provide
odd and even conformal GPD moments, respectively, which are always even polynomials in $\eta$. Note also
that compared to the CPWs of a DA, entering in the CPWE (\ref{varphi-DA_CPWE}), the normalization of $(-1)^n p_n(x,\eta)$ for $\eta=1$ differs by the factor
\begin{equation}
\frac{(-1)^n\, p_n(u-\bu)}{6 u\bu\, C^{3/2}_n(u-\bu)}=
\frac{4}{6}\; \frac{2^{n} \Gamma\big(n+\frac{5}{2}\big)}{\Gamma\big(\frac{3}{2}\big)\Gamma(n+3)}
=\frac{1}{3}\frac{2^{n+1} \Gamma\big(n+\frac{5}{2}\big)}{\Gamma\big(\frac{3}{2}\big)\Gamma(n+3)},
\end{equation}
where the inclusion of the  factor $(-1)^n$ takes care on the negative argument
$-x=-(u-\bu)$ of the Gegenbauer polynomials in $p_n(x)$.

Since the support of integral CPWs is restricted to the interval $u,v\in [0,1]$,
the convolution of these CPWs with the hard quark amplitude,
\begin{equation}
p_n(x,\xi)\stackrel{x}{\otimes}\! \qT\!\left(\!\frac{\xi+x}{2\xi},v\Big|\!\cdots\!\right)\stackrel{v}{\otimes} 6 v\bv C_k^{3/2}(v-\bv) =
(-1)^n\xi^{-n-1}\, \qT_{nk}
\end{equation}
is  up to a factor $(-1)^n\,\xi^{-n-1}$ defined as
\begin{subequations}
\label{^qT_jk}
\begin{eqnarray}
\qT_{nk}(\cdots ) &\!\!\! =\!\!\! &
\frac{2^{n+1}\ \Gamma\big(n+\frac{5}{2}\big)}{\Gamma\big(\frac{3}{2}\big)\Gamma(n+3)} \times 3 \times c_{nk}(\cdots )
\,
\label{^qT_jk-1}
\end{eqnarray}
with
\begin{eqnarray}
c_{nk}(\cdots ) &\!\!\! =\!\!\! & 2 u\bu C_n^{3/2}(u-\bu) \stackrel{u}{\otimes}\! \qT(u,v|\cdots)\stackrel{v}{\otimes} 2 v\bv C_k^{3/2}(v-\bv)\,,
\label{^qc_jk}
\end{eqnarray}
\end{subequations}
and where the prefactor $2^{n+1} \Gamma(n+5/2)/\Gamma(3/2)\Gamma(n+3)$ is
associated with the Clebsch--Gordon coefficient in the CPWE of CFFs.  The factor 3 results from the normalization of the DA.
As in the form factor coefficients (\ref{c_{nk}}), these normalization factors are pulled out in the $c_{nk}$ coefficients
which in LO approximation are normalized to one.

For the vector valued GPD $\mbox{\boldmath $F$}$ in the flavor singlet sector we utilize for the CPWs the vector
\begin{eqnarray}
\mbox{\boldmath $p$}_n (x, \eta) = {\eta^{-n-1}}   \left(
{
p_n^{\Sigma}
\atop
-\eta\, p_n^{\rm G}
}
\right)\!\!\left(\!\frac{x}{\eta}\!\right)\,, \quad  \mbox{\boldmath $p$}_n (x) \equiv \mbox{\boldmath $p$}_n (x, \eta=1)\,,
\end{eqnarray}
where $p_n^\Sigma \equiv p_n$ is already defined in (\ref{eq:p_n}) and the gluonic CPWs are expressed by Gegenbauer polynomials with index $\nu=5/2$
\begin{eqnarray}
\label{eq:pG_n}
p^{\rm G}_n(x)
= \theta(1-x^2)\frac{2^{n} \Gamma\big(n+\frac{5}{2}\big)} {\Gamma\big(\frac{3}{2}\big)\Gamma(n+3)}\, \frac{3}{n+3} (1-x^2)^2\,  C_{n-1}^{5/2}(-x)\,.
\end{eqnarray}
This implies that the gluonic entries are  evaluated from
\begin{subequations}
\label{^GT_jk}
\begin{eqnarray}
{^\text{G}}T_{nk}(\cdots) &\!\!\! =\!\!\! & \frac{2^{n+2}\ \Gamma\big(n+\frac{5}{2}\big)}{\Gamma\big(\frac{3}{2}\big)\Gamma(n+4)} \times 3\times {^\text{G}}c_{nk}(\cdots)\,
\label{^GT_jk-1}
\end{eqnarray}
with
\begin{eqnarray}
{^\text{G}}c_{nk}(\cdots)  &\!\!\! =\!\!\! & 12 (u\bu)^2 C_{n-1}^{5/2}(u-\bu)
\stackrel{u}{\otimes}\! {^\text{G}}T(u,v|\cdots)\stackrel{v}{\otimes} 2 v\bv C_k^{3/2}(v-\bv)\,,
\label{^Gc_jk}
\end{eqnarray}
\end{subequations}
and where again the ${^\text{G}}c_{nk}$ coefficients are in LO approximation normalized to one.
Note that the  prefactor for gluons in (\ref{^GT_jk-1}) is $2/(n+3)$ times the prefactor  for quarks in (\ref{^qT_jk-1}).
Passing from $x$ to $u$,  we also pulled out here, as in the quark case, an overall normalization factor $(-1)^n \xi^{-n-1}$.
Let us add that the integral conformal GPD moments are calculated as in \cite{Kumericki:2007sa} from
\begin{eqnarray}
\label{Fn} \mbox{\boldmath $F$}_{n}(\eta,t,\mu^2)=\frac{
\Gamma(3/2)\Gamma(n+1)}{2^{n}  \Gamma(n+3/2)}
\frac{1}{2}\int_{-1}^1\! dx\; \eta^{n-1} \left(\!\!
\begin{array}{cc}
\eta\, C_n^{3/2} & 0 \\
0 & \frac{3}{n}\,  C_{n-1}^{5/2}
\end{array}
\!\!\right)\!\!\left(\frac{x}{\eta}\right)
 \mbox{\boldmath $F$}(x,\eta,t,\mu^2)\,.
\end{eqnarray}
This definition ensures that in the forward limit ($\eta=0,t=0$)  the entries of  GPD $ \mbox{\boldmath $H$}_{n}$   are  given by the odd Mellin moments
of unpolarized quark and gluon PDFs:
$$
\mbox{\boldmath $H$}_{n}(\eta=0,t=0,\mu^2)= \int_0^1\!dx\, x^n  \left( \Sigma(x,\mu^2) \atop g(x,\mu^2)\right)\;\;\mbox{with}\;\;\
\Sigma(x,\mu^2)= \sum_{q=u,d,\cdots} \left[q(x,\mu^2) + \overline{q}(x,\mu^2)\right]
$$
for $n\in \{1,3,\cdots\}$.
To complete the description of our conventions, let us note that the evolution of the conformal GPD moments (\ref{Fn}) reads
\begin{eqnarray}
\mu\frac{d}{d\mu} \mbox{\boldmath $F$}_{n}(\eta,t,\mu^2)  = \left[\frac{\alpha_s(\mu)}{2\pi} \mbox{\boldmath $\gamma$}^{(0)}_{n}
+ O(\alpha_s^2) \right]\cdot\mbox{\boldmath $F$}_{n}(\eta,t,\mu^2)  \quad\mbox{for}\quad n\in\{1,3,5,\cdots\}\,.
\end{eqnarray}
At LO  it is governed by the anomalous dimension matrix
\begin{subequations}
\label{eq:andim}
\begin{eqnarray}
\mbox{\boldmath $\gamma$}_n^{(0)}=
\left(
\begin{array}{cc}
{^{\Sigma \Sigma}}\gamma^{(0)}_n & {^{\Sigma {\rm G}}}\gamma^{(0)}_n\\
{^{{\rm G} \Sigma }}\gamma_{n}^{(0)} & {^{{\rm G}{\rm G}}}\gamma^{(0)}_n
\end{array}\right)
\,.
\end{eqnarray}
In accordance with the LO order kernel (\ref{Vb^{(0)}}), the LO entries  coincide with those known from deep inelastic scattering.
The quark-quark entry is given in (\ref{eq:gamma0}) and the three other entries read
\begin{eqnarray}
\label{Def-LO-AnoDim-QG-V}
{^{\Sigma\g}\!\gamma}_{n}^{(0)} &\!\!\!=&\!\!\! -2n_f
\frac{4 + 3\,n + n^2 }{( n + 1 )( n + 2 )( n + 3)}
\,,
\\
\label{Def-LO-AnoDim-GQ-V}
{^{\g \Sigma}\!\gamma}_{n}^{(0)} &\!\!\!=&\!\!\!
-2\CF\frac{4 + 3\,n + n^2 }{n( n + 1 )( n + 2 )}
\,,
\\
\label{Def-LO-AnoDim-GG-V} {^{\g\g}\!\gamma}_{n}^{(0)}
&\!\!\!=&\!\!\!  \CA \left(4S_1( n + 1 ) + \frac{4}{( n + 1 )( n + 2)} - \frac{12}{n( n + 3)}  \right)+ \beta_0
\,,
\end{eqnarray}
\end{subequations}
More information on anomalous dimensions is given
in App. \ref{app:ad}.

An integral  CPWE for TFFs does not converge in the physical region and, thus, we need for them a Mellin-Barns integral representation.
To pass from the CPWE (\ref{F(Q^2)}) for form factors to those of quark TFFs,
we can perform a Sommerfeld-Watson transform, intuitively written as%
\footnote{Here we used that ${\rm Res}_{j=n}1/\sin(\pi j) =  (-1)^n/\pi$ for $n=\{0,1,2,\cdots\}$, which generates the $(-1)^n$  factor
of $\sigma (-\xi)^{-n-1}$. The factor $1/\sin(\pi j)$ compensates the exponential growth of the CPW for $j\to \infty$ while the
continuation of $\xi$ to $-\xi+i \epsilon$  yields
$\frac{\sigma - e^{-i \pi j}}{\sin(\pi j)} = i \pm\left\{{\tan \atop \cot }\right\}\left(\frac{\pi\,j}{2}\right)$ for $\sigma = \left\{{+1 \atop -1}\right\}$.
}
\begin{equation}
\sum_{n=0\atop {\rm even}}^\infty  \varphi_n\, {\rm T}_{nk}  \Rightarrow \sum_{n=0}^\infty \left[\sigma (-\xi)^{-n-1}+(\xi- i\epsilon)^{-n-1}\right] F_n\,  {\rm T}_{nk}
\Rightarrow \frac{1}{2 i} \int_{c-i\infty}^{c+i\infty}\! dj\,\xi^{-j-1}\, \frac{\sigma-e^{-i \pi j}}{\sin(\pi j)}  F_j\,  {\rm T}_{jk}\,.
\end{equation}
Then, the CPW amplitudes (\ref{eq:Tnk(Q,mu0)}), containing also the evolution operator, must be continued in such a manner that they are  bounded for $j\to \infty$, where Carlson`s theorem assures us that this continuation is unique \cite{Car14}.
Furthermore, all singularities lie on the l.h.s.~of the final integration contour, which is parallel to the imaginary axis.
Taking into account the overall normalization, we can write in analogy to the CFF notation from \cite{Kumericki:2007sa} the TFFs (\ref{eq:tffF^A},\ref{eqs:tffF^S}) as Mellin-Barnes integrals.

Flavor non-singlet TFFs (\ref{eq:tffF^A}) or charge parity odd quark ones evolve autonomously. Furthermore,
restricting us to those with definite signature, i.e., $\text{A}\in\{\qmi,3^{(\pm)},\cdots\}$, we can represent them as
\begin{subequations}
\label{eq:tffqF_M-MBI}
\begin{eqnarray}
\label{eq:tffqF_M-MBI-a}
\tffF^\text{A}_{\rm  M}(\xB,t,\cQ^2)
&\!\!\! \stackrel{\rm Tw-2}{=}\!\!\! &
 \frac{C_F  f_{\rm M}}{N_c \cQ}
  \frac{1}{2 i} \int_{c- i \infty}^{c+ i \infty}\! dj\, \xi^{-j-1}
\left[
  i \pm \left\{{\tan \atop \cot }\right\}\!\left(\!\frac{\pi\,j}{2}\!\right)
  \right]
\qquad\qquad\mbox{for}\quad  \sigma\!\left(\!F^\text{A}\!\right) = \left\{ +1 \atop -1 \right\}
  \nonumber  \\[0.3cm]
&&\phantom{\frac{C_F  f_{\rm M}}{N_c \cQ} \frac{1}{2 i}}
\times \left[
{^\sigma}{\rm T}_{jk}(\cQ^2,\cQ_0^2)
\stackrel{k}{\otimes}
\varphi_{{\rm  M},k}(\cQ_0^2)
\right]
F^\text{A}_j(\xi,t,\cQ_0^2)\,,
\end{eqnarray}
where in accordance with the signature definition (\ref{eq:sigma4F}) one chooses the $\tan(\pi j/2)$ and  $-\cot(\pi j/2)$ function for $\sigma\!\left(\!F^\text{A}\!\right) =+1$ and $\sigma\!\left(\!F^\text{A}\!\right) =-1$, respectively. The CPW amplitudes
\begin{eqnarray}
\label{eq:tffqF_M-MBI-b}
{^\sigma}{\rm T}_{jk}(\cQ^2,\cQ_0^2) =
{^\sigma}T_{j+m,k+l}\!\left(\!\alpha_s(\muR),\frac{\cQ^2}{\muF^2},\frac{\cQ^2}{\muphi^2},\frac{\cQ^2}{\muR^2}\!\right)
\stackrel{l,m}{\otimes} E_{k+l,k}(\muphi,\cQ_0)\, {^\sigma}E_{j+m,j}(\muF,\cQ_0)\,,
\end{eqnarray}
\end{subequations}
having pre-superscript $\sigma$,
are obtained by analytic continuation of those where $n=j$ is odd for $\sigma =+1$ and where it is even for
$\sigma =-1$, respectively. Here on the l.h.s.~their residual factorization and renormalization scale
dependencies are again not indicated, and the input scales for DA and GPD are chosen to be the same.
In LO approximation they are trivially given as products of LO evolution operators, valid for
both signatures. At NLO the signature must be set in both the hard coefficients and the flavor non-singlet anomalous dimensions.
As mentioned in the preceding section, the summation in (\ref{eq:tffqF_M-MBI-b}) can be numerically precalculated.

In an analogous fashion, we can write down the Mellin-Barnes integral for the  flavor singlet TFF (\ref{eqs:tffF^S}), which has even signature:
\begin{subequations}
\label{eq:tffSF_M-MBI}
\begin{eqnarray}
\label{eq:tffSF_M-MBI-a}
\tffF_{{\rm  V}^0}^{\SigG}(\xB,t,\cQ^2)
&\!\!\! \stackrel{\rm Tw-2}{=}\!\!\! &
\frac{C_F f_{{\rm  V}^0}}{N_c \cQ}
 \frac{1}{2 i} \int_{c- i \infty}^{c+ i \infty}\! dj\, \xi^{-j-1}\left[ i+ \tan\!\left(\!\frac{\pi\,j}{2}\!\right)\right]
\\[0.3cm]
&&\phantom{\frac{C_F  f_{\rm M}}{N_c \cQ}  \frac{1}{2 i} \int_{c- i \infty}^{c+ i \infty}}
\qquad
\times
\left[
\mbox{\boldmath ${\rm T}$}_{jk}(\cQ^2,\cQ_0^2)
\stackrel{k}{\otimes}
\varphi_{{\rm  V}^0,k}(\cQ_0^2)
\right]
\cdot \mbox{\boldmath $F$}_j(\xi,t,\cQ_0^2)
\nonumber
\end{eqnarray}
with
\begin{eqnarray}
\mbox{\boldmath ${\rm T}$}_{jk}(\cQ^2,\cQ_0^2)  =
\left[\mbox{\boldmath $T$}_{j+m,k+l}\!\left(\!\alpha_s(\muR),\frac{\cQ^2}{\muF^2},\frac{\cQ^2}{\muphi^2},\frac{\cQ^2}{\muR^2}\right)
\stackrel{l}{\otimes} E_{k+l,k}(\muphi,\cQ_0) \right]\stackrel{m}{\otimes}\mbox{\boldmath$\cE$}_{j+m,j}(\muF,\cQ_0)
\quad
\end{eqnarray}
and the  vector
\begin{eqnarray}
\label{^ST_{jk}}
\mbox{\boldmath $T$}_{jk} = \frac{2^{j+1}\ \Gamma\big(j+\frac{5}{2}\big)}{\Gamma\big(\frac{3}{2}\big)\Gamma(j+3)} \left({{^\Sigma c}_{jk} , \frac{2}{C_F (j+3)}{^{\rm G} c}_{jk} }\right) \times 3\,,\quad
{^{\Sigma}c}_{jk} = \frac{1}{n_f}\, {^+}c_{jk} +  {^\pS c}_{jk}\,.
\end{eqnarray}
\end{subequations}

Finally, for the conformal moments ${^\text{A}}c_{jk}$, appearing in the hard coefficients (\ref{^qT_jk},\ref{^GT_jk},\ref{eq:tffSF_M-MBI}),
we adopt the analogous perturbative expansion as in Eqs.~(\ref{eq:T-pQCD}),
\begin{eqnarray}
\label{^Ac_{jk}}
{^\text{A}}c_{jk}\!\left(\alpha_s(\muR),
\frac{\cQ^2}{\muF^2},\frac{\cQ^2}{\muphi^2},\frac{\cQ^2}{\muR^2}
\right)
&\!\!\!=\!\!\! &
\alpha_s(\muR) \: {^\text{A}}c_{jk}^{(0)}
+ \frac{\alpha^2_s(\muR)}{2\pi}\:
{^\text{A}}c_{jk}^{(1)}\!\left(\!\frac{\cQ^2}{\muF^2},\frac{\cQ^2}{\muphi^2},\frac{\cQ^2}{\muR^2}\!\right) + O(\alpha_s^3)\,.
\end{eqnarray}
Furthermore, the perturbative expansion of anomalous dimensions is inherited from those of the evolution kernel (\ref{eq:V-pQCD}), i.e., it is the same
as in \cite{Kumericki:2007sa}. The  conformal moments read to LO as
\begin{subequations}
\label{c_jk-pQCD}
\begin{eqnarray}
\label{c^{(0)}_jk}
{^\pm}c^{(0)}_{jk} =  {^\g}c^{(0)}_{jk} = 1
\quad\mbox{and}\quad
{^\pS}c^{(0)}_{jk} =0.
\end{eqnarray}
At NLO  we have for the quark contribution
\begin{eqnarray}
{^\pm}c^{(1)}_{jk}\!\left(\!\frac{\cQ^2}{\muF^2},\frac{\cQ^2}{\muphi^2},\frac{\cQ^2}{\muR^2}\!\right)  &\!\!\! =\!\!\! &
- \frac{1}{2}\ln\frac{\cQ^2}{\muF^2} \gamma_j^{(0)}- \frac{1}{2}
\ln\frac{\cQ^2}{\muphi^2}  \gamma_k^{(0)}  +
\frac{\beta_0}{2}  \ln\frac{\cQ^2}{\muR^2}+ \cdots,
\end{eqnarray}
in the pure   singlet quark channel
\begin{eqnarray}
{^\pS}c^{(1)}_{jk}\!\left(\!\frac{\cQ^2}{\muF^2}\!\right) &\!\!\! =\!\!\! &
 -\frac{1}{2}\ln\frac{\cQ^2}{\muF^2}\,  \frac{1}{\CF}\, \frac{2}{j+3}\, {^{{\rm G}\Sigma} \gamma}^{(0)}_j + \cdots
\,,
\end{eqnarray}
and for the gluons
\begin{eqnarray}
{^{\rm G}{c}}^{(1)}_{jk}\!\left(\!\frac{\cQ^2}{\muF^2},\frac{\cQ^2}{\muphi^2},\frac{\cQ^2}{\muR^2}\!\right)&\!\!\!=\!\!\!&
-\frac{1}{2} \ln\frac{\cQ^2}{\muphi^2} \gamma^{(0)}_k
-\frac{1}{2}\ln\frac{\cQ^2}{\muF^2} \left( {^{\rm GG} \gamma}^{(0)}_j +
\frac{C_F}{n_f}\, \frac{j+3}{2}\, {^{\Sigma {\rm G}} \gamma}_j^{(0)}\right)
+\frac{\beta_0}{2}  \ln\frac{\cQ^2}{\muR^2} +\cdots\,.
\nonumber
\\
\end{eqnarray}
\end{subequations}
Note that due to the change of normalization when going from ${^\pS}T_{jk}$ to ${^\pS}c_{jk}$ and ${^\g}T_{jk}$ to ${^\g}c_{jk}$  in (\ref{^ST_{jk}})
the off-diagonal entries of the anomalous dimensions (\ref{eq:andim}) are accompanied in the pure singlet contribution by a factor $2/\CF\times 1/(j+3)$
and in the gluonic one by a factor $\CF/2n_f\times (j+3)$. The  color factors are changed as in the corresponding  momentum fraction expressions  (\ref{eq:logSigma}) and (\ref{^GcalT^(1)}).

\subsubsection{Analytic continuation of integral conformal moments}
\label{sec:prel-MB-analytic}

As mentioned in the previous section, these complex valued  conformal moments must satisfy a bound for $j\to \infty$, which implies
that their  continuation from integer values is unique. Most of the conformal moments, we need in the NLO expressions, have been
already evaluated in a different context and with various methods, see Refs.~\cite{Melic:2002ij,Kumericki:2007sa}.  However,
in  hard DVMP amplitudes, known at NLO accuracy,  we  encounter a new class of functions that calls for a  more powerful method.
It is of crucial importance for us that a method exists which allows to solve this continuation problem on general grounds.
The method we propose to use is based on DR technique and allows to perform this mapping purely {\em numerically} and, moreover,
it can be utilized to link conformal moments of certain functions via a standard Mellin transform directly to harmonic sums.
This will be used to evaluate some of our more intricate conformal moments analytically, which we could not achieve
by utilizing other methods. To our best knowledge this method has not  been used so
far for the evaluation of CPW amplitudes, however, it is well known from the SO(3) PWE of scattering amplitudes and carries there
the name Froissart–-Gribov projection.

For the sake of a compact presentation let us introduce integral CPWs
\begin{eqnarray}
\widehat{p}^{\frac{3}{2}}_n(u) =  2 u\bu C_n^{(3/2)}(u-\bu) \quad\mbox{and}\quad  \widehat{p}^{\frac{5}{2}}_n(u) =  12 (u\bu)^2\, C_{n-1}^{(5/2)}(u-\bu)
\end{eqnarray}
in which overall normalization factors are absorbed in the definition of $T_{jk}$ , see (\ref{^qT_jk-1},\ref{^GT_jk-1},\ref{^ST_{jk}}). As already shown implicitly in the preceding section, the map from the momentum faction representation
to conformal moments (\ref{^Ac_{jk}}) takes then the simple form
\begin{equation}
 {^\text{A}}c_{nk} = \widehat{p}_n^{\nu(\text{A})}(u) \stackrel{u}{\otimes}
 {^\text{A}}T(u,v)\stackrel{v}{\otimes}  \widehat{p}_k^\frac{3}{2}(v)\quad \mbox{with} \quad c^{(0)}_{nk} =1\,,
\label{^Ac_{nk}}
\end{equation}
where for quark-quark (gluon) channels  Gegenbauer polynomials with index $\nu=3/2$ ($\nu=5/2$) have to be taken.
For our purpose it is now more appropriate to generate these polynomials
by differentiation w.r.t.~$u$ applied $n$ and $n-1$  times  to the function $(u\bu)^{n+1}$, respectively, i.e.,
we utilize the Rodrigues formula \cite{AbrSte},
\begin{equation}
\label{hp^A_n}
\widehat{p}^{\nu(\text{A})}_n(u) = \left\{
\begin{array}{c}
{\displaystyle \frac{(n+2)}{(-1)^n\, n!}\, \frac{d^{n}(u\bu)^{n+1}}{du^{n}}}
\\
{\displaystyle \frac{(n+2)_2}{(-1)^{n-1}\,(n-1)!}\,, \frac{d^{n-1}(u\bu)^{n+1}}{du^{n-1}}}
\end{array}
 \right\}
\quad\mbox{with} \quad
\left\{
\begin{array}{l}
\nu(\text{A}) =3/2 \mbox{ for quarks}  \\
\nu(\text{A})=5/2 \mbox{ for gluons}
\end{array}
\right. \,,
\end{equation}
where the Pochhammer symbol, defined as usual as ratio of two Euler $\Gamma$ functions
\begin{eqnarray}
\label{Pochhammer}
(m)_a= \frac{\Gamma(m+a)}{\Gamma(m)} =  m\times \cdots \times (m+a-1)\quad\mbox{for}\quad a \in \{1,2,\cdots\}
\quad\mbox{and}\quad (m)_0= 1  \,,
\end{eqnarray}
has the value $(n+2)_2 = (n+2)(n+3)$.

For the hard scattering amplitudes in (\ref{^Ac_{nk}}) with definite signature
we now utilize the single $u$-variable DR (\ref{^AT(u,v)-DR-r}). In the following we prefer  the equivalent form
\begin{equation}
{^\text{A}}T(u,v|\cdots) = \int_0^1\!\frac{dy}{1-u y}\,  \frac{2\,{^\text{A}}t\!\left(\frac{y}{2-y},v|\cdots\!\right)}{2-y}
\,,
\label{^AT(u,v)-DR}
\end{equation}
which is obtained from (\ref{^AT(u,v)-DR-r}) by the variable transformation $r=y/(2-y)$.
Plugging this representation and the CPWs (\ref{hp^A_n}) into the CPW amplitudes (\ref{^Ac_{nk}}), reshuffling the differential operators by partial integration to act on the dispersion kernel, and symbolically performing the $u$ integration, yields the desired representation
\begin{equation}
\label{T2cjk}
 {^\text{A}}c_{jk}(\cdots) = \widetilde{p}_j^{\nu(\text{A})}(y) \stackrel{y}{\otimes} \frac{2\, {^\text{A}}t\!\left(\frac{y}{2-y},v|\cdots\!\right)}{2-y}
  \stackrel{v}{\otimes} \hat{p}^{\frac{3}{2}}_k(v)
\,,
\end{equation}
where the conformal moments of the dispersion integral kernel are given as integrals over $u$,
\begin{subequations}
\begin{equation}
\label{tp^A_n}
\widetilde p^{\nu(\text{A})}_j(y) = \left\{
\begin{array}{c}
{\displaystyle (j+2)\, y^j \int_0^1\! du\,  \frac{(u \bu)^{j+1}}{(1-u y)^{j+1}}}
\\
{\displaystyle (j+2)_2\, y^{j-1} \int_0^1\! du\,  \frac{(u \bu)^{j+1}}{(1-u y)^{j}}}
\end{array}
 \right\}
\quad\mbox{with} \quad
\left\{
\begin{array}{l}
\nu(\text{A}) =3/2 \mbox{ for quarks}  \\
\nu(\text{A})=5/2 \mbox{ for gluons}
\end{array}
\right. \,.
\end{equation}
The reader may recognize that these functions are nothing but hypergeometric functions,
\begin{eqnarray}
\label{tp^Q}
{\widetilde p}_j^{3/2}(y) &\!\!\! \equiv \!\!\!& \frac{\Gamma(j+2)\Gamma\!\left(j+3\right)}{\Gamma(2j+4)}\, y^{j}\, {_2F_1}\left({{j+1, j+2}\atop {2j+ 4}}\Big|y\right)\,,
\\
\label{tp^G}
{\widetilde p}_j^{5/2}(y) &\!\!\! \equiv \!\!\!&
\frac{\Gamma(j+2)\Gamma\!\left(j+4\right)}{\Gamma(2j+4)}\, y^{j-1}\,
{_2F_1}\left({{j, j+2}\atop {2j+ 4}}\Big|y\right)\,,
\end{eqnarray}
\end{subequations}
which can also  be expressed in terms of associated Legendre functions of the second kind \cite{AbrSte}.
These functions may be viewed as the `dual' CPWs that generalize the common Mellin moments.
The integral representation (\ref{tp^A_n}) obviously tells us that in our case, i.e., $0\le y \le 1$,
the CPWs are bounded for $j\to \infty$ and, consequently, also the conformal moments (\ref{T2cjk}).
Having at hand a numerical routine for hypergeometric functions, the formula (\ref{T2cjk}) can be employed for
the numerical evaluation of conformal moments for complex valued $j$.

A more convenient representation, is obtained if we  rewrite the conformal moments (\ref{T2cjk}) in terms of a common Mellin transform.
To do so, we insert the integral representation (\ref{tp^A_n}) into (\ref{T2cjk}) and  introduce the new integration variable $w= y u\bar{u}/(1-u y)$,
which yields
\begin{subequations}
\label{^Ac_{jk}-Mellin}
\begin{eqnarray}
\label{^Ac_{jk}-Mellin-1}
{^{\text{A}}}c_{jk}(\cdots) = \int_{0}^1\! dw\, w^j\,{^{\text{A}}} m^{\nu({\text{A})}}_{k}(w|\cdots)\,,
\end{eqnarray}
where the quark  and gluon coefficients read
\begin{eqnarray}
\label{^Qm_{jk}-Mellin}
{^{\text{A}}}m^{\frac{3}{2}}_{k}(w|\cdots) &\!\!\! = \!\!\! & \int_w^1\!du\,   \frac{2(j+2)u\bu}{2 u\bu + w (u-\bu)}\,
 {^{\text{A}}}t\!\left(\!\frac{w}{2 u\bu + w (u-\bu)},v|\cdots\!\right)\stackrel{v}{\otimes} \widehat{p}^{\frac{3}{2}}_k(v)\,,
\nonumber\\
\\
\label{^Gm_{jk}-Mellin}
{^\g}m^{\frac{5}{2}}_{k}(w|\cdots) &\!\!\! = \!\!\! &  \frac{1}{w}\int_w^1\!du\,   \frac{2(j+2)_2\, u^2\bu^2}{2 u\bu + w (u-\bu)}\,
 {^\g}t\!\left(\!\frac{w}{2 u\bu + w (u-\bu)},v|\cdots\!\right)\stackrel{v}{\otimes} \widehat{p}^{\frac{3}{2}}_k(v)\,.
\end{eqnarray}
\end{subequations}
These formulae can be utilized for the analytical evaluation of conformal moments from the imaginary part of the hard scattering amplitude in NLO approximation, presented in Sec.~\ref{sec:anatomy-quark}--\ref{sec:anatomy-gluon}.
Otherwise, one may simply perform a two-dimensional integration.

\subsection{Mixed representations}
\label{sec:prel-mixed}

Although we will only present the NLO results  in momentum fraction representation, including the explicit expressions
for the imaginary part of TFFs, and in the CPWE for complex valued $j$ and integral $k$, we should at least mention here that these representations
can be combined in various manners. There is possibly even some need for doing so, e.g., if one is interested to provide predictions from a GPD model that is given in momentum fraction representation. Also if the CPWE of a DA converges only slowly one may prefer to switch to the momentum fraction
representation. We will present in the next section the NLO results in such a manner that once one is interested in a mixed representation one
can easily recover it from the collection of formulae, given below.

\begin{itemize}
\item Supposing that the integral CPWE for DAs can be truncated, it might be practical to combine this expansion with the
momentum fraction representation of GPDs. The hard coefficients can be analytically calculated,  e.g.,  for $k \in \{0,2,4\}$, from
\begin{eqnarray}
{^\text{A}}\widehat{T}_k(u|\cdots) = 3\times  {^\text{A}}T(u,v|\cdots) \stackrel{v}{\otimes} \widehat{p}^{\frac{3}{2}}_k(v)\,.
\end{eqnarray}
The analogous approach might be used directly for the evaluation of the imaginary part,
\begin{eqnarray}
\label{^Awidehat{t}_k(r|cdots)}
{^\text{A}}\widehat{t}_k(r|\cdots) =  3\times {^\text{A}}t(r,v|\cdots) \stackrel{v}{\otimes} \widehat{p}^{\frac{3}{2}}_k(v)\,,
\end{eqnarray}
which leads to simpler analytical functions.
\item  CPWE of GPDs and momentum fraction representation for DA,
\begin{eqnarray}
{^\text{A}}T_j(v|\cdots) = T_{j}^{\nu(\text{A})}\; {^\text{A}}c_j(v|\cdots)\,, \quad
{^\text{A}}c_j(v|\cdots)=  \widetilde p_j^{\nu(\text{A})}(y) \stackrel{y}{\otimes}  \frac{2\,{^\text{A}}t\!\left(\frac{y}{2-y},v|\cdots\!\right)}{2-y}\,,
\end{eqnarray}
where $T^{3/2}_j=\frac{2^{j+1}\ \Gamma\left(j+\frac{5}{2}\right)}{\Gamma\left(\frac{3}{2}\right)\Gamma(j+3)}$ and
 $T^{5/2}_j=\frac{2^{j+2}\ \Gamma\left(j+\frac{5}{2}\right)}{\Gamma\left(\frac{3}{2}\right)\Gamma(j+4)}$.
\item  CPWE of GPDs and integral CPWE of DA,
\begin{eqnarray}
\label{eq:ima2CM}
{^\text{A}}T_{jk}(\cdots) = T^{\nu(\text{A})}_{j}\times 3\times {^\text{A}}c_{jk}(\cdots)\,, \quad
{^\text{A}}c_{jk} &\!\!\!=\!\!\!&
{^\text{A}}c_j(v|\cdots) \stackrel{v}{\otimes} \widehat{p}^{(3/2)}_k(v)
\\
&\!\!\!=\!\!\!&\widetilde{p}_j^{\nu(\text{A})}(y)\;  \stackrel{y}{\otimes} \frac{2\, {^\text{A}}\widehat{t}_k\!\left(\frac{y}{2-y}\!\right)}{2-y}\,,
\nonumber
\end{eqnarray}
where, alternatively, the  Mellin transform (\ref{^Ac_{jk}-Mellin}) can be employed.
\end{itemize}
Let us add that instead of a slowly convergent integral CPWE for rather broad or narrow DAs one may be interested to have a complex valued expansion, too,
which can be alternatively used to the momentum fraction representation.  This is indeed possible, however, we will not present technical details here.

\section{Anatomy of next-to-leading order corrections}
\label{sec:NLO}
\setcounter{equation}{0}

The  NLO corrections to the hard DVMP amplitudes are known in momentum fraction representation.
In the flavor non-singlet channel they were obtained by analytic continuation \cite{Belitsky:2001nq,Diehl:2007hd}
from  diagrammatical result for the pion form factor \cite{Melic:1998qr} (and references therein).
This  finding in the flavor non-singlet channel can be used for all DVMP channels since the {\em two}
$\gamma_5$ matrices, arising from two intrinsic parity odd operators, are irrelevant.
Furthermore, the hard scattering amplitudes for  DV$\!V^0_L$P to NLO accuracy in the pure singlet quark and  gluon-quark channel
were diagrammatically evaluated in \cite{Ivanov:2004zv}.
For the non-singlet case also the integral conformal moments were evaluated,
where the most intricate part was only given in terms of an integral \cite{Mueller:1998qs}.
The  NLO  evolution kernel in the non-singlet channel was also obtained
some time ago by means of the extension rule \cite{Dittes:1988xz} from
the diagrammatical result \cite{Dittes:1983dy,Sarmadi:1982yg,Mikhailov:1984ii},
while the singlet kernels were constructed from the anomalous dimensions
\cite{Belitsky:1999hf}, obtained from the understanding of conformal symmetry
breaking in the modified minimal subtraction scheme
($\overline{\rm MS}$) \cite{Mueller:1993hg,Belitsky:1998gc}.
Thus, the full NLO formalism is available to leading twist accuracy
for all flavor non-singlet and DV$\!V^0_L$P processes.

In the following we present compact expressions for all the  hard scattering amplitudes that are known to NLO accuracy in
the momentum fraction representation, cf.~Sec.~\ref{sec:prel-MF}, their imaginary parts, cf.~Sec.~\ref{sec:prel-DR},
and their conformal moments, cf.~Sec.~\ref{sec:prel-MB}. For the sake of a compact presentation, we comment in
Sec.~\ref{sec:NLO-generic}  on the general structure of NLO corrections
and introduce building blocks for all the three representations in a one-to-one correspondence.
In Sec.~\ref{sec:NLO-result} we present then the  NLO corrections in terms of these building blocks.

\subsection{Generic structure of NLO corrections}
\label{sec:NLO-generic}

In our presentation of the NLO corrections in the channel
$A\in\{\sigma=\pm,\pS,\g\}$
we will decompose these channels w.r.t.~color structure.
In the momentum fraction representation we write the NLO approximation
of the perturbative expansion (\ref{eq:T-pQCD}) with the LO coefficient
(\ref{eq:T^{(0)}}) as
\begin{eqnarray}
\label{AT^{(1)}}
{^A T}\!\left(u,v|\cdots\right)
= \alpha_s\,\frac{1-\delta_{A,\pS}}{\bu\bv}
+ \frac{\alpha^2_s}{2\pi}\,
{^A T}^{(1)}\!\left(u,v|\cdots\right) + O(\alpha_s^3)\;\; \mbox{with}\;\;
{^A}T^{(1)} = \sum_{\text{c}} C_{\text{c}}\, {^A}T^{(1,\text{c})}\,,
\end{eqnarray}
where color factors (combinations) can take the values
$$C_{\text{c}}\in\{\CF,\CA,\CG=\CF-\CA/2,\beta_0\} \mbox{ with }
\CF= \frac{4}{3},\; \CA=3,\; \CG=-\frac{1}{6},  \mbox{ and }\beta_0 = -11 +
\frac{2n_f}{3}.
$$
Both the imaginary and real part of the hard scattering amplitude
can be easily evaluated. The relevant terms are listed
in App. \ref{app:ImRe}.
In this section we present only the full NLO expressions for the imaginary
parts of the hard scattering amplitudes,
written in the form  (\ref{eqs:DRtffFqC},\ref{eqs:DRF^S_V}) with the
perturbative expansion (\ref{eq:tT-pQCD})
and  the color decomposition of the NLO contribution analogous to (\ref{AT^{(1)}})
\begin{eqnarray}
\label{At^{(1)}}
{^A t}\!\left(r,v|\cdots\right)
=\alpha_s \frac{(1-\delta_{A,\pS})\, \delta(1-r)}{\bv}
+ \frac{\alpha^2_s}{2\pi}\,
{^A t}^{(1)}\!\left(r,v|\cdots\right) + O(\alpha_s^3)\;\;\mbox{with}\;
{^{A}}t^{(1)}
= \sum_{\text{c}} C_{\text{c}}\, {^A}t^{(1,\text{c})}.
\end{eqnarray}
As above we use for shortness the variable $r=\xi/x $.
The conformal moments of (\ref{AT^{(1)}}), see the perturbative expansion
(\ref{^Ac_{jk}}), inherit the color decomposition,
\begin{eqnarray}
\label{c^(1)_jk}
{^A c}_{jk}\!\left(\cdots\right)
=\alpha_s\, (1-\delta_{A,\pS})
+ \frac{\alpha^2_s}{2\pi}\,
{^A c}^{(1)}_{jk}\!\left(\cdots\right) + O(\alpha_s^3)\;\;\mbox{with}\;\;
{^A}c^{(1)}_{jk} = \sum_{\text{c}} C_{\text{c}}\,
{^A}c^{(1,\text{c})}_{jk}\,.
\end{eqnarray}
At LO these moments are consistently normalized to one for both quarks and
gluons.

Additionally, the NLO corrections (\ref{AT^{(1)}}) can be decomposed in $u$
and $v$  separable and non-separable contributions,
\begin{eqnarray}
\label{eq:ATc-dec}
{^A}T^{(1,c)}(u,v) = \sum_{i,j} a^c_{ij}\, f_i(u) f_j(v)  +
\Delta{^A}T^{(1,c)}(u,v) \quad \mbox{with}\quad \Delta{^A}T^{(1,c)}(u,v)=
\sum_i a^c_{i}\, f_i(u,v)\,,
\end{eqnarray}
where  $f_i(u)$ are certain single variable functions and
$\Delta{^A}T^{(1,\cdots)}(u,v)$ denotes the non-separable part in channel $A$ with color
structure $c$. In the following we call such an additive term 'addendum'.
They arise, e.g., from crossed ladder Feynman  diagrams, and their
origin is considered in  App. \ref{app:diag}. They can be further
decomposed into a set of  functions $f_i(u,v)$, depending on two variables.
In the next two sections we introduce the building blocks for separable
and non-separable functions $f_i(u)$ and $f_i(u,v)$, respectively, give
their imaginary parts, and evaluate their conformal moments.
We group theses building blocks w.r.t.~their analytic properties,
where the most singular terms are removed from the non-separable
building blocks.  Such an ordering provides insight into the
qualitative features of  NLO corrections, which is explicitly spelled out
in Sec.~\ref{sect-numerics}.

\subsubsection{Building blocks for separable NLO terms}
\label{sec:NLO-blocks1}

First we introduce  the building blocks for separable contributions to the
NLO hard scattering amplitudes.
The evaluation of their imaginary parts is straightforward and is together
with
the evaluation of their real parts systematized for the general case in
App. \ref{app:ImRe}, listed  there in Tab.~\ref{tab:ReImf}.
Most of the conformal moments are already known
\cite{Mueller:1998qs,Melic:2002ij,Kumericki:2007sa}. We will evaluate the
missing ones from the imaginary parts by means the mapping technique, discussed in
Sec.~\ref{sec:prel-MB-analytic}.
We will list the building blocks, their imaginary parts, and the
corresponding conformal moments in  Tabs. \ref{subT-1}--\ref{subT-Li}.

\begin{itemize}
\item{\em Most singular building blocks.}
\end{itemize}
\vspace{-2mm}
We recall that the LO coefficient ${^A}T^{(0)}(u,v)$, given in
(\ref{eq:T^{(0)}}), consist of two factorized poles $1/\bu\bv$ at the
cross-over point
$u=1$ and endpoint $v=1$. Surely, the imaginary part of $1/\bu$ yields then
a Dirac delta-function in the corresponding coefficient (\ref{t^{(0)}})
for the imaginary part. At higher orders of the  perturbative expansion
these poles appear, too, and moreover, they are partially accompanied
by logarithmic $[1,\infty]$-cuts along the positive real axis, starting
at one and ending at infinity.  Such a logarithmical enhancement implies
large perturbative corrections in the
vicinity of the  cross-over point and/or the endpoint region.
The most singular function that appears at NLO is a pole that is accompanied
by a squared logarithm. Thus, we consider here the building blocks
\begin{equation}
\label{eq:uv-func-g}
\frac{1}{\bu-i\epsilon}\,, \quad
\frac{\ln(\bu-i\epsilon)}{\bu-i\epsilon}\,,  \quad
\frac{\ln^2(\bu-i\epsilon)}{\bu-i\epsilon}\quad  \mbox{(analogous for $u\to
v$)},
\end{equation}
which we denote as the {\em most singular} ones.
Their values on the cut is governed by the $\bu-i\epsilon$-prescription,
inherited from Feynman`s causality prescription, and they are
generalized functions in the mathematical sense \cite{GelShi64}. Obviously,
the first term appears at LO and was treated above.

Most singular building blocks are generated for  (non-negative integral) $p$
values by differentiation of
$\ln^{p+1}(\bu - i\epsilon)= \left[\ln|\bu| - i
\pi\theta(-\bu)\right]^{p+1}$  w.r.t.~$u$, see App. \ref{app:ImRe}.
In particular their imaginary parts for the cases   of interest $p\in\{1,2\}$
read as follows
\begin{eqnarray}
\label{eq:uv-func-g-IM_0}
\im  \frac{\ln(\bu-i\epsilon)}{\bu-i\epsilon} = \pi \frac{d
\theta(-\bu)\ln(-\bu)}{du} \quad\mbox{and}\quad
\im \frac{\ln^2(\bu-i\epsilon)}{\bu-i\epsilon} = \pi
\frac{d\theta(-\bu)\ln^2(-\bu)}{du} -2\pi \zeta(2) \delta(\bu),
\nonumber
\end{eqnarray}
which can be also expressed in terms of more common $+$-prescriptions
(\ref{[[]]_+}).
If we switch to momentum fraction variables, used in the convolution
integrals (\ref{eqs:DRtffFqC},\ref{eqs:DRF^S_V}) for the imaginary part of
TFFs,
we define the $+$-prescriptions as in (\ref{{}_+}). They explicitly read as
follows
\begin{subequations}
\label{f+}
\begin{eqnarray}
\label{f+-1}
\int_{\xi}^{1}\!\frac{dx}{x}\,\left\{f\!\left(\!\frac{\xi}{x}\!\right)\right\}_+
\tau(x) = \int_{\xi}^{1}\!\frac{dx}{x}\, f\!\left(\!\frac{\xi}{x}\!\right)
\left[\tau(x)-\tau(\xi)\right] + c_f(\xi) \tau(\xi)\,,
\end{eqnarray}
with the $\xi$-dependent subtraction terms interest
\begin{eqnarray}
c_f(\xi) = \ln\frac{1-\xi}{2 \xi}\;\;\mbox{for}\;\;f=\frac{1}{1-r}
\quad\mbox{and}\quad
c_f(\xi) =\frac{1}{2}\ln^2\frac{1-\xi}{2
\xi}-\zeta(2)\;\;\mbox{for}\;\;f=\frac{\ln\frac{1-r}{2r}}{1-r}  \,.
\end{eqnarray}
\end{subequations}
Note that the difference between our $\{\cdots\}_+$-prescription and the more
common $[\cdots]_+$-prescription, used in inclusive processes, is
just the (finite) subtraction term $c_f(\xi) \delta(1-r)$.

\begin{table}[t]
\begin{center}
\begin{tabular}{|ccccl|}
\hline
$\frac{1}{\bu}$   & $\Leftrightarrow$ & $\delta(1-r)$ & $\Leftrightarrow$ &
${\footnotesize \left\{ \begin{array}{l}
1
\\
1
\end{array}\right. }$
\\
$\frac{\ln \bu}{\bu}$ & $\Leftrightarrow$ & $\left\{
\frac{1}{1-r}\right\}_+$
& $\Leftrightarrow$ &
${\footnotesize \left\{ \begin{array}{l}
-2 S_1(j + 1) + \frac{1}{(j+1)_2}
\\
-2 S_1(j+1)+1+ \frac{4 (j+1)_2-2}{(j)_4}
\end{array}\right. }$
\\
$\frac{\ln^2 \bu}{\bu}$ & $\Leftrightarrow$ & $\left\{
\frac{2\ln\frac{1-r}{2r}}{1-r}\right\}_+$
& $\Leftrightarrow$ &
${\footnotesize \left\{ \begin{array}{l}
\left[2 S_1(j+1)-\frac{1}{(j+1)_2}\right]^2+\frac{2
(j+1)_2+1}{\left[(j+1)_2\right]^2} \\
\left[2 S_1(j+1)-1-\frac{4 (j+1)_2 -2 }{(j)_4}\right]^2  -1 +
\frac{2j(j+3)+9}{[j(j+3)]^2}+\frac{2 (j+1)_2+1}{[(j+1)_2]^2}
\end{array}\right. }$\\\hline
\end{tabular}
\end{center}
\caption{\small Substitution rules for the most singular building blocks
(left column), their imaginary parts (middle column), and
conformal moments (right column) for quarks (upper lines) and gluons (lower
lines). The $\left\{\cdots\right\}_+$-definitions and first order harmonic
sum are specified in (\ref{f+}) and (\ref{Sq}), respectively, and
the Pochhammer symbol $(\cdots)_a$ is defined in (\ref{Pochhammer}).
\label{subT-1}
}
\end{table}

The conformal moments of the most singular building blocks
(\ref{eq:uv-func-g})  can be easily generated from $\bu^{-\beta}$ by taking
derivatives w.r.t.~$\beta$ at $\beta=1$.
Utilizing Rodrigues formulae (\ref{hp^A_n}) one arrives at a closed
expression, see, e.g., App. B of \cite{Mueller:1998qs},
\begin{eqnarray}
\frac{\ln^p\bu}{\bu} \stackrel{u}{\otimes}  \widehat{p}^{\nu}_n(u)  =
\frac{
\Gamma\big(n+\frac{3}{2}+\nu\big)\,(-1)^p}{\Gamma\big(\nu-\frac{1}{2}\big)\Gamma\big(n+\frac{5}{2}-\nu\big)
}\,
\frac{d^p}{d\beta^p}\,
\exp\left\{\ln\frac{\Gamma\big(\nu-\frac{1}{2}-\beta\big)\Gamma\big(n+\frac{5}{2}-\nu+\beta\big)}{\Gamma(1+\beta)\Gamma\big(n+\frac{3}{2}+\nu-\beta\big)}\right\}
\Bigg|_{\beta=0},
\end{eqnarray}
which for $\nu \in \{3/2,5/2\}$ and  $p=0$ is normalized to one.
The analytic continuation of the r.h.s.~can be done in an obvious manner,
simply replace integral $n$ by complex valued $j$.  For integral $p\ge 1$
powers of the logarithmic derivative of Euler's $\Gamma$ function are
generated. We express them by the first order harmonic sum $S_1(j+1)$,
defined in
the standard manner, e.g., for any order $q$ as
\begin{eqnarray}
\label{Sq}
\quad S_q(z) = \frac{(-1)^{q-1}}{(q-1)!} \left[\psi^{(q-1)}(z+1)-
\psi^{(q-1)}(1)\right]\quad \mbox{with}\quad \psi^{(q-1)}(z+1) =
\frac{d^q}{d z^q} \ln \Gamma(z+1)\,.
\end{eqnarray}

Finally, the most singular building blocks  (\ref{eq:uv-func-g}), their
imaginary parts, and the corresponding conformal moments are collected as
substitution rules in Tab.~\ref{subT-1}. The logarithmic enhancement of
the pole at $u=1$ causes the need for $+$-prescriptions and is also encoded in the
logarithmic growth of the conformal moments at large $j$ since the harmonic
sum behaves for large $j$ as
\begin{eqnarray}
\label{S_1-asym}
S_1(j+1) = \ln(j+1) + \gamma_E + O(1/(j+1))\,,\quad \mbox{where}\quad
\gamma_E = 0.5772\cdots\,.
\end{eqnarray}

\begin{itemize}
\item{\em Building blocks with logarithmical $[1,\infty]$-cuts.}
\end{itemize}
\vspace{-2mm}
\begin{table}[t]
\begin{center}
\begin{tabular}{|ccccl|}
\hline
$\frac{\ln \bu}{u} $ & $\Leftrightarrow$ & $-\frac{1}{1+r}$ &
$\Leftrightarrow$ &
${\footnotesize \left\{ \begin{array}{l}
\frac{ -1}{(j+1)_2}
\\
\frac{-2 (j+1)_2-2}{(j)_4}
\end{array}\right. }$
\\
$\frac{\ln^2 \bu}{u}$  & $\Leftrightarrow$ & $-\frac{2}{1+r}
\ln\frac{1-r}{2r}$ & $\Leftrightarrow$ &
${\footnotesize \left\{ \begin{array}{l}
\frac{4 S_1(j+1)}{(j+1)_2} -\frac{(j+1)_2 +1}{[(j+1)_2]^2}
\\
\frac{8\, [j (j+3)+3]\, S_1(j+1)-6 j (j+3)- 22 }{(j)_4} -\frac{8\, [j
(j+3)+3]\,[2 j (j+3)+3]}{[(j)_4]^2}
\end{array}\right. }$
\\
$\frac{\ln \bu+u}{u^2}$  & $\Leftrightarrow$ & $-\frac{2r}{(1+r)^2}$ &
$\Leftrightarrow$ &
${\footnotesize \left\{ \begin{array}{l}
\frac{(j+1)_2}{2}
\left[S_2\!\Big(\!\frac{j+1}{2}\!\Big)-S_2\!\Big(\!\frac{j}{2}\!\Big)
\right]-1
\\
-\frac{2}{(j+1)_2}
\end{array}\right. }$
\\
$\frac{\ln^2 \bu}{u^2}$ & $\Leftrightarrow$ &
$-\frac{4r}{(1+r)^2}\ln\frac{1-r}{2r}$ & $\Leftrightarrow$ &
${\footnotesize \left\{ \begin{array}{l}
-- \\
\frac{8\, S_1(j+1)-6}{(j+1)_2}-\frac{4}{[(j+1)_2]^2}
\end{array}\right. }$
\\\hline
\end{tabular}
\end{center}
\caption{\label{subT-2}
\small Substitution rules among subtracted log functions (left column),
their imaginary parts (middle column), and
conformal moments (right column) for quarks (upper lines) and gluons (lower
lines), presented in terms of
Pochhammer`s symbols (\ref{Pochhammer}) and harmonic sums (\ref{Sq}).
}
\end{table}
In the  NLO expressions we  also  encounter terms  which possess only
logarithmical $[1,\infty]$-cuts. Since the LO pole at $u=1$ (or $u=0$) is
absent, such terms can be in general
considered as rather harmless. They can be expressed by means of the
following building blocks (analogously for $u\to v$)
\begin{equation}
\label{eq:uv-func-3}
\frac{\ln(\bu-i\epsilon)}{u}\,,\quad \frac{\ln^2(\bu-i\epsilon)}{u} \,,
\left[\frac{\ln\bu}{u^2}\right]^{\rm sub} \equiv
\frac{\ln(\bu-i\epsilon)+u}{u^2}\,,
\quad  \frac{\ln^2(\bu-i\epsilon)}{u^2}\,,
\end{equation}
where  terms proportional to $1/u^2$  may  occur in the original NLO expressions only in the gluon-quark channel,
see the convolution formula (\ref{T^G-pQCD_TcoV}).
Note that $\ln(\bu-i\epsilon)/(u-i\epsilon)^2$ possesses also a pole at $u=0$, whose
imaginary part is taken according to  Feynman`s causality prescription.
This pole is removed in $[\ln(\bu-i \epsilon)+u]/u^2$ by subtraction.  Thus, the
imaginary part of all building blocks (\ref{eq:uv-func-3}) is simply
determined by the logarithmical cut. Hence, from the imaginary part of $\ln^p(\bu - i\epsilon)$,
see (\ref{imf^p_1+(u|a,0)}) with $a=0$ for $p\in\{1,2\}$,   we find the
imaginary parts of subtracted functions and
setting $u= (1+r)/2r$ as well as taking into account the prefactor $1/2\pi
r$, see (\ref{^sigmat}), the following correspondences emerge
\begin{eqnarray}
\label{subT`u2im2-ln}
\im \left[\frac{\ln^p \bu}{u^a}\right]^{\rm sub}\!\! = -\pi\theta(u-1)
\frac{\ln^{p-1}(u-1)}{u^a}  \;\;\Rightarrow \;\;  - \frac{\theta(r)}{2
r}\,\frac{(2r)^a \ln^{p-1}\!\frac{1-r}{2r}}{(1+r)^a}  \;\;\mbox{for}\;\;
p\in\{1,2\}\,,
\quad
\end{eqnarray}
where the superscript $^{\rm sub}$ is superfluous if no poles are present.

The conformal moments of the building blocks (\ref{eq:uv-func-3}) can be
essentially read off for the quark channel from Appendix C in \cite{Melic:2002ij} and for the
gluonic one from  appendix C.1 in \cite{Kumericki:2007sa}, where the
normalization factors $1/2 N_j$ and $1/12 N^{(5/2)}_{j-1}$, explicitly shown there,
must be neglected. The quark conformal moments of $(\ln\bu+u)/u^2$ were given
in \cite{Mueller:1998qs} only in terms of a hypergeometric function $_3F_2$
with unit argument. On the other hand they can be easily evaluated from its
imaginary part by adopting the Mellin moment technique
(\ref{^Ac_{jk}-Mellin}). Performing the integral (\ref{^Qm_{jk}-Mellin})
provides the conformal moments (\ref{^Ac_{nk}}) in terms of Mellin
moments (\ref{^Ac_{jk}-Mellin-1}) %
\footnote{Obviously, the $v$-convolution can be here ignored in all of these
equations.},
\begin{subequations}
\label{<lnbu/u^2>_j}
\begin{eqnarray}
\label{<lnbu/u^2>_j-1}
\left[\frac{\ln\bu}{u^2}\right]^{\rm sub} \stackrel{u}{\otimes}
\widehat{p}^{\frac{3}{2}}_n(u) = -1-
\frac{(n+1)_2}{2} \int_{0}^1\! dw\, \frac{4\ln w}{1+w}\,  w^{n+1}\,,
\end{eqnarray}
where the remaining integral represents the difference of two second order
harmonic sums (\ref{Sq}),
\begin{eqnarray}
\label{<lnbu/u^2>_j-2}
- \int_{0}^1\! dw\, \frac{4\ln w}{1+w}\,  w^{n+1} =
S_2\!\Big(\!\frac{n+1}{2}\!\Big)-S_2\!\Big(\!\frac{n}{2}\!\Big) \,,
\end{eqnarray}
\end{subequations}
with half integer argument. Note that it is popular to express such combinations for integer $n$ by a
sign alternating sum $S_{-p}$, i.e.,
\begin{eqnarray}
\label{Smp}
\Delta S_p \!\Big(\!\frac{n+1}{2}\!\Big) \equiv
S_p\!\Big(\!\frac{n+1}{2}\!\Big)-S_p\!\Big(\!\frac{n}{2}\!\Big) = (-1)^{n+1}
2^p \left[S_{-p}(n+1) + \left(1-2^{1-p}\right) \zeta(p)\right]\,.
\end{eqnarray}
Since for the analytic continuation of $S_{-p}$ one has to fix the signature
first, which may provide some confusion, we prefer to present our results in
terms of harmonic sums with half integer arguments, however,
for shortness we will denote their difference (\ref{Smp}) with the
symbol $\Delta S_p$.

Finally, we list the building blocks (\ref{eq:uv-func-3}), the substitution
rules (\ref{subT`u2im2-ln}), and the corresponding conformal moments in
Tab.~\ref{subT-2}. Compared to the LO pole $1/\bu$, their mild
logarithmical behavior in the vicinity of $u=1$ and for their imaginary
parts at $r=1$
implies that their  conformal moments vanish in the limit $j\to \infty$ as
$1/j^2$ or $(\ln j)/j^2$.

\begin{itemize}
\item{\em Building blocks with dilog`s.}
\end{itemize}
\vspace{-2mm}
We also encounter in the NLO hard scattering amplitudes terms that contain the dilog (or
Spence) function $\Li(u+i\epsilon)$, where  causality implies
the $u+i\epsilon$-prescription.  This function  behaves in the vicinity of
$u=0$ as $u + O(u^2) $ and it contains a logarithmical $[1,\infty]$-cut,
i.e.,
\begin{equation}
\label{imLi(u)}
\im \Li(u+i\epsilon) = \pi \theta(u-1) \ln u.
\end{equation}
We need the following two building blocks (analogously for $u\to v$)
\begin{equation}
\label{eq:uv-func-4}
\frac{\Li(u+i\epsilon)}{u}\quad \mbox{and} \quad
\frac{\Li(u+i\epsilon)}{u^2}\;\;\mbox{or}\;\;\left[\frac{\Li(u+i\epsilon)}{u^2}\right]^{\rm
sub} \equiv  \frac{\Li(u+i\epsilon)-u}{u^2}\,,
\end{equation}
where the single pole in $\Li(u)/u^2$  is  subtracted.
Furthermore, dilog functions appear also accompanied by poles at $u=1$.
Although $u=1$ is a branch point we can nevertheless subtract these poles.
To keep track on the most singular pieces in the hard scattering amplitudes, we
introduce the following subtracted building blocks (analogously for $u\to v$)
\begin{equation}
\label{eq:uv-func-5}
\left[\frac{\Li(u)}{\bu}\right]^{\rm sub} \equiv
\frac{\Li(u)-\zeta(2)}{\bu} \quad\mbox{and}\quad
\left[\frac{\Li(u)}{\bu^2}\right]^{\rm sub} \equiv
\frac{\Li(u)-\zeta(2)-\bu\ln\bu +\bu}{\bu^2}\,,
\end{equation}
which possess harmless logarithmical singularities in the vicinity of $u=1$
and approach a constant at $u=0$. For all of our subtracted building
blocks
we can easily evaluate their imaginary parts from (\ref{imLi(u)}), yielding
with $u= (1+r)/2r$ and the prefactor $1/2\pi r$ the substitution rules
\begin{subequations}
\label{subT`u2im2-Li}
\begin{eqnarray}
\label{subT`u2im2-Li-1}
\im \left[\frac{\Li(u)}{u^a}\right]^{\rm sub} = \pi \theta(1-u) \frac{\ln
u}{u^a}
\quad \Rightarrow \quad \frac{\theta(r)}{2r} \frac{(2r)^a
\ln\frac{1+r}{2r}}{(1+r)^a}
\end{eqnarray}
for (\ref{eq:uv-func-4}) and
\begin{eqnarray}
\left[\frac{\Li(u)}{\bu}\right]^{\rm sub} \Rightarrow
\quad  -\frac{\theta(r)}{2r} \frac{2r \ln\frac{1+r}{2r}}{1-r}
\quad\mbox{and}\quad
\left[\frac{\Li(u)}{\bu^2}\right]^{\rm sub} \Rightarrow \quad
\frac{\theta(r)}{2r} \frac{2r \left[2r \ln\frac{1+r}{2r} -
1+r\right]}{(1-r)^2}
\nonumber \\
\label{subT`u2im2-Li-2}
\end{eqnarray}
\end{subequations}
for (\ref{eq:uv-func-5}), where in the last rule also the imaginary part of
the accompanying $\ln\bu$ function is taken into account, cf.
(\ref{subT`u2im2-ln}).

\begin{table}[t]
\begin{tabular}{|ccccl|}
\hline
$\frac{\Li(u)}{u} $ & $\Leftrightarrow$ & $\frac{\ln\frac{1+r}{2r}}{1+r}$ &
$\Leftrightarrow$ &
${\footnotesize \left\{ \begin{array}{l}
-\frac{1}{2}\left[
S_2\!\Big(\!\frac{j+1}{2}\!\Big)-S_2\!\Big(\!\frac{j}{2}\!\Big)\right]+
\frac{(j+1)_2+1}{[(j + 1)_2]^2}
\\
\textstyle\frac{1}{2}
\left[S_2\!\Big(\!\frac{j+1}{2}\!\Big)-S_2\!\Big(\!\frac{j}{2}\!\Big)\right]+\frac{18-j
(j+3)}{2j^2 (j+3)^2}-\frac{2+(j+1)_2}{2[(j+1)_2]^2}
\end{array}\right. }$
\\
$\frac{\Li(u)-u}{u^2}$  & $\Leftrightarrow$ &
$\frac{2r\ln\frac{1+r}{2r}}{(1+r)^{2}}$ & $\Leftrightarrow$ &
${\footnotesize \left\{ \begin{array}{l}
1+ (j+1)_2 \Big\{\frac{1}{4}\!\left[
S_3\!\Big(\!\frac{j+1}{2}\!\Big)-S_3\!\Big(\!\frac{j}{2}\!\Big)\right]+
\left[S_2\!\Big(\!\frac{j+1}{2}\!\Big)-S_2\!\Big(\!\frac{j}{2}\!\Big)\right]
\\
\phantom{(1)^{j}} \times \left[S_1(j+1)-\frac{1}{2}\right] +4 (-1)^{j}
\left[S_{-2,1}(j+1)+\frac{5\zeta(3)}{8}\right]\!\Big\}
\\
\frac{j (j+3)}{2} \left[S_2\!\Big(\!\frac{j+1}{2}\!\Big) -
S_2\!\Big(\!\frac{j}{2}\!\Big)\right]+ \frac{2 + 3(j+1)_2}{[(j+1)_2]^2}-1
\end{array}\right. }$
\\
$\frac{\Li(u)-\zeta(2)}{\bu}$  & $\Leftrightarrow$ & $-
\frac{\ln\frac{1+r}{2r}}{1-r}$  & $\Leftrightarrow$ &
${\footnotesize \left\{ \begin{array}{l}
-\frac{(j+1)_2+1}{[(j + 1)_2]^2}
\\
-\frac{18-j(j+3)}{2j^2 (j+3)^2}-\frac{2+5 (j+1)_2}{2[(j+1)_2]^2}
\end{array}\right. }$
\\
$\frac{\Li(u)-\zeta(2)-\bu \ln\bu + \bu}{\bu^2}$  & $\Leftrightarrow$ &
$\frac{2r\ln\frac{1+r}{2r}-1+r}{(1-r)^2}$  & $\Leftrightarrow$ &
${\footnotesize \left\{ \begin{array}{l}
2(j+1)_2\left[ S_3(j+1)-\zeta(3) \right]+1-\frac{1}{(j+1)_2}
\\ --
\end{array}\right. }$
\\\hline
\end{tabular}
\caption{\small Substitution rules among subtracted dilog functions (left
column), their imaginary parts (middle column), and
conformal moments (right column) for quarks (upper line) and gluons (lower
line), presented in terms of
Pochhammer`s symbol (\ref{Pochhammer}) and harmonic sums
(\ref{Sq},\ref{Sm21}).
\label{subT-Li}}
\end{table}
The  conformal moments of the first building block in (\ref{eq:uv-func-4})
and (\ref{eq:uv-func-5}) for quarks  can be found in appendix~C of
\cite{Melic:2002ij}. Analogously as for (\ref{<lnbu/u^2>_j}), for the
remaining ones in the quark sector we obtain from the integral
transformation (\ref{^Qm_{jk}-Mellin}) of the corresponding imaginary parts
(\ref{subT`u2im2-Li}) the Mellin moments
\begin{subequations}
\label{<Li(u)/u2>_j <Li(u)/bu2>_j-3/2}
\begin{eqnarray}
\label{<Li(u)/u2>_j-3/2}
\left[\frac{\Li(u)}{u^2}\right]^{\rm sub} \stackrel{u}{\otimes}
\widehat{p}^{\frac{3}{2}}_j(u)
&\!\!\!= \!\!\! & 1+
2(j+1)_2\!\! \int_0^1\!\!dw \frac{\left[1-\frac{1}{2}\ln w+
2\ln(1\!+\!w)\right]\ln w +2 \Li(-w)+ \zeta(2)}{1+w} w^{j+1},
\nonumber\\
\\
\label{<Li(u)/bu2>_j-3/2}
\left[\frac{\Li(u)}{\bu^2}\right]^{\rm sub} \stackrel{u}{\otimes}
\widehat{p}^{\frac{3}{2}}_j(u) &\!\!\!= \!\!\! &
1-\frac{1}{(j+1)_2}+(j+1)_2\!\! \int_0^1\!\!dw \frac{\ln^2 w}{1-w} w^{j+1},
\end{eqnarray}
\end{subequations}
where we utilized integration by parts. The  Mellin integrals yield third
order  harmonic sums
$$\quad \int_0^1\!\!dw \frac{\ln^2 w}{1-w} w^{j+1}  =-2S_3(j+1) + 2
\zeta(3)\,,  \quad\mbox{see also (\ref{Sq})},$$
while the remaining integral in (\ref{<Li(u)/u2>_j-3/2}) can be read off from
\cite{Vermaseren:1998uu,Blumlein:1998if},
\begin{eqnarray}
&&\!\!\!\!\!\! \int_0^1\!\!dw \frac{\left[1-\frac{1}{2}\ln w+
2\ln(1\!+\!w)\right]\ln w +2 \Li(-w)+ \zeta(2)}{1+w} w^{j+1}
=(-1)^{j+1}
\\
&&\times\!\left\{\!
S_{-3}(j+1)+\frac{3\zeta(3)}{4}+\left[2S_1(j+1)-1\right]\!\left[\!
S_{-2}(j+1)+\frac{\zeta(2)}{2}\! \right]
-2\left[\! S_{-2,1}(j+1)+\frac{5\zeta(3)}{8}\! \right]\! \right\},
\nonumber
\end{eqnarray}
represented as combination of harmonic sums  with negative order. Here, we
have
\begin{eqnarray}
\label{Sm21}
(-1)^{j+1}\left[S_{-2,1}(j+1)+\frac{5\zeta(3)}{8}\right] = \int_0^1\!dw\,
\frac{\zeta(2)-\Li(w)}{1+w} w^{j+1} \,.
\end{eqnarray}
For gluons the integral transformation (\ref{^Gm_{jk}-Mellin}) of
the corresponding imaginary parts (\ref{subT`u2im2-Li}) of (subtracted)
dilog building blocks  leads after integration by parts to integral representations
of rational functions and/or  second order harmonic sums (\ref{Smp}).

Finally, we list our results for the building blocks
(\ref{eq:uv-func-4},\ref{eq:uv-func-5}), their imaginary parts
(\ref{subT`u2im2-Li}), and conformal moments  as substitution rules in
Tab.~\ref{subT-Li}. Again we can consider them as rather harmless. They have only
logarithmical $[1,\infty]$-cuts on the $u$-axis, a vanishing or
constant behavior of their imaginary parts at $r=1$, and vanishing conformal
moments in the limit $j\to \infty$.

\pagebreak[3]
\begin{itemize}
\item{\em Peculiarities at $u\to \infty$ and for $n=0$.}
\end{itemize}
\vspace{-2mm}
We add that only in the pure singlet quark contribution the functions
$$\ln(\bu-i\epsilon)\,,\quad \ln^2(\bu-i\epsilon)\,, \quad\mbox{and}\quad
\Li(u+i\epsilon)$$
appear, which do not vanish in the limit $u\to\infty$.
Their imaginary parts are obtained from (\ref{eq:uv-func-3}) and
(\ref{subT`u2im2-Li-1}) with $a=0$ and they are singular at $r=0$.
Their conformal moments for $n\ge 1$ are easily
calculated, e.g., using Rodrigues formula, and their analytic continuation
yields the substitution rules
\begin{eqnarray}
\label{<lnbu>_j}
\ln\bu \Rightarrow \frac{-1}{j(j+3)}, \quad  \ln^2\bu \Rightarrow
\frac{6}{j^2 (j+3)^2}+\frac{4 S_1(j+3)+3}{j (j+3)} +\frac{1}{(j+1)_2},
\quad
\Li(u) \Rightarrow \frac{2(j+1)_2+2}{[(j)_4]^2} .
\nonumber\\
\end{eqnarray}
Note that the integral conformal moments for $n=0$ are finite. Nevertheless,
the $j=0$ poles in (\ref{<lnbu>_j}) are the correct results
for the analytic  continuation of odd integral conformal moments.
Since the second order pole at $j=0$  cancel at the end, we list specific
combinations of these building blocks in
Tab.~\ref{subT-pS}, given below.  We add that in the gluon-quark channel such
$j=0$ poles appear, too, see Tabs. \ref{subT-1}--\ref{subT-Li}, where the
second order pole will also disappear in the final NLO result.

\pagebreak[3]
\begin{itemize}
\item{\em Exploiting symmetry.}
\end{itemize}
\vspace{-2mm}
As explained in Sec.~\ref{sec:prel-DR-TFF}, we can exploit symmetry to
express the hard scattering amplitudes with definite signature
in such a manner that they are holomorphic except for discontinuities on the
positive axis. This can be achieved by means of a $u\to \bu$  transformation which maps possible terms with poles at
$u=0$ and logarithmical $[-\infty,0]$-cuts  along the negative axis, e.g.,
\begin{equation}
\label{eq:uv-func-2}
\frac{\ln^p u}{u}\,, \quad
\frac{\ln^p u}{\bu}\,, \quad  \left[\frac{\ln u}{\bu^a}\right]^{\rm sub}\,,
\quad \left[\frac{\Li(\bu)}{\bu^a}\right]^{\rm sub} \,, \quad
\left[\frac{\Li(\bu)}{u^a}\right]^{\rm sub}
\end{equation}
to those in (\ref{eq:uv-func-3}), having poles at $u=1$ and logarithmical
$[1,\infty]$-cuts
along the positive axis.  According to (\ref{^sigmat}) and (\ref{Im^GT}), in
such a $u\to \bu$ or $r\to -r$  map a
signature factor $-\sigma$ and $\sigma$ has to be included in the resulting quark
and gluon building blocks, respectively.
Conformal moments are getting decorated with a factor $(-1)^j$, which is
replaced in the quark-quark and gluon-quark channel by $-\sigma$ and $\sigma$,
respectively.
For building blocks that depend on $v$, the momentum fraction of the meson
DA, an additional $-1$ factor appears only for anti-symmetric DAs, which we
do not consider here.  Note that the corresponding factor $(-1)^k$ in the conformal
moments of (anti-)symmetric DAs can be set to $+1$ ($-1$).

Obviously, we can always eliminate functions that have cuts along the
positive and negative real axes. In the NLO hard scattering amplitudes we only
encounter the term $\ln u \ln\bu$, decorated with some rational function.
Utilizing the formula
\begin{equation}
\label{eq:Li2-identity}
\Li(u+i \epsilon) +\Li(\bu+i \epsilon) + \ln(u -i \epsilon) \ln(\bu-i\epsilon) - \Li(1) \simeq 0 \quad\mbox{with}\quad \Li(1)=\zeta(2)=\frac{\pi^2}{6} \,,
\end{equation}
getting an identity in the limit $\epsilon \to 0$,
we can split $\ln(u -i \epsilon) \ln(\bu-i\epsilon)$ in two terms that are expressed by dilog  functions.
Note that for $u\ge 1$ ($u<0$) the $u+i\epsilon$  ($\bu+i\epsilon$)
prescription in the dilog  is  consistent with the $\bu-i \epsilon$ ($u-i
\epsilon$) one in the log. Finally, we employ then a $u\to \bu$ map and  subtract possible poles in an
appropriate manner to get rid of  $\Li(\bu)$ and poles at $u=0$.

\subsubsection{Building blocks for non-separable NLO terms}
\label{sec:NLO-blocks2}

All non-separable addenda (\ref{eq:ATc-dec}) in the various channels will be expressed in terms of
\begin{eqnarray}
\label{L_{ab}}
\frac{1}{u^a \bv^b} \times \frac{\Li(\bv)  -\Li(\bu)   +\ln v \ln \bu  -\ln u \ln \bu}{u-v}
\end{eqnarray}
and its derivatives w.r.t.~the $v$ variable, where the poles at $u=0$ and $v=1$ are of first and/or second order.  Note that the representation of non-separable terms is not unique, since one might use another combination of dilog and log functions. Moreover,  the accompanying rational function can be chosen differently, e.g.,
 $$
 \frac{1}{v}\, \frac{\Li(\bv)  -\cdots}{u-v} = \frac{1}{u}\, \frac{\Li(\bv)  -\cdots}{u-v} + \frac{\Li(\bv)  -\cdots}{u v}\,.
 $$

To clarify the analytic properties of the building blocks (\ref{L_{ab}}) and to simplify their treatment, yielding the representation
that is given below in (\ref{Delta T^{cdots}(u,v)}), we study the auxiliary function
\begin{subequations}
\label{eqs:L}
\begin{eqnarray}
\label{L}
L(u,v) =
\Li(\bv)  -\Li(\bu)   +\ln v \ln \bu  -\ln u \ln \bu
\end{eqnarray}
and its derivatives
\begin{eqnarray}
\label{Lu,Lv,Luv}
L_u(u,v)  & \!\!\! \equiv  \!\!\!  & \frac{\partial}{\partial u} L(u,v) =  -\frac{\ln\bu}{u} - \frac{\ln v }{\bu} \,,
\quad
L_v(u,v) \equiv \frac{\partial}{\partial v} L(u,v)=\frac{\ln \bu }{v} + \frac{\ln v }{\bv}\,,
\nonumber \\
L_{u,v}(u,v)  & \!\!\! \equiv  \!\!\!  & \frac{\partial^2}{\partial u \partial v} L(u,v)= -\frac{1}{\bu v}
\, .
\end{eqnarray}
First we note that the function $L(u,v)$ vanishes in the vicinity of $u=v$ as $(u-v)$ and, thus, our building blocks (\ref{L_{ab}}) exist also on the line $u=v$.
Furthermore, by means of the dilog identity (\ref{eq:Li2-identity})
we can express the $L$-function (\ref{L}) also as
 \begin{eqnarray}
\label{L-1}
L(u,v) = \Li(\bv)  +\Li(u)   +\ln \bu \ln v   -\zeta(2) \,.
\end{eqnarray}
Hence, this function (\ref{L},\ref{L-1})  posses due to the $\ln \bu$ and $\Li(u)$ terms a cut $[1,\infty]$ on the real  $u$-axis and due to the $\ln v$ and $\Li(\bv)$ terms  a cut $[-\infty,0]$ on the real $v$-axis, while it is holomorphic in the vicinity of $u=0$ and $v=1$.  At these points the function has the values
\begin{equation}
L(u=0,v) = \Li(\bv)-\zeta(2) \quad\mbox{and}\quad L(u,v=1) = \Li(u)-\zeta(2)\,,
\end{equation}
respectively. Finally, the identity (\ref{eq:Li2-identity}) tells us also that the representation
$$L(u,v) = \Li(u) -\Li(v)  +\ln \bu \ln v   -\ln\bv \ln v$$
holds true. Comparing this formula with the definition (\ref{L}), we realize that
the $u\leftrightarrow v$ exchange arises from a simultaneous $u\to\bu$ and $v\to \bv$ exchange, i.e., we have the $u\leftrightarrow \bv$ symmetry relation
 \begin{eqnarray}
\label{L-sym}
L(v,u)=L(\bu,\bv)\,.
\end{eqnarray}
\end{subequations}
More details on the $L$  function are given in App. \ref{app:H0}.

Since the $L(u,v)$-function is holomorphic in the vicinity of $u=0$ and $v=1$, we can straightforwardly
subtract the poles in the building blocks (\ref{L_{ab}}).
We will heavily utilize, e.g., in the pure  singlet quark and gluon-quark channel, the pole subtracted expression
 \begin{subequations}
 \label{eqs:Lsub}
\begin{eqnarray}
\label{eq:Lsub-u1v1}
\left[
\frac{1}{u\bv}\,\frac{L(u,v)}{u-v}
\right]^{\rm sub} &\!\!\! \equiv \!\!\!&
\frac{L(u,v)}{u(u-v)\bv} +\frac{L(u,v=1)}{u\bu\bv}+\frac{L(u=0,v)}{u v\bv}-\frac{L(u=0,v=1)}{u\bv}\,,
\end{eqnarray}
which is symmetric under $u\leftrightarrow \bv$-reflection. To shorten the notation in the flavor non-singlet channel  we also introduce
the associated building blocks
\begin{eqnarray}
\label{eq:Lsub-v1}
\left[
\frac{1}{\bv}\,\frac{L(u,v)}{u-v}
\right]^{\rm sub} &\!\!\! \equiv \!\!\!&
\frac{L(u,v)}{(u-v)\bv} + \frac{L(u,v=1)}{\bu\bv}\,,
\\
\label{eq:Lsub-v2}
\left[
\frac{1}{\bv^2}\,\frac{L(u,v)}{u-v}
\right]^{\rm sub} &\!\!\! \equiv \!\!\!&
\frac{L(u,v)}{(u-v)\bv^2} +\frac{(\bu+\bv)L(u,v=1)}{\bu^2 \bv^2}-\frac{L_v(u,v=1)}{\bu\bv}\,,
 \end{eqnarray}
 and their $u\leftrightarrow \bv$-reflected analog, see the symmetry relation (\ref{L-sym}),
 \begin{eqnarray}
 \label{eq:Lsub-u1}
\left[\frac{1}{u}\, \frac{L(u,v)}{u-v}\right]^{\rm sub} &\!\!\! \equiv \!\!\!& \frac{L(u,v)}{u(u-v)} + \frac{L(u=0,v)}{u v} \,,
 \\
  \label{eq:Lsub-u2}
 \left[\frac{1}{u^2}\,\frac{L (u,v)}{u-v} \right]^{\rm sub} &\!\!\! \equiv \!\!\!&
\frac{L(u,v)}{u^2(u-v)} + \frac{(u+v) L(u=0,v)}{u^2 v^2}+ \frac{L_u (u=0,v)}{u v}  \,.
\end{eqnarray}
 \end{subequations}
These non-separable building blocks can now be  considered as rather harmless, where the reshuffled
subtraction terms, separable in the $u$ and $v$ variables,  contain only one pole in $u$ or $v$ that is accompanied with a rather harmless
function in $v$ or $u$.

Finding such a representation (\ref{L_{ab}}), where the poles are now subtracted, and the associated differential operator, which we generically call $\vec{\cal D}^{\cdots,ab}_v$, labeled by
the (negative) powers $a$ and $b$ of the accompanying $u$ and $\bv$ factors for the color structure  $^{\cdots}$ in a given channel,
is now a straightforward algebraic procedure. It leads us to the following simple form of the addenda (\ref{eq:ATc-dec})
\begin{eqnarray}
\label{Delta T^{cdots}(u,v)}
\Delta T^{\cdots}(u,v) = \sum_{a,b} \vec{\cal D}^{\cdots,ab}_v \left[\frac{1}{u^a\, \bv^b}\frac{L(u,v)}{u-v}\right]^{\rm sub}
\end{eqnarray}
in a given channel. Note that $\vec{\cal D}^{\cdots,ab}_v$ can be a second order differential, first order differential, or simply
a multiplication operator.

Since we have removed all poles in the subtracted building blocks (\ref{eqs:Lsub}), their
imaginary parts follow simply from the imaginary part of the $L$ function (\ref{L-1}) and the associated subtraction terms, i.e., we can simply
apply the rules (\ref{subT`u2im2-ln},\ref{subT`u2im2-Li-1}) for log and dilog functions,
$$
\frac{L(u,v)}{u-v} \Rightarrow  \frac{ \theta(r)\ln\frac{1+r}{2 r v}}{1+r -2r v}\,,\quad \ln \bu \Rightarrow - \frac{\theta(r)}{2 r}\,,
\quad \ln^2\bu \Rightarrow - \frac{2\theta(r)\ln\frac{1-r}{2r}}{2 r}\,,
$$
and set  in the remaining rational functions $u=(1+r)/2r$.
Plugging the explicit expressions (\ref{eqs:L}) into (\ref{eqs:Lsub}), we obtain the following substitution rule
\begin{subequations}
\label{subsL`u2im}
\begin{eqnarray}
\label{subsL`u2im-u}
\left[\frac{1}{u^a\, \bv^b}\frac{L(u,v)}{u-v}\right]^{\rm sub}
&\Leftrightarrow&   \left[\frac{\theta(r)\,(2 r)^a}{(1+r)^a\, \bv^b}\, \frac{\ln\frac{1+r}{2 r v}}{1+r -2r v} \right]^{\rm sub}
\\
 &&\equiv  \frac{\theta(r)\, (2 r)^a}{(1+r)^a\, \bv^b}\left[
\frac{\ln\frac{1+r}{2 r v}}{1+r -2r v} - \sum_{i=0}^{b-1}\frac{(-\bv)^i}{i!} \frac{\partial^i}{\partial v^i} \frac{\ln\frac{1+r}{2 r v}}{1+r -2r v} \Big|_{v=1}
\right],\qquad
\nonumber
\end{eqnarray}
where in the $b=0$ case no subtraction appears and for $b\in\{1,2\}$ the subtraction terms read
 \begin{eqnarray}
\frac{\ln\frac{1+r}{2 r v}}{1+r -2r v} \Big|_{v=1} =  \frac{\ln\frac{1+r}{2 r}}{1-r}\,, \quad
 \frac{\partial}{\partial v} \frac{\ln\frac{1+r}{2 r v}}{1+r -2r v} \Big|_{v=1} = \frac{2 r\ln\frac{1+r}{2 r}-1+r}{(1-r)^2}\,.
 \end{eqnarray}
\end{subequations}
The substitution (\ref{subsL`u2im}) provides then the imaginary part $\Delta t^{\cdots}(r,v)$ of the addenda $\Delta T^{\cdots}(u,v)$, where the differential operator in (\ref{Delta T^{cdots}(u,v)}) remains the same.

We also write the conformal moments of the addenda (\ref{eq:ATc-dec}) in one-to-one correspondence to the representation (\ref{Delta T^{cdots}(u,v)}) as
\begin{equation}
\label{Delta^{A}c^{cdots}_{jk}-def}
\Delta ^{\text{A}}c^{\cdots}_{jk} = \sum_{a,b}\Delta ^{\text{A}}c^{\cdots,ab}_{jk}\quad\mbox{with}\quad
\Delta ^{\text{A}}c^{\cdots,ab}_{jk} = \widehat{p}^{\text{A}}_j(u)\stackrel{u}{\otimes}
\left[\frac{1}{u^a\, \bv^b}\frac{L(u,v)}{u-v}\right]^{\rm sub}\stackrel{v}{\otimes}\vec{\cal D}^{^\dag\cdots,ab}_v\, \widehat{p}^{\frac{3}{2}}_k(v)\,,
\end{equation}
where $\vec{\cal D}^{^\dag\cdots,ab}_v $ is the adjoint differential (or simply a multiplication) operator. To perform the analytic continuation
in $j$, we can utilize the representation (\ref{T2cjk}) of conformal moments in terms of associated Legendre functions. Plugging the
imaginary parts  (\ref{subsL`u2im})  into (\ref{T2cjk}) and switching to the variable $y=2r/(1+r)$ we obtain the non-separable conformal
moments (\ref{Delta^{A}c^{cdots}_{jk}-def}) of the addenda,
\begin{equation}
\label{Delta^{A}c^{cdots}_{jk}}
\Delta ^{\text{A}}c^{\cdots}_{jk} = \sum_{a,b}   \widetilde{p}_j^{\nu(\text{A})}(y) \stackrel{y}{\otimes}
\frac{y^a}{\bv^b}\left[
\frac{-\ln(y v)}{1-y v} + \sum_{i=0}^{b-1}\frac{(-\bv)^i}{i!} \frac{\partial^i}{\partial v^i} \frac{\ln(y v)}{1- y v} \Big|_{v=1}
\right]\stackrel{v}{\otimes}  \vec{\cal D}^{^\dag\cdots,ab}_v\,  \widehat{p}^{\frac{3}{2}}_k(v)\,.
\end{equation}
This formula allows us to evaluate the conformal moments for complex valued $j$ and non-negative integer $k$ numerically. Furthermore, we will choose the second order differential operator as the defining one for Gegenbauer polynomials with index $3/2$
\begin{subequations}
\label{dif-op}
\begin{equation}
\label{eq:eigenvalue-C^{3/2}_k}
v\bv\frac{d^2}{dv^2}\, \widehat p_k^{3/2}(v) = -(k+1)_2\; \widehat p_k^{3/2}(v)\,,
\end{equation}
and take the following set of first order differential operators
\begin{eqnarray}
\label{dif-op1}
v \bv \frac{d}{dv}\,  \widehat p_k^{3/2}(v) &\!\!\!=\!\!\!&
\frac{(k+1)_2}{2(2 k+3)}\, \widehat p_{k-1}^{3/2}(v) -\frac{ (k+1)_2 }{2(2 k+3)}\,\widehat p_{k+1}^{3/2}(v)\,,
\\
\label{dif-op2}
(v \bv)^2 \frac{d}{dv}\,  \frac{\widehat p_k^{3/2}(v)}{v\bv}  &\!\!\!=\!\!\!&
\frac{(k+2)_2}{2(2 k+3)}\, \widehat p_{k-1}^{3/2}(v) -\frac{ (k)_2 }{2(2 k+3)}\,\widehat p_{k+1}^{3/2}(v)\,,
\\
\label{dif-op3}
(v-\bv) v \bv\frac{d}{dv}\,  \frac{\widehat p_k^{3/2}(v)}{v\bv}  &\!\!\!=\!\!\!& 2k\, \widehat p_k^{3/2}(v)+ \sum_{l=0}^{k-1}\left[1+(-1)^{k-l}\right](2 l + 3)\widehat p_l^{3/2}(v) \,,
\qquad\quad
\end{eqnarray}
which allows us to replace the differential operator in (\ref{Delta^{A}c^{cdots}_{jk}}) by two Gegenbauer polynomials or in the case of (\ref{dif-op3}) by a finite sum over them.
\end{subequations}
We add that $k$ can be analytically continued in complete analogy to the procedure that we used for $j$ by means
of a double dispersion integral. An example is given in App.~\ref{app:L}.

To improve the  efficiency of the numerical evaluation, we calculated the non-separable conformal moments in terms of harmonic sums, where  (\ref{Delta^{A}c^{cdots}_{jk}-def},\ref{eq:eigenvalue-C^{3/2}_k}) tells us that only those of the building blocks (\ref{eqs:Lsub}) are needed.
We denote these moments as
\begin{eqnarray}
L^{\nu,a,b}_{nk} = \widehat p_n^{\nu}(u) \stackrel{u}{\otimes} \left[\frac{1}{u^a\, \bv^b}\frac{L(u,v)}{u-v}\right]^{\rm sub}\!\!\! \stackrel{v}{\otimes}  \widehat p_k^{3/2}(v) \quad\mbox{with}\quad \nu\in\{3/2,5/2\}
\end{eqnarray}
and reduce their evaluation to the $a=1,b=1$ and $\nu=3/2$ case. This task can be done for any given $k$ by a straightforward calculation, in the
following we derive a closed expression for them.

To do so let us first derive a set of algebraic reduction formulae.  For the quark case $\nu=3/2$ we can exploit the $u\leftrightarrow \bv$ reflection symmetry which implies the  relation
\begin{subequations}
\begin{eqnarray}
\label{L_{nk}-sym}
L_{nk}^{\frac{3}{2},a,b} = (-1)^{n-k} L_{kn}^{\frac{3}{2},b,a}\,.
\end{eqnarray}
The gluonic conformal moments are most easily obtained by decomposing the gluonic CPW in terms of quark ones,
$$
\widehat{p}_n^{\frac{5}{2}}(u) =  \frac{(n+2)_2}{2(2 n+3)}\,  \widehat{p}_{n-1}^{\frac{3}{2}}(u) -\frac{(n)_2}{2(2 n+3)}\,  \widehat{p}_{n+1}^{\frac{3}{2}}(u),
$$
which implies that the gluonic conformal moments of (\ref{eq:Lsub-u1v1}) are given as combination of two shifted quark conformal moments,
\begin{eqnarray}
\label{L^{frac{5}{2},a,b}_{jk}}
L_{nk}^{\frac{5}{2},a,b}=   \frac{(n+2)_2}{2(2 n+3)} L_{n-1,k}^{\frac{3}{2},a,b} - \frac{(n)_2}{2(2 n+3)} L_{n+1,k}^{\frac{3}{2},a,b}\,.
\end{eqnarray}
Furthermore, to link the $a=0,b=1$ case to the $a=1,b=1$ one, we might employ the algebraic relation among the corresponding building blocks (\ref{eq:Lsub-u1v1},\ref{eq:Lsub-v1}),
$$
\left[\frac{1}{\bv}\, \frac{L(u,v)}{u-v} \right]^{\rm sub}=  v \left[\frac{1}{u \bv}\, \frac{L(u,v)}{u-v} \right]^{\rm sub}  +
\frac{ \Li(u)- \zeta(2) u}{u\bu} + \frac{\ln\bu \ln v}{u\bv},
$$
a recurrence relation among Gegenbauer polynomials, written as
$$
v \, \widehat p_k(v) = \frac{k+1}{2(2k+3)} \widehat p_{k+1}(v)  +\frac{1}{2}  \widehat p_{k}(v)  +\frac{k+2}{2(2k+3)} \widehat p_{k-1}(v)\,,
$$
and the moments of the additional subtraction terms, listed in Tab.~\ref{subT-2} and \ref{subT-Li}, which yields
\begin{eqnarray}
\label{L_{jk}^{frac{3}{2},0,1}-rec}
L_{jk}^{\frac{3}{2},0,1} = \frac{k+1}{2(2k+3)}  L_{j,k+1}^{\frac{3}{2},1,1} + \frac{1}{2}  L_{jk}^{\frac{3}{2},1,1} +
\frac{k+2}{2(2k+3)} L_{j,k-1}^{\frac{3}{2},1,1} + \frac{(-1)^k}{(j+1)_2 (k+1)_2}\,.
\end{eqnarray}
Note that the $k=0$ case deserves special considerations. It is contained in the quoted recurrence relation, i.e., a Kronecker delta contribution $\delta_{k0}$  does finally not appear. To evaluate the $a=0$, $b=2$  case we use the algebraic relation
$$
\left[\frac{1}{\bv^2}\, \frac{L(u,v)}{u-v} \right]^{\rm sub}=  \frac{1}{\bv}\left[\frac{1}{u \bv}\, \frac{L(u,v)}{u-v} \right]^{\rm sub}-\left[\frac{1}{u \bv}\, \frac{L(u,v)}{u-v} \right]^{\rm sub} +
\frac{\Li(u)-u \zeta(2)}{u\bu \bv}+\frac{\ln\bu\ln v}{u \bv^2}
$$
and the following expansion in terms of a finite sum
$$
\frac{1}{\bv}\,\widehat p_k(v) =   2v C_k^{3/2}(1)  + \sum _{l=0}^{k} (2 l+3)\, \frac{(l+1)_2-(k+1)_2}{(l+1)_2}\, \widehat p_l(v)\,, \quad
C_k^{3/2}(1) = (k+1) (k+2)\,.
$$
Consequently, we can evaluate $L_{jk}^{\frac{3}{2},0,2} $ for fixed non-negative integer $k$ as a finite sum over the conformal moments $L_{jk}^{\frac{3}{2},1,1} $ and some additional separable terms,
\begin{eqnarray}
\label{L_{nk}^{frac{3}{2},0,2}-rec}
L_{nk}^{\frac{3}{2},0,2} &\!\!\!=\!\!\!& -\sum _{l=0}^{k} (2 l+3)\, \frac{(k-l)(k+l+3)}{(l+1)_2} L_{nl}^{\frac{3}{2},1,1} -   L_{nk}^{\frac{3}{2},1,1}-\frac{(j-k)(j+k+3)\Delta S_2\!\big(\!\frac{j+1}{2}\!\big)}{2(j+1)_2}+
\frac{ 1+ (-1)^k}{(j+1)_2}
\nonumber\\
&&- (k+1)_2 \left[2S_3(j+1)-2\zeta(3)-\frac{\Delta S_2\!\big(\!\frac{j+1}{2}\!\big)}{2}+ (-1)^k\frac{\Delta S_2\!\big(\!\frac{k+1}{2}\!\big)}{2 (j+1)_2}+\frac{\zeta(2)}{(1+j) (2+j)}\right]
\,.
\end{eqnarray}
\end{subequations}

We come now to the remaining non-trivial task, i.e., to the evaluation of $L_{jk}^{\frac{3}{2},1,1}$ in terms of harmonic sums. First  we  mapped the
symmetric building block (\ref{eq:Lsub-u1v1}), i.e., $a=1$, $b=1$, into the mixed representation%
\footnote{We do not include here the  normalization factor 3 that ensures the normalization of the DA.} (\ref{^Awidehat{t}_k(r|cdots)})
for fixed non-negative integer $k$, denoted as
\begin{eqnarray}
\label{L_k^{1,1}(r)-def}
L_k^{1,1}(r) = \int_0^1\!dv\,  \left[\frac{2r}{(1+r)\bv}\, \frac{\ln\frac{1+r}{2r v}}{1+r - 2 r v}\right]^{\rm sub} 2v\bv C_k^{\frac{3}{2}}(v-\bv)\,.
\end{eqnarray}
In this representation it is given as linear combination  of subtracted polylog functions
\begin{eqnarray}
\label{L_k^{1,1}(r)}
L_k^{1,1}(r)= \frac{2 - y}{2} \sum_{l=0}^k\frac{(-1)^{k-l}\, \Gamma(k+l+3)}{l!(l+1)!\, \Gamma(k-l+1)} \left[\frac{\Li(1-y)}{y^{l+1}}\right]^{\rm sub}\Bigg|_{\textstyle y=\frac{2r}{1+r}}\,.
\end{eqnarray}
As a side remark we note that this new result allows us to present all NLO addenda in the mixed
representation for the imaginary parts of the hard scattering  amplitudes  in a closed form, while the real parts might be restored by means of the DR.
This offers an alternative method to the `step-by-step' procedure \cite{Diehl:2007hd}.
From (\ref{L_k^{1,1}(r)}) we calculated then the quark Mellin kernels (\ref{^Qm_{jk}-Mellin}), which are given in closed form in terms of (integrated)
$\ln(w)/(1+w)$ functions.
Finally, by means of the Mellin transform (\ref{^Ac_{jk}-Mellin-1}) we find a rather simple functional form in terms of second order harmonic sums (\ref{Smp}), see also the Mellin transform (\ref{<lnbu/u^2>_j}),
\begin{eqnarray}
\label{Delta c^{frac{3}{2},1,1}_{jk}}
L_{jk}^{\frac{3}{2},1,1} = -(-1)^k\,
\frac{
(j+1)_2\, \Delta S_2\!\big(\!\frac{j+1}{2}\!\big) - (k+1)_2\, \Delta S_2\!\big(\!\frac{k+1}{2}\!\big)
}{
2(j-k)(j+k+3)
}\,.\qquad\qquad
\end{eqnarray}
These conformal moments can be used also for complex valued $k$ and they are finite for $j=k$. Implied by our subtraction procedure they
are numerically less important, e.g., $L_{00}^{\frac{3}{2},1,1}\approx -0.10$,
and behave in the limit $j\to \infty$ or $k\to \infty$ as $1/j^2$ or as $1/k^2$.

The associated building blocks for the $a=0,b=1$ or  $a=1,b=0$ cases
can be now easily calculated from the recurrence relations (\ref{L_{jk}^{frac{3}{2},0,1}-rec}) and the symmetry relation (\ref{L_{nk}-sym}). Thus,
replacing the factor $(-1)^{j+k}$ by a signature factor allows us also to give the results for complex valued $n$, i.e., $j$, and complex valued $k$.
The conformal moments for the $a=0, b=2$ building block can be straightforwardly obtained from (\ref{L_{nk}^{frac{3}{2},0,2}-rec}) for complex valued $j$
and non-negative integer $k$, where the sum over $k$ can be reduced, e.g., to
$$\sum_{l=0}^{k}\frac{(-1)^l (2l+1)
\left[\Delta S_2\!\big(\!\frac{j+1}{2}\!\big)-\Delta S_2\!\big(\!\frac{l}{2}\!\big)\right]}{4(j-l+1)(j+l+2)}.$$
Note that this finite sum might be represented as a  (double) Mellin transform, which defines the function for complex valued $k$. Such an integral can be also expressed in terms of higher order hypergeometric functions.
The integral conformal moments for the $a=2,b=0$  case  follow again from the symmetry relation (\ref{L_{nk}-sym}), where we can in addition utilize the identity
\begin{eqnarray}
\label{identity-sums}
&&\!\!\!\!\!(-1)^n \left[S_3(k+1)-\zeta(3)\right]-(-1)^n \sum_{l=0}^{n}\frac{(-1)^l (2l+1)
\left[\Delta S_2\!\big(\!\frac{k+1}{2}\!\big)-\Delta S_2\!\big(\!\frac{l}{2}\!\big)\right]}{4(k-l+1)(k+l+2)}
\\
&&\qquad=
\frac{\left[S_2\big(\!\frac{n+1}{2}\!\big) -S_2\big(\!\frac{n}{2}\!\big)\right]  \left[S_1(n+1)-S_1(k + 1)\right]
+4 (-1)^{n}\left[S_{-2,1}(n+1)+\frac{5\zeta(3)}{8}\right]}{2}
\nonumber\\
&&\qquad\qquad + \frac{S_3\big(\!\frac{n+1}{2}\!\big)-S_3\big(\!\frac{n}{2}\!\big)}{8} -\sum_{l=0}^{k}\frac{(2l+1)\left[\Delta S_2\!\big(\!\frac{n+1}{2}\!\big)-\Delta S_2\!\big(\!\frac{l}{2}\!\big)\right]}{4(n-l+1)(n+l+2)}
+\frac{\Delta S_2\!\big(\!\frac{k+1}{2}\!\big)- \Delta S_2\!\big(\!\frac{n+1}{2}\!\big)}{4 (n-k)}
\nonumber
\end{eqnarray}
to transform the finite sum over $n$ in one over $k$. Finally, the non-separable conformal moments for gluons follow from the quark ones by means of (\ref{L^{frac{5}{2},a,b}_{jk}}).

\begin{table}[t]
\begin{tabular}{|ccccl|}
\hline
$\left[\frac{1}{u\bv}\, \frac{L(u,v)}{u-v}\right]^{\rm sub}$ &
$\!\!\!\Leftrightarrow\!\!$ &
$\frac{-2 r}{1+r}\, \frac{\frac{2r}{1-r} \ln\frac{1+r}{2 r}+\frac{\ln v}{\bv}}{1+r-2r v}$ &
$\!\!\Leftrightarrow\!\!\!$ &
${\footnotesize \left\{ \begin{array}{l}
-(-1)^k (k+1)_2\, \Delta S_2\!\big(\!\frac{j+1}{2},\frac{k+1}{2}\!\big)  - \frac{(-1)^k}{2} \Delta S_2\!\big(\!\frac{j+1}{2}\!\big)
\\
\frac{(-1)^k (k+1)_2}{2}\Big[\frac{(j)_2}{2 j+3} \Delta S_2\!\big(\!\frac{j+2}{2},\frac{k+1}{2}\!\big)
- \frac{(j+2)_2}{2 j+3}  \Delta S_2\!\big(\!\frac{j}{2},\frac{k+1}{2}\!\big)
\Big]
\\
\phantom{\frac{(-1)^k(j)_4}{2(2 j+3)}\Big[ }+\frac{(-1)^k}{2} \Delta S_2\!\big(\!\frac{j+1}{2}\!\big) - (-1)^k\, \frac{3(j+1)_2+2}{[(j+1)_2]^2}
\end{array}\right. }$
\\\hline
$\left[\frac{1}{\bv}\, \frac{L(u,v)}{u-v}\right]^{\rm sub}$  &
$\!\!\!\Leftrightarrow\!\!$ &
$-\frac{\frac{2r}{1-r} \ln\frac{1+r}{2 r}+\frac{\ln v}{\bv}}{1+r-2r v}$  &
$\!\!\Leftrightarrow\!\!\!$ &
${\footnotesize  \begin{array}{l}
\frac{(-1)^k (k+1)_2}{2} \Big[
\frac{-k-3}{2k+3} \Delta S_2\!\big(\!\frac{j+1}{2},\frac{k+2}{2}\!\big)+
\frac{-k}{2k+3} \Delta S_2\!\big(\!\frac{j+1}{2},\frac{k}{2}\!\big)
\\
\phantom{\frac{(-1)^k (k+1)_2}{2} }
+ \Delta S_2\!\big(\!\frac{j+1}{2},\frac{k+1}{2}\!\big)\Big] + \frac{(-1)^k}{(j+1)_2 (k+1)_2}
\end{array} }$
\\\hline
$\left[\frac{1}{\bv^2}\, \frac{L(u,v)}{u-v}\right]^{\rm sub}$  &
$\!\!\!\Leftrightarrow\!\!$ &
$\frac{\frac{2 r}{1-r}\left[\frac{2 r}{1-r}\ln\frac{1+r}{2 r}-1\right]-\frac{\ln v+\bv}{\bv^2}}{1+r-2 v r}$  &
$\!\!\Leftrightarrow\!\!\!$ &
${\footnotesize  \begin{array}{l}
a_{jk}\Big[S_3(j+1)-\zeta(3)
-\sum_{l=0}^{k}\frac{(-1)^l (2l+1)}{2} \Delta S_2\!\big(\!\frac{j+1}{2},\frac{l}{2}\!\big)\Big]
\\
-\frac{(-1)^k(k+1)_2}{2}\!\Bigg[2\Delta S_2\!\big(\!\frac{j+1}{2},\frac{k+1}{2}\!\big)
-\frac{\Delta S_2\big(\!\frac{j+1}{2}\!\big)-\Delta S_2\big(\!\frac{k+1}{2}\!\big)}{k+2}
\\
\phantom{-\frac{(-1)^k(k+1)_2}{2}\!\Bigg[}  -\frac{(j+1)_2-1}{(j+1)_2} \left(\Delta S_2\big(\!\frac{k+1}{2}\!\big)-\frac{2}{(k+1)_2} \right)\Bigg]
\end{array}}$
\\
\hline
\end{tabular}
\caption{\small Substitution rules among subtracted non-separable functions (\ref{eqs:Lsub}) [left
column], their imaginary parts w.r.t.~the $u$-variable (middle column), and
conformal moments (right column) for quarks (upper lines) and gluons (lower
lines), presented in terms of
Pochhammer`s symbol (\ref{Pochhammer}) and harmonic sums
(\ref{Sq},\ref{Sm21}), where $\Delta S_2\!\big(\!\frac{j+1}{2}\!\big)=S_2\!\big(\!\frac{j+1}{2}\!\big)-S_2\!\big(\!\frac{j}{2}\!\big)$ and
$\Delta S_2\!\big(\!\frac{j+1}{2},\frac{k+1}{2}\!\big)=
\frac{1}{a_{jk}}\left[\Delta S_2\!\big(\!\frac{j+1}{2}\!\big)-\Delta S_2\!\big(\!\frac{k+1}{2}\!\big)\right]$ with
$a_{jk}= 2[(j+1)_2-(k+1)_2]$.
\label{subT-L}}
\end{table}
We finally summarize our findings for the equivalent representations of non-separable building blocks, obtained as described and used
in the next section to express the non-separable addenda (\ref{eq:ATc-dec}),  in Tab.~\ref{subT-L}.
For shortness the conformal moments are given  in terms of the function
\begin{eqnarray}
\label{Delta S_2}
\Delta S_2\!\Big(\!\frac{j+1}{2},\frac{k+1}{2}\!\Big)=
\frac{\Delta S_2\!\Big(\!\frac{j+1}{2}\!\Big)-\Delta S_2\!\Big(\!\frac{k+1}{2}\!\Big)}{2(j-k)(j+k+3)}\,, \quad
\Delta S_2\!\Big(\!\frac{j+1}{2},\frac{j+1}{2}\!\Big)=- \frac{\Delta S_3\!\Big(\!\frac{j+1}{2}\!\Big)}{2j+3}\,.
\end{eqnarray}

\subsection{Next-to-leading corrections}
\label{sec:NLO-result}

In Sec.~\ref{sec:anatomy-quark}, \ref{sec:anatomy-pS} and \ref{sec:anatomy-gluon} we will list the NLO corrections for the flavor non-singlet,  pure singlet quark-quark, and gluon-quark channel, respectively.  We give for each channel first the color decomposition (\ref{AT^{(1)}}) and list the separate terms in momentum fraction representation using our building blocks, where we will group the leading singularities together with factorization and renormalization logarithms.
In the momentum fraction representation the factorization logarithms are proportional to the convolution with the LO evolution kernel, see
(\ref{qT^(1)},\ref{T^PS},\ref{^GT^(1)}).
The imaginary parts, given in terms $r= x/\xi$, and the conformal moments of separable functions follow then from the  substitution rules that are listed in
tables \ref{subT-1}--\ref{subT-Li}, where references to specific functions are given. The non-separable terms are presented in terms of the building blocks (\ref{eqs:Lsub}) and differential operators,
where latter are given by the adjoint operators that appear on the l.h.s.~of the differential equations (\ref{dif-op}). Their imaginary parts and conformal moments  are obtained from the substitution rules listed in Tab.~\ref{subT-L}, see also (\ref{subsL`u2im}), where for moments we also utilize  the differential equations (\ref{dif-op}).
We add that the NLO expressions for the evolution kernels in momentum fraction are derived in Ref.~\cite{Belitsky:1999hf} and match our conventions,
see App.~\ref{app:def}. The NLO expressions for the evolution operator in terms of conformal moments can be simply taken from
\cite{Kumericki:2007sa}.

\subsubsection{Flavor non-singlet channel}
\label{sec:anatomy-quark}

The NLO contributions in the flavor non-singlet channel can be read off from \cite{Melic:1998qr}.
The color factor decomposition of the corresponding coefficient (\ref{qT^(1)}) can be chosen to be
\begin{eqnarray}
\label{T^{(1)}-1}
\qT^{(1)}\!\left(\!u,v\Big|\frac{\cQ^2}{\muF^2},\frac{\cQ^2}{\muphi^2},\frac{\cQ^2}{\muR^2}\right)
&\!\!\! =\!\!\! &
\CF \, \qT^{(1,\text{F})}\!\left(\!u,v\Big|\frac{\cQ^2}{\muF^2},\frac{\cQ^2}{\muphi^2}\right)
+ \beta_0 \, \qT^{(1,\beta)}\!\left(\!u,v\Big|\frac{\cQ^2}{\muR^2}\right)
+\,
\CG \, \qT^{(1,\text{G})}(u,v) \,,
\nonumber\\
\end{eqnarray}
and as discussed in Sec.~\ref{sec:prel-MF-definitions}, they are symmetric under $(u,\muF) \leftrightarrow  (v,\muphi)$ exchange.
The $\qT^{(1,\text{F})}(u,v)$ and $\qT^{(1,\beta)}(u,v)$ functions are entirely expressed by  separable building blocks that are most singular, listed in
Tab.~\ref{subT-1}, and those from Tab.~\ref{subT-2}. The $\qT^{(1,\text{G})}(u,v)$ function has besides such singularities also logarithmical cuts on the {\em negative} $u$- and $v$-axis and it contains a non-separable piece. Due to the subtraction procedure, introduced in Sec.~\ref{sec:NLO-blocks2}, its explicit form is rather lengthy
\begin{eqnarray}
\label{Delta T^{(1,G)}}
\Delta\qT^{(1,\text{G})}(u,v) & \!\!\! =  \!\!\!  &
\left[\!\frac{\bu u}{\bv} + \frac{\bv v}{\bu} + \frac{(u-v)^3}{\bu \bv}\right]  \!
\frac{ \Li(\bv) - \Li(\bu) + \ln v \ln \bu  - \ln u \ln \bu  }{(u-v)^3} + \frac{\bu \ln v + \bv^2}{\bu\bv^2\,(u-v)}
\nonumber \\ &&\!\!
+ \frac{ 2\bv \ln\bu + 2 v \ln v }{\bv (u - v)^2} - \frac{\ln \bu  \ln v +\Li(\bv) }{\bu \bv^2 }-
\frac{(\bu-u) \left[\Li(u)  -\zeta(2) \right] +\bu \ln \bu }{\bu^2 \bv} \,.
\qquad
 \phantom{\Bigg\}}
\end{eqnarray}
This addendum possesses  $1/(u-v)^n$ terms with up to $n=3$, however, it is finite at $u=v$.
As desired, $\Delta\qT^{(1,\text{G})}(u,v)$  has only logarithmical cuts on the positive $u$-  and negative $v$-axis.
We write the separate terms in the color decomposition (\ref{T^{(1)}-1}) as follows.\\

\noindent
\textit{Momentum fraction representation.}
\begin{subequations}
\label{qT^{(1)}}
\begin{eqnarray}
\label{T^{(1,F)}}
\qT^{(1,\text{F})}(u,v)
& \!\!\! =  \!\!\!  &
\left[ \ln\frac{\cQ^2}{\muF^2}+ \frac{1}{2}\ln(\bu \bv) +1 \right] \frac{3+ 2\ln \bu }{2 \bu \bv}
-\frac{23}{6 \bu \bv} -\frac{\ln \bu }{2u \bv} + \{\muF \to \muphi, u \leftrightarrow v\},
\qquad\quad\phantom{\Bigg|}\\
\label{T^{(1,beta)}}
\qT^{(1,\beta)}(u,v)
& \!\!\! =  \!\!\!  &
\left[
\frac{1}{2}\ln\frac{\cQ^2}{\muR^2} + \ln\bu  - \frac{5}{6}
\right]  \frac{1}{2\bu \bv} + \{u \leftrightarrow v\},
\phantom{\Bigg|}
\\
\label{T^{(1,G)}}
\qT^{(1,\text{G})}(u,v)
& \!\!\! =  \!\!\!  &
\left[
   \ln\bu \frac{\ln v }{\bv}+ \ln \bv  - \frac{7}{6} - \zeta(2) + 2  \Li(v)- 2\Li(\bv)  - \ln u \ln v  \right] \frac{1}{\bu\bv}
\\
&&\!\!\!\!  +\!\left[ \frac{\Li(\bv)- \Li(v) +\zeta(2)}{\bv}+\frac{\ln v}{\bv }-1\right] \frac{1}{\bu\bv}  + \Delta T^{(1,\text{G})}(u,v) + \{u \leftrightarrow v\} \, .
\nonumber
\phantom{\Bigg|}
\end{eqnarray}
The addendum $\Delta T^{(1,\text{G})}(u,v)$, see (\ref{Delta T^{(1,G)}}), can be expressed by means of a differential operator that acts on the non-separable
building block (\ref{eq:Lsub-v1}) and the building block (\ref{eq:Lsub-v2}),
\begin{eqnarray}
\label{Delta T^{(1,G)}-1}
\Delta T^{(1,\text{G})}(u,v)
&\!\!\! =\!\!\! &
\left[\frac{\vec{\partial}^2}{\partial v^2}  -\frac{2}{v\bv} \right]v\bv \left[\frac{1}{\bv}\frac{L(u,v)}{u-v}\right]^{\rm sub}\!\!
-\frac{\vec{\partial}}{\partial v}\, \bv  \left[\frac{1}{\bv}\frac{L(u,v)}{u-v}\right]^{\rm sub}\!\!
 +\left[\frac{1}{\bv^2}\, \frac{L(u,v)}{u-v}\right]^{\rm sub} \!\!\!.
\qquad
\end{eqnarray}
Note that to avoid boundary term in  a partial integration, we introduced an oversubtraction  for the second order derivative.
The $u\leftrightarrow v$-reflected addendum can be conveniently  written in terms of the variables $\bv$  and $\bu$ as
\begin{eqnarray}
\label{Delta T^{(1,G)}-bvbu}
\Delta T^{(1,\text{G})}(\bv,\bu) &\!\!\!=\!\!\! &
\left[\frac{\vec{\partial}^2}{\partial v^2}  -\frac{2}{v\bv} \right]  v\bv \left[\frac{1}{u}\frac{L(u,v)}{u-v}\right]^{\rm sub}
+\frac{\vec{\partial}}{\partial v}\, v \left[\frac{1}{u}\frac{L(u,v)}{u-v}\right]^{\rm sub}
\\
&&\!\phantom{\left[\frac{\vec{\partial}^2}{\partial v^2}  - \frac{2}{v\bar{v}} \right] }
 +\frac{1}{v}\! \left[\frac{1}{u}\, \frac{L(u,v)}{u-v}\right]^{\rm sub}\!\! -\frac{\Li(u)+\ln \bu + [\ln\bu+ u]\ln v }{u^2 v}\,.
\phantom{\Bigg]}
\nonumber
\end{eqnarray}
Here, the last term subtracts the pole contribution at $v=0$.
Note that the addendum with definite signature is then obtained from
\begin{eqnarray}
\Delta^\sigma\!T^{(1,\text{G})}(u,v) = \Delta  T^{(1,\text{G})}(u,v) - \sigma \Delta  T^{(1,\text{G})}(v,\bu)\,.
\end{eqnarray}
\end{subequations}

\noindent
\textit{Imaginary parts of} (\ref{qT^{(1)}}) \textit{from quark exchange.}\\
(positive momentum fraction $x\ge \xi$, i.e., poles at $u=1$ and $u$-cuts $[1,\infty]$)
\begin{subequations}
\begin{eqnarray}
\label{t^{(1,F)}}
t^{(1,\text{F})}(r,v)
& \!\!\! =  \!\!\! &
\left[\ln\frac{\cQ^2}{\muF^2}+ \frac{1}{2}\ln\bv + 1 \right]\!\left[\frac{3}{2}\delta(1-r)+ \left\{\frac{1}{1-r}\right\}_+\right] \frac{1}{\bv}
+\left\{\frac{\frac{3}{4}+\ln\frac{1-r}{2r}}{1-r}\right\}_+ \frac{1}{\bv}
\nonumber\\
&&\!\!\!\!\!
+\left[\left(\ln\frac{\cQ^2}{\muphi^2}+ \frac{1}{2}\ln\bv +1\right) \delta(1-r) +
\frac{1}{2}\left\{\frac{1}{1-r}\right\}_+\right] \frac{3+ 2\ln\bv }{2\bv}
\\
&&\!\!\!\!\!
-\left[\frac{23}{3}+\frac{\bv}{2v}\ln\bv  \right]\frac{\delta(1-r)}{\bv} + \frac{1}{1+r}\; \frac{1}{2\bv}\,,
\nonumber
\end{eqnarray}
\begin{eqnarray}
\label{t^{(1,beta)}}
t^{(1,\beta)}(r,v)
& \!\!\! =  \!\!\!  &
\left[\ln\frac{\cQ^2}{\muR^2} - \frac{5}{3} + \ln \bv \right]\frac{\delta(1-r)}{2 \bv}   + \left\{\frac{1}{1-r}\right\}_+ \frac{1}{2\bv}\,,
\end{eqnarray}
\begin{eqnarray}
\label{t^{(1,G)}}
t^{(1,\text{G})}(r,v)
& \!\!\! =  \!\!\!  &
  \left\{\frac{1}{1-r}\right\}_+ \frac{\ln v }{\bv^2}+
\left[
2\zeta(2)-\frac{7}{3}+ \frac{v-\bv}{\bv}\left[\Li(\bv)-\Li(v)+\zeta(2)\right] \right.
 \\
&&  \left.  \phantom{-\frac{13}{3}+2\zeta(2) ++\; }
+ \frac{\ln v - \bv}{\bv}  \right]\frac{\delta(1-r)}{\bv} -  \frac{2\ln\frac{1 + r}{2 r}-1+r}{(1 - r)^2\, \bv}  + \Delta t^{(1,\text{G})}(r,v)\,.
\phantom{\Bigg)}
\nonumber
\end{eqnarray}
The imaginary part of the addendum (\ref{Delta T^{(1,G)}-1}) can be written in a compact form as
\begin{eqnarray}
\label{Delta t^1G}
\Delta t^{(1,\text{G})}(r,v) & \!\!\! = \!\!\! &
-\left[\frac{v-\bv}{\bv^2}+\frac{\partial}{\partial v}\frac{v\,\partial}{\partial v}\right] \left[\frac{2r\bv \ln\frac{1+r}{2 r}+(1-r)\ln v}{(1-r)(1+r-2r v)}\right] + \frac{2r \ln\frac{1+r}{2 r }-1+r}{(1-r)^2\bv}
.\qquad
\end{eqnarray}
\end{subequations}

\noindent
\textit{Imaginary parts of}  (\ref{qT^{(1)}}) \textit{from antiquark exchange}.\\
(negative momentum fraction $x\le -\xi$, i.e., $u$-cuts $[-\infty,0]$)
\begin{subequations}
\begin{eqnarray}
\label{tb^{(1,F)}}
\label{tb^{(1,beta)}}
\overline{t}^{(1,\text{F})}(r,v) & \!\!\! =  \!\!\!  & \overline{t}^{(1,\beta)}(r,v) \equiv 0
\phantom{\Bigg|}\\
\label{tb^{(1,G)}}
\overline{t}^{(1,\text{G})}(r,v)
& \!\!\! =  \!\!\!  &
-\left[ \ln\frac{1 + r}{2 r v}  + r  \ln\frac{\bv}{v} + r \right]\frac{2}{(1+r)^2\, \bv }  +\Delta t^{(1,\text{G})}(r,v)\, .
\phantom{\Bigg|}
\end{eqnarray}
The addendum, following  from $\Delta T^{(1,\text{G})}(v,u)$ by means of (\ref{Delta T^{(1,G)}-bvbu}), reads
\begin{eqnarray}
\label{Delta tb^1G}
\Delta \overline{t}^{(1,\text{G})}(r,v) & \!\!\! = \!\!\! &
\frac{\partial}{\partial v}\frac{v\,\partial}{\partial v}
\left[\frac{2r\bv \ln\frac{1+r}{2r\bv}}{(1+r)(1+r-2r\bv)}\right] -
\frac{4r}{(1+r)^2}\frac{\ln\frac{1+r}{2r\bv}}{1+r-2r\bv} +\frac{2 r}{(1+r)^2\bv} .
\qquad
\end{eqnarray}
\end{subequations}

\noindent
\textit{Conformal moments of} (\ref{qT^{(1)}}).
\begin{subequations}
\label{c^{(1)}}
\begin{eqnarray}
 \label{c^{(1,F})}
c^{(1,\text{F})}_{jk} &\!\!\! =\!\!\!&
\left[-\ln\frac{\cQ^2}{\muF^2}+S_ 1(j+1)+S_ 1(k+1)-1-\frac{1}{2(j+1)_2}-\frac{1}{2(k+1)_2}\right]
 \frac{\gamma_j^{(0,\text{F})}}{2}
\\
&&-\frac{23}{6}+ \frac{3 (j+1)_2+1}{2[(j+1)_2]^2}
+ \{j \leftrightarrow k, \muF\to \muphi \}
\,,
\nonumber \\
{\ }
\nonumber \\
\label{c^{(1,beta)}}
c^{(1,\beta)}_{jk}
&\!\!\! =\!\!\!& \frac{1}{4}\ln\frac{\cQ^2}{\muR^2}
- S_1(j+1)-\frac{5}{12}+\frac{1}{2(j+1)_2}
 + \{ j \leftrightarrow k\}
\, , \qquad
 \\
 {\ }
\nonumber \\
\label{c^{(1,G)}}
 c^{(1,\text{G})}_{jk}
 &\!\!\! =\!\!\!&
 \left[2 S_1(j+1) -\frac{1}{(j+1)_2}\right] \left[
 1+(-1)^k-(-1)^k (k+1)_2\frac{\Delta S_2\big(\!\frac{k+1}{2}\!\big)}{2}
  \right] +\zeta(2) -\frac{7}{6}
  \\
 &&\!\!\!\!\!
 +\left[(-1)^k\, \mathbb{S}_3(k+1)+
 \frac{(-1)^k\, \Delta S_2\big(\!\frac{k+1}{2}\!\big)}{2(k+1)_2}  - S_3(k+1) + \zeta(3)-\frac{(k+1)_2-1}{2[(k+1)_2]^2} \right] 2(k+1)_2
 \nonumber\\
&&- \frac{2\left[1+(-1)^k\right]\left[(k+1)_2+1\right]}{\left[(k+1)_2\right]^2} - \frac{(-1)^{j+k}}{(j+1)_2(k+1)_2}
+ \Delta c^{(1,\text{G})}_{jk} + \{ j \leftrightarrow k\}\,,
 \nonumber
\end{eqnarray}
where
\begin{eqnarray}
\label{gamma_j^{(0,F)}}
\gamma_j^{(0,F)} =  4 S_{1}(j + 1) - 3 - \frac{2}{(j + 1 )_2}
\end{eqnarray}
is apart from the color factor the anomalous dimension (\ref{eq:gamma0}) and we use here the shorthand
\begin{eqnarray}
\mathbb{S}_3(n)  &\!\!\! =\!\!\!&
\frac{S_3\big(\!\frac{n}{2}\!\big)-S_3\big(\!\frac{n-1}{2}\!\big)}{8} +
\frac{\left[S_2\big(\!\frac{n}{2}\!\big) -S_2\big(\!\frac{n-1}{2}\!\big)\right]  S_1(n)}{2}
-2 (-1)^{n}
\left[S_{-2,1}(n)+\frac{5\zeta(3)}{8}\right].
\qquad
\end{eqnarray}
This auxiliary function is finite at $n=0$ and it vanishes like $1/n^4$ for $n\to \infty$.
The conformal moments of the addendum  (\ref{Delta T^{(1,G)}-1})
are obtained as described above and they read for complex $j$ and non-negative integer $k$ as following
\begin{eqnarray}
\label{Deltac^{(1,G)}_jk}
\Delta c^{(1,\text{G})}_{jk} &\!\!\!=\!\!\!&
a_{jk} \left[S_3(j+1)-\zeta(3)+\frac{(-1)^k (k+1)\Delta S_2\big(\!\frac{j+1}{2},\frac{k+1}{2}\!\big)  }{2}
-\sum _{l=0}^k \frac{(2 l+1) (-1)^l\Delta S_2\big(\!\frac{j+1}{2},\frac{l}{2}\!\big)}{2}
\right]
\nonumber\\
&&+\frac{(-1)^k (k+1)_2}{2} \sum_{b=0}^{2} \frac{(-1)^b(2k+3b)\left[4+3b(3-b)+2k b+2 (k+1)^2\right]
\Delta S_2\big(\!\frac{j+1}{2},\frac{k+b}{2}\!\big)}{[3+(-1)^b](2k+3)}
\nonumber\\
&&+
\frac{(-1)^k\left[(j+1)_2-1\right]\left[(k+1)_2\,\Delta S_2\big(\!\frac{k+1}{2}\!\big)-2\right]}{2(j+1)_2}-\frac{2 (-1)^k}{(j+1)_2 (k+1)_2}
\,,
\end{eqnarray}
where $a_{jk}= 2(j-k)(j+k+3).$
In the case that the first argument is $k$  we write this addendum as, see identity (\ref{identity-sums}),
\begin{eqnarray}
\label{Deltac^{(1,G)}_kj}
\Delta c^{(1,\text{G})}_{kj} &\!\!\!=\!\!\!&
a_{kj}(-1)^j \Bigg[\mathbb{S}_3(j+1)
-\frac{S_1(k+1) \Delta S_2\big(\!\frac{j+1}{2}\!\big)}{2}
-\sum _{l=0}^k \frac{(2 l+1)\Delta S_2\big(\!\frac{j+1}{2},\frac{l}{2}\!\big)}{2}
\Bigg]
\\
&&-\frac{(-1)^j (k+1)_2}{2} \sum_{b=0}^{2} \frac{(2k+3b)\left[4+3b(3-b)+2k b+2 (k+1)^2\right]
\Delta S_2\big(\!\frac{j+1}{2},\frac{k+b}{2}\!\big)}{[3+(-1)^b](2k+3)}
\nonumber\\
&&-\left[(k+1)^2+2 +\frac{(j+1)_2}{(k+1)_2}\right] \frac{(-1)^j \Delta S_2\big(\!\frac{j+1}{2}\!\big)}{2}
-\frac{(-1)^j (k+1)\Delta S_2\big(\!\frac{k+1}{2}\!\big)}{2}  +\frac{(-1)^j}{(k+1)_2}
\,.
\nonumber
\end{eqnarray}
\end{subequations}
The conformal moments $c^{(1,\text{F})}_{jk}$, $c^{(1,\beta)}_{jk}$, and $\Delta c^{(1,\text{G})}_{jk}$  are independent on the signature while the  $(-1)^j$ factors for complex valued $j$
in $c^{(1,\text{G})}_{jk}$ and $\Delta c^{(1,\text{G})}_{kj}$ must be  replaced by $-\sigma$.

\subsubsection{Pure  singlet quark channel}
\label{sec:anatomy-pS}
The pure singlet contribution  arises from six contributing Feynman diagrams, see Fig.~\ref{f:contributions}b). Only two of them have to be
evaluated and the rest is obtained using $u\to\bu$ and $v\to \bv$
symmetries.  Our diagrammatical evaluation confirms the result in \cite{Ivanov:2004zv}.
In order to obtain a representation that contains only a branch cut $[1,\infty]$ on the real $u$-axis  for  $0\le v \le 1$,
we employ the known symmetry properties of this contribution:
the result is antisymmetric under $u\to\bu$, symmetric under $v\to\bv$ and
antisymmetric under $(u,v)\to(\bu,\bv)$.
 The non-separable contributions are collected in
\begin{eqnarray}
\label{Delta{^pS}T^{(1)}}
\Delta{^\pS T}^{(1)}(u,v) &\!\!\! =\!\!\! &
\frac{u\bu+u\bv-v\bv}{u\bv}\, \frac{\Li(u)+\Li(\bv) + \ln\bu \ln v -\zeta(2)}{(u - v)^2}  + \frac{\bv \ln\bu +u \ln v }{u\bv(u-v)}
\phantom{\Bigg|}
\nonumber\\
&&+\frac{\Li(u)+\ln\bu \ln v}{u\bv} + \frac{\Li(\bv)-\zeta(2)}{u v}\,.
\phantom{\Bigg|}
\end{eqnarray}
As in the preceding section, this function is finite on the line $u=v$ and the pole at $u=0$ is subtracted, while a pole at $v=1$ remains.
It can be  expressed by the building block
(\ref{eq:Lsub-u1}), which makes the analytical properties of the addendum  obvious.
Our results read as follows. \\

\noindent
{\textit{Momentum fraction representation.}}
\begin{subequations}
\label{{^pS}T^{(1,F)}-0}
\begin{eqnarray}
\label{{^pS}T^{(1)}}
{^\pS T}^{(1)}(u,v) &\!\!\! = \!\!\!&
\left[ \ln\frac{\cQ^2}{\muF^2} + \frac{1}{2} \ln \bu  + \ln(v \bv)- 1 \right] \frac{\bu-u}{u v\bv} \ln \bu -\frac{2 \Li(u) }{v\bv}
\\
&&\!\!\!\!\!
- \left[\frac{1}{2v\bv} +  \frac{\ln v }{\bv} + \frac{\ln\bv }{v} \right]\frac{\ln \bu}{u} + \Delta{^\pS}T^{(1)}(u,v)\, ,
\phantom{\Bigg|}
\nonumber \\
\label{Delta{^pS}T^{(1)}-1}
\Delta{^\pS T}^{(1)}(u,v) &\!\!\! = \!\!\!&
\frac{1}{v\bv}\frac{\partial}{\partial v} v^2 \bv \left[\frac{1}{u}\,\frac{L(u,v)}{u-v}\right]^{\rm sub}
.
\end{eqnarray}
\end{subequations}
The first two terms on the r.h.s.~of (\ref{{^pS}T^{(1)}}) diverge logarithmically in the limit $u\to \infty$, but the terms proportional to $\ln^2\bu$ and $\Li(u)$  cancel each other, leaving a constant that vanishes  by antisymmetrization. The remaining divergent term is contained
in  $(\bu-u)\ln(\bu)/u$, which is nothing  but the convolution of the LO  evolution kernel in the gluon-quark channel with the LO hard scattering amplitude, see
(\ref{eq:logSigma}). The substitution rules for these functions are given in Tab.~\ref{subT-pS}, where
the $j=0$ pole is absorbed in the anomalous dimension of the gluon-quark channel.
\begin{table}[t]
\begin{center}
\begin{tabular}{|ccccl|}
\hline
$\frac{\bu-u}{u}\ln\bu $ & $\Leftrightarrow$ & $ \frac{1}{r(1+r)}$ & $\Leftrightarrow$ &
$
 -\frac{{^{{\rm G} \Sigma}\gamma}_{j}^{(0,{\rm F})}}{2(j+3)}\,,
$
\\
$\frac{\bu-u}{2u}\ln^2\bu- 2\Li(u) $  & $\Leftrightarrow$ & $\frac{\ln\frac{1-r}{1+r}}{r(1+r)} -\frac{\ln\frac{1+r}{2r}}{1+r} $ & $\Leftrightarrow$ &
$
\left[S_1(j+1)-1\right] \frac{{^{{\rm G} \Sigma}\gamma}_{j}^{(0,{\rm F})}}{j+3} -\frac{(j+1)_2+1}{[(j+1)_2]^2}
$
\\\hline
\end{tabular}
\end{center}
\caption{\small Quark building blocks which diverge for $u\to\infty$ [left column], their imaginary parts
(\ref{eq:uv-func-3},\ref{subT`u2im2-Li-1}) [middle column], and
conformal moments (\ref{<lnbu>_j}) [right column], where ${^{{\rm G} \Sigma}\gamma}_{j}^{(0,{\rm F})}$ is defined in (\ref{{^{g Sigma}gamma}^{(0,F)}_j}).
\label{subT-pS}}
\end{table}
\\

\noindent
{\textit{Imaginary part of} (\ref{{^pS}T^{(1,F)}-0}).}
\begin{subequations}
\label{{^Sigma t}^{(1)}}
\begin{eqnarray}
\label{^Sigma t^{(1,F)}}
{^\pS t}^{(1)}\!\!\left(\!r,v\Big|\frac{\cQ^2}{\muF^2}\right) &\!\!\!=\!\!\!&
\left[\ln\frac{\cQ^2}{\muF^2} + \ln(v\bv) + \ln\frac{1-r}{1+r} - 1\right]  \frac{1}{r(1+r)v\bv}
-\frac{\ln\frac{1+r}{2r}}{(1+r)v\bv}
\\
&&\!\!\!\!\!+ \left[\frac{1}{2v\bv} +  \frac{\ln v }{\bv} + \frac{\ln\bv }{v} \right] \frac{1}{1+r}
+ \Delta{^\pS}t^{(1,\text{F})}(r,v)\, ,
\phantom{\Bigg|}\nonumber\\
\label{Delta^pS t^(1,F)}
\Delta{^\pS}t^{(1)}(r,v) &\!\!\!=\!\!\!&
\frac{1}{v\bv}\frac{\partial}{\partial v} v \bv\left[\frac{2r v}{1+r}\, \frac{\ln\frac{1+r}{2r v}}{1+r - 2 r v}\right]
.
\end{eqnarray}
\end{subequations}

\noindent
{\textit{Conformal moments of} (\ref{{^pS}T^{(1,F)}-0}).}
\begin{subequations}
\label{^Sigma c^{(1,F)}-0}
\begin{eqnarray}
\label{^Sigma c^{(1,F)}}
{^\pS c}^{(1)}_{jk}
&\!\!\!=\!\!\!&
\left[-\ln\frac{\cQ^2}{\muF^2}+2 S_ 1(j+1) + 2S_ 1(k+1)-1\right] \frac{{^{\text{G}\Sigma}\gamma}_j^{(0,\text{F})}}{j+3}
\\ &&\!\!\!\!\!\!
- \left[\frac{1}{2}+\frac{1}{(j+1)_2}+\frac{1}{(k+1)_2}\right]\frac{2}{(j+1)_2} + \Delta{^\pS}c^{(1)}_{jk}
\,,
\nonumber\\
{}
\nonumber\\
\label{Delta {^pS}c^{(1,F)}_{jk}}
\Delta{^\pS}c^{(1)}_{jk} &\!\!\!=\!\!\!&
\frac{(k)_4\left[\Delta S_{2}\big(\!\frac{j+1}{2},\frac{k}{2}\!\big)- \Delta S_{2}\big(\!\frac{j+1}{2},\frac{k+2}{2}\!\big)\right]}{2(2k+3)},
\phantom{\Bigg]}
\end{eqnarray}
where we extracted the color factor from  the  anomalous dimension (\ref{Def-LO-AnoDim-GQ-V}),
\begin{eqnarray}
\label{{^{g Sigma}gamma}^{(0,F)}_j}
\frac{{^{\text{G}\Sigma}\gamma}_j^{(0,\text{F})}}{j+3} = -\frac{4 +2 (j + 1 )_2}{(j)_4},
\end{eqnarray}
\end{subequations}
 in the gluon-quark channel. Note that if we express (\ref{Delta{^pS}T^{(1)}-1}) in terms of the building block (\ref{eq:Lsub-u1v1}) the addendum (\ref{Delta {^pS}c^{(1,F)}_{jk}}) follows straightforwardly  from utilizing  the differential operator (\ref{dif-op2}) and the corresponding conformal moments, given
 in Tab.~\ref{subT-L}. Thereby, an artificial $\delta_{k0}$ term appears only in intermediate steps.

\subsubsection{Gluon-quark channel}
\label{sec:anatomy-gluon}

For the gluon-quark contribution we take the results from Ref. \cite{Ivanov:2004zv}
and rewrite them in a compact form, using
symmetry under  $u\leftrightarrow \bu$ and $v\leftrightarrow \bv$, in such a
manner that the net results have the desired analytic properties%
\footnote{To shorter the expression we will allow for one pole contribution at $u=0$.},
where we prefer functions symmetric  under $v \leftrightarrow  \bv$.
The LO contribution ${^\text{G}T}^{(0)}$ is defined in (\ref{eq:T^{(0)}})
and the NLO part (\ref{^GT^(1)}) can be decomposed as
\begin{eqnarray}
\label{^G T^{(1)}}
{^\text{G}T}^{(1)}\left(\!u,v\Big| \frac{\cQ^2}{\muF^2},\frac{\cQ^2}{\muphi^2},\frac{\cQ^2}{\muR^2}\right)
= \CA {^\text{G}T}^{(1,\text{A})}\!\left(\!u,v\Big| \frac{\cQ^2}{\muF^2}\right)
  + \CF  {^\text{G}T}^{(1,\text{F})}\!\left(\!u,v\Big| \frac{\cQ^2}{\muF^2},\frac{\cQ^2}{\muphi^2}\right) +
\frac{ \beta_0 }{2\bu  \bv} \ln \frac{\muF^2}{\muR^2}
\, .
\nonumber\\
\end{eqnarray}
The term  proportional to $\beta_0$, arising from the gluon self-energy insertion, is given by $\ln(\muF^2/\muR^2)$ times the LO amplitude,
see  (\ref{eq:T^{(0)}},\ref{t^{(0)}},\ref{c^{(0)}_jk}). Its imaginary part and conformal moments  follow from
$$
\frac{ \beta_0 }{2\bu  \bv} \ln \frac{\muF^2}{\muR^2}
\;\; \Leftrightarrow \;\;
\frac{ \beta_0 \delta(1-r)}{2  \bv} \ln \frac{\muF^2}{\muR^2}
\;\; \Leftrightarrow \;\;
\frac{ \beta_0}{2} \ln \frac{\muF^2}{\muR^2}\,.
$$
We introduce two addenda for the parts proportional to $\CA$ and $\CG$,
\begin{subequations}
\label{Delta^G T^{(1)}}
\begin{eqnarray}
\label{Delta^G T^{(1,A)}}
\Delta {^\text{G}T}^{(1,\text{A})}(u,v) &\!\!\! = \!\!\!&
\frac{\bu-u}{4 v\bv}
\left[
\frac{\Li(u) + \Li(\bv)+ \ln\bu \ln v - \zeta(2)}{(u - v)^2} + \frac{u \ln v + \bv \ln\bu }{u (u-v)\bv}
\right]
\\
&&\!\!\!  + \frac{(3-4v) \Li(u) }{4 u v\bv }  + \frac{ (1-4v) \left[ \Li(u)-\zeta(2) \right] }{4 \bu v \bv}
+ \frac{ \ln\bu }{u}\,  \frac{\ln v +1}{2 v \bv }  + \frac{\ln v }{2 v\bv^2},
\nonumber\\
\label{Delta^G T^{(1,F)}}
\Delta {^\text{G}T}^{(1,\text{F})}(u,v) &\!\!\! = \!\!\!&
\frac{u\bv -(u-v)^2}{2u\bv}\, \frac{\Li(u)+\Li(\bv)+\ln\bu \ln v -\zeta(2)}{(u-v)^3}
+ \frac{u \ln v  + \bv \ln\bu }{2u (u - v)^2 \bv}
\quad\phantom{\Bigg|}\nonumber\\
&&
+ \frac{\ln \bu + u}{4 u (u-v) v}  + \frac{\ln v +\bv}{4(u-v)\bv^2}
-\frac{\Li(u)-\zeta(2)}{2u\bu \bv} - \frac{\Li(\bv)-\bv \zeta(2)}{2u v \bv}.
\end{eqnarray}
\end{subequations}
As before they  are finite at $u=v$, posses only logarithmical cuts on the positive $u$-axis,  and can be expressed by
means of  differential operators in terms of the building block (\ref{eq:Lsub-u1v1}).  Both addenda posses still poles at $v=0$ and/or $v=1$. They can be  straightforwardly removed, which, however, would yield more cumbersome expressions. \\

\noindent
{\textit{Momentum fraction representation.}}
\begin{subequations}
\label{eqs:^GT^{(1)}}
\begin{eqnarray}
\label{^G T^{(1,A)}}
{^\text{G}T}^{(1,A)}(u,v)&\!\!\!=\!\!\!&
\left[ \ln\frac{\cQ^2}{\muF^2}+\frac{\ln \bu }{2}
+\frac{3\ln(v\bv)}{4} -\frac{3}{2} \right]\!
\left[1+\frac{\bu^2}{u^2} \right]\frac{ \ln \bu }{2 \bu v\bv}
 \\
&&\!\!\!\!\!
+
\left[\frac{\ln\bu}{2}- \frac{\ln(v \bv) }{4} -\frac{3}{2}\right]\frac{\ln\bu }{u\, v\bv}
+
\left[1 + \zeta(2) - \frac{v^2 \ln v+ \bv^2 \ln\bv }{2v\bv}\right]\frac{1}{4\bu v\bv}
\nonumber\\
&& \!\!\!\!\!
-\left[\frac{(\bu-u) \Li(u) + u\, \zeta(2) + \bu  \ln^2 \bu }{u \bu} + \left[2+  \ln(v \bv)\right] \frac{\ln\bu + u }{4 u^2}
\right] \frac{1}{2 v\bv}  + \Delta{^\text{G}T}^{(1,\text{A})}(u,v),
\nonumber
\end{eqnarray}
\begin{eqnarray}
\label{^G T^{(1,F)}}
{^\text{G}T}^{(1,F)}(u,v)&\!\!\!=\!\!\!&
\left[
\ln\frac{\cQ^2}{\muF^2}+  \frac{\ln \bu }{2} - \frac{1}{\bu}- (1-2v \ln v - 2\bv \ln\bv ) \frac{u}{2\bu}
\right]\frac{(-1)\ln \bu }{4 u^2 v\bv} - \frac{31}{16 \bu v\bv}
\nonumber\\
&&\!\!\!\!\!+ \left[
\ln\frac{\cQ^2}{\muphi^2} +   \frac{\ln \bu }{2 u} + \frac{\ln \bv }{2}  + \frac{1}{4}
\right]  \frac{3 + 2 \ln \bv }{2\bu \bv}
\\
&&\!\!\!\!\!+
\left[
\frac{v^2 \ln v + \bv^2 \ln \bv }{4 v \bv} -
\frac{(\bv -v)\left[\Li(v) - \Li(\bv)\right]+\zeta(2)}{2}  \right]\frac{1}{2 \bu v \bv}
+ \Delta{^\text{G}T}^{(1,\text{F})}(u,v)
\, .
\phantom{\Bigg|}
\nonumber
\end{eqnarray}
Note that $\ln \bu /u^2$, appearing in the first term on the r.h.s.~of (\ref{^G T^{(1,A)}}), contains a  pole at $u=0$.
The addenda, explicitly given in (\ref{Delta^G T^{(1)}}), read in terms of the building block (\ref{eq:Lsub-u1v1}) as
\begin{eqnarray}
\label{Delta^G T^{(1,A)}-1}
\Delta{^\text{G}T}^{(1,\text{A})}(u,v) &\!\!\!=\!\!\!&
\frac{1}{v\bv}\frac{\partial}{\partial v}
\frac{v \bv(\bv-v)}{4}
\left[\frac{1}{u\bv}\frac{L (u,v)}{u-v}\right]^{\rm sub},
\\
\label{Delta^G T^{(1,F)}-1}
\Delta {^\text{G}T}^{(1,\text{F})}(u,v)&\!\!\!=\!\!\!&
\left[\frac{\partial^2}{\partial v^2} - \frac{2}{v\bv}\right]\frac{v \bv}{4}
\left[\frac{1}{u \bv}\, \frac{L(u,v)}{u-v} \right]^{\rm sub}.
\end{eqnarray}
\end{subequations}

\noindent
\textit{Imaginary parts of} (\ref{eqs:^GT^{(1)}}).
\begin{subequations}
\label{eqs:{^Gt}^{(1)}}
\begin{eqnarray}
\label{^G t^{(1,A)}}
{^\text{G} t}^{(1,\text{A})}(r,v)
&\!\!\!=\!\!\!&
\left[ \ln\frac{\cQ^2}{\muF^2} + \frac{3\ln(v\bv)}{4}  -\frac{3}{2} \right]
\left[ \left\{\frac{1}{1-r}\right\}_+ -\delta(1-r) + \frac{1-r}{(1+r)^2} \right]\frac{1}{2v\bv}
\\
&&\!\!\!\!\!
+\left[
\left\{\frac{\ln\frac{1-r}{2 r}}{1-r}\right\}_+  -  \frac{(1+3r)
\ln\frac{1-r}{2r}}{(1+r)^2} + \frac{3}{1+r} + \frac{\ln( v\bv )}{2(1+r)}
\right] \frac{1}{2v\bv}
\nonumber \\
&&\!\!\!\!\!
+ \left[1+ \zeta(2) - \frac{v^2 \ln v + \bv^2 \ln\bv }{2v\bv}\right] \
 \frac{\delta(1-r)}{4v\bv}
\nonumber \\
&&\!\!\!\!\!
+ \left[\frac{2\ln\frac{1-r}{1+r}}{1+r}-\frac{2r \ln\frac{1+r}{2 r}}{1-r^2} + \left[2 + \ln(v \bv) \right]\frac{ r  }{2 (1 + r)^2}  \right]
\frac{1}{2v\bv} + \Delta{^\text{G}t}^{(1,\text{A})}(r,v)
\,, \nonumber\\
\label{^G t^{(1,F)}}
{^\text{G} t}^{(1,\text{F})}(r,v)&\!\!\!=\!\!\!&
\Bigg[
\ln\frac{\cQ^2}{\muF^2} \delta(1-r) +\ln\frac{\cQ^2}{\muF^2}\, \frac{2r}{(1+r)^2}  +
\frac{3- 2 v \ln v - 2\bv \ln\bv}{2}
\\
&&
\phantom{\ln\frac{\cQ^2}{\muF^2} \delta}
\times \left(\left\{\frac{1}{1-r}\right\}_+ -\frac{1}{1+r}\right)
-\frac{35}{4}  \delta(1-r) + \frac{2r \ln\frac{1 - r}{2 r} -2r}{(1+r)^2}
\Bigg]  \frac{1}{4v\bv}
\nonumber\\
&& \!\!\!\!\!
+
\left[
\ln\frac{\cQ^2}{\muphi^2} \delta(1-r) + \frac{1}{2} \left\{\frac{1}{1-r}\right\}_+
+ \frac{1+2 \ln\bv }{4}\delta(1-r)  -\frac{1}{2(1+ r)}
\right]
\frac{3+2 \ln\bv }{2\bv}
\nonumber \\ && \!\!\!\!\!
+ \left[
 \frac{v^2 \ln v + \bv^2 \ln\bv }{4v \bv} - \frac{(\bv - v) \left[ \Li(v) - \Li(\bv) \right] + \zeta(2)}{2}
\right] \frac{\delta(1 - r)}{2 v\bv} + \Delta{^\text{G}t}^{(1,\text{F})}(r,v)
\,.
\phantom{\Bigg|}
\nonumber
\end{eqnarray}
Note that we  utilized symmetry under $r\to -r$ (or $u\to \bu$) to reexpress the Dirac function $\delta(1+r)$,
which stems from the  remaining $u=0$ pole [see also discussion below (\ref{eq:uv-func-3})],
$$
\frac{\ln \bu }{u^2 } \;\;\Rightarrow \;\; -\frac{2 r }{(1+r)^2}-\delta(1+r) \;\;\Rightarrow \;\;  -\frac{2 r }{(1+r)^2}-\delta(1-r)\,.
$$
The imaginary parts of the addenda (\ref{Delta^G T^{(1,A)}-1}) and (\ref{Delta^G T^{(1,F)}-1}) are
\begin{eqnarray}
\label{Delta^G t^{(1,A)}}
\Delta {^\g t}^{(1,\text{A})}(r,v)&\!\!\!=\!\!\!&
\frac{1}{4v\bv} \frac{\partial}{\partial v} \frac{2 r v (\bv-v)  }{1+r} \! \left[
\frac{ \ln\frac{1+r}{2r v}}{1+r-2 r v} -  \frac{\ln\frac{1+r}{2r}}{1-r}\right]
, \phantom{\Bigg|}\\
\label{Delta^G t^{(1,F)}}
\Delta {^\g t}^{(1,\text{F})}(r,v)&\!\!\!=\!\!\!&
\frac{1}{2}\left[\frac{\partial^2}{\partial v^2} - \frac{2}{v\bv}\right]\frac{r v}{1+r}\! \left[ \frac{\ln\frac{1+r}{2r v}}{1 + r - 2 v r}-\frac{\ln\frac{1+r}{2r}}{1 - r}\right].
\phantom{\Bigg|}
\end{eqnarray}
\end{subequations}

\noindent
\textit{Conformal moments of} (\ref{eqs:^GT^{(1)}}).
\begin{subequations}
\label{^G c^{(1)}}
\begin{eqnarray}
\label{^G c^{(1,A)}}
{^\text{G} c}^{(1,\text{A})}_{jk} &\!\!\!=\!\!\!&
\left[
-\ln\frac{\cQ^2}{\muF^2} + S_ 1(j+1) + \frac{3}{2} S_ 1(k+1) +   \frac{1}{2} +   \frac{1}{(j+1)_2}
\right] \frac{^{\g\g}\gamma_j^{(0,\text{A})}}{2}
\phantom{\Bigg|}\\  &&
- \frac{3\left[2S_1(j+1) + S_1(k+1) -6\right]}{j (j+3)} + \frac{8 + 4\zeta(2) - (k+1)_2\, \Delta S_2\big(\frac{k+1}{2}\big)}{8}
\nonumber \\ &&
 - \frac{\Delta S_2\big(\frac{j+1}{2}\big)}{2} -\frac{10(j+1)_2 + 4}{[(j+1)_2]^2}  + \Delta{^\text{G}c}^{(1,\text{A})}_{jk}
\,,
\nonumber
\end{eqnarray}
\begin{eqnarray}
\label{^G c^{(1,F)}}
{^\text{G} c}^{(1,\text{F})}_{jk} &\!\!\!=\!\!\!&
\left[
-\ln\frac{\cQ^2}{\muphi^2}+S_ 1(j+1)+S_ 1(k+1) - \frac{3}{4}-\frac{1}{2(k+1)_2}-\frac{1}{(j+1)_2}
\right] \frac{\gamma_k^{(0,\text{F})}}{2}
 \\ &&\!\!\!\!\!
+\left[-\ln\frac{\cQ^2}{\muF^2}+ 3 S_ 1(j+1) - \frac{1}{2}  + \frac{2 S_1(j+1)-1}{(k+1)_2} -\frac{1}{(j+1)_2}\right]
\frac{j+3}{2}\frac{{^{\Sigma\text{G}}\gamma}_j^{(0,{n_f})}}{2}
\nonumber \\ &&\!\!\!\!\!
-\left[35 - \left[(k+1)_2+2\right] \Delta S_2\Big(\frac{k+1}{2}\Big) - \frac{4}{[(k+1)_2]^2}\right]\frac{1}{8}
\nonumber \\ &&\!\!\!\!\!
+\left[\frac{\left[(k+1)_2+2\right] S_ 1(j+1)}{(k+1)_2} +1 \right] \frac{1}{(j+1)_2}+ \Delta{^\text{G}c}^{(1,\text{F})}_{jk}
\,.
\nonumber
\end{eqnarray}
where $\gamma_k^{(0,\text{F})}$  is defined in (\ref{gamma_j^{(0,F)}}),
\begin{eqnarray}
\label{^Sigma gamma_j^{(0,n_f)}}
{^{\Sigma\g}}\gamma_j^{(0,n_f)} = - \frac{4 +2(j + 1 )( j + 2 ) }{(j + 1 )( j + 2 )( j + 3)}
\end{eqnarray}
can be read off from (\ref{Def-LO-AnoDim-QG-V}), and
\begin{eqnarray}
\label{^gg gamma_j^{(0,A)}}
{^{\g\g}}\gamma_j^{(0,\text{A})} = 4S_1(j + 1 ) + \frac{4}{(j + 1 )( j + 2)} - \frac{12}{j(j + 3)}
\end{eqnarray}
is the part proportional to  $\CA$  of the anomalous dimension (\ref{Def-LO-AnoDim-GG-V}) in the gluon channel.
The addenda follow from (\ref{Delta^G t^{(1,A)}},\ref{Delta^G t^{(1,F)}}) and they can be
cast with a little bit of algebra in the  form
\begin{eqnarray}
\label{Delta{g}c^{(1,A)}_{jk}}
\Delta{^\g}c^{(1,\text{A})}_{jk} &\!\!\!=\!\!\!&
 - \left[\frac{\Delta S_2\big(\!\frac{j+1}{2}\!\big)}{2(k+1)_2}+
\frac{(k-1)\Delta S_{2}\big(\!\frac{j+1}{2},\frac{k}{2}\!\big) +  (k+4)\, \Delta S_{2}\big(\!\frac{j+1}{2},\frac{k+2}{2}\!\big)}{2k+3}\right]
\frac{(k)_4}{4}
\\
&&\qquad
+\frac{(k+1)_2\,S_1(k+1)-2}{(j+1)_2(k+1)_2}\,,
\nonumber
\\
\label{Delta{g}c^{(1,F)}_{jk}}
\Delta{^\g}c^{(1,\text{F})}_{jk} &\!\!\!=\!\!\!& \Bigg[
 \frac{\Delta S_2\big(\!\frac{j+1}{2}\!\big)}{2(k+1)_2}
-\frac{(k-1)_2\,\Delta S_{2}\big(\!\frac{j+1}{2},\frac{k}{2}\!\big)- (k+3)_2\,  \Delta S_{2}\big(\!\frac{j+1}{2},\frac{k+2}{2}\!\big)}{2(2k+3)}
\Bigg] \frac{(k+1)_2 [(k+1)_2 +2]}{4}
\nonumber\\
&&\qquad .-\frac{(k+1)_2 +2}{2(j+1)_2(k+1)_2} \,,
\end{eqnarray}
\end{subequations}
where the finite sum, appearing in $\Delta{^\g}c^{(1,\text{A})}_{jk}$  could be performed, cf.~(\ref{dif-op3}).

\section{Estimates of radiative NLO corrections}
\label{sect-numerics}
\setcounter{equation}{0}

In specific model estimates the size of NLO corrections were reported to be large for DV$\!\pi^+$P \cite{Belitsky:2001nq} and  DV$\!V^0_L$P in the small-$\xB$
region \cite{Ivanov:2004zv}. A comprehensive study, restricted to GPD models that are build with Radyushkin's factorized double distribution ansatz (RDDA) \cite{Radyushkin:1997ki,Musatov:1999xp}, was performed in \cite{Diehl:2007hd} at a rather large input scale square $\cQ_0^2 = 16\, \GeV^2$
with three active flavors, where the authors also reported rather large corrections.  Numerical model studies were also given for the DVCS amplitude, including
the consistent treatment of evolution effects \cite{Belitsky:1999sg,Freund:2001rk,Kumericki:2007sa}, where NLO corrections are more moderate, see also \cite{Moutarde:2013qs}.
After all these studies, mainly restricted to one class of GPD models that is not entirely favored from GPD phenomenology, we have the desire to understand radiative corrections on a generic level.  The basic idea is to identify terms that vanish if the GPD does not evolve, which is presumable the case in the valence region ($\xB \sim 0.3$). Furthermore, we analytically calculate the TFFs in the large-$x_B$ ($x_B \gtrsim 0.5$) and small-$x_B$ ($x_B \ll 0.1$) region. For shortness we will in the following only discuss radiative corrections at the input scale, thereby, we concentrate us on DV$V_L$P processes, however, our results can be easily adopted to DV$PS$P processes, too.

In Sec.~\ref{sec:NLOestimates-model} we recall for later use a flexible GPD model, based on the CPWE as it is outlined in Sec.~\ref{sec:prel-MB-TFF}.
In Sec.~\ref{sect-NLOcorrections-generic}  we present the technicalities for a generic analysis of NLO radiative corrections in momentum fraction space.
In the remaining Sec.~\ref{sec:NLOestimates} we  discuss the NLO corrections in both the flavor non-singlet and singlet channel, where we illuminate generic properties and model dependencies with numerical predictions from our GPD model.  Finally, we compare our DV$\!V_{L}$P results with the NLO corrections in DVCS, providing for the latter also a generic understanding.

\subsection{GPD models and evaluation of TFFs}
\label{sec:NLOestimates-model}

A flexible GPD model, which is/will be employed for global fitting \cite{Kumericki:2009uq,Aschenauer:2013qpa,Lautenschlager:2013}, can be easily set up in terms of conformal GPD moments.
We adopt the common PDF terminology for the parton species, see (\ref{F^q-deomomposition}).
For simplicity, we take a universal functional form for the various antiquarks and specify the flavor content of the sea
at the input scale
\begin{eqnarray}
H^{\bar{q}}(\cdots) = \frac{1}{2}S_q H^{\rm sea}(\cdots)\quad\mbox{with}\quad S_u =S_d= \frac{2}{5} S\,,\quad\mbox{and}\quad  S_s=\frac{1}{5}\,.
\end{eqnarray}
Here, $S_q$ are the sea quark  asymmetry parameters, which  we took from the MRST  parameterization in \cite{Martin:1998sq}, and
we equate the sea quark and antiquark distributions, i.e.,  $q^{\rm sea} = \bar{q}$. Furthermore, the quark distribution $q=q^{\rm val}+q^{\rm sea}$
is the sum of valence and sea quarks.  Hence, we have at the input scale for charge even and odd quark GPDs:
\begin{eqnarray}
\label{H-parton decompositon}
H^\qpl(x,\cdots) &\!\!\!=&\!\!\!   {H}^{q_{\rm val}}(x,\cdots)+\frac{2}{5} H^{\rm sea}(x,\cdots) \quad\mbox{for}\quad q\in\{u,d\}\\
H^\spl(x,\cdots) &\!\!\!=&\!\!\! \frac{1}{5} H^{\rm sea}(x,\cdots) \,,
\qquad\qquad\mbox{and}\quad
H^\qmi(x,\cdots) = {H}^{q_{\rm val}}(x,\cdots) \,.
\nonumber
\end{eqnarray}
To overcome the quark-gluon mixing in the charge even sector, we switch to the group theoretical basis  (\ref{eq:SU(n)}) and
build with the flavor singlet quark  and gluon GPDs  the vector valued GPD (\ref{eq:gpdFvec}).

Our PDF models are formulated in terms of Mellin moments, which we also dress with $t$-dependence.
For sea quark and gluon PDFs we utilize a simple, however, realistic model that is described in \cite{Kumericki:2007sa,Kumericki:2009uq}.
For both quark and gluon GPDs we use the ansatz
\begin{eqnarray}
\label{KMP-ans-generic}
H_j(\eta=0,t,\mu_0^2) &\!\!\!=\!\!\!& \frac{N}{\left(1-\frac{t}{M^2}\right)^p}\, \frac{\Gamma(2-\alpha+j)\Gamma(3-\alpha+\beta)
}{(1-\alpha(t)+j)\Gamma(2-\alpha)\Gamma(2-\alpha+j+\beta)} \,,
\end{eqnarray}
where the normalization $N=H_{j=1}(\eta=0,t=0,\mu_0^2)$ is the momentum fraction, $\alpha(t)= \alpha + \alpha^\prime t $ is the effective
`pomeron' trajectory with  $\alpha \approx 1.1$ and $\alpha^\prime \approx 0.15$. The residual $t$-dependence is parameterized by a $p$-pole ansatz
with the cut-off mass $M$.  At $t=0$ the Mellin moments (\ref{KMP-ans-generic}) are obtained from the  sea quark  or gluon PDF parameterizations
(see (\ref{Fn}) and below)
\begin{eqnarray}
\left\{{q^{\rm sea} \atop g }\right\}(x,\mu_0^2)=N\frac{\Gamma(3-\alpha+\beta)}{\Gamma(2-\alpha) \Gamma(1+\beta)}\, x^{-\alpha} (1-x)^\beta\,.
\end{eqnarray}

For valence quarks we take a model which has been discussed in Ref.~\cite{Bechler:2009me,Kumericki:2011zc}.
It is based on generic arguments and a simple PDF ansatz, where the $t$-dependence is now entirely contained in the leading `Regge' trajectory
$\alpha(t)= \alpha+ \alpha^\prime t$:
\begin{eqnarray}
\label{KLMSPM-ans-generic}
H^{q_{\rm val}}_j(\eta=0,t,\mu_0^2) &\!\!\!=\!\!\!& N^q\, \frac{\Gamma(1-\alpha(t)+j)\Gamma(1+p-\alpha+j)\Gamma(2-\alpha+\beta)
}{
\Gamma(1-\alpha)\Gamma(1+p-\alpha(t)+j)\Gamma(2-\alpha+j+\beta)}
\\
&&\times \left[(1-h)+
h \frac{\Gamma(2-\alpha+\beta+\delta\beta)\Gamma(2-\alpha+j+\beta)}{\Gamma(2-\alpha+j+\beta+\delta\beta)\Gamma(2-\alpha+\beta)}\right].
\nonumber
\end{eqnarray}
Here, the normalization $N^q=2\,(1)$ gives now the number $u$ ($d$) valence quarks,
$p$  determines the large $-t$ behavior, $\beta$ and $\delta\beta$ the large $j$
behavior, and $h$ is a phenomenological parameter%
\footnote{In the case that large-$x$ counting rules \cite{Brodsky:1994kg} are not spoiled by non-leading terms in a $1-x$
expansion, $h\times N^q$ might be interpreted as the probability for a quark to have opposite helicity to that of the longitudinally polarized proton.}.
The valence quark PDFs are then given by an inverse Mellin transform,
\begin{eqnarray}
\label{KLMSPM-ans-genericPDF}
q^{\rm val}(x,\mu_0^2) &\!\!\!=\!\!\!& N^q\, \frac{\Gamma(2-\alpha+\beta)}{\Gamma(1-\alpha)\Gamma(1+\beta)} x^{-\alpha} (1-x)^\beta
\\
&&\times\left[
(1-h) + h \frac{\Gamma(1+\beta)\Gamma(2-\alpha+\beta+\delta\beta)}{\Gamma(2-\alpha+\beta)\Gamma(1+\beta+\delta\beta)} (1-x)^{\delta\beta}
\right],
\nonumber
\end{eqnarray}
where the $t$-dependent PDF analog can be expressed in terms of hypergeometric  $_2F_1$-functions. Note
that such a Mellin moment--to--momentum fraction GPD modeling conveniently allows to implement form factor data or lattice predictions.
This is basically the inverse procedure as chosen in \cite{Diehl:2004cx,Guidal:2004nd,Diehl:2013xca}.

\begin{figure}[t]
\begin{center}
\includegraphics[width=15cm]{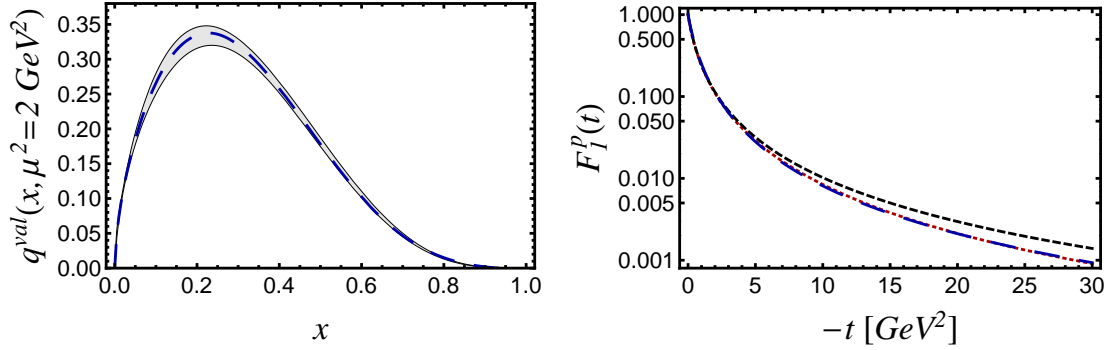}
\end{center}
\vspace{-0.7cm}
\caption{
\small
A simple GPD model (long dashed), based on the ansatz (\ref{KLMSPM-ans-generic}),  versus Alekhins LO PDF parameterization \cite{Alekhin:2002fv} (grayed area) [left panel] and Kelly's \cite{Kelly:2004hm} (dotted)
[Sachs (short dashed)] form factor parameterization [right panel].
\label{KLMSPM-fig-genmod}}
\end{figure}

We add as side remark that the  model%
\footnote{A slightly different version is used in \cite{Bechler:2009me}.}
(\ref{KLMSPM-ans-genericPDF}) with generic parameter values  provides reasonable results for the isotriplet part of $\widetilde H$ (also used for $\widetilde E$) in \cite{Bechler:2009me}. To adopt here our GPD $H$ to Alekhin`s LO PDF parameterization \cite{Alekhin:2002fv} we choose the Regge intercept  $\alpha= 0.43$, $\beta=
3.2$, $\delta\beta=2.2$, and $h= -1$.  For the `Reggeon' slope parameter we take the typically value $\alpha^\prime=0.85$ and to match with the form factor data we chose  $p= 2.12$.  Note that $\beta$, $\delta\beta$, and $p$ only  differ slightly from the canonical values 3, 2, and 2, respectively, and that $\alpha(t)= 0.43+0.85 t$ is essentially the $\rho/\omega$ trajectory.
 In Fig.~\ref{KLMSPM-fig-genmod} we illustrate that our results (long dashed) for the valence
GPD  $H^{\rm val} = (4/9) H^{u^{\rm val}} + (1/9)H^{d^{\rm val}}$ (left panel) and the electromagnetic form factor $F_1^p$ (right panel)
are  consistent with  Alekhin`s PDF (band) and Kelly's parameterization (dotted) \cite{Kelly:2004hm}, respectively.

To parameterize the degrees of freedom that
can be accessed in hard exclusive reactions, one might expand the conformal moments in terms of  $t$-channel SO(3)-PWs~\cite{Polyakov:1998ze}, expressed by Wigner rotation matrices. We denote them as $\hat d_n(\eta)$ and normalize them for $\eta=0$ to one, i.e.,  $\hat d_n(\eta=0) =1$.
 Depending on the GPD type, these SO(3)-PWs are either
Gegenbauer polynomials with index $\nu=1/2$, i.e., Legendre polynomials, or with index $\nu=3/2$~\cite{Diehl:2003ny}.
Since we will not discuss in the following the $D$-term contribution, which might be understood as an integral part of the  SO(3)-PWs that
completes polynomiality \cite{Kumericki:2007sa}, we can restrict ourselves in the following to Gegenbauer polynomials with index $\nu=3/2$.
An effective GPD model at a given input scale $\cQ_0$ is provided by taking into account three  SO(3)-PWs, e.g., for integral $n\ge 4$:
\begin{subequations}
\label{mod-nnlo}
\begin{eqnarray}
\label{KLMSPM-mod-nnlo}
F_n(\eta,t) =\hat d_n(\eta) f_n^{n+1}(t) +\eta^2 \hat d_{n-2}(\eta) f_n^{n-1}(t) + \eta^4 \hat d_{n-4}(\eta) f_n^{n-3}(t)
\quad\mbox{for}\quad n\in\{4,5,6,\cdots\}\,.
\end{eqnarray}
In the simplest version of such a next-next-leading (nnl) SO(3)-PW model, one might introduce just
two additional parameters by setting the non-leading SO(3)-PW amplitudes to:
\begin{eqnarray}
\label{KLMSPM-mod-nnlo1}
f_j^{j+1-\nu}(t) = s_\nu F_j(t) \; \; \mbox{for}\; \; \nu\in\{2,4\}
\quad \mbox{and}\quad
f_j^{j-\nu}(\eta,t) = 0\; \; \mbox{for}\; \; \nu\in\{6,8,\cdots\},
\end{eqnarray}
where $F_j(t)\equiv f_j^{j+1}(t)$ are the Mellin moments of a skewless GPD, e.g., specified in (\ref{KMP-ans-generic}) and (\ref{KLMSPM-ans-generic}), and the proper
choice of the complex valued SO(3)-PWs is given by representing the Gegenbauer polynomials with index $\nu=3/2$ by the following hypergeometric function
\begin{eqnarray}
\label{hat{d}_j}
\hat{d}_j(\eta) =
\frac{\Gamma\big(\frac{3}{2}\big) \Gamma(j+3)}{2^{j+1} \Gamma\big(j+\frac{3}{2}\big)}\, \frac{2\eta^{j+1}}{1+\eta}\,
_2F_1\!\left(\!{-j-1,j+2 \atop 2}\bigg|\frac{\eta -1}{2 \eta }\!\right)\,.
\end{eqnarray}
\end{subequations}

The TFFs are evaluated by means of Mellin-Barnes integrals (\ref{eq:tffqF_M-MBI}) and (\ref{eq:tffSF_M-MBI}),
where the constraint (\ref{KLMSPM-mod-nnlo}) is taken into account by a shift of the integration variable in
the Mellin-Barnes integral. For technical details see Sec.~3.2 of \cite{Kumericki:2009uq}.  Inserting the ansatz (\ref{mod-nnlo}) into (\ref{eq:tffqF_M-MBI}) yields  our model for the flavor non-singlet TFF (\ref{eq:tffqF_M-MBI-a}), which reads  at the input scale $\cQ=\cQ_0$ as follows
\begin{eqnarray}
\label{tffF_NS-model}
\tffF^\text{A}_{\rm  M}(\xB,t,\cQ_0^2)
&\!\!\! \stackrel{\rm Tw-2}{=}\!\!\! &
 \frac{3C_F  f_{\rm M}}{N_c \cQ_0}\,
  \frac{1-\frac{\xB}{2}}{2 i} \int_{c- i \infty}^{c+ i \infty}\! dj\,
\left[
  i \pm \left\{{\tan \atop \cot }\right\}\!\left(\!\frac{\pi\,j}{2}\!\right)
  \right] (3+2j)
\qquad\qquad\mbox{for}\; \sigma = \left\{ +1 \atop -1 \right.
  \nonumber  \\[0.3cm]
&&
\times {_2F_1}\!\left(\!{-j-1,j+2 \atop 2}\bigg|\frac{\xB-1}{\xB}\!\right)
F^\text{A}_j(t)  \sum_{{\nu=0 \atop {\rm even}}}^4  s_{\nu}
\frac{2^{\nu} \big(j+\frac{3}{2}\big)_{\nu}}{(j+2)_{\nu}} \sum_{{k=0 \atop {\rm even}}}^\infty {^\sigma\! c}_{j+\nu,k}(\cdots )\, \varphi_{\text{M},k}
\,,  \nonumber \\
\end{eqnarray}
where $F_j(t)$ and $\varphi_{\text{M},k}$ (with $\varphi_{\text{M},0}=1$) are the moments of our skewless GPD and meson DA at the input scale, respectively, and the conformal moments (\ref{^Ac_{jk}}) are specified to NLO accuracy in (\ref{c^(1)_jk},\ref{c^{(1)}}). We recall that for signature even and odd TFFs $c< 1$
and $c< 0$, respectively, is required while a lower bound for $c$ arises from the requirement that all singularities lie on the
l.h.s.~of the integration path. Analogously, one can write down the flavor singlet TFFs (\ref{eq:tffSF_M-MBI}), c.f.~the treatment of CFFs in the small-$\xB$ region in Sec.~3.2 of \cite{Kumericki:2009uq}.

In the small-$\xB$ region the TFFs can be easily evaluated. Shifting in (\ref{tffF_NS-model}) the integration path to the l.h.s., we pick up the leading `Regge' pole at $j=\alpha(t)-1$:
\begin{eqnarray}
\label{tffF_NS-model-smallxB}
\tffF^\text{A}_{\rm  M}(\xB,t,\cQ_0^2)
&\!\!\! \stackrel{\xB\to 0}{=}\!\!\! &
 \frac{3C_F  f_{\rm M}}{N_c \cQ_0}\,\pi
\left[
  i \mp \left\{{\cot \atop \tan }\right\}\!\left(\!\frac{\pi\,\alpha(t)}{2}\!\right)
  \right] \left(\frac{\xB}{2}\right)^{-\alpha(t)}
  \qquad \qquad \qquad\mbox{for}\; \sigma = \left\{ +1 \atop -1 \right.
 \\[0.3cm]
&&\!\!\!\!\! \times \frac{2^{\alpha(t)}\ \Gamma\big(\alpha(t)+\frac{3}{2}\big)}{\Gamma\big(\frac{3}{2}\big)\Gamma(\alpha(t)+2)}
 \sum_{{\nu=0 \atop {\rm even}}}^4 s_{\nu}  \frac{2^{\nu} \big(\alpha(t)+\frac{3}{2}\big)_{\nu}}{(\alpha(t)+2)_{\nu}}
 \sum_{{k=0 \atop {\rm even}}}^\infty {^\sigma\!c}_{\alpha(t)+\nu-1, k}(\cdots)\,  \varphi_{\text{M},k}\,
 {\rm Res}F^{\text{A}}_{j=\alpha(t)-1}(t)\,,
   \nonumber
\end{eqnarray}
where ${\rm Res}F^{\text{A}}_{j=\alpha(t)-1}(t)$ is the residue of the skewless GPD. Obviously, the normalization of the TFF is controlled by
both the SO(3)-PW and meson DA  parameters. For example, restricting us to three lowest CPW amplitudes and to LO accuracy the TFF is in the
small-$\xB$ region proportional to
$$
\sum_{{\nu=0 \atop {\rm even}}}^4 s_{\nu}  \frac{2^{2\nu} \big(\alpha(t)+\frac{3}{2}\big)_{\nu}}{(\alpha(t)+2)_{\nu}}
 \sum_{{k=0 \atop {\rm even}}}^4{^\sigma\!c}^{(0)}_{\alpha(t)+\nu-1, k}\,  \varphi_{k} =
\left(1+ \frac{2^{4} \big(\alpha(t)+\frac{3}{2}\big)_{2}}{(\alpha(t)+2)_{2}}  s_2 +
 \frac{2^{8} \big(\alpha(t)+\frac{3}{2}\big)_{4}}{(\alpha(t)+2)_{4}}  s_4 \right) (1+\varphi_2+\varphi_4)\,.
$$
Furthermore, since  the two non-leading  mesonic CPW amplitudes in $[1+\varphi_2(\cQ^2)+\varphi_4(\cQ^2)]$  evolve with different strength,
we can use them to control the evolution of the overall normalization, where their sum  at the input scale can be fixed.
Rather analogously,  the two non-leading SO(3)-PWs evolve differently, giving us an additional handle to control the evolution flow, too.

\subsection{Generic properties of NLO corrections}
\label{sect-NLOcorrections-generic}

Having rather simple analytic formulae for the hard scattering amplitudes in terms of our building blocks at hand, presented in Sec.~\ref{sec:NLO-result},
we can easily understand,  even quantify, NLO corrections in an analytic manner. The {\em leading} singularities,
listed in Tab.~\ref{subT-1} and grouped together with factorization and renormalization logarithms, play a key role.
In the momentum fraction representation we count them as the $1/\bu$ (and $1/\bv$) poles that are also combined with a logarithmical $[1,\infty]$-cut (at NLO squared logarithms can appear), i.e., their imaginary parts are given in terms of $+$-prescriptions. In the language of CPWs they are identified as the first order harmonic sum $S_1(j+1)$ [or $S_1(k+1)$], up to the second power, which grow logarithmically  at large $j$ [or $k$], see (\ref{S_1-asym}).
Furthermore, the strength of NLO terms proportional to the LO pole is counted by their residue (times $\alpha_s/2\pi$) while all other contributions can be considered as moderate or small.

Let us remind that the NLO corrections in the flavor non-singlet channel, given in Sec.~\ref{sec:anatomy-quark}, have been already intensively discussed for the pion form factor, where the signature $\sigma=-1$ applies \cite{Field:1981wx,Dittes:1981aw,Khalmuradov:1984ij,Braaten:1987yy,Melic:1998qr,Braun:1999uj,Bakulev:2004cu}. It was found that the leading singular terms yield a logarithmical enhancement  in the endpoint region or a
logarithmical enhancement of higher CPWs \cite{Mueller:1998qs}. Consequently, if the DA has a concave shape, sizeable corrections show up and their size is directly related to  the endpoint behavior of the DA, i.e., it increases with the flatness of the DA. Note, however, that contributions from separate singular terms may cancel each other and that such a generic counting may not hold true if the DA has a more intricate shape.

To allow for a straightforward, compact,  and rather generic discussion, we express the first order harmonic sums,
appearing in the  conformal moments (\ref{c^{(1)}}) of the hard scattering amplitude (\ref{qT^{(1)}}), in terms of anomalous dimensions (\ref{gamma_j^{(0,F)}}).
Hence, in momentum fraction representation all leading singularities can be expressed as convolution of the LO hard scattering amplitude with
the LO evolution  operator,
\begin{subequations}
\label{1/buV}
\begin{eqnarray}
\label{1/buV^1}
\frac{-\gamma^{(0,\text{F})}_j}{2} =  2 S_{1}(j+1) - \frac{3}{2}- \frac{1}{(j+ 1)_2}
 & \Leftrightarrow & \frac{2\ln\bu+3}{2\bu} =\frac{1}{\bu^\prime}\otimes \frac{V^{(0)}(u^\prime,u)}{\CF}\,,
\end{eqnarray}
see, e.g., (\ref{T^q-TcoV}). For the square of anomalous dimensions we obtain the equivalent representation
\begin{eqnarray}
\label{1/buV^2}
\frac{\left(\gamma^{(0,\text{F})}_j\right)^2}{4} & \Leftrightarrow &
\left[\frac{1}{\bu^{\prime\prime}} \stackrel{u^{\prime\prime}}{\otimes}  \frac{V^{(0)}(u^{\prime\prime},u^{\prime})}{\CF}\stackrel{u^{\prime}}{\otimes}  \frac{V^{(0)}(u^{\prime},u)}{\CF}  \right]\!\!(u) = \frac{\left(2\ln \bu +3\right)^2 }{4\bu}
+\frac{\ln\bu}{u}+\frac{\Li(u)-\zeta(2)}{\bu}\,,
\nonumber
\\
\end{eqnarray}
\end{subequations}
where the two last terms on the r.h.s.~posses only a logarithmic $[1,\infty]$-cut and are considered as harmless in the endpoint region.
However, note that the sum of their  conformal moments is given by  $-[2(j+1)_2+1]/[(j+1)_2]^2$, see Tab.~\ref{subT-2} and \ref{subT-Li}, and yields $-5/4$ for the asymptotic DA, while the second Gegenbauer moment is already suppressed by a factor $5/36\approx 0.15$.

To quantify the relative NLO corrections in a mostly  model independent manner, we define now two numbers in terms of
the convolution integrals (\ref{1/buV}), which absorb the leading singularities,
\begin{subequations}
\label{ell_phi}
\begin{eqnarray}
\label{ell_phi1}
\frac{-\gamma^{(0,\text{F})}_k}{2} \quad & \Rightarrow & \quad
\ell^{\prime}_\varphi \equiv  \frac{ \int_0^1\!dv\, \frac{3 +2 \ln\bv }{2\bv}\, \varphi(v,\mu^2)}{ \int_0^1\!dv\,\frac{1}{\bv} \varphi(v,\mu^2)}\,, \quad
\\
\label{ell_phi2}
\frac{\left(\gamma^{(0,\text{F})}_k\right)^2}{4}  \quad & \Rightarrow & \quad
\ell^{\prime\prime}_\varphi  \equiv
\frac{ \int_0^1\!dv\, \left[\frac{\left(3+ 2\ln \bv \right)^2 }{4\bv} +\frac{\ln\bv}{v}+\frac{\Li(v)-\zeta(2)}{\bv}\right] \varphi(v,\mu^2)}{ \int_0^1\!dv\,\frac{1}{\bv} \varphi(v,\mu^2)}\,.
\end{eqnarray}
\end{subequations}
The two numbers $\ell^{\prime}_\varphi$ and $\ell^{\prime\prime}_\varphi$  characterize the DAs with respect to their behavior under evolution, which allows us in return to judge the size of radiative corrections in dependence on the behavior of the DA under evolution.
Furthermore, for a given class of DAs we can easily provide some bounds for $\ell^{\prime}_\varphi(\mu^2)$ and $\ell^{\prime\prime}_\varphi(\mu^2)$. Our reference DA is the asymptotic one $\varphi^{\rm asy}=6 v\bar{v}$ for which $\ell^{\prime}_\varphi$ and $\ell^{\prime\prime}_\varphi$ vanish. If we choose a  broader/narrower DA, the value of both $-\ell^{\prime}_\varphi(\mu^2)$  and $\ell^{\prime\prime}_\varphi(\mu^2)$ will become positive/negative.
In the following we compare two rather extreme models, which should provide a good feeling for the possible range of results.
The first model DA $\varphi^{\rm broad}=8 \sqrt{v\bar{v}}/\pi$, e.g., suggested in an AdS/QCD model \cite{Brodsky:2003px}. The second model DA assumes
equal-momentum sharing $\varphi^{\rm narrow}= \delta(v-1/2)$. These two yield the estimates
\begin{eqnarray}
\label{DA-constraint}
-1\,  (\mbox{narrow})\, \lesssim    -\ell^{\prime}_\varphi(\mu^2)  \lesssim 1\, (\mbox{broad})\quad
\mbox{and} \quad
-1\,  (\mbox{narrow})\, \lesssim    \ell^{\prime\prime}_\varphi(\mu^2)  \lesssim 4\, (\mbox{broad})\,.
\end{eqnarray}
Finally, we can express the relative NLO corrections in terms of $\ell^{\prime}_\varphi$, $\ell^{\prime\prime}_\varphi$, the residuum of the pole at $u=1$, counted at LO as one, and more harmless terms. In the latter higher CPWs are suppressed, see discussion below (\ref{1/buV^2}). Thus,  we will take into account the contribution of the lowest CPW,
\begin{eqnarray}
\label{DA-approximation}
\frac{ \int_0^1\!dv\, f(v)\varphi(v,\mu^2)}{ \int_0^1\!dv\,\frac{1}{\bv} \varphi(v,\mu^2)} \approx
\frac{ f_0}{1+ \sum_{k>0} \varphi_k(\mu^2)} \sim f_0 \quad\mbox{with}\quad  \big|{\textstyle\sum_{k>0}}\, \varphi_k(\mu^2)\big| \le \frac{1}{3}\,,
\end{eqnarray}
and neglect the higher order ones, which is for our purpose sufficient.

Analogously as for DAs, we can study the relative NLO corrections to the imaginary part of TFFs.  Replacing in (\ref{ell_phi}) the DA convolution integrals by a quark GPD ones and taking the imaginary part of both numerator and denominator immediately yields according to tables \ref{subT-1}--\ref{subT-Li}:
\begin{subequations}
\label{ell_F}
\begin{eqnarray}
\label{ell_F1}
\frac{-\gamma^{(0,\text{F})}_j}{2} \; & \Leftrightarrow & \;
\ell^{\prime}_F(\xi,t)  \equiv
\int_\xi^1\!dx\left\{\frac{1}{x-\xi}\right\}_+ \frac{F(x,\xi,t,\mu^2)}{F(\xi,\xi,t,\mu^2)} + \frac{3}{2}\,,
\\
\label{ell_F2}
\frac{\left(\gamma^{(0,\text{F})}_j\right)^2}{4}
\; & \Leftrightarrow & \;
\ell^{\prime\prime}_F(\xi,t)  \equiv
 \int_\xi^1\!dx\left[\left\{\frac{2\ln\frac{x-\xi}{2\xi} +3}{x-\xi}\right\}_+
- \frac{1}{x+\xi} - \frac{\ln\frac{x+\xi}{2\xi}}{x-\xi} \right]\frac{F(x,\xi,t,\mu^2)}{F(\xi,\xi,t,\mu^2)}
 + \frac{9}{4} \,.
 \nonumber\\
\end{eqnarray}
\end{subequations}
These functions characterize the evolution of the quark GPD as function of the two kinematical variables $\xi$ and $t$.
Analogously, one may define such quantities for gluon GPDs. In doing so, one should keep in mind
that first order harmonic sums appear in the gluonic  anomalous dimensions (\ref{Def-LO-AnoDim-GG-V}), however,
do not appear in the mixed channels. Moreover, a perturbative `pomeron' pole at $j=0$ appears in both the gluon-quark  and gluon-gluon anomalous dimensions, see (\ref{Def-LO-AnoDim-GG-V},\ref{Def-LO-AnoDim-GQ-V}), which drives the evolution in the small-$\xi$ region. We postpone the discussion of defining appropriate quantities in the flavor singlet channel to Sec.~\ref{sec:estimates-singlet}.

To derive a quark GPD analog of the DA constraints (\ref{DA-constraint}) for the quantities $\ell^{\prime}_F$ and $\ell^{\prime\prime}_F$, we consider first the
convolution of the GPD in the large- and then in the small-$\xi$ region, which b.t.w.~would allow us to solve analytically the LO evolution
equation in these both limits.

\begin{itemize}
\item {\em GPD convolution integrals in the large $\xB$ region.}
\end{itemize}
\vspace{-2mm}
In the convolution integrals (\ref{eq:imtffFqC},\ref{eq:imtffF^S_V}) for the imaginary part of a TFF,
$$
\im \tffF(\xB,t,\cQ^2) \propto \int^1_\xi\frac{dx}{x} F(x,\xi,\cdots)\, t(r=\xi/x,v)\,,
$$
the argument $\xi/x$ of the hard scattering amplitude remains for large $\xi$  in the vicinity of $1$.
The leading $\xB\to 1$ behavior is governed by the most singular terms, see Tab.~\ref{subT-1},
where $\delta (1-x/\xi)$  simply gives the GPD $F(\xi,\xi,\cdots)$ on the cross-over line.  Obviously,
the regular part of the hard scattering amplitude $t(r)$  tends for $r\to 1$ to a constant and, thus, the integration gives
an additional $(1-\xi)$ suppression  factor.
To evaluate the remaining convolution integrals (\ref{f+}), containing a singular $+$-prescription,
we suppose that for large $\xi$  the GPD  behaves in the outer region as
$$
F(x \geq \xi,\xi,t,\mu^2) \simeq   F(\xi,\xi,t,\mu^2)   \left(\frac{1-x}{1-\xi}\right)^\beta + \cdots\,,
$$
where the ellipses stand for terms that die out faster than $(1-x)^\beta$. Guided by RDDA, we might even consider $\beta$ as the parameter that characterizes the large $x$-behavior of the corresponding PDF. The convolution integrals are
now straightforward to calculate,
\begin{subequations}
\label{conv-largexB}
\begin{eqnarray}
\label{conv1-largexB}
\int_\xi^1\!\frac{dx}{x} \left\{ \frac{x}{x-\xi} \right\}_+ \frac{F(x,\xi,\cdots)}{F(\xi,\xi,\cdots)} &\!\!\!\stackrel{\xi\to 1}{\approx}\!\!\!&
-\ln\frac{2\xi}{1-\xi} - S_1(\beta) \,,
\\
\label{conv2-largexB}
\int_\xi^1\!\frac{dx}{x} \left\{ \frac{2x\ln\frac{x-\xi}{2\xi}}{x-\xi}  \right\}_+ \frac{F(x,\xi,\cdots)}{ F(\xi,\xi,\cdots) }
&\!\!\!\stackrel{\xi\to 1}{\approx}\!\!\!\!&
\left[\ln\frac{2\xi}{1-\xi} + S_1(\beta)\right]^2 +  S_2(\beta)-2\zeta(2)\,.
\end{eqnarray}
\end{subequations}
As one realizes, the result is expressible by the GPD $F(\xi,\xi,\cdots)$ on the cross-over line, where
the subtraction procedure (\ref{f+}) causes a logarithmical enhancement effect and the regularized integral a constant, which
depends on the large $\xi \le x$ behavior of the GPD, parameterized by the PDF parameter $\beta$.
We add that this large-$\xB$ discussion can be repeated in terms of the Mellin-Barnes integral  along the lines of Sec.~3.4 in \cite{Bechler:2009me}.

For the large-$\xB$ asymptotics we obtain from the convolution integral (\ref{conv1-largexB}) that the quantity
 $\ell_F^\prime$, defined in  (\ref{ell_F1}), behaves as
 \begin{subequations}
\label{evol-x_large}
\begin{eqnarray}
\label{evol-x_large1}
 \ell^{\prime}_F(\xi,t,\mu^2) \stackrel{\xi \to 1}{\approx} -\ln\frac{2\xi}{1-\xi} - S_1(\beta) + \frac{3}{2} \approx
 -\ln\frac{2}{1-\xi} - S_1(\beta) + \frac{3}{2}   <0\,.
\end{eqnarray}
For  a realistic  $\beta \gtrsim 3$ value the result is negative and decreases with growing $\xi$.
Furthermore,  we find  from (\ref{ell_F}) and
(\ref{conv-largexB}) that $\ell^{\prime\prime}_F$  can be  practically expressed by the square $ \ell^{\prime\,2}_F$,
\begin{eqnarray}
\label{evol-x_large2}
\ell^{\prime\prime}_F(\xi,t,\mu^2) \stackrel{\xi \to 1}{\approx} \ell^{\prime\,2}_F(\xi,t,\mu^2) -\zeta(2) \sim  \ell^{\prime\,2}_F(\xi,t,\mu^2)\,.
\end{eqnarray}
\end{subequations}

\pagebreak[3]
\begin{itemize}
\item {\em GPD convolution integrals in the small-$\xB$ region.}
\end{itemize}
\vspace{-2mm}
The `Regge'  asymptotics of the TFF (\ref{tffF_NS-model-smallxB}), calculated
with a given GPD model in terms of conformal moments, tells us that in contrast to the large-$\xB$ region
the radiative corrections at small-$\xB$
can not be read off from the hard scattering amplitude only and that it depends on the skewness parameters.
Of course,  these asymptotics can be analytically discussed in momentum fraction representation, too.
Such NLO discussions were given for PDF-like cases in  \cite{Ivanov:2004zv}  and \cite{Diehl:2007hd},
in what follows we incorporate the skewness dependence.

Going along the line of \cite{Kumericki:2009uq}, we parameterize a realistic GPD, e.g., for a quark GPD as
$$
F(x\geq \xi,\xi,t,\cQ^2) \stackrel{x\to 0}{=}  x^{-\alpha(t)} r(\xi/x,t,\cQ^2)  + \cdots\,,
$$
where $\alpha(t) > 0$ is the  leading `Regge' trajectory, $r(x,t,\cQ^2)$ is a residue function that factorizes further,
and the ellipsis indicates less singular terms. Note that $r(\eta=0,t,\cQ^2)$ is the residue of the skewless GPD and, hence,
the normalized ratio
$$
 \frac{F(x\geq \xi,\eta=\xi,t,\cQ^2)}{F(x\geq \xi,\eta=0,t,\cQ^2)} = \frac{r(\xi/x,t,\cQ^2)}{r(\eta=0,t,\cQ^2)}
$$
quantifies the skewness effect and controls thus the normalization of the TFF in the `Regge' asymptotics.
Performing a variable transformation $x \to \xi/x$ in the convolution integral (\ref{eqs:DRtffFqC}),
we obtain a more convenient representation for the imaginary part of a quark TFF,
\begin{eqnarray}
\label{tff-x_small}
\im \tffF(\xB,t,\cQ^2)  \stackrel{\xi\to 0}{\propto} \xi^{-\alpha(t)} \lim_{\xi\to 0}\int^1_\xi\!\frac{dx}{x}\, x^{\alpha(t)}\, r(x,t,\cQ^2)\, t(x,v)\stackrel{v}{\otimes} \varphi(v)
\quad \mbox{for} \quad \alpha(t)>0 \,.
\end{eqnarray}
If $t(x,v)/x$ is regular at $x=0$ we can take the limit $\xi\to 0$ and calculate then the integral. Otherwise, we consider the lower limit in the convolution integral as a regulator, split the integral by means of a subtractions procedure, calculate the singular part exactly,
and finally take in the regularized integral the limit $\xi\to 0$.

Let us suppose that $t(x,v)$ behaves in the vicinity of $x=0$ as $x^{-p}$, where the two cases $p\in\{0,1\}$ with $\alpha(t)-p > -1$ are of interest.
Then, we find the following relative contribution
\begin{eqnarray}
\label{x^(-p)-integrals}
\int^1_\xi\!\frac{dx}{x^{p+1}}\, \frac{F(\xi/x,\xi,\cdots)}{F(\xi,\xi,\cdots) }
\stackrel{\xi\to 0}{=}
\frac{1-\xi^{\alpha(t)-p}}{\alpha(t)-p}\, \frac{r(x=0,\cdots)}{r(x=1,\cdots)} +
\int^1_0\!\frac{dx}{x}\, x^{\alpha(t)-p}\, \frac{r(x,\cdots)-r(0,\cdots)}{r(1,\cdots)}\,.
\quad
\end{eqnarray}
The subtraction term is proportional to the residue $r(x=0,\cdots)$ of a forward GPD, where the regularized integral, in which we set the lower limit to zero, depends on the skewness ratio $r(x,\cdots)$.

Terms such as (\ref{x^(-p)-integrals}) with $p=1$ and an effective `pomeron' trajectory $\alpha(t)\sim 1$ appear in the pure  singlet quark channel and
with $p=0$ and $\alpha(t)-1$ in the gluon-quark channel. They have been viewed as a source of big  corrections, e.g., exemplified for the generic `pomeron' intercept $\alpha(t)=1$ \cite{Ivanov:2004zv}, which implies that the corrections (\ref{x^(-p)-integrals}) are logarithmically enhanced,
$$
\int^1_\xi\!\frac{dx}{x^{p+1}}\, \frac{F(\xi/x,\xi,\cdots)}{F(\xi,\xi,\cdots) }
\stackrel{\xi\to 0}{=}
\left[\ln\frac{1}{\xi} +  \int^1_0\!\frac{dx}{x}\left\{ \frac{r(x,\cdots)}{r(0,\cdots)}-1\right\}\right]  \frac{r(x=0,\cdots)}{r(x=1,\cdots)} \approx
\ln\frac{1}{\xi} \,  \frac{r(x=0,\cdots)}{r(x=1,\cdots)}\,.
$$
Surely, in the `soft' regime $\alpha(t)-p <0$, the ratio (\ref{x^(-p)-integrals}) diverges in the small-$\xB$ asymptotics, too,
$$
\int^1_\xi\!\frac{dx}{x^{p+1}}\, \frac{F(\xi/x,\xi,\cdots)}{F(\xi,\xi,\cdots) }
\stackrel{\xi\to 0}{=}
\frac{\xi^{\alpha(t)-p}}{p-\alpha(t)}\, \frac{r(x=0,\cdots)}{r(x=1,\cdots)} \,.
$$
Hence, in both scenarios the naive application of the pQCD formalism might be spoiled and a BFKL inspired framework might be considered as more appropriate. Fortunately, evolution tells us that the effective `pomeron' trajectory increases with growing $\cQ^2$. Hence, once we have reached the  `hard' regime $\alpha(t)-p >0$ the NLO corrections are finite in the small-$\xB$ asymptotic,
$$
\int^1_\xi\!\frac{dx}{x^{p+1}}\, \frac{F(\xi/x,\xi,\cdots)}{F(\xi,\xi,\cdots) }
\stackrel{\xi\to 0}{=}
\frac{1}{\alpha(t)-p}\, \frac{r(x=0,\cdots)}{r(x=1,\cdots)} +
\int^1_0\!\frac{dx}{x}\, x^{\alpha(t)-p}\, \frac{r(x,\cdots)-r(0,\cdots)}{r(1,\cdots)}\,.
$$
However, they  are maybe enhanced by a factor  $1/\left(\alpha(t) -p\right)$, see also \cite{Diehl:2007hd}.
Obviously, in the `hard' regime the net value depends on the skewness ratio $r(x,\cdots)$, too.

We add that the case $p=0$ and $\alpha(t) = \alpha + \alpha^\prime t \lesssim 0$ can also appear for a `Reggeon' trajectory  at larger $-t$ values in the flavor non-singlet channel. However, since the cross section (\ref{dX^V}) will vanish in this limit,
large or even huge relative radiative corrections  are irrelevant for phenomenology. Note that the approximations are not applicable for $\alpha(t) \ge 0$.

The small-$\xi$ asymptotics of popular (quark) GPD models, based on RDDA, is given by
\begin{eqnarray}
\label{RDDA-r}
\frac{r(\xi/x,\cdots)}{r(0,\cdots)} = {_2F_1}\left({\alpha(t)/2,\alpha(t)/2+1/2 \atop b+3/2} \Big| \frac{\xi^2}{x^2} \right)
\mbox{ with }
\frac{r(1,\cdots)}{r(0,\cdots)} =
\frac{\Gamma(2 b+2) \Gamma(b-\alpha(t)+1)}{2^{\alpha(t)} \Gamma(b+1) \Gamma(2 b-\alpha(t)+2)}\,.
\nonumber\\
\end{eqnarray}
This skewness ratio is governed by the positive profile function parameter $b$ and it decreases with growing $b$ (narrowing the profile function),
reaching the value 1 for $b\to \infty$.
Plugging (\ref{RDDA-r}) into (\ref{tff-x_small}) and performing the integration yields a simple functional form
\begin{eqnarray}
\label{x^(-1)-integrals-RDDA1}
\int^1_\xi\!\frac{dx}{x^{2}}\,\frac{F^{\mbox{\tiny RDDA}}(\xi/x,\xi,\cdots)}{F^{\mbox{\tiny RDDA}}(\xi,\xi,\cdots) }
\stackrel{\xi\to 0}{=}
\frac{2\left(b-\alpha(t)+1\right)}{\left(\alpha(t) -1\right) \left(2b-\alpha(t)+2\right)}\!\left[1-
\frac{\Gamma(b+1)  \Gamma(2 b-\alpha(t)+3)} {\Gamma(2 b+2)  \Gamma(b-\alpha(t)+2)}\, \xB^{\alpha(t)-1}\right].
\nonumber\\
\end{eqnarray}
This result exemplifies that not necessarily a numerical enhancement occurs in the `hard' scenario. Namely,  for small positive $b$, which, however, is phenomenologically disfavored, the $1/\left(\alpha(t)-1\right)$ factor is partially neutralized by the prefactor $\left(\alpha(t)-1-b\right) \sim \left(\alpha(t) -1\right)$.

\pagebreak[3]
\begin{itemize}
\item {\em Constraints for $\ell_F^{\prime}$ and $\ell_F^{\prime\prime}$.}
\end{itemize}
\vspace{-2mm}
For RDDA based GPD models the `Regge' asymptotics of the convolution integrals (\ref{ell_F}) is obtained from straightforward calculations,
\begin{subequations}
\label{conv-smallxB}
\begin{eqnarray}
\label{conv1-smallxB}
\ell^\prime_F(\xi|b,\alpha(t)) &\!\!\!\stackrel{\xi\to 0}{=}\!\!\!&
-\frac{\gamma^{(0,\text{F})}_{\alpha(t)-1}}{2} +S_1(\alpha(t)+1) - S_1(2b-\alpha(t)+1) +S_1(b-\alpha(t))  \,,
\\
\label{conv2-smallxB}
\ell^{\prime\prime}_F(\xi|b,\alpha(t)) &\!\!\!\stackrel{\xi\to 0}{=}\!\!\!&
\ell^{\prime\,2}_F(\xi|b,\alpha(t)) +\frac{1}{2}\Big[S_2(2b-\alpha(t)+1)-S_2(\alpha(t)+1) -2 S_2(b-\alpha(t)) \Big]
\\
&&\phantom{\ell^{\prime\,2}_F(\xi|b,\alpha(t))} - \frac{1}{2} \left[
\frac{\Delta S_1\!\big(\!\frac{2b-\alpha(t)+1}{2}\!\big)}{2}- \frac{\Delta S_1\!\big(\!\frac{\alpha(t)+1}{2}\!\big)}{2}-1- \frac{1}{\alpha(t) [1+\alpha(t)]}\right]
\nonumber\\
&&\phantom{\ell^{\prime\,2}_F(\xi|b,\alpha(t))\frac{1}{2}}\,\times
\left[\frac{\Delta S_1\!\big(\!\frac{2b-\alpha(t)+1}{2}\!\big)}{2}- \frac{\Delta S_1\!\big(\!\frac{\alpha(t)+1}{2}\!\big)}{2}\right]  \,,
\nonumber
\end{eqnarray}
\end{subequations}
where we used for shortness the notation (\ref{Smp}).
Our quantities (\ref{conv-smallxB}) for a minimalist GPD model, which is set up in the Mellin-Barnes representation or `dual' parametrization \cite{Polyakov:2002wz,Polyakov:2007rw},
are formally obtained by setting $b=\alpha(t)$ \cite{Musatov:1999xp,Polyakov:2008aa,Kumericki:2009uq}. They are then entirely expressed by
the first and second power of $-\gamma^{(0,\text{F})}_{\alpha(t)-1}/2$, i.e., by the anomalous dimension (\ref{gamma_j^{(0,F)}})
taken at  $j=\alpha(t)-1$:
\begin{subequations}
\label{conv1-smallxB-minimalist}
\begin{eqnarray}
\label{conv1-smallxB-minimalist1}
\ell^\prime_F(\xi|\alpha(t),\alpha(t)) &\stackrel{\xi\to 0}{=}&
\frac{3}{2} + \frac{1}{\alpha(t) [\alpha(t) + 1 ]}  -2 S_{1}(\alpha(t))\,,
\\
\label{conv1-smallxB-minimalist2}
\ell^{\prime\prime}_F(\xi|\alpha(t),\alpha(t)) &\stackrel{\xi\to 0}{=}&
\ell^{\prime\,2}_F(\xi|\alpha(t),\alpha(t)).
\end{eqnarray}
\end{subequations}
For the `Reggeon' case $0 < \alpha(t) < 1$  the quantity $\ell_F^{\prime}(\xi,t)$ and trivially also $\ell_F^{\prime\prime}$ are positive,
they vanish for the `pomeron' case $\alpha(t)=1$, while for small $j+1=\alpha(t)$ their sizes are governed by
the $j=-1$ pole.
These results apply for a whole class of specific GPD models \cite{Shuvaev:1999fm,Shuvaev:1999ce} in which `Regge' poles are implemented in the complex $j$-plane rather in the complex angular momentum plane \cite{Kumericki:2009ji}. Thus, it can be trivially obtained from the Mellin-Barnes integral,
see above (\ref{tffF_NS-model-smallxB}) with $s_2=s_4=0$. Furthermore,  in this model class the skewness ratio (\ref{RDDA-r}) takes the value
$$
\frac{F(x,\eta=x,t=0)}{ F(x,\eta=0,t=0)}={_2F_1}\left({\alpha/2,\alpha/2+1/2 \atop \alpha+3/2} \Big| 1 \right) =
\frac{2^\alpha \Gamma\big(\alpha + \frac{3}{2}\big)}{\Gamma\big(\frac{3}{2}\big) \Gamma(\alpha+2)}\,,
$$
which we consider here us an upper bound for the set of our GPD models%
\footnote{This ratio has been viewed as a GPD `property'  \cite{Shuvaev:1999ce} in the small $x$-region.  In \cite{Kumericki:2009ji} it has been clarified that such a statement arises from an oversimplified mathematical treatment, which can be defended  by the assumption that `Regge' poles lie in the complex conformal spin plane than the  angular momentum one \cite{Martin2009zzb}.}.

If the parameter $b$ increases, the skewness ratio (\ref{RDDA-r}) decreases and the quantity $\ell_F^{\prime}$ will monotonously grow,
reaching for a forward  GPD  with $ 0<\alpha(t)$ the `Regge' asymptotic value
\begin{subequations}
\label{conv-smallxB-PDF}
\begin{eqnarray}
\label{conv1-smallxB-PDF}
\lim_{b\to\infty}\ell^\prime_F(\xi|b,\alpha(t)) &\!\!\!\stackrel{\xi\to 0}{=}\!\!\!&
\frac{3}{2} + \frac{1}{\alpha(t)}  - S_{1}(\alpha(t)) - \ln(2)\,,
\end{eqnarray}
where $\ell^{\prime\prime}_F(\xi|b,\alpha(t))$ is bounded from above by $\ell^{\prime\, 2}_F(\xi|b,\alpha(t))$,
\begin{eqnarray}
\lim_{b\to\infty}\ell^{\prime\prime}_F(\xi|b,\alpha(t)) &\!\!\!\stackrel{\xi\to 0}{=}\!\!\!&  \ell^{\prime\, 2}_F(\xi|\infty,\alpha(t))
 -\frac{S_2(\alpha(t)+1) +\zeta(2)}{2}  - \frac{\Delta S_1\!\big(\!\frac{\alpha(t)+1}{2}\!\big)}{2}
\\
&&\phantom{\ell^{\prime\,2}_F(\xi|\infty,\alpha(t))}
\times \left[ 1+\frac{1}{\alpha(t) [1+\alpha(t)]} + \Delta S_1\!\!\left(\!\frac{\alpha(t)+1}{2}\!\right)\right].
\qquad
\nonumber
\end{eqnarray}
For the `pomeron' case $\alpha(t)=1$ both of them do not vanish anymore, however, their values $\ell_F^{\prime} \sim 1$ and $\ell_F^{\prime\prime} \sim -1$
can be considered as rather small. With decreasing $\alpha(t)$, both quantities will grow and $\ell_F^{\prime\prime}$ will change sign, i.e., for `Reggeon' exchange we have the inequality
\begin{eqnarray}
\label{conv2-smallxB-PDF}
0 \lesssim \ell^{\prime\prime}_F(\xi|b=\infty,\alpha(t)) <  \ell^{\prime\, 2}_F(\xi|b=\infty,\alpha(t)) \quad\mbox{for small $\xi$ and $0 <\alpha(t)  \lesssim0.8$}\,.
\end{eqnarray}
\end{subequations}

In conclusion, we can  state that the value of $-\ell_F^{\prime}(\xi,t)$ will be positive in the large-$\xi$ region and turns negative for common valence GPDs
in the small-$\xi$ asymptotics.  Obviously, we have at least one node
$\ell^{\prime}_F(\xi=\overline{\xi},t,\mu^2)=0$ and,  thus we have analogously as for PDFs also for the class of popular GPD models one
stable point $\overline{\xi}$ at which the GPD does not evolve. Note, however, that the value of $\overline{\xi}$ may depend on $t$.
We may consider  (\ref{evol-x_large1})  and  (\ref{conv1-smallxB-PDF})  as an upper and lower bound, which yields with
common  $\beta \gtrsim 3$ the constraint
\begin{eqnarray}
\label{ell^{prime}_F-constraint}
-\frac{3}{2}-\frac{1}{\alpha(t) } + S_1(\alpha(t)) + \ln(2)  \lesssim - \ell^{\prime}_F(\xi,t)
\lesssim  \ln\frac{2}{1-\xi} + S_1(\beta)  -\frac{3}{2} \quad\mbox{for}\quad 0 <\alpha(t) < 1
\,.
\end{eqnarray}
Moreover, in  both limits we have in addition the bound  $\ell^{\prime\prime}_F(\xi,t) \lesssim \ell^{\prime 2}_F(\xi,t)$, specified further in (\ref{evol-x_large2},\ref{conv2-smallxB-PDF}). For the `pomeron' case  quark evolution  plays no crucial role in the small-$\xi$ region and we might
roughly set
\begin{eqnarray}
\label{ell_F-estimates_pomeron}
0 \le \lim_{\xi\to 0}\ell^{\prime}_F(\xi,t)  \lesssim 1 \quad\mbox{and}\quad
 0 \le -\lim_{\xi\to 0}\ell^{\prime\prime}_F(\xi,t)  \lesssim 1 \quad\mbox{for}\quad \alpha(t)\sim 1
\,.
\end{eqnarray}

\subsection{Generic and model dependent features}
\label{sec:NLOestimates}

As alluded to above, we will now analyze the NLO radiative corrections, stemming from the hard scattering amplitude, at fixed photon virtuality
\begin{subequations}
\label{settings}
\begin{eqnarray}
\label{scale-setting-Q_0}
\cQ^2=\cQ_0^2=4\,\GeV^2,
\end{eqnarray}
where at LO and NLO the GPD and DA models are taken the same. Of course, we are aware that such considerations, which only sketch the qualitative features
of radiative NLO corrections, are not entirely realistic. An appropriate method, which is beyond the scope of this article, would be the quantification of reparameterization effects that arise from LO and NLO fits to experimental data.  Furthermore, we will  quote NLO corrections with the scale setting prescription
\begin{eqnarray}
\label{scale-setting}
\muF = \muphi = \muR = \cQ_0\,,
\end{eqnarray}
where we consistently take at LO and NLO the phenomenological values of the running coupling
\begin{eqnarray}
\label{alpha_s-setting}
\alpha^{\rm LO}_\text{s}(\mathcal \cQ_0=2\,\GeV)=0.34 \quad\mbox{and}\quad \alpha^{\rm NLO}_\text{s}(\mathcal \cQ_0=2\,\GeV)=0.29
\end{eqnarray}
\end{subequations}
with four active quarks. We will also shortly discuss `optimal' scale setting prescriptions.

\begin{table}[t]
\begin{center}
\begin{tabular}{|c|c|cccccccccc|}
   \hline
   parton & Eq. & $N$ & $\alpha(0)$ & $\alpha^\prime$ & $\beta$ & $\delta\beta$ & $h$ &$p$  &$M^2$ & $s_2$ & $s_4$ \\
  \hline\hline
  $u^{\rm val}$ &(\ref{KLMSPM-ans-generic})  & $2$ & $0.43$ & $0.85$ & $3.2$ & $2.2$ & $-1$  &$2.12$ & - & $-0.26$ & 0.04 \\
  $d^{\rm val}$ &(\ref{KLMSPM-ans-generic}) & $1$ & $0.43$ & $0.85$ & $3.2$ & $2.2$ & $-1$  &$2.12$ & - & $-0.26$ & 0.04 \\
  $q^{\rm sea}$ & (\ref{KMP-ans-generic}) & 0.152 & 1.158 & 0.15 & 8 & -& - & 2 & 0.446  & $-0.442$ & 0.089  \\
  G             & (\ref{KMP-ans-generic}) & 0.448 & 1.247 & 0.15 & 6 & -& -& 2 & 0.7 & $-2.309$ & 0.812\\
   \hline
\end{tabular}
\end{center}
\caption{Model parameters for valence quark, sea quark, and gluon GPDs with nnlo-PWs, where Regge slope  and cut-off mass parameters are
given in units of $\GeV^{-2}$ and $\GeV^{2}$, respectively.}
\label{tab:models}
\end{table}
The size of radiative corrections to the imaginary part of TFFs will be discussed  using analytic expressions.
To visualize the relative NLO corrections, we  employ as in \cite{Kumericki:2007sa} the ratio of the TFF at NLO
to that at LO
\begin{subequations}
\label{deltaKdeltavarphi}
\begin{align}
\label{deltavarphi}
  \frac{\tffF^\text{NLO}_{\text{M}}(\xB,t,\cQ^2)}{\tffF^\text{LO}_{\text{M}}(\xB,t,\cQ^2)} =
  K_{\text{M}}(\xB,t,\cQ^2) \exp\left\{ i\delta\phi_{\text{M}}(\xB,t,\cQ^2) \right\},
\end{align}
which we parameterize by the  modulus ratio $K$ and the phase difference $\delta\phi_{\text{M}}$.
Obviously, both  the deviation of the $K$ ratio from one,
\begin{align}
\label{deltaK}
  \delta K_{\text{M}}(\xB,t,\cQ^2)=K_{\text{M}}(\xB,t,\cQ^2)-1,
\end{align}
\end{subequations}
and  the phase difference $\delta\phi_{\text{M}}$ quantify the relative size of radiative corrections. For our numerical illustration
we utilize a next-to-next-leading (nnl) SO(3)-PW model, where the parameters for the various parton species are listed in
Tab.~\ref{tab:models} and the DA is chosen to be narrow, specified by the  CPW amplitudes
\begin{eqnarray}
\label{DA-narrow}
 \varphi_0 =1 \,, \quad \varphi_2 = -\frac{1}{4} \,, \quad \varphi_4 = \frac{1}{30}\,,
\;\;\mbox{and}\;\; \varphi_k = 0\;\; \mbox{for}\;\; k\in\{6,8,\cdots\}\,.
\end{eqnarray}
We will also utilize a minimalist model, i.e., we take only the leading (l) SO(3)-PW, i.e., $s_2=s_4=0$, and the so-called asymptotic  DA. For this DA only
$\varphi_0 =1$ differs from zero and the explicit factorization scale dependence of the hard scattering amplitudes drops out at NLO, however, note that this DA
evolves at the considered order \cite{Mikhailov:1984ii,Mueller:1993hg}.

\subsubsection{Flavor non-singlet channel}

{F}rom the associated color factors of the most singular terms
one may conjecture that such terms are related to the factorization/renormalization procedure or a reminiscence of Sudakov suppression \cite{Li:1992nu},
see also the discussion for the pion-to-photon transition form factor in \cite{Musatov:1997pu}.
Utilizing our generic findings of Sec.~\ref{sect-NLOcorrections-generic}, we first analyze the NLO corrections to the imaginary part of the TFF (\ref{eqs:DRtffFqC}), where we consider the three different color structures (\ref{qT^{(1)}}) to the NLO corrections (\ref{T^{(1)}-1}), separately.

\pagebreak[3]
\begin{itemize}
\item {\em The $\CF=4/3$  part.}
\end{itemize}
\vspace{-2mm}
We express the conformal moments (\ref{c^{(1,F})}) in terms of anomalous dimensions
$$
\frac{4}{3} \left[\ln\frac{\cQ^2}{\muF^2} -\frac{\gamma_j^{(0,\text{F})}}{4} - \frac{\gamma_k^{(0,\text{F})}}{4} - \frac{1}{2}\right]
 \frac{(-1)\gamma_j^{(0,\text{F})}}{2}
-\frac{46}{9} + \frac{6 (j+1)_2+2}{3[(j+1)_2]^2}
+ \{j \leftrightarrow k, \muF\to \muphi \},
$$
where the color factor $\CF=4/3$ is now included.   From the substitution rules (\ref{ell_phi},\ref{ell_F}) we find the ratio for the imaginary part
of the corresponding NLO term to that of the LO one,
\begin{eqnarray}
\label{F^{(1,CF}}
\frac{\im \tffF^{(1,{\rm F})}_{\text{M}} }{\im \tffF^{\rm LO}_{\text{M}} } &\!\!\! = \!\!\!  &
\frac{2}{3} \left[2\ln\frac{\cQ^2}{\muF^2} + \frac{\ell^{\prime\prime}_F(\xi,t)}{\ell^{\prime}_F(\xi,t)}+2\ell^{\prime}_\varphi-1\right]\ell^{\prime}_F(\xi,t)
 + \frac{2}{3}  \left[2\ln\frac{\cQ^2}{\muphi^2} + \frac{\ell^{\prime\prime}_\varphi}{\ell^{\prime}_\varphi} -1 \right] \ell^{\prime}_\varphi -\frac{163}{18}
+ \cdots\,,
\nonumber\\
\end{eqnarray}
in units of $\alpha_s(\cQ_0\approx 2\, \GeV)/2\pi \approx 0.05$. The ellipsis stands for rather harmless  contributions, e.g., we neglected $(6 (j+1)_2+2)/3[(j+1)_2]^2$, which corresponds in momentum fraction to
\begin{eqnarray}
\label{F^{(1,CF}-remaining}
\frac{6 (j+1)_2+2}{3[(j+1)_2]^2} \quad\Leftrightarrow\quad  \frac{2}{3}\int_\xi^1\!\frac{dx}{x}\left[\frac{2}{1+x}+\frac{\ln\frac{1+x}{2x}}{1-x}\right] \frac{F(\xi/x,\xi,\cdots)}{F(\xi,\xi,\cdots)}\,.
\end{eqnarray}
Furthermore, according to the procedure (\ref{DA-approximation}) we took the expression $(6 (k+1)_2+2)/3[(k+1)_2]^2$ for $k=0$ into account, which gives  a comparable small positive correction $7/6$ from the lowest CPW of the meson DA. Hence, the large negative constant $-2\times 46/9 = - 92/9$ from the residue of the pole at $v=1$ slightly decreases to $-163/18 \simeq -9$. The constant part, which is independent on the kinematical variables
\begin{eqnarray}
\label{ell^2-term}
\frac{\alpha_s}{2\pi}\left\{\frac{2}{3}\left[\frac{\ell^{\prime\prime}_\varphi}{\ell^{\prime}_\varphi} -1 \right] \ell^{\prime}_\varphi -\frac{163}{18}\right\}
\quad\mbox{e.g., for}\quad \muphi^2=\cQ^2\,,
\end{eqnarray}
contains also the meson DA in terms of the convolution integrals (\ref{ell_phi}) with the LO evolution kernel.
Since they vanish for the asymptotic DA, we have for this specific choice a sizeable negative contribution of $-45\%$.
Furthermore, it will only slightly change if the DA gets narrow since
$$
(\ell^{\prime\prime}_\varphi/\ell^{\prime}_\varphi -1) \ell^{\prime}_\varphi \approx  (\ell^{\prime}_\varphi -1) \ell^{\prime}_\varphi
\quad \mbox{with}\quad  0 \lesssim \ell^{\prime}_\varphi \lesssim 1 \quad \mbox{for a narrow DA}.
$$
For a broader DA, both
$\ell^{\prime\prime}_\varphi$ and $-\ell^{\prime}_\varphi$ are positive, yielding the surprising result that the size of relative corrections will decrease.
For the AdS/QCD model they are, e.g., reduced to -30\% or so.

For our class of popular GPD models we can suppose that a stable point
$\xB= \overline{x}_{\rm B}$ exists, at which the GPD does not evolve, and that it lies in the valence quark region.
Setting $\ell^{\prime}_F=\ell^{\prime\prime}_F=0$ and taking $\muphi^2=\cQ_0^2$ in (\ref{F^{(1,CF}}),
we can immediately state from  our discussion that relative NLO corrections  are negative and are  of the order
\begin{equation}
\label{NLO-CF_estimate}
-0.45\, (\mbox{narrow \& asymptotic DA}) \lesssim
\frac{\alpha_s}{2 \pi} \frac{\im \tffF^{(1,\text{F})}(\overline{x}_{\rm B},t,\cQ^2)}{\im \tffF^{\rm LO}(\overline{x}_{\rm B},t,\cQ^2)}
\lesssim  -0.3\, (\mbox{broad DA})
\,.
\end{equation}

Clearly, in the large-$\xB$ region the ratio (\ref{F^{(1,CF}}) is dominated by the positive $\ell^{\prime\prime}_F(\xi,t) \approx \ell^{\prime\,2}_F(\xi,t)$ term (\ref{evol-x_large}), providing a (squared) logarithmical grow
that overcompensates the sizeable negative constant.  Thus, the shape of the DA  influences the strength of the linear term $(2\ell^{\prime}_\varphi-1)\ell^{\prime}_F(\xi,t)$. Hence, the shape of the DA plays some role in the transition region from the valence to the large-$\xB$ region. Strictly spoken, as in, e.g., deep inelastic scattering or DVCS, the perturbative expansion breaks down in the $\xB \to 1$ limit. Nevertheless,  such
logarithmical corrections are absorbed by  a slight reparametrization of the $\beta$-parameter at the input scale. Note that evolution leads to a growth of this parameter with increasing $\cQ^2$, i.e., to a suppression of the large-$\xB$ region.

In the small-$\xB$ region the positive $\ell_F^{\prime}$ and $\ell_F^{\prime\prime}$ terms, e.g., evaluated  from RDDA in (\ref{conv-smallxB}),  are relatively small for a  $\rho/\omega$-pole at low $-t$ where $\alpha(t\sim 0) \sim  0.5$.  Thus, for such values the large negative constant in (\ref{F^{(1,CF}}) dominates and the relative NLO contribution is still be negative, e.g., of the order  -20\% or so.
However, for growing  $-t$  the value of $\alpha(t)$ decreases. For our class of GPD models both  $\ell_F^{\prime} \sim 1/\alpha(t)$  and  $\ell_F^{\prime\prime}\sim \ell_F^{\prime\, 2}$ will increase and the relative NLO corrections (\ref{F^{(1,CF}})
may become positive and sizeable for $0 < \alpha(t) \ll 0.5$,
$$
\sim \frac{\alpha_s}{2\pi} \frac{2}{3} \left\{ \left[ \frac{1}{\alpha(t)} +2\ell^{\prime}_\varphi-\frac{1}{2}+\cdots\right]\!\left(\frac{1}{\alpha(t)} +\frac{1}{2}+\cdots\!\right)
 +  \left[\frac{\ell^{\prime\prime}_\varphi}{\ell^{\prime}_\varphi} -1 \right] \ell^{\prime}_\varphi -\frac{163}{12}
+ \cdots\right\}.
$$
As explained in Sec.~\ref{sect-NLOcorrections-generic}, the full result in the small-$\xB$ region also strongly depends on the remaining terms (\ref{F^{(1,CF}-remaining}) and GPD model details. In particular the case with small $\alpha(t)$ might be considered to be of academic interest only.

\begin{itemize}
\item {\em The $\beta_0=-11+ 2n_f/3$  part.}
\end{itemize}
\vspace{-2mm}
In the term  proportional to $\beta_0$ (\ref{c^{(1,beta)}})  [or (\ref{T^{(1,beta)}})] the large-$j$ and -$k$ behavior  [or end-point singularities]
are logarithmical enhanced, too, and we may write this expression as
$$
\frac{\beta_0}{2}\left\{\ln\frac{\cQ^2}{\muR^2} -
\frac{\gamma_{j}^{(0,\text{F})}}{2} - \frac{\gamma_{k}^{(0,\text{F})}}{2} -\frac{14}{3}\right\}
\quad\mbox{or}\quad
\frac{\beta_0}{2}\left\{\ln\frac{\cQ^2}{\muR^2}+ \frac{3 +2 \ln\bu }{2\bu\bv} +  \frac{3+2\ln\bv }{2\bu\bv} -\frac{14}{3\bu\bu}\right\}.
$$
Together with the sizeable  $\beta_0$ ($=-25/3$ for $n_f$=4) it provides for $\muR^2=\cQ^2$ rather large positive corrections
in the vicinity of $u=1$ or $v=1$.
Employing our definitions (\ref{ell_phi1},\ref{ell_F1}) we can immediately write down the relative correction to the associated imaginary part
\begin{eqnarray}
\label{F^{(1,beta}}
\frac{\im \tffF^{(1,\beta)}_{\text{M}}}{\im \tffF^{\rm LO}_{\text{M}}} =  \frac{25}{6} \left\{
-\ln\frac{\cQ^2}{\muR^2}
-\ell_F^\prime(\xi,t)+ \frac{14-3\ell_\varphi^\prime }{3}  \right\}.
\end{eqnarray}
Setting  $\muR^2=\cQ^2$ and taking our class of popular models, we find that in the valence quark region, i.e., more precisely
for $\xB=\overline{x}_{\rm B}$, the  NLO corrections are as sizable than the LO contribution,
\begin{equation}
\label{NLO-beta_estimate}
0.8\, \mbox{(narrow)} \lesssim \frac{\alpha_s}{2\pi} \frac{\im \tffF^{(1,\beta)}_{\text{M}}(\overline{x}_{\rm B},t,\cQ^2)}{\im \tffF^{\rm LO}_{\text{M}}(\overline{x}_{\rm B},t,\cQ^2)} \sim 1\, \mbox{(asymptotic)}   \lesssim  1.2\,  \mbox{(broad)}
\quad\mbox{for}\quad \cQ^2 \approx 4\,\GeV^2 \,.
\end{equation}
As expected, a narrow (broad) DA provides smaller (larger) NLO corrections than the asymptotic one.
Outside this region these corrections are determined by the behavior of $\ell_F^\prime(\xi,t)$, i.e.,
they will increase further  in the large $\xB$-region, cf.~(\ref{evol-x_large1}) and they will decrease in the small $\xB$-region, cf.~(\ref{conv1-smallxB}).

The reader may realize that our estimates are too naive and probably overestimate the true NLO corrections. If we change from
LO to NLO we have also to change the value of $\alpha_s^{\rm LO}$ to $\alpha_s^{\rm NLO}$, which means that the (relative) NLO  should be defined as
$$
\frac{\im \tffF^{\rm NLO}_{\text{M}}}{\im \tffF^{\rm LO}_{\text{M}}} -1 =  \frac{\alpha_s^{\rm NLO}(\cQ)-\alpha_s^{\rm LO}(\cQ)}{\alpha_s^{\rm LO}(\cQ)} +
\frac{\alpha_s^{\rm NLO}(\cQ)}{\alpha_s^{\rm LO}(\cQ)}\,  \frac{\alpha_s^{\rm NLO}(\cQ)}{2\pi}\,  \frac{\im \tffF^{(1)}_{\text{M}}}{\im \tffF^{(0)}_{\text{M}}}\,.
$$
Clearly, the change of $\alpha_s$ affects the term proportional to $\beta_0$ and it
reduces the naive estimate (\ref{NLO-beta_estimate}) of about $30\%$, e.g.,  to a relative $\sim 50\%$ effect for the equal momentum sharing DA.

\pagebreak[3]
\begin{itemize}
\item {\em `Optimal' scale setting prescriptions.}
\end{itemize}
\vspace{-2mm}
It is very popular to seek for an `optimal' renormalization scale setting prescription \cite{Belitsky:2001nq,Anikin:2004jb,Brodsky:2005vt,Ivanov:2004zv}.
The  Brodsky--Lepage--Mackenzie (BLM) scale setting prescription proposes to eliminate the $\beta_0$ contribution \cite{Brodsky:1982gc},
which, e.g., yields the momentum fraction dependent scale
\begin{subequations}
\begin{eqnarray}
\mu_{\rm BLM}^2(\bu,\bv,\cQ^2)= e^{-\frac{5}{3}}\, \bu \bv \cQ^2 =  e^{-\frac{14}{3}} \times e^{\frac{3}{2} +\ln\bu}  \times  e^{\frac{3}{2} +\ln\bv }\, \cQ^2 \,.
\end{eqnarray}
To avoid a complex valued scale in DVMP expressions or two different ones for the imaginary and real part, see discussion in \cite{Brodsky:2005vt}, we use as above a global scale setting prescription, written in terms of the functional $\ell_F^\prime(\xi,t)$.  To illustrate the analogy between large $\xB$-behavior and end-point behavior once more, we may  also write the BLM scale as a functional of the meson DA,
\begin{eqnarray}
\mu_{\rm BLM}^2(\xi,t,\cQ^2) =  e^{-\frac{14}{3}}\times  e^{\ell_\varphi^\prime}\times e^{\ell_F^\prime(\xi,t)} \;  \cQ^2\,.
\end{eqnarray}
In the valence region and for the asymptotic DA we find a very small value $\mu_{\rm BLM}^2 \approx 0.01 \cQ^2$, which
decreases (increases) for a broader (narrower) DA,
\begin{eqnarray}
 0.003\, e^{\ell_F^\prime(\xi,t)} \lesssim  \frac{\mu_{\rm BLM}^2(\xi,t,\cQ^2)}{\cQ^2}
\lesssim  0.03\, e^{\ell_F^\prime(\xi,t)}\,.
\end{eqnarray}
\end{subequations}
Clearly, for experimental accessible photon virtualities of a few $\GeV^2$ the renormalization scale is pushed deeply into the non-perturbative region.
Hence, one has to give a non-perturbative model prescription for the behavior of $\alpha_s(\mu_{\rm BLM})$ in the infrared region, e.g., one conjectures that the coupling constant freezes \cite{Shirkov:1997wi} and that the perturbative expansion of the TFF remains meaningful, e.g., as in \cite{Brodsky:1997dh,Bakulev:2004cu}.

\begin{itemize}
\item {\em The $\CG=-1/6$ proportional part.}
\end{itemize}
\vspace{-2mm}
The term proportional to $\CG=\CF-\CA/2=-1/6$   (\ref{T^{(1,G)}}) is  suppressed by $1/\NC^2$ relatively to  the terms proportional to $\CF$ or $\beta_0$,
which strongly suppresses the logarithmical enhancement in the endpoint region%
\footnote{It seems to obvious that not all of this enhancement effects can be associated with the factorization or renormalization logarithms.
}.
As before we replace in the conformal moments (\ref{c^{(1,G)}}--\ref{Deltac^{(1,G)}_kj}) the corresponding harmonic sum $S_1(j+1)$ by the anomalous dimension.
However, in contrast to the two other color structures, the coefficients of these logarithmical terms
posses a more intricate dependence. We quote  the result that provides the correct $j$-asymptotics for the asymptotic DA
$$
\frac{-1}{6}\left\{ \left[2\zeta(2)-2 + \cdots\right]\frac{\gamma_{j}^{(0,\text{F})}}{2}    +  \frac{\gamma_{k}^{(0,\text{F})}}{2} -\frac{10}{3}-\zeta(2)  +6 \zeta(3) + \cdots \right\},
$$
where the ellipses stand for $k$-dependent terms, which, however, are of less numerical importance.
Utilizing once more the substitution (\ref{ell_phi},\ref{ell_F}), this expression translates into small corrections
\begin{equation}
\label{NLO-G-estimate}
\frac{\im \tffF^{(1,\text{G})}_{\text{M}}}{\im \tffF^{\rm LO}_{\text{M}}} =  \frac{1}{6} \left\{
\left[2\zeta(2)-2 + \cdots\right] \ell_F^\prime(\xi,t)  +  \ell_\varphi^\prime  + \frac{10}{3}+ \zeta(2) -6 \zeta(3) + \cdots  \right\}.
\end{equation}
For the valence region, where GPD evolution effects are considered as small, we find for the asymptotic DA a small negative relative correction
$(\alpha_s/2\pi)\{\ell_\varphi^\prime + 10/3+ \zeta(2) -6 \zeta(3)\}/6 \sim -0.02$, which moderately depends on the DA.  This small negative correction  decreases logarithmically
in the large-$\xB$ region, see (\ref{evol-x_large1}). Since singular terms are absent in the
antiquark contribution (\ref{tb^{(1,G)}}), the difference between the signature even and odd case is in the valence and large-$\xB$ region small, too.
However, in the small-$\xB$ region the size of NLO corrections may differ in the two cases.
For instance, for  our minimalist model we find in the `Regge' asymptotics
\begin{eqnarray}
\label{NLO-G-estimate1}
\frac{\alpha_s}{2\pi} \frac{\im \tffF^{(1,\text{G})}_{\text{M}}}{\im \tffF^{\rm LO}_{\text{M}}} &\!\!\! \stackrel{\xB\to 0}{=} \!\!\!&
\frac{\alpha_s}{2\pi}\,\frac{2}{3}\Bigg\{
  [\zeta(2)-1] S_1(\alpha(t))- \frac{1}{12}-\zeta(2)  +\frac{3\zeta(3)}{2}-\frac{2+3 \alpha(t)[1+\alpha(t)]}{4 \alpha^2(t)[1+\alpha(t)]^2}(1-\sigma )
\nonumber\\
&&\phantom{\frac{\alpha_s}{2\pi}\,\frac{2}{3}\Bigg\{}
+ \frac{\sigma}{8} \left[S_3\!\left(\alpha(t)/2\right)-S_3\!\left(\alpha(t)/2-1/2\right) \right] +\cdots
\Bigg\},
\end{eqnarray}
where we neglected numerically small contributions. Clearly, for even signature  the $j-1=\alpha(t)$ poles vanish and in the odd  signature case
they can perhaps cause rather large corrections for small positive $\alpha(t)$ values.

\begin{itemize}
\item {\em Total contribution.}
\end{itemize}
\vspace{-2mm}
Summing up the three separate estimates (\ref{F^{(1,CF}},\ref{F^{(1,beta}},\ref{NLO-G-estimate1}),
\begin{subequations}
\begin{eqnarray}
\label{(1)2(0)-net-NS}
\frac{\im \tffF^{(1)}_{\text{M}}}{\im \tffF^{(0)}_{\text{M}}} &\!\!\! = &\!\!\!
\left[2\ln\frac{\cQ^2}{\muF^2} +\frac{\ell^{\prime\prime}_F(\xi,t)}{\ell^{\prime}_F(\xi,t)}
-9+\frac{n_f+\ell^{\prime}_\varphi}{2}+\frac{\pi^2-9}{12}+\cdots\right]\!\! \frac{2\ell^{\prime}_F(\xi,t)}{3} +
\left[2\ln\frac{\cQ^2}{\muphi^2} + \frac{\ell^{\prime\prime}_\varphi}{\ell^{\prime}_\varphi}-1\right]\!\! \frac{2\ell^{\prime}_\varphi}{3}
\nonumber\\
&&- \frac{33-2n_f}{6} \ln\frac{\cQ^2}{\muR^2}
+16 (1-\ell^{\prime}_\varphi/3)  - \frac{14n_f}{9} (1-3 \ell^{\prime}_\varphi/14)
+\frac{42+\pi^2-36 \zeta(3)}{36}  + \cdots,
\nonumber\\
\end{eqnarray}
allows us to judge the net size of radiative corrections in the flavor non-singlet channel,
\begin{eqnarray}
\label{NLO-net-NS}
\frac{\im \tffF^{\rm NLO}_{\text{M}}(\xB,t,\cQ^2)}{\im \tffF^{\rm LO}_{\text{M}}(\xB,t,\cQ^2)}
&\!\!\! =\!\!\!& \frac{\alpha_s^{\rm NLO}(\muR)}{\alpha_s^{\rm LO}(\muR)}+ \frac{\alpha_s^{\rm NLO}(\muR)}{\alpha_s^{\rm LO}(\muR)}
\frac{\alpha_s^{\rm NLO}(\muR)}{2\pi}\, \frac{\im \tffF^{(1)}_{\text{M}}(\xB,t,\cQ^2|\muR,\muF,\muphi)}{\im \tffF^{(0)}_{\text{M}}(\xB,t,\cQ^2|\muR,\muF,\muphi)} +O(\alpha_s^3)\,.\nonumber\\
\end{eqnarray}
For $\xB=\overline{x}_{\rm B}$ the large positive $\beta_0$-part is partially canceled by both the negative $\CF$-part and the reparametrization of the strong coupling, while the $\CG$-part plays practically no role. This provides for the settings (\ref{settings}) a  moderate $\sim 25\%$ net correction for the asymptotic DA,
\begin{equation}
\label{NLO_estimate}
0.1\, \mbox{(narrow DA)} \lesssim  \frac{\im \tffF^{\rm NLO}_{\text{M}}(\overline{\xB},t,\cQ^2_0)}{\im \tffF^{\rm LO}_{\text{M}}(\overline{\xB},t,\cQ^2_0)}-1 \sim 0.25\, \mbox{(asymptotic  DA)}   \lesssim  0.5\,  \mbox{(broad  DA)}
\,,
\end{equation}
\end{subequations}
which is getting smaller for a narrow DA and larger for a broader DA. Independent of the shape of the DA,
in the large-$\xB$ region the relative NLO corrections are dominated by the $\ell_F^2 \approx \ell_F^{\prime\prime}$  term, arising from the $\CF$-part. Hence, they are positive. However, the increase will be strengthened by the linear $\ell_F^{\prime}$ term, which only slightly depends on the DA. As before, we observe in the small-$\xB$ region that a `pomeron' behavior provides similar NLO corrections as in the valence region. A rather flat behavior will strongly increase the relative NLO corrections.
For our GPD model the small-$\xB$ asymptotics of the TFF as function of $\alpha(t)$ can be obtained from (\ref{tffF_NS-model-smallxB}),
\begin{subequations}
\label{NLO-net-NS-smallxB}
\begin{eqnarray}
\label{NLO-net-NS-smallxB1}
\frac{\im \tffF^{(1)}_{\text{M}}}{\im \tffF^{(0)}_{\text{M}}}
&\!\!\! \stackrel{\xB\to 0}{=}\!\!\!&
\sum_{{\nu=0 \atop {\rm even}}} \hat{s}_{\nu}(t)
  \Bigg\{
{ \textstyle - \frac{39-2n_f-2\zeta(2)}{6}} \left[1-\frac{\frac{4}{3} \ell^\prime_\varphi - \frac{1}{3}\gamma^{(0,\text{F})}_{\alpha(t)-1+\nu}+\frac{4}{3}\ln\frac{\cQ^2}{\muF^2}
}{\frac{39-2n_f-2\zeta(2)}{6}
}\right]\gamma^{(0,\text{F})}_{\alpha(t)-1+\nu}
\\
&&\phantom{\sum_{{\nu=0 \atop {\rm even}}} \hat{s}_{\nu}(t) }+
{ \textstyle \frac{309-28 n_f+3\zeta(2)-18\zeta(3)}{18}}\left[
1-\frac{\frac{18-n_f}{3} \ell^\prime_\varphi -\frac{2}{3}\ell^{\prime\prime}_\varphi
-\frac{4}{3}\ell^\prime_\varphi \ln\frac{\cQ^2}{\muphi^2} +\frac{33-2n_f}{6}\ln\frac{\cQ^2}{\muR^2}
}{
\frac{309-28 n_f+3\zeta(2)-18\zeta(3)}{18} }\right]
\nonumber\\
&&\phantom{\sum_{{\nu=0 \atop {\rm even}}} \hat{s}_{\nu}(t) }
+\frac{\frac{17}{6}-\frac{\sigma}{2} -\frac{1}{3} \zeta(2)}{(\alpha(t)+\nu)_2}+\frac{1-\frac{\sigma}{3} }{[(\alpha(t)+\nu)_2]^2} +
\frac{\sigma\Delta S_3\big(\!\frac{\alpha(t)+\nu}{2}\!\big)}{12} +\cdots
\Bigg\},
\nonumber
\end{eqnarray}
where  for shortness we introduced relative skewness parameters
\begin{eqnarray}
\label{hat{s}_nu(t)}
\hat{s}_\nu(t)  = \frac{2^{2\nu} \big(\alpha(t)+\frac{5}{2}\big)_{\nu}}{(\alpha(t)+2)_{\nu}}\bigg/
\sum_{{\mu=0 \atop {\rm even}}} s_\mu\frac{2^{2\mu} \big(\alpha(t)+\frac{5}{2}\big)_{\mu}}{(\alpha(t)+2)_{\mu}}
\quad\mbox{with}\quad s_0\equiv 1\,.
\end{eqnarray}
\end{subequations}
For the minimalist model, i.e.,  $s_{\nu}=0$ for $\nu> 0$, we find
small/moderate contributions  $\approx 17\%\, [24\%]$ for $\alpha(t) \approx 0.5\, [0.6]$ for signature even [odd].
As explained they  diverge for $\alpha(t)\to 0$.

\begin{figure}[t]
\begin{center}
\includegraphics[width=\textwidth]{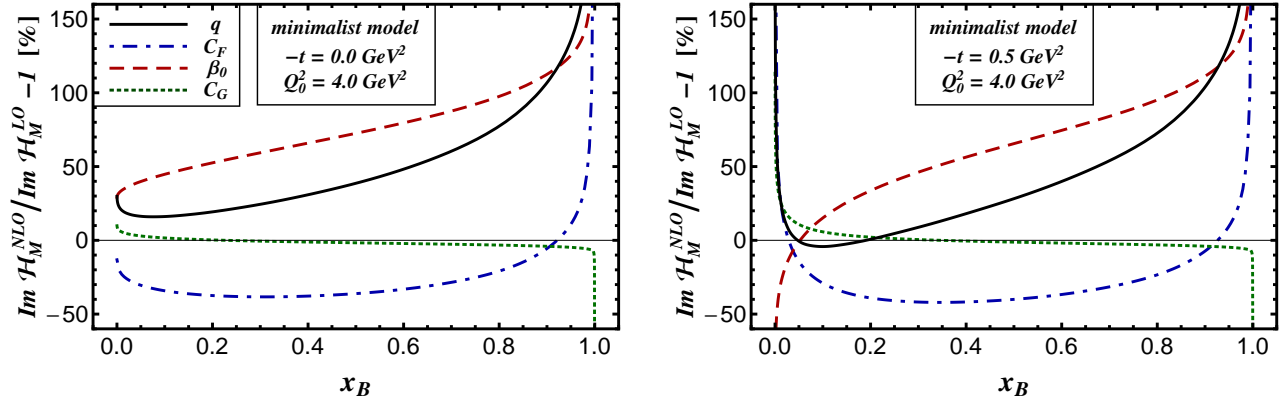}
\end{center}
\vspace{-8mm}
\caption{
\small
Relative NLO corrections to the imaginary part of the TFF $\tffH^{u^{(-)}}_\text{M}$ (solid) broken down to the
$\CF$ (dash-dotted),  $\beta_0$ (dashed), and $\CG$ (dotted) parts at $t=0\, \GeV^2$ (left panel) and $t=-0.5\, \GeV^2$ (right panel) at the initial scale
$\cQ_0^2= 4\,\GeV^2$ with four active quarks.}
\label{Fig-TFF-beta0CFCG}
\end{figure}
In Fig.~\ref{Fig-TFF-beta0CFCG} we display this generic behavior of relative NLO corrections (solid curves) for the imaginary part of
the signature even TFF $\tffH^{\umi}_\text{M}$, which is build from the valence quark ansatz (\ref{KLMSPM-ans-generic}). It is evaluated from our
minimalist  model (leading (l) SO(3)-PW for valence GPD and asymptotic DA) at the input scale $\cQ^2_0=4\,\GeV^2$ for $t=0$ (left panel) and $t=-0.5\,\GeV^2$ (right panel).
Our generic estimates are numerically confirmed.  In the valence region we have
large negative, positive, and small corrections for the $\CF$ (dash-dotted curve), $\beta_0$ (dashed curve), and $\CG$ (dotted curve) parts, respectively,
which finally yields the net-result of  a moderate positive $\sim 30\%$ correction.
Furthermore, the reader may easily convince himself that the large-$\xB$ behavior is consistent with our estimates. The analytical values for the small-$\xB$ asymptotics (\ref{tffF_NS-model-smallxB}),
$$
-0.10\; (\CF\mbox{-part})\,, \quad 0.29\;  (\beta_0\mbox{-part})\quad   0.11 \;  (\CG\mbox{-part})\quad\Rightarrow\quad 0.3\; (\mbox{net part})\,,
$$
are in agreement with the numerical values, which can be read off from the left panel in Fig.~\ref{Fig-TFF-beta0CFCG}. If we
lower the value of $\alpha(t=-0 \,\GeV^2) = 0.43$ to  $\alpha(t=-0.5\,\GeV^2) = 0.005$, the corrections are huge in the small-$\xB$ asymptotics, see right panel.

We add that the NLO corrections to the real part, obtained from the DR-integral, may also require to calculate the $D$-term contribution (\ref{eq:tffD}).
The discussion for the size of NLO corrections can be adopted from the previous one, e.g.,
simply by replacing $\ell^{\prime}_F(\xi,t,\muF^2) \to \ell^{\prime}_d(t,\muF^2)$, where the function $d(u-\bu,t,\muF^2)$ plays the role of another
(generalized) DA. Note that the replacement $\ell^{\prime}_F(\xi,t,\muF^2)\to \ell^{\prime}_\varphi(\muphi^2)$ gives the result for the elastic form factor.

\begin{itemize}
\item {\em Model dependency.}
\end{itemize}
\vspace{-2mm}
Let us now illustrate the generic features and model dependency of the relative NLO corrections to both the modulus and the phase of flavor non-singlet TFFs,
see (\ref{deltaKdeltavarphi}).
\begin{figure}[t]
\begin{center}
\includegraphics[width=\textwidth]{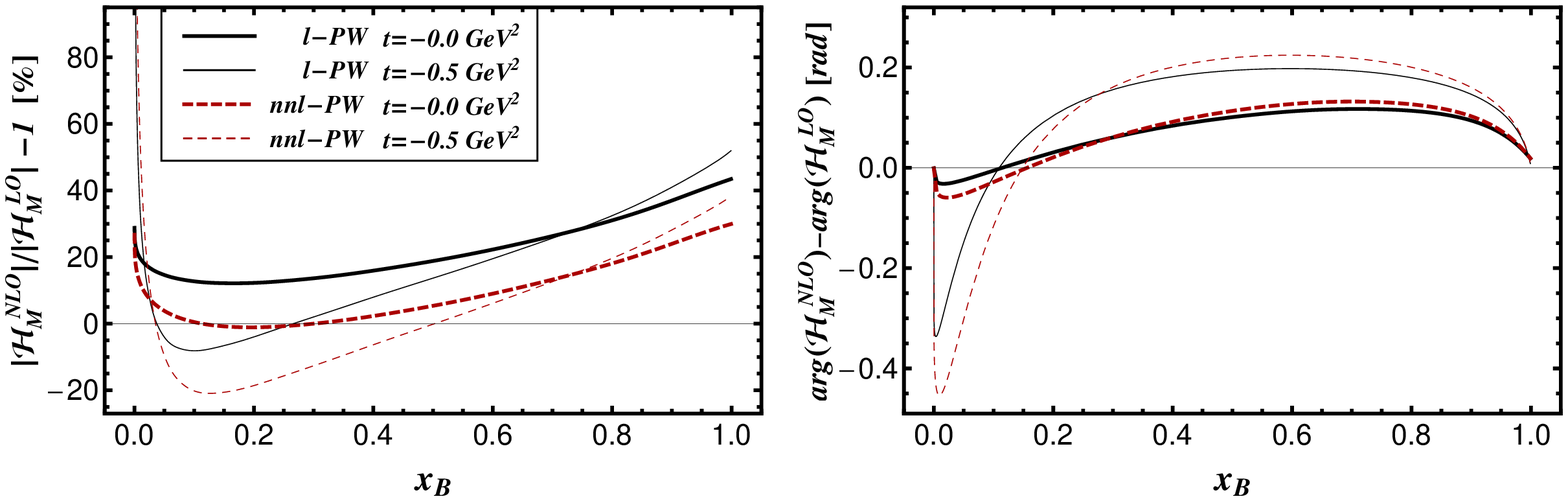}
\end{center}
\caption{
\small
Relative NLO corrections to the modulus (left panel) and phase (right panel) for charge odd TFF $\tffH_M^{u^{(-)}}$
for our minimalist (l-SO(3) PW, asymptotic DA) model (solid curves) and a nnl-SO(3) PW model with a narrow DA (dashed curves) at $t=0\,\GeV^2$ (thick curves) and $t=-0.5\,\GeV^2$ (thin curves) at the initial scale $\cQ_0^2=4\,\GeV^2$.}
\label{Fig-TFF-valence}
\end{figure}
In Fig.~\ref{Fig-TFF-valence}
the relative NLO corrections for the modulus (left panel) and the phase (right panel) of the TFF  $\tffH^{u^{(-)}}_{\text{M}}$
are displayed  for our minimalist model as solid curves for fixed $-t=0$ (thick curves) and $-t=0.5\, \GeV^2$ (thin curves) as function of $\xB$ %
\footnote{This allows for a simple comparison with the NLO corrections to the imaginary
part, displayed in Fig.~\ref{Fig-TFF-beta0CFCG}. However, note that the experimental accessible variable $t-t_{\rm min}(\xB,\cQ^2)$
and the DVMP requirement $-t \ll \cQ^2$ yields an upper bound on $\xB$.}.

The DR (\ref{DR}) implies that the NLO corrections to the real part are governed by those of the imaginary part. In the $\xB\to 1$ limit the modulus is determined by the real part, i.e., by the net contributions to the imaginary part over the whole $\xB$-region. Thus, the relative corrections to the modulus in the large-$\xB$ region are smaller in comparison  with those for the imaginary part, shown as solid curves in Fig.~\ref{Fig-TFF-beta0CFCG}, and they remain finite in the $\xB\to 1$ limit (which is experimentally not reachable).   In the small $\xB$-region the relative corrections to the modulus are governed
for $\alpha(t)>0$ by those from the imaginary part, since the  TFF follows simply from its imaginary part by multiplication with a well-known factor
\begin{eqnarray}
\tffF(\xB,t,\cQ_0^2) \stackrel{\xB\to 0}{=} \left\{
{
i -  \cot(\pi \alpha(t)/2)
\atop
i+  \tan(\pi \alpha(t)/2)
}\right\} \im \tffF(\xB,t,\cQ_0^2) \quad \mbox{for} \quad
\sigma= \left\{
{
+1
\atop
-1
}
\right.\,.
\end{eqnarray}
This formula is a consequence of the `Regge'-behavior we assumed and it can be easily derived from the DR or the Mellin-Barnes integral.
Hence, if radiative corrections in the small-$\xB$ asymptotics do not alter the overall sign of the TFF, the phase difference
$\delta \phi_{\text{M}}$ vanishes and
$$
\delta K_{\text{M}}(\xB,t,\cQ^2)
\stackrel{\xB\to 0}{=}
\frac{\im\tffF^{\rm NLO}_{\text{M}}(\xB,t,\cQ^2)}{\im\tffF^{\rm LO}_{\text{M}}(\xB,t,\cQ^2)}-1
\quad\mbox{for}\quad \delta \phi_{\text{M}}(\xB,t,\cQ^2) \stackrel{\xB\to 0}{=}  0
$$
is given by the NLO corrections to the imaginary part. As above the values of TFFs in the small-$\xB$ asymptotics,
obtained numerically, are reproduced from analytic expressions.

In the valence region we have rather moderate NLO
corrections, which for the moduli are comparable or even smaller than those of the imaginary parts, see solid curves in Figs.~\ref{Fig-TFF-beta0CFCG} and \ref{Fig-TFF-valence} (left panel).  As observed in the DVCS case \cite{Belitsky:1999sg}, we also realize from the right panel in Fig.~\ref{Fig-TFF-valence} that the NLO corrections to the phase are rather small over the whole $\xB$-region.

For our nnl SO(3)-PW model with an asymptotic DA we find similar corrections (not shown). As explained above, if we use a narrow DA, e.g., (\ref{DA-narrow})
with $\ell_\varphi^\prime \approx 0.54$,
the size of radiative NLO corrections to the imaginary part decreases and thus also the corrections to the modulus, while the corrections to the phase are
essentially unchanged. These more generic expectations are illustrated by the dashed lines in Fig.~\ref{Fig-TFF-valence}, where again we set $-t=0$ (thick curves) and $-t=0.5\, \GeV^2$ (thin curves).

\subsubsection{Flavor singlet channel (even signature and even intrinsic parity)}
\label{sec:estimates-singlet}

In the flavor singlet channel the gluon and quark contributions are rather strongly related in the small-$\xB$ region. Thus, and in view  of the group theoretical decomposition, cf.~TFFs (\ref{eqs:tffV0}--\ref{eqs:tffV0-SU}),  it is appropriate to study both contributions together. We introduce the ratios
of the  quark  or gluon contribution to the imaginary part of the flavor singlet TFF  in  LO accuracy as
\begin{subequations}
\begin{eqnarray}
\label{R^A}
 \textbf{R}^{\text{A}}(\xB,\cdots) &\!\!\!\equiv\!\!\! & \frac{\im \tffF^{(0,\text{A})}(\xB,\cdots)}{\im \tffF^{(0,\text{S})}(\xB,\cdots)}
\quad \mbox{with}\quad \text{A} \in\{\Sigma,\g\} \,.
\end{eqnarray}
Trivially, their sum $\textbf{R}^\Sigma+ \textbf{R}^\g =1$ is model independent. The ratios can be directly expressed by
GPDs
 \begin{eqnarray}
 \label{R^A-1}
  \textbf{R}^{\text{A}}(\xB,\cdots) &\!\!\!=\!\!\! &
 \frac{F^{\text{A}}(\xi,\xi,\cdots)}{\left[F^{\Sigma} + \frac{n_f}{\CF \xi}F^\g\right](\xi,\xi,\cdots)}\times
 \left\{
 {1 \atop \frac{n_f}{\CF \xi}}
 \right\}\quad\mbox{for}\quad \text{A} =\Bigg\{ {\Sigma \atop \g}\,.
 \qquad
\end{eqnarray}
In the small-$\xB$ region the sea quarks dominate in the  flavor singlet quark GPD. Hence, the $\textbf{R}^{\text{A}}$-ratios in the small-$\xB$ region
are for our model, set up in Sec.~\ref{sec:NLOestimates-model}, approximately given by
\begin{eqnarray}
 \label{R^A-2}
  \textbf{R}^{\text{A}}(\xB,\cdots) &\!\!\!\stackrel{\xB \to 0}{\approx}\!\!\! &
 \frac{\xi^{-\alpha^\text{A}}
\mbox{$\displaystyle  \sum_{{\nu=0 \atop {\rm even}}}$} s^\text{A}_{\nu}  \frac{2^{2\nu} \big(\!\alpha^\g+\frac{5}{2}\!\big)_{\nu}}{(\alpha^\g+2)_{\nu}}
{\displaystyle  \left\{{1 \atop \frac{2n_f}{\left(\alpha^\g+2+\nu\right)\CF} } \right\}} {\rm Res}F_j^{\text{A}} |_{j=\alpha^\text{A}-1}
}{
{\displaystyle  \sum_{{\nu=0 \atop {\rm even}}}}  \frac{2^{2\nu} \big(\!\alpha^\g+\frac{5}{2}\!\big)_{\nu}}{(\alpha^\g+2)_{\nu}}\!
\left[
{\scriptstyle s^{\rm sea}_{\nu}\, \xi^{-\alpha^{\rm sea}}} {\rm Res}F_j^{\Sigma} |_{j=\alpha^{\rm sea}-1} +\frac{2n_f s^\g \xi^{-\alpha^\g}}{\left(\alpha^\g+2+\nu\right)\CF}   {\rm Res}F_j^\g |_{j=\alpha^\g-1}
 \right]
}\,,
\!\!\!{ {\displaystyle \mbox{for}\;\; \text{A} =\Bigg\{ {\Sigma \atop \g}} \atop  \phantom{\frac{\Bigg\{}{\Big\{}}}\,.
\end{eqnarray}
\end{subequations}
Note that with our choice $\alpha^\g-\alpha^{\rm sea} =0.089$ the ratio $\textbf{R}^\g$ approaches  slowly $1$ in the $\xB\to 0$  limit, while $\textbf{R}^\Sigma$  slowly vanishes.

\pagebreak[3]
\begin{itemize}
\item {\em Large-$\xB$ region.}
\end{itemize}
\vspace{-2mm}
In the large $\xB$-region we rely on the standard scenario in which sea quark and gluon contributions die out faster than the valence ones and so the large-$\xB$ behavior in the flavor singlet channel is governed by the valence quark content, as  defined in (\ref{H-parton decompositon}).  Furthermore, since the conformal moments (\ref{^Sigma c^{(1,F)}-0}) in the pure singlet contribution die out at large $j$, the pure singlet NLO contribution can be neglected in these kinematics. Consequently, the characteristic size of NLO corrections in these kinematics arises from the remaining quark part and it can be already read off from Figs.~\ref{Fig-TFF-beta0CFCG} and \ref{Fig-TFF-valence}. Nevertheless, we add that the gluon contribution in the large-$\xB$ region is governed by a $\CA \left[\ln\frac{1}{1-\xB} + \cdots \right]^2$ term, arising from the $+$-prescription (\ref{conv2-largexB}) that enters in
(\ref{^G t^{(1,A)}}). Note that such a squared contribution is absent in the  term proportional to $\CF$  (\ref{^G t^{(1,F)}}). Finally, we can conclude that the relative NLO corrections stemming from the gluonic $t$-channel exchange must be positive in the (very) large-$\xB$ region, too.

\pagebreak[3]
\begin{itemize}
\item {\em A specific model estimate for the valence region.}
\end{itemize}
\vspace{-2mm}
In the valence region the generic picture may become more diffuse.
As for valence quarks it can be easily shown that in pure gluo-dynamics the LO evolution suppresses the gluon GPD in the large-$\xB$ region and  enhances
it in the small-$\xB$ region.
Hence, a stable point $\xB=\overline{x}_{\rm B}$ must exist, which, however, not necessarily lies in the valence region.
On the other hand, for a minimalist sea  quark GPD model with $\alpha(t)>1$  the quantity $\ell^\prime_F(\xB,t)$ can be also negative
in the small-$\xB$ region, see (\ref{conv-smallxB}). Thus, we can not necessarily assume that $\ell^\prime_F(\xB,t)$ vanishes at some given $\xB=\overline{x}_{\rm B}$ in the quark-quark channel. Moreover, in the small-$\xB$ region the evolution of the quark GPD is driven by the gluonic one while in the large $\xB$-region the driving force is the valence quark distribution,
see discussion above that can be adopted to evolution. In the valence region we expect that both  mixing and model dependence
play an important role.

For the sake of simplifying the discussion,
let us consider here a model scenario in which we have a stable point in both, the quark-quark and gluon-quark channel, which is possible
for $n_f\ge 4$. Let us recall that the LO anomalous dimensions are related to each other in a supersymmetric Yang-Mills theory \cite{Dok77,BukFroKurLip85,Belitsky:1998gu}, which imply a QCD relation,
\begin{subequations}
\label{gamma-susy}
\begin{eqnarray}
\label{gamma-susy1}
^{\g\g}\gamma^{(0,\text{A})}_j+ \frac{\beta_0}{3} &\!\!\!=\!\!\!&  \frac{3+j}{j}\, \frac{1}{^{\g\Sigma}\gamma^{(0,\text{F})}_j}\, \gamma^{(0,\text{F})}_j\; ^{\Sigma\g}\gamma^{(0,nf)}_j  + \frac{3}{j}\; ^{\Sigma\g}\gamma^{(0,nf)}_j + \frac{2n_f-6}{9}
\\
\label{gamma-susy2}
&& \mbox{with}\quad  \frac{3+j}{j}\, \frac{^{\Sigma\g}\gamma^{(0,nf)}_j}{^{\g\Sigma}\gamma^{(0,\text{F})}_j} =1\,,
\end{eqnarray}
\end{subequations}
that can also be  formulated in  momentum fraction representation. The first term on the r.h.s.~of (\ref{gamma-susy1}) can be understood as a  projection onto the quark evolution operator and will be set to zero. Since the quark-gluon anomalous dimension (\ref{^Sigma gamma_j^{(0,n_f)}}) is negative and $n_f-3$ is positive, the remaining two terms can add to zero.  Hence, as required, we can have a vanishing gluon anomalous dimension, i.e., (another) stable point in the gluon-quark channel. As we will also exemplify below, this model scenario is not necessarily true for all popular GPD models.

Apart from the pure singlet contribution, we can use in the quark-quark channel the estimate from our considerations in the flavor non-singlet channel,
given in (\ref{NLO-net-NS}) with $\ell^{\prime\prime}_F=\ell^\prime_F=0$ of the previous section.
These quark corrections are  positive and the estimate (\ref{NLO_estimate}), weighted with $\textbf{R}^\Sigma$, tells us that they can be roughly quoted as $\textbf{R}^\Sigma\times (1-2\ell_\varphi^\prime/3)\times 30\%$.
To obtain an estimate for the pure singlet quark  contribution, we follow the procedure of the previous section
and start with the conformal moments (\ref{^Sigma c^{(1,F)}-0}) in terms of anomalous dimensions (\ref{gamma_j^{(0,F)}})
\begin{eqnarray}
 \label{{^pS c}^{(1)}_{jk}}
 {^\pS c}^{(1)}_{jk} &\!\!\!=\!\!\!&
\left[\ln\frac{\cQ^2}{\muF^2} - \frac{\gamma^{(0,\text{F})}_j}{2} - \frac{\gamma^{(0,\text{F})}_k}{2} -\frac{5}{2}-
\frac{1}{(k+1)_2}\right]\frac{(-1) ^{\text{G}\Sigma}\gamma_j^{(0,\text{F})}}{j+3} - \frac{2}{(j+1)_2 (k+1)_2}
+ \Delta{^\pS}c^{(1)}_{jk} .
\nonumber\\
\end{eqnarray}
The gluon-quark anomalous dimension (\ref{{^{g Sigma}gamma}^{(0,F)}_j}), weighted with  $(-1)/(j+3)$, reads in momentum fraction representation as
\begin{eqnarray}
\label{j02x-quark}
\frac{(-1) ^{\text{G}\Sigma}\gamma_j^{(0,\text{F})}}{j+3} = \frac{4}{j (j+3)}-\frac{2}{(j+1)_2} \quad\Rightarrow\quad  \int_\xi^1\!\frac{dx}{x} \frac{2}{x(1+x)} \frac{F^\Sigma(\xi/x,\xi,\cdots)}{F^\Sigma(\xi,\xi,\cdots)}\,,
\end{eqnarray}
see corresponding expressions in Tab.~\ref{subT-pS}. As already noted this integral is unimportant in the large-$\xB$ region, however,  it
is sharply peaked due to the $1/x^2$ factor at the lower integral boundary.
Thus, we can not exclude that even for $\xB\sim 0.3$ or so (i.e., $\xi \sim 0.15$ or so) the integral is of order one and so we take now such model dependent contributions, stemming from the $j=0$ pole, into account\footnote{For the purpose of tracing large NLO corrections it would be still justified to neglect $1/(j+1)_2$ terms in the valence region,
which we will, however, include.}. Setting the quark anomalous dimension to zero, projecting as above onto the lowest CPW of the DA in terms
proportional to $1/(k+1)_2$, and  employing the substitution rules  (\ref{ell_phi},\ref{j02x-quark}),  we obtain from (\ref{{^pS c}^{(1)}_{jk}}) the  estimate
\begin{equation}
\label{NLO-pS-estimate-valence}
\frac{\im \tffF^{(1,\pS)}_{\text{V}^0} }{\im \tffF^{(0,\text{S})}_{\text{V}^0}}
\stackrel{\xB \approx\overline{x}_{\rm B}}{\approx}
\!\int^1_{\overline{\xi}}\!\frac{dx}{x} \left[\frac{2\ln\frac{\cQ^2}{\muF^2} - 6+2\ell_\varphi^\prime -2x}{x(1+x)}\right]\!\!
\frac{F^\Sigma(\overline{\xi}/x,\cdots)}{F^\Sigma(\overline{\xi},\cdots)}\,  \textbf{R}^\Sigma(\overline{x}_{\rm B},\cdots)
\end{equation}
for the valence region. The $-2x$ term in the square brackets stems from the expression $-2/(k+1)_2(j+1)_2$  in  the
conformal moments (\ref{{^pS c}^{(1)}_{jk}}) and can be as above considered as a less important contribution. Furthermore,
we can safely neglect the non-separable expressions, which vanish for $k=0$ and can be considered as small for $k\ge 2$.
For the scale setting prescription  $\mu_F = \cQ$  one realizes that we have a negative contribution
for our class of models. If the  integral (\ref{j02x-quark}) is  of order one or so, this
negative NLO correction is of the order $- (1-\ell_\varphi^\prime/3)\textbf{R}^\Sigma \times 30\% $ or so. In this scenario they
would (partially) cancel those of
the remaining quark part, quoted above as $(1-2\ell_\varphi^\prime/3)\textbf{R}^\Sigma\times 30\%$,  or even overwhelm them. Note that the pure  singlet quark estimate varies w.r.t.~DA dependence only on the $30\%$ level or so.

Our specific model assumptions allow us to express the harmonic sums $S_1(j+1)$ in the gluon-quark channel (\ref{^G c^{(1)}}) by quark anomalous dimensions, which we finally set to zero, see the anomalous dimension relation (\ref{gamma-susy}). Analogously as above, we find our specific estimate in the momentum fraction,
\begin{eqnarray}
\label{NLO-gluon-estimate-valence}
\frac{\im \tffF^{(1,\g)}_{\text{V}^0}}{\im \tffF^{(0,\text{S})}_{\text{V}^0}}
&\!\!\!\stackrel{\xB \approx\overline{x}_{\rm B}}{\approx} \!\!\! &
\Bigg\{\!
\frac{5 \pi ^2}{18}+\frac{185-89 \ell_\varphi^\prime }{24}
-\frac{23 \ln\frac{\cQ^2}{\muF^2}+(33-2n_f)\ln \frac{\muF^2}{\muR^2}}{6}
 +\frac{2}{3}\!\left[2\ln\frac{\cQ^2}{\muphi^2}+\frac{\ell^{\prime\prime}_\varphi}{\ell^\prime_\varphi}-1\!\right] \ell^\prime_\varphi
\nonumber\\
&&\phantom{\Bigg\}} +\left[\frac{26 \pi ^2}{18}-\frac{149}{12}+\frac{19}{12} \ell_\varphi^\prime\right]
\int^1_{\overline{\xi}}\!\frac{dx}{x}\,
\frac{x}{(1+x)^2}\,
\frac{F^\g(\overline{\xi}/x,\cdots)}{ F^\g(\overline{\xi},\cdots)}
\\
&&\phantom{\Bigg\}}
+6\left[\ln\frac{\cQ^2}{\muF^2} -2  +\ell_\varphi^\prime\right]\int^1_{\overline{\xi}}\!\frac{dx}{x}\,
\frac{1+\frac{13}{18} x}{(1+x)^2}\,
\frac{F^\g(\overline{\xi}/x,\cdots)}{ F^\g(\overline{\xi},\cdots)} \Bigg\} \textbf{R}^\g(\overline{\xi},\cdots)\,.
\nonumber
\end{eqnarray}
To quantify this estimate, we use again the settings (\ref{settings}).
The upper line in the braces on the r.h.s.~provides a positive relative NLO correction,
\begin{equation}
\label{NLO-imG-valence-1}
0.3  \textbf{R}^\g\;\mbox{(narrow DA)} \lesssim  0.52 \left[1 + 0.42 \ell_\varphi^\prime +0.06 \ell^{\prime\prime}_\varphi \right]\textbf{R}^\g
 \lesssim  0.9 \textbf{R}^\g \; \mbox{(broad DA)}
\end{equation}
for DA dependence see (\ref{DA-constraint}) and discussion around (\ref{ell^2-term}). The middle line contains a less important contribution, having the prefactor
$$0\; \mbox{(broad DA)} \lesssim  \frac{\alpha_s}{2\pi} 1.8(1+ 0.9\ell_\varphi^\prime) \lesssim 0.2\; \mbox{(narrow DA)}$$
in front of a harmless convolution integral, which provides an additional suppression and so this contribution can be safely ignored in our estimate.
The lower line contains the $j=0$ pole in the convolution integral and has a rather large negative prefactor in front
$$
-0.9\; \mbox{(broad DA)} \lesssim  \frac{\alpha_s}{2\pi}\, 12(-1+\ell_\varphi^\prime/2) \sim -0.6\; \mbox{(asymptotic DA)}\lesssim  -0.3\; \mbox{(narrow DA)}.
$$
As in the quark case, the value of the convolution integral is model dependent and because of the $j=0$ pole it is now much more sensitive to the
$\overline{x}_{\rm B}$ value. We expect that this negative contribution can not compensate the positive one from the upper line and so diminishes the
size of the estimate (\ref{NLO-imG-valence-1}).

Let us summarize the situation in the valence region. The NLO correction in the pure  singlet quark channel is negative, however, model dependent.
The assumption that we have a stable point $\xB=\overline{x}_{\rm B}$ in both the quark-quark and gluon-quark channel, allows us to give a more detailed estimate.
Namely, the correction in the remaining quark-quark channel are positive and of order $(1-\textbf{R}^\g )\times 30\%$ (for asymptotic DA) or so, and the gluonic corrections are also positive and of
the order of $\textbf{R}^\g  \times 50\%$ (for asymptotic DA) or so. Hence, the net result is
$30\% + \textbf{R}^\g \times 20\%$  for the asymptotic DA, which will increase, i.e., up to $50\% + \textbf{R}^\g \times 40\%$, and decrease, i.e., up to $10\% + \textbf{R}^\g \times 20\% $, for a broader (narrower) DA, respectively. Further  GPD model dependent contributions, which are associated with the $j=0$ poles, in both the pure singlet quark and gluon-quark channel will decrease these estimates,
where compared to the asymptotic DA the cancelation will be more pronounced  for a broader and weaker for a narrower DA.

\pagebreak[3]
\begin{itemize}
\item {\em Small-$\xB$ region.}
\end{itemize}
\vspace{-2mm}
For the analysis of the NLO corrections in the small-$\xB$ region it is realistic to take an effective `pomeron' trajectory $\alpha(t)\sim 1$.
Thus, in contrast to the `Reggeon' case, negative poles in the $j$-plane, i.e., quark convolution integrals such as in (\ref{F^{(1,CF}-remaining})
may contribute to some extent, however, they are harmless in the group theoretical part of the quark sector. Furthermore, the estimate
(\ref{ell_F-estimates_pomeron}), given  in Sec.~\ref{sect-NLOcorrections-generic}, tells us that  evolution effects in the quark-quark channel can be neglected.
Hence, if going from the valence region to the small-$\xB$ one the  estimate (\ref{NLO-net-NS}) will only slightly change. We recall that this NLO correction, e.g., a $\sim 30\%$ effect for the asymptotic DA, has to be translated to the net contribution in the flavor singlet channel. As we will see now, it becomes then a rather unimportant correction.

We may adopt the estimates (\ref{ell_F-estimates_pomeron}) also for the pure  singlet quark part and the gluon-quark channel, which immediately yields the conclusion that the $j=0$ integrals (\ref{x^(-p)-integrals}) that appear in (\ref{NLO-pS-estimate-valence},\ref{NLO-gluon-estimate-valence}) may give the dominant contribution,
\begin{subequations}
\label{NLO-S-estimate-small1}
\begin{eqnarray}
\label{NLO-pS-estimate-small1}
\frac{\im \tffF^{(1,\Sigma)}_{\text{V}^0} }{\im \tffF^{(0,\text{S})}_{\text{V}^0}}
&\!\!\! \stackrel{\xB \to 0 }{\approx} \!\!\!& - 6 \left[1-\frac{1}{3} \ell_\varphi^\prime -\frac{1}{3}\ln\frac{\cQ^2}{\muF^2} \right]
\int^1_{\xi}\!\frac{dx}{x^2}\,
\frac{F^\Sigma(\xi/x,\cdots)}{F^\Sigma(\xi,\cdots)}\,  \textbf{R}^\Sigma(\xi,\cdots) + \cdots\,,
\\
\label{NLO-G-estimate-small1}
\frac{\im \tffF^{(1,\g)}_{\text{V}^0}}{\im \tffF^{(0,\text{S})}_{\text{V}^0}}
&\!\!\! \stackrel{\xB \to 0 }{\approx} \!\!\!&
-12\left[1-\frac{1}{2}\ell_\varphi^\prime -\frac{1}{2}\ln\frac{\cQ^2}{\muF^2}\right]\int^1_{\xi}\!\frac{dx}{x}
\frac{F^\g(\xi/x,\cdots)}{ F^\g(\xi,\cdots)}  \textbf{R}^\g(\xi,\cdots)+ \cdots\,.
\end{eqnarray}
\end{subequations}
At our input scale $\cQ_0$ the `pomeron' is hard $\alpha(t) >1$, which is consistent with phenomenological findings in hard exclusive processes and for $t=0$ in
inclusive processes.  As discussed in Sec.~\ref{sect-NLOcorrections-generic}, the ratios (\ref{NLO-S-estimate-small1}) will then remain finite in the $\xB\to 0$ limit, however they are enhanced by relative large $1/(\alpha(t)-1)$ factors.
Since the prefactors in front of the integrals in (\ref{NLO-S-estimate-small1}) also depend on the factorization scale one might be attempted to partially remove them by choosing a low factorization scale \cite{Ivanov:2004zv,Diehl:2007hd}, e.g.,
$$
0.05\; \mbox{(broad DA)} \quad \lesssim \quad\muF^2/\cQ^2  = e^{\ell_\varphi^\prime-2}  \approx 0.14\;\mbox{(asymptotic DA)}\quad \lesssim \quad0.36\;\mbox{(narrow DA)}
$$
removes the gluon contribution and reduces the pure  singlet quark one. However, as the reader will realize at such low scale, e.g., $0.14 \cQ^2_0 \approx 0.5\GeV^2 $ for the asymptotic DA, we can not trust pQCD evolution. Furthermore, we recall the well-known fact that it is the $j=0$ pole of the gluon anomalous dimension that causes the strong gluon evolution at small $\xB$ and drives so also the quark singlet, yielding an increase of the effective `pomeron' intercepts with growing $\cQ^2$. A compromising factorization scale choice  $1\,\GeV^2 < \muF^2 < \cQ_0^2 $ will only partially remove the $j=0$ poles in the hard scattering amplitudes, however, it also softens the `pomeron' behavior of the GPDs at $\muF^2$. Consequently, the value of the convolution integrals will increase again.  The upshot is that whatever we do the (truncated) factorization scale independence tells us that we can not avoid this $j=0$ pole contribution by `optimizing' the factorization scale.

To understand the transition from the valence to the small-$\xB$ region in more detail and its interplay with the skewness dependence, we approximatively calculate the $\xB\to 0$ asymptotics of relative NLO corrections for our GPD model, specified in Sec.~\ref{sec:NLOestimates-model}, in analogy to
the flavor non-singlet result (\ref{NLO-net-NS-smallxB}).  In very good approximation the results for the relative NLO corrections are
\begin{subequations}
\label{NLO-estimate-small2}
\begin{eqnarray}
\label{NLO-pS-estimate-small2}
\frac{\im \tffF^{(1,\pS)}_{\text{V}^0} }{\im \tffF^{(0,\text{S})}_{\text{V}^0}}
&\!\!\!\stackrel{\xB \to 0}{\approx} \!\!\! &\!\!\! -\!\sum_{{\nu=0 \atop {\rm even}}} \hat{s}^{\rm sea}_{\nu}
\Bigg\{\!
\frac{12+6(\alpha^{\rm sea}+\nu)_2}{(\alpha^{\rm sea}+\nu)_3}\,
\frac{1- \frac{1}{3}\ell_\varphi^\prime+  \frac{1}{6}\gamma^{(0,\text{F})}_{\alpha^{\rm sea}-1+\nu}-\frac{1}{3}\ln\frac{\cQ^2}{\muF^2}}{\alpha^{\rm sea} -1+\nu} +
\frac{1}{(\alpha^{\rm sea}+\nu)_2}
\\
&&\phantom{\sum_{{\nu=0 \atop {\rm even}}} s_{\nu} \Bigg\{}
 -12\delta_{\nu,0}
\frac{1-\frac{1}{3}\ell_\varphi^\prime-\ln\frac{\cQ^2}{\muF^2}}{\alpha^{\rm sea} -1}
\frac{\Gamma\big(\frac{1}{2}\big) \Gamma(2+\alpha^{\rm sea})}{2^{1+2\alpha^{\rm sea}}\,\Gamma\big(\frac{3}{2}+\alpha^{\rm sea} \big)}
\frac{\xB^{\alpha^{\rm sea} -1}F^\Sigma_j|_{j=0}}{{\rm Res}F^\Sigma_j|_{j=\alpha^{\rm sea}-1}}
\Bigg\} \textbf{R}^\pS(\xi,t)
\nonumber
\end{eqnarray}
for the pure singlet quark part and
\begin{eqnarray}
\label{NLO-G-estimate-small2}
\frac{\im \tffF^{(1,\g)}_{\text{V}^0} }{\im \tffF^{(0,\text{S})}_{\text{V}^0}}
&\!\!\!\stackrel{\xB \to 0}{\approx}\!\!\! &
\sum_{{\nu=0 \atop {\rm even}}} \hat{s}^\g_{\nu}(t)
\Bigg\{
-\frac{36+18(\alpha^{\rm sea}+\nu)_2}{(\alpha^{\rm sea}+\nu)_3}\, \frac{1-\frac{1}{2}\ell_\varphi^\prime+\frac{1}{4} \gamma^{(0,\text{F})}_{\alpha^\g -1+\nu}
 -\frac{1}{2}\ln\frac{\cQ^2}{\muF^2}}{\alpha^\g -1+\nu}
\\&&
\phantom{\sum_{{\nu=0 \atop {\rm even}}}}
+\frac{55}{12}\left[
1-\frac{7}{22} \ell_\varphi^\prime +\frac{9}{110} \gamma^{(0,\text{F})}_{\alpha^\g -1+\nu}-\frac{18}{55} \ln\frac{\cQ^2}{\muF^2}
\right] \gamma^{(0,\text{F})}_{\alpha^\g -1+\nu}
\nonumber\\&&
\phantom{\sum_{{\nu=0 \atop {\rm even}}}}
+\frac{5 [37+8 \zeta(2)]}{24}\left[1-\frac{21  \ell_\varphi^\prime -\frac{16}{5}  \ell_\varphi^{\prime\prime} +
\frac{92}{5} \ln\frac{\cQ^2}{\muF^2} -\frac{32}{5} \ell_\varphi^\prime \ln\frac{\cQ^2}{\muphi^2}}{37+8 \zeta(2)}\right]
\nonumber\\
&&\phantom{\sum_{{\nu=0 \atop {\rm even}}}}
-\frac{181-104\zeta(2)}{12 (\alpha^\g+\nu)_2}\left[1-\frac{35\ell_\varphi^\prime  -8\gamma^{(0,\text{F})}_{\alpha^\g -1+\nu}+
16 \ln\frac{\cQ^2}{\muphi^2}}{181-104\zeta(2)}\right] +\cdots
\nonumber\\
&&\phantom{\sum_{{\nu=0 \atop {\rm even}}}} - 6\delta_{\nu,0}
\frac{1-\frac{1}{2}\ell_\varphi^\prime-\frac{1}{2}\ln\frac{\cQ^2}{\muF^2}}{\alpha^\g -1}
\frac{\Gamma\big(\frac{1}{2}\big) \Gamma(3+\alpha^\g)}{2^{2\alpha^\g}\,\Gamma\big(\frac{3}{2}+\alpha^\g \big)}
\frac{\xB^{\alpha^\g -1}F^\g_j|_{j=0}}{{\rm Res}F^\g_j|_{j=\alpha^\g-1}}
\Bigg\}\textbf{R}^\g(\xi,t)
\nonumber
\end{eqnarray}
for the gluon part, respectively. The relative skewness parameters read with $s^{\rm sea}_0=s^\g_0=1$
\begin{eqnarray}
\hat{s}^{\rm sea}_{\nu} =  \sum_{{\mu=0 \atop {\rm even}}}
\frac{s^{\rm sea}_{\nu}}{s^{\rm sea}_{\mu}}\,
\frac{2^{2\nu} \big(\alpha^{\rm sea}+\frac{5}{2}\big)_{\nu} (\alpha^{\rm sea}+2)_{\mu}}{
2^{2\mu} \big(\alpha^{\rm sea}+\frac{5}{2}\big)_{\mu} (\alpha^{\rm sea}+2)_{\nu}
}
\;\;\mbox{and}\;\;
\hat{s}^\g_\nu = \sum_{{\mu=0 \atop {\rm even}}}
\frac{s^\g_\nu}{s^\g_{\mu}}\, \frac{2^{2\nu} \big(\alpha^\g+\frac{5}{2}\big)_{\nu} (\alpha^\g+2)_{\mu+1}
}{
2^{2\mu} \big(\alpha^\g+\frac{5}{2}\big)_{\mu} (\alpha^\g+2)_{\nu+1}
}.
\quad
\end{eqnarray}
\end{subequations}
The $j=0$ pole contribution induce
the first and last term in the braces on the r.h.s.~of both ratios (\ref{NLO-pS-estimate-small2},\ref{NLO-G-estimate-small2}), apparently  only in the leading SO(3)-PW, i.e., $\nu=0$.   To proceed, we expand the ratios in a Laurent series at $\alpha=1$, where the first terms in the braces
contain pieces,
$$
\frac{\gamma_{\alpha-1}^{(0,\text{F})}}{\alpha-1}= -\frac{5}{2}+\frac{2 \pi ^2}{3} + O(\alpha-1)\approx 4.1 + O(\alpha-1)\,,
$$
which decrease the net result of the constant terms  by a factor of two or so. Taking also into account the remaining quark part (\ref{NLO-net-NS-smallxB}), having no pole contribution, we can quote the relative NLO corrections  for the minimalist GPD model as
\begin{subequations}
\label{NLO-S-estimate-small3}
\begin{eqnarray}
\frac{\im \tffF^{(1,\Sigma)}_{\text{V}^0} }{\im \tffF^{(0,\pS)}_{\text{V}^0}}
&\!\!\!\approx\!\!\! &  \left\{\!-4 \frac{1-\frac{1}{3}\ell_\varphi^\prime}{\alpha^{\rm sea} -1}\left[1-\xB^{\alpha^{\rm sea} -1}\right]+ 12.3-6.1 \ell_\varphi^\prime +
0.7 \ell_\varphi^{\prime\prime}+ O(\alpha^{\rm sea}-1)\! \right\}\!\textbf{R}^\Sigma(\xi,t)\,,
\\
\frac{\im \tffF^{(1,\g)}_{\text{V}^0} }{\im \tffF^{(0,\text{S})}_{\text{V}^0}} &\!\!\!\approx\!\!\! & \left\{\! -12 \frac{1-\frac{1}{2}\ell_\varphi^\prime}{\alpha^\g -1}\left[1-\xB^{\alpha^\g -1}\right] + 10.8 -9.4 \ell_\varphi^\prime  +0.7 \ell_\varphi^{\prime\prime} + O(\alpha^\g-1)\!\right\}\! \textbf{R}^\g(\xi,t)\,,\qquad
\end{eqnarray}
\end{subequations}
Clearly, in this model, e.g., as specified in Tab.~\ref{tab:models}, the corrections are governed by the $j=0$ poles,
$$
-4 \frac{1-\frac{1}{3}\ell_\varphi^\prime}{\alpha^{\rm sea} -1}\approx -25.3 + 8.4 \ell_\varphi^\prime
\quad \mbox{and} \quad
-12 \frac{1-\frac{1}{2}\ell_\varphi^\prime}{\alpha^\g -1} \approx -48.6 + 24.3 \ell_\varphi^\prime\,,$$
however, the positive constants in the Laurent expansion (\ref{NLO-S-estimate-small3}) already diminish them.  Hence, we can trust in this special case the $j=0$ pole approximation%
\footnote{Of course, (\ref{NLO-pS-estimate-small1}) gives with (\ref{x^(-1)-integrals-RDDA1}) and $b=\alpha^{\rm sea}$ the same $j=0$ contribution
as shown in (\ref{NLO-pS-estimate-small2}). The same holds true for analogous gluonic expressions.}
(\ref{NLO-S-estimate-small1}) only on a qualitative level. Taking further SO(3)-PWs will modify $\hat{s}^{\Sigma}_0$ and $\hat{s}^{\g}_0$,
i.e., the residues of the corresponding $j=0$ poles,  and the constant term in the Laurent expansion. Thus, terms proportional to
$\gamma^{(0,\text{F})}_{\alpha^\g -1+\nu}$ with $\nu\ge 2$ are getting important. From the given numbers one can easily imagine that with a special choice of the $s_\nu$-parameters one can essentially cancel the $j=0$ contributions. Hence, one can not generally conclude from  model dependent findings, see \cite{Ivanov:2004zv,Diehl:2007hd}, that NLO corrections are necessarily large in the small-$\xB$ region.

\begin{figure}[t]
\begin{center}
\includegraphics[width=\textwidth]{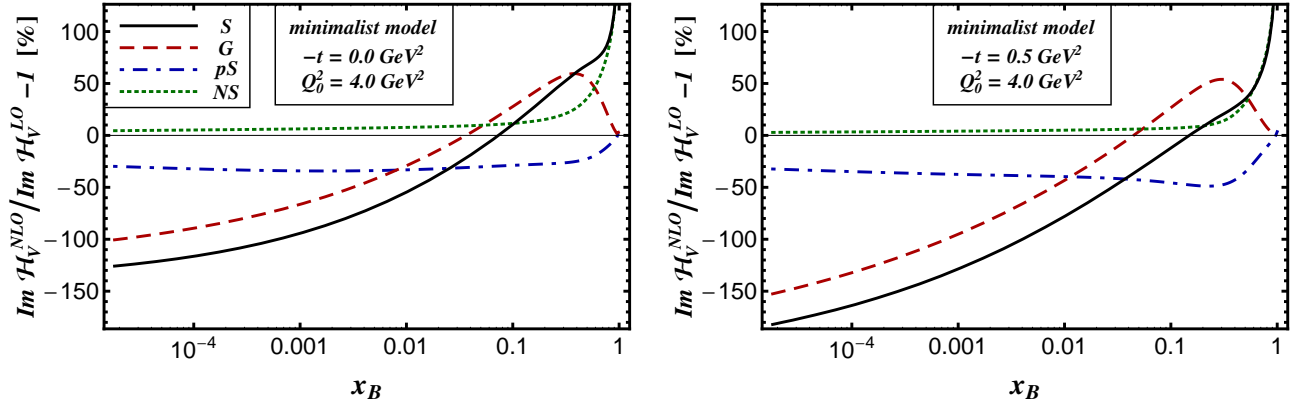}
\end{center}
\vspace{-8mm}
\caption{\small
Relative NLO corrections to the imaginary part of the flavor singlet TFF
$\tffF^{\SigG}_\text{V}$ (solid) broken down to the
gluon (dashed), pure singlet quark  (dash-dotted) and `non-singlet' quark
(dotted) at $t=0\, \GeV^2$ (left panel) and $t=-0.5\, \GeV^2$ (right
panel) at the initial scale
$\cQ_0^2= 4\,\GeV^2$.
}
\label{Fig-S-Im}
\end{figure}
Our discussion of the imaginary part for the flavor singlet TFF, evaluated
from the minimalist GPD model, is
visualized in Fig.~\ref{Fig-S-Im} for $t=0$ (left panel) and $t=
-0.5\,\GeV^2$ (right panel). Here we display the relative NLO corrections,
evaluated analogously to (\ref{NLO-net-NS}), that arise from the gluonic
NLO coefficients (dashed curves), the pure singlet quark (short
dash-dotted curves), the remaining quark part (dotted curves), and the net
contribution (solid curves).  Clearly, the large-$\xB$ asymptotic arises
from the
valence content, compare with Fig.~\ref{Fig-TFF-beta0CFCG}, where the
contributions from the `non-singlet' hard scattering amplitude dominate the net
result.
As stated, in the valence region the non-singlet contribution is
moderately positive and the pure singlet is negative, while it turns out
that in our model
the gluonic one yields a rather sizeable positive correction. In the
small-$\xB$ region the gluons dominate in our model and their
contributions are essentially governed by the $j=0$ pole. These large
corrections increase further with growing $-t$ since the `pomeron' pole at
$j=0.247 + 0.15 t $ gets slightly softer.
Note that the shape of the curves in the small-$\xB$ region is governed by
the functional forms as it arises
from the $j=0$ pole contributions, e.g., shown in
(\ref{NLO-S-estimate-small3}), and $\textbf{R}$-ratios (\ref{R^A-2}). For
our model the quark content vanishes in the $\xB\to 0$ limit  while the
gluon ones is as sizable as  $-144\%$ [$-233\%$] for $t= -0$ $[t=-0.5
\GeV^2]$.

\begin{figure}[t]
\begin{center}
\includegraphics[width=\textwidth]{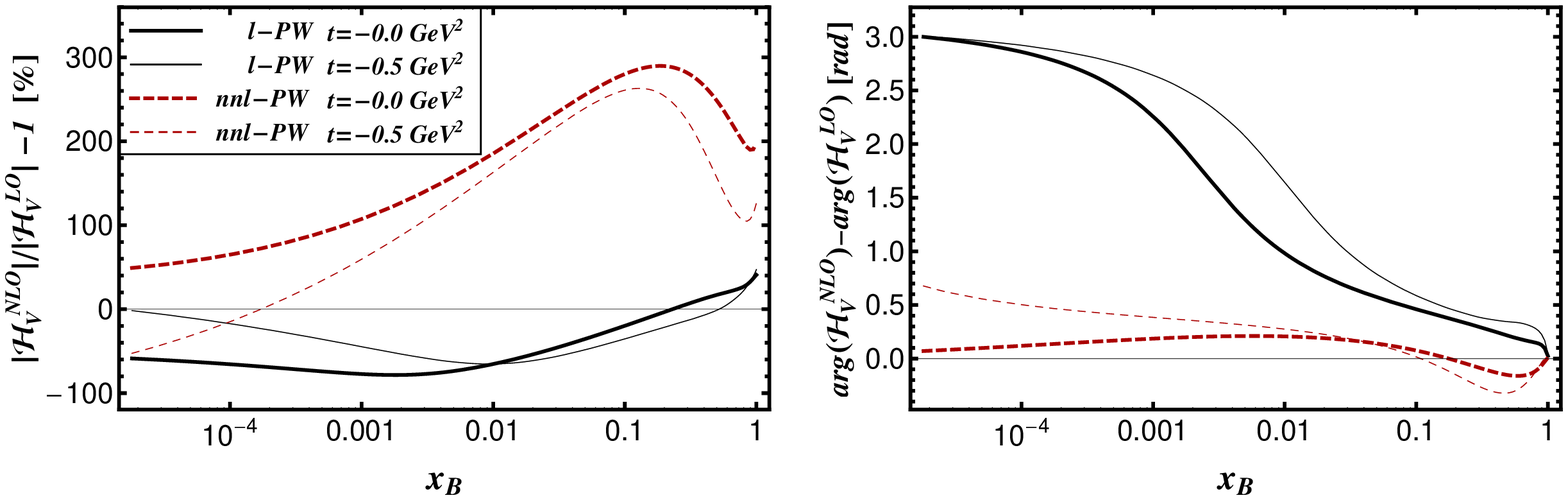}
\end{center}
\caption{
\small
Relative NLO corrections to the modulus (left panel) and phase (right panel) for the flavor singlet TFF $\tffH^{\text{S}}_{V}$
for our minimalist (l-SO(3) PW, asymptotic DA) model (solid curves) and a nnl-SO(3) PW model with a narrow DA (dashed curves) at $t=0\,\GeV^2$ (thick curves) and $t=-0.5\,\GeV^2$ (thin curves) at the initial scale $\cQ_0^2=4\,\GeV^2$.}
\label{Fig-TFF-smallx}
\end{figure}

\pagebreak[3]
\begin{itemize}
\item {\em Model dependency.}
\end{itemize}
\vspace{-2mm}
Finally, let us also demonstrate in Fig.~\ref{Fig-TFF-smallx} for the
modulus (left panel) and the phase change (right panel), defined as in
(\ref{deltaKdeltavarphi}),
that the  NLO corrections in the flavor singlet sector are rather model
dependent.
We display again the minimalist model (solid curves) and the nnl-SO(3) PW
model
as specified in Tab.~\ref{tab:models} (dashed curves) with the narrow DA
(\ref{DA-narrow}) for $t=-0\,\GeV^2$ (thick curves) and $t=-0.5\,\GeV^2$
(thin curves). For the minimalist model the corrections to the modulus are
smaller than
$100\%$ and they become negative in the small $\xB$ region.
As we have discussed the reduction of relative NLO corrections to the
modulus in the large-$\xB$ region is a naturally consequence of
analyticity, i.e., the validity of the DR.
Compared to Fig.~\ref{Fig-S-Im}, the relative NLO corrections to the modulus in the small-$\xB$
region are looking rather mild (solid curves). Note that this is caused by the negative size of the
NLO contribution, dominated by the $j=0$ pole, which overcompensates the
positive LO contribution and induces a phase difference of $\pi$ in the
small-$\xB$ asymptotic. Entirely different features
appear for our nnl-SO(3) PW model. Here the size of NLO corrections is for
$t= 0$ positive in the small-$\xB$ region and gets for a softer `pomeron'
behavior at
$t=-0.5\, \GeV^2$ negative, where the phase difference slightly increases.
On the other hand we have now huge corrections to the moduli around the
valence region, see dashed curves, which are caused by strong evolution effects in the gluon sector.

\subsubsection{Comparison of DVMP and DVCS NLO  corrections}
We add that an estimate of radiative corrections for DVCS can be
generically done in an analog manner, providing us a deeper understanding of various model dependent studies \cite{Belitsky:1999sg,Freund:2001rk,Kumericki:2007sa,Moutarde:2013qs}. To adopt it to our notation here, we write the quark conformal moments (127,128) form \cite{Kumericki:2007sa} as
\begin{subequations}
\label{^sigmac_j^{DVCS}}
\begin{eqnarray}
^\sigma\!c_j^{\rm DVCS} &\!\!\! = \!\!\!& 1 + \frac{\alpha_s(\muR)}{2\pi} \CF\, ^\sigma\!c_j^{(1,\text{F})} + O(\alpha_s^2)\,,
\\
^\sigma\!c_j^{(1,\text{F})} &\!\!\! = \!\!\!&
\bigg[ \ln\frac{\cQ^2}{\muF^2}-\frac{ \gamma^{(0,\text{F}) }_j}{4}-\frac{3(j+1)_2+1}{2(j+1)_2}\bigg]
\frac{(-1)\gamma^{(0,\text{F}) }_j}{2} -\frac{27}{8}+\frac{(11+2 \sigma )(j+1)_2+4}{4[(j+1)_2]^2}\,,\qquad\qquad
\end{eqnarray}
\end{subequations}
where signature $\sigma =+1$ and $\sigma =-1$ applies for the twist-two CFFs ${\cal F} \in \{{\cal H},{\cal E}\}$  and ${\cal F} \in \{\widetilde{\cal H},\widetilde{\cal E}\}$, respectively. Already the similarity of the most singular terms with the TFF estimate (\ref{(1)2(0)-net-NS}) implies that the NLO corrections have similar features. This is not the case in the flavor singlet channel, where the conformal moments read as follows
\begin{subequations}
\label{cS-DVCS}
\begin{eqnarray}
^{\sigma}\!\mbox{\boldmath $c$}_j^{\rm DVCS} &\!\!\! = \!\!\!& (1,0) + \frac{\alpha_s(\muR)}{2\pi}
 \left(\CF\, ^\sigma\!c_j^{(1,\text{F})},\frac{2n_f}{j+3}\, ^\sigma\!c_j^{(1,n_f)}\right) + O(\alpha_s^2)\,,
\\
^{\sigma}\!c_j^{(1,n_f)} &\!\!\! = \!\!\!&
\left[ \ln\frac{\cQ^2}{\muF^2}-\frac{\gamma^{(0,\text{F}) }_j}{2}-\frac{5}{2}\right] \frac{j+3}{2}\,
\frac{(-1) ^{{\Sigma\g}^\sigma}\!\gamma^{(0,nf)}_j}{2}+\frac{1+\sigma}{2(j+1)_2}\,,
\end{eqnarray}
where we use the notation $^{{\Sigma\g}^+}\!\gamma^{(0,nf)}_j \equiv ^{{\Sigma\g}}\!\gamma^{(0,nf)}_j$ and
$^{{\Sigma\g}^-}\!\gamma^{(0,nf)}_j= -2j/(j+1)_2$\,.
\end{subequations}

\begin{itemize}
\item {\em Flavor non-singlet channel.}
\end{itemize}
\vspace{-2mm}
Replacing  the anomalous dimensions in (\ref{^sigmac_j^{DVCS}}) by (\ref{ell_F}) yields the generic estimate
\begin{subequations}
\begin{eqnarray}
\label{NLO-DVCS-estimate}
\frac{\im\, ^\text{NS}\tffF^{\rm NLO}_\gamma}{\im\, ^\text{NS}\tffF^{\rm LO}_\gamma} =1+
\frac{\alpha^{\rm NLO}_s(\muR)}{2\pi}\,\frac{2}{3}\left\{ \left(2\ln\frac{\cQ^2}{\muF^2} +\frac{\ell^{\prime\prime}_F(\xi,t)}{\ell^{\prime}_F(\xi,t)}-3 \right)\ell^{\prime}_F(\xi,t)-\frac{27}{4} + \cdots\right\}
\end{eqnarray}
for the DVCS flavor non-singlet channel.  As in the TFF estimate (\ref{(1)2(0)-net-NS}), the behavior in the large-$\xB$ region  is dictated by a
logarithmical growth of $-\ell^{\prime}_F(\xi,t)$, where
the squared term $\ell^{\prime\prime}_F(\xi,t) \approx \ell^{\prime\, 2}_F(\xi,t)$ is universal.  Furthermore, neglecting evolution effects, we find in the valence region a moderate NLO correction
\begin{eqnarray}
\frac{\im\, ^\text{NS}\tffF^{\rm NLO}_\gamma}{\im\, ^\text{NS}\tffF^{\rm LO}_\gamma}-1 \sim \frac{\alpha_s}{2\pi} (-9/2) \approx -25\%
\quad\mbox{for}\quad \cQ_0^2= 4\,\GeV^2\,,
\end{eqnarray}
which size is comparable with our findings for flavor non-singlet TFFs, however, it is now negative.
In the small-$\xB$ region non-singular terms, which contain a pole at $j=-1$ yield large relative NLO corrections, however, are phenomenological unimportant.
The exact $\xB\to 0$ asymptotics for a minimal GPD model can be trivially found from the conformal moments,
\begin{eqnarray}
\frac{\im\, ^\text{NS}\tffF^{\rm NLO}_\gamma}{\im\, ^\text{NS}\tffF^{\rm LO}_\gamma} &\!\!\! \stackrel{\xB\to 0}{=} \!\!\!& 1+ \frac{\alpha_s}{2\pi}\,
\frac{2}{3} \Bigg\{4\left[ S_1(\alpha(t))\right]^2- \frac{2S_1(\alpha(t))}{\alpha(t)  [1+\alpha(t) ]} -9
+\frac{ (4 +\sigma ) \alpha(t) [1+ \alpha(t)]+2}{ \alpha^2(t) [1+\alpha(t)]^2}\Bigg\}.\qquad\quad
\end{eqnarray}
\end{subequations}
Compared to the flavor non-singlet TFFs, given in (\ref{NLO-net-NS-smallxB}), we have rather similar features, which, however, differ in some details.
Again caused by the first and second order pole at $\alpha(t)=0$, they become huge and positive for smaller values of $\alpha(t)$,
they vanish for $\alpha(t)\sim 0.5$, while for $\alpha(t) \sim  1$ we find that they are negative and smaller than in the valence region.
The differences of even and odd signature sector are not essential.

\begin{itemize}
\item {\em Flavor singlet channel for even signature.}
\end{itemize}
\vspace{-2mm}
In the large-$\xB$ region the signature even singlet contribution is as in DVMP entirely determined by the valence quark content and is positive.
Hence, its process independent features may be read off from (\ref{NLO-DVCS-estimate}).
In the valence region both the quark and gluonic component yields a negative correction, which looks rather harmless.
For GPD models that only evolve weakly  in the valence region we can immediately quote the crude approximation
\begin{eqnarray}
\frac{\im\, ^\text{S}\tffF^{\rm NLO}_\gamma}{\im\, ^\text{S}\tffF^{\rm LO}_\gamma}-1 \stackrel{\xB=\overline{x}_{\rm B}}{\approx} \frac{\alpha_s}{2\pi} \left\{
-\frac{9}{2} -  \frac{5}{4}\, \frac{n_f\, F^\g(\xi,\xi,\cdots)}{F^\Sigma(\xi,\xi,\cdots)} + \cdots
\right\}
\quad\mbox{for}\quad \cQ_0^2= 4\,\GeV^2\,.
\end{eqnarray}
The gluonic induced correction is  determined by the  ratio $n_f F^\g(\xi,\xi,\cdots)/F^\Sigma(\xi,\xi,\cdots)$. This model dependent
ratio may be of order one and so the gluonic component gives a  moderate contribution to the quark induced one, which decreases the flavor
non-singlet estimate of $\sim -25\%$ further.  Since a $j=0$ pole is absent in the conformal moments
(\ref{cS-DVCS}), also the corrections in the small-$\xB$ region  possess now different qualitative features than for DVMP.
For the minimalist GPD  model they are roughly given in the presence of a `pomeron' pole by
\begin{eqnarray}
\frac{\im\, ^\text{S}\tffF^{\rm NLO}_\gamma}{\im\, ^\text{S}\tffF^{\rm LO}_\gamma}-1 \stackrel{\xB\to 0}{\approx} \frac{\alpha_s}{2\pi} \left\{
-2 -  \frac{4}{3} \frac{n_f\, {\rm Res} F^\g_j|_{j=\alpha^\g-1}}{{\rm Res} F^\Sigma_j|_{j=\alpha^{\rm sea}-1}} \left(\frac{\xB}{2}\right)^{\alpha^{\rm sea}-\alpha^\g}
+\cdots\right\}\,,
\end{eqnarray}
where different coefficients appear in other models. To describe  DVCS or deep inelastic scattering data from the H1 and ZEUS
collaboration in LO approximation, one has usually $\alpha^\g > \alpha^{\rm sea} >1$ at our input scale and so the NLO corrections are sizeable and negative.
In contrast to DVMP a `hard' gluon yields now sizeable corrections. However, we should emphasize that the rather large LO value of
$\alpha^\g$ arise simply from the fact that the role of the $j=0$ pole in the evolution operator
is then less important and so the pQCD $\cQ^2$-evolution can be brought in agreement with the observed one in deep inelastic scattering.
At NLO the gluon PDF/GPD is also directly controlled which leads in general to a smaller value of  $\alpha^\g$.
Taking the PDF reparameterization consistently into account in building GPD models, one may expect smaller radiative corrections for DVCS.
What matters for phenomenology is not the estimated size of radiative corrections of some GPD models, rather how pQCD  connects the
different processes of interest in the small-$\xB$ region. This is a kind of fine tuning problem and only after data are successfully described
on can quantify reparameterization effects.

\section{Summary and conclusions}

We have systematized the theoretical framework for the perturbative
treatment of the DVMP process. In particular we have expressed
the differential cross section in terms of physically motivated TFFs.
We have also considered the case that the polarization
of the final nucleon state is observed, which offers at least in principle the possibility for a complete measurement of the TFFs.
Furthermore, we have recalculated the pure  singlet quark part
for DV\!$V_L^0$P and presented a  set of NLO formulae which allows to easily
implement the radiative corrections, which are classified into
flavor non-singlet contributions with defined signature and flavor singlet contributions.
Also, we have considered besides the common momentum fraction representation
the dispersive approach and conformal partial wave expansion in terms of a Mellin-Barnes integral.
We  have evaluated the imaginary parts of NLO hard scattering amplitudes and their conformal moments, which
are needed for implementing the radiative corrections into an existing GPD fitting code.
In this way, we represented the conformal moments by common Mellin moments,
which allowed us to calculate them analytically in terms of rational
functions and harmonic sums.
We also pointed out that the analytic continuation can be numerically
performed by means of single or double dispersion relations.
Our presentation of the NLO corrections to the hard scattering amplitude,
essentially given in terms of building blocks, can be easily adapted to
mixed representations. This is possibly useful for the case that
the meson DA is  much broader or narrower than the asymptotic one.

NLO corrections to the hard scattering amplitudes for phenomenologically
important DV$\!V_L$P and  DV$\!P\!S$P reactions were presented
in such a manner that they match existing conventions,
used in DVCS and for the evolution operator or kernels.
Here, only for  DV$\!\eta$P, having {\em odd} $t$-channel charge parity
and  {\em odd} intrinsic parity,
the NLO contributions to the hard scattering amplitudes for both the singlet quark DA,
i.e., its pure singlet part $\gamma^\ast (q\bar{q})^{(-)}\to (q\bar{q})^\pS$, and  gluonic DA  are still missing.
These reactions were measured or are planned to be measured in
near future at COMPASS II and JLAB@12 $\GeV$.
Consequently, for the more experimental challenging  DV$\!h_0$P  reaction
($h_0=1^{+-}$), having  {\em even} $t$-channel charge parity
and  {\em odd} intrinsic parity, both the pure  singlet quark-quark
[$\gamma^\ast(q\bar{q})^\pS \to (q\bar{q})^{(-)}$] and the gluon-quark
[$\gamma^\ast gg\to (q\bar{q})^{(-)}$] channel remain unknown at NLO.
We also add that for DV$\!f_0$P ($f_0=0^{++}$), having {\em odd} $t$-channel charge parity
and {\em even} intrinsic parity, all NLO ingredients including the corresponding
pure singlet quark and gluon scalar meson DA contributions are obtained from those of  DV$\!V_L^0$P by an exchange
of the in- and out-momentum fraction variables.
Whether these reactions can be accessed in  high luminosity experiments at JLAB, remains so far unclear to us.
Finally, we emphasize that the NLO formulae can be also utilized  for crossed processes, e.g.,
for exclusive Drell-Yan processes $\pi N \to N^\prime \gamma^\ast_L$ \cite{Muller:2012yq}.
So far, however, it remains questionable if such reactions can be observed in planned Drell-Yan measurements at COMPASS II.

Furthermore, we comprehensively analyzed the role of radiative NLO corrections
that arises from the hard scattering part, where for the first time we also gave an analytical discussion that is based on generally expected GPD properties. We observed that perturbative corrections in the large-$\xB$ asymptotics appear in the same manner in both DVCS and DVMP. We add that the resummation of so-called soft and collinear contributions, proposed for DVCS in \cite{Altinoluk:2012nt}, seems to be only useful in this region.   For GPD models that evolve only weakly in the valence region we found that the corrections in the valence region in both DVCS and DVMP are more or less harmless, however, they have different sign. In the presence of `pomeron' behavior the size of radiative corrections in the small-$\xB$ region is very model dependent, since it is crucially governed by the effective `pomeron' intercepts of both the quark singlet and gluon trajectories and also in a more moderate manner on the skewness effect. Hence,  only a fit to data can tell us  how large reparametrization effects are.

To get a handle on GPDs from present (and future) longitudinal DVMP and transversal DVCS measurements in a reliable manner, one certainly should in the first place utilize the collinear framework, where at least factorization was proven. In such analyses evolution must definitely be consistently included.
We add that one usually assumes that the scheme (and framework) dependent meson DA is known, but even in the case of the pion DA rather different
models were proposed \cite{Efremov:1978fi,Lepage:1980fj,Chernyak:1983ej,Bakulev:2005cp,Bakulev:2001pa,Brodsky:2003px,Cloet:2013tta,Radyushkin:2009zg} which, unfortunately, can be discriminated only partially by present experimental pion form factor
\cite{Farrar:1979aw,Volmer:2000ek,Horn:2006tm,Huber:2008id}
and pion-to-photon transition form factor \cite{Behrend:1990sr,Gronberg:1997fj,Aubert:2009mc,Uehara:2012ag} data and lattice results \cite{Braun:2006dg}. For light vector meson DAs only QCD sum rule results \cite{Ball:1996tb, Ball:1998sk}, see also references therein, and AdS/QCD model predictions \cite{Brodsky:2003px} are known to us.

Based on present phenomenological experience in the description of DVCS and DV$\!V^0_L$P processes in the small-$\xB$ region \cite{Kumericki:2009uq,Meskauskas:2011aa}  we are rather optimistic that a global description of these processes is reachable at NLO.
One arrives at the same conclusion if one considers the results based on the handbag approach \cite{Goloskokov:2005sd,Goloskokov:2007nt} and their confrontation with DVCS  data \cite{Meskauskas:2011aa,Kroll:2012sm}. Analogously, the phenomenological findings in the handbag approach \cite{Goloskokov:2005sd,Goloskokov:2007nt} and its confrontation with DVCS \cite{Meskauskas:2011aa,Kumericki:2011zc,Kroll:2012sm}, which  indicates that a description of such processes might be feasible in the valence region.  The description of the DV$\!V^0_L$P data from the CLAS collaboration in the large-$\xB$ region requires a separate study. In the case of DV$\!\pi^+$P we have also phenomenological constraints for the pion DA that arise from the electromagnetic pion and photon-to-pion transition form factor. A simultaneous  description of these form factor and DV$\!\pi^+$P data
in the pQCD framework remains an interesting problem.

\subsection*{Acknowledgements}
We are indebted to K.~Kumeri{\v c}ki for many discussions and numerical implications of DVMP formulae, which were used by us for cross checks.
We also like to thank H.~Avakian, N.~Kivel, P.~Kroll, S.V.~Mikhailov, A.V.~Pimikov, and N.~G.~Stefanis for discussions one some selected topics.
This work was supported in part by the Joint Research Activity
\emph{Study of Strongly Interacting Matter} (acronym HadronPhysics3,
Grant Agreement No.~283286) under the Seventh Framework Program of the European Community;
by the Croatian Ministry of Science, Education and Sport under the contract no.~098-0982930-2864;
and by the German Ministry of Science and Education (BMBF grant OR 06RY9191 and 05P12WRFTE).

\newpage
\appendix

\renewcommand{\theequation}{\Alph{section}.\arabic{equation}}

\section{Conventions}
\label{app:def}
\setcounter{equation}{0}
In order to use the results from the literature
or to compare them to our results,
it is important to be aware of the different conventions used.
The definitions of  GPDs and DAs uniquely determine the  hard scattering amplitudes,
evolution kernels and their anomalous dimensions.
One often encounters these elements separately in the literature.
In this section we spell out our definitions, adopted from  \cite{Mankiewicz:1997bk,Belitsky:1999hf}, and show how ours,
those in \cite{Diehl:2003ny}, and in \cite{Belitsky:2005qn} are connected to each other.

\subsection{GPD definitions}
\label{app:def-GPD}
GPDs are defined as expectation values of renormalized light-ray operators, sandwiched between a polarized in- and out-proton state \cite{Mueller:1998fv,Radyushkin:1996nd,Ji:1996nm}.
The most often encountered definitions for twist-two GPDs read%
\footnote{We ignore a gauge link along the light-cone in bi-local operators, which is absent in axial gauge $A^+=0$, and indicate the renormalization
procedure by a subscript $(\mu^2)$.}
in the notation of \cite{Mankiewicz:1997bk,Belitsky:1999hf}
for  unpolarized partons (parity even operators)
\begin{subequations}
\label{eq:defGPDV}
\begin{eqnarray}
 q(x, \eta, t,\mu^2)&\!\!\! =\!\!\!&  \int \frac{d \kappa}{2 \pi}\: e^{i x (P\cdot n)\kappa}
 \langle s_2,p_2|\bar{q}(-\kappa n)\gamma^+ q(\kappa n)|p_1,s_1\rangle_{(\mu^2)},
\label{eq:defquarkGPDV} \\
 G(x, \eta, t,\mu^2)&\!\!\! =\!\!\!& \frac{4}{P\cdot n} \int \frac{d \kappa}{2 \pi}\: e^{i x (P\cdot n) \kappa}
 \langle s_2, p_2|G^{+\mu}_{a}(-\kappa n) G_{a \mu}^{\;\;\:+} (\kappa n)|p_1,s_1\rangle_{(\mu^2)},
\label{eq:defgluonGPDV}
\end{eqnarray}
\end{subequations}
and for  polarized ones (parity odd operators)
\begin{subequations}
\label{eq:defGPDA}
\begin{eqnarray}
 \Delta q(x, \eta, t,\mu^2)&\!\!\! =\!\!\!& \int \frac{d \kappa}{2 \pi}\: e^{i x(P\cdot n)\kappa}
 \langle s_2, p_2|\bar{q}(-\kappa n)\gamma^+ \gamma_5 q(\kappa n)|p_1,s_1\rangle_{(\mu^2)},
\label{eq:defquarkGPDA} \\
\Delta G(x, \eta, t,\mu^2)&\!\!\! =\!\!\!& \frac{4}{P\cdot n} \int \frac{d \kappa}{2 \pi}\: e^{i x (P\cdot n) \kappa}
\langle s_2, p_2|G^{+\mu}_{a}(-\kappa n) i \epsilon^{\perp}_{\mu\nu} G_{a}^{\nu +} (\kappa n)|p_1,s_1\rangle_{(\mu^2)}.
\label{eq:defgluonGPDA}
\end{eqnarray}
\end{subequations}
Here
\begin{equation}
P=p_1 + p_2 \;, \quad \Delta = p_2 - p_1 \;, \quad  t\equiv \Delta^2\;, \qquad \eta = - \frac{\Delta\cdot n}{P\cdot n}\;,
\quad \epsilon^{\perp}_{\mu\nu} = \epsilon_{\mu\nu\alpha\beta}\, n^{\ast\alpha} n^\beta\;,\quad \epsilon^{0123}=1,
\label{eq:defkin} \nonumber
\end{equation}
$n^\mu$ and $n^{\ast\mu}$ are (conventional) light-like vectors with $n\cdot n^{\ast} =1$, where $a^+ \equiv a\cdot n$ projects on the +-component.
The quark GPD definitions are chosen in such a manner that they coincide with the PDF definitions in terms of light-ray
operators  in the forward limit $p_2\to p_1\equiv p$, i.e., $P=2p$, $s_2\to s_1\equiv s$, while for the gluon  an additional factor $1/x$ appears in
the PDF definition, see, e.g., \cite{Brock:1994er}. We note that the momentum fraction $\eta$ is often equated with $\xi$, which plays in DVMP the role of a
scaling variable and that this variable depends on the conventional choice of the light-cone vector. The sign in the $\eta$
definition is not chosen uniformly in the literature. Here, and in the following $\eta$ is taken to be non-negative. Furthermore, often one denotes with $P$
the average of $p_1$ and $p_2$.
The target spin content of unpolarized (\ref{eq:defGPDV}) and polarized (\ref{eq:defGPDA})  GPDs  is parameterized in terms of form factors
\begin{subequations}
\label{eq:FtoHE}
\begin{eqnarray}
q &\!\!\! =\!\!\! &
\bu_2\!\left[\frac{\gamma^+}{P^+}  H^q + \frac{  i \sigma^{+\mu}\Delta_\mu}{P^+\,2 M} E^q \right]\! u_1\,,
\quad
\phantom{\Delta}\;\; G=
\bu_2\!\left[\frac{\gamma^+}{P^+}  H^\g + \frac{  i \sigma^{+\mu}\Delta_\mu}{P^+\,2 M} E^\g \right]\! u_1\,,
\\
\Delta q &\!\!\! =\!\!\! &
\bu_2\!\left[\frac{\gamma^+ \gamma_5}{P^+} \widetilde{H}^q  + \frac{\Delta^+\, \gamma_5}{P^+\, 2M} \widetilde{E}^q \right]\! u_1\,,
\quad
\Delta G =
\bu_2\!\left[\frac{\gamma^+ \gamma_5}{P^+} \widetilde{H}^\g  + \frac{\Delta^+\, \gamma_5}{P^+\, 2M} \widetilde{E}^\g \right]\! u_1\,,
\end{eqnarray}
\end{subequations}
where Dirac spinors $u_i\equiv u(p_i,s_i)$ are normalized as $\bu(p,s)\gamma^\mu u(p,s) = 2 p^\mu$. Hence, in the forward limit the form factors of the target helicity flip contributions vanish and $q=H^q,\g=H^\g$ and $\Delta q=\widetilde{H}^q, \Delta\g=\widetilde{H}^\g$ GPDs reduce as shown in (\ref{eq:Fq2pdf},\ref{eq:Fg2pdf}) to unpolarized and polarized PDFs, respectively.
These definitions agree with those in \cite{Diehl:2003ny} and  in \cite{Belitsky:2005qn}.
See also remarks in \cite{Diehl:2003ny} about some mismatches with other definitions.

Note that GPDs and PDFs from operator definitions have the support $x\in [-1,1]$ and, thus, a quark GPD/PDF
contains both quark and antiquark contributions:
\begin{subequations}
\label{eq:Fq2pdf}
\begin{eqnarray}
q(x\ge 0,\eta=0,t=0,\mu^2) &\!\!\! = \!\!\!& q(x,\mu^2)\,,
\phantom{\Delta}
\quad
q(x\le 0 ,\eta=0,t=0,\mu^2) = -\bar{q}(-x,\mu^2)\,,
\phantom{\Delta\Delta}
\\
\Delta q(x\ge 0,\eta=0,t=0,\mu^2) &\!\!\! = \!\!\!&  \Delta q(x,\mu^2)\,,
\quad
\Delta q (x\le 0 ,\eta=0,t=0,\mu^2) = \Delta  \bar{q}(-x,\mu^2)\,.
\phantom{\Delta\Delta}
\end{eqnarray}
\end{subequations}
In the forward limit the gluon GPD $(\Delta)\g$) as defined in (\ref{eq:defgluonGPDV},\ref{eq:defgluonGPDA})  yields the standard unpolarized
(polarized) gluon PDF
$g$ ($\Delta g$) in both regions $x \ge 0$ and $x\le 0$, where, however, gluon PDFs have an additional  $x$  factor:
\begin{equation}
\label{eq:Fg2pdf}
G(x\ge 0,\eta=0,t=0,\mu^2) = x g(x,\mu^2)\,,\quad
\Delta G(x\ge 0,\eta=0,t=0,\mu^2) =  x \Delta g(x,\mu^2)\,.
\end{equation}
Because of Bose symmetry gluon GPDs (PDFs) have definite symmetry behavior under $x\to -x$ reflection,
\begin{equation}
\label{eq:G-symmetry}
G(-x,\cdots) = G(x,\cdots) \quad\mbox{and}\quad\Delta G(-x,\cdots) =  -\Delta G(x,\cdots)\,.
\end{equation}

Gluon GPDs have  even charge parity, while a quark GPD basis (\ref{gpdF^{q(pm)}}) with definite charge parity reads explicitly as
\begin{subequations}
\label{eq:Fqpm}
\begin{align}
F^{\qplmi}(x, \cdots)& \equiv
F^{q}(x, \cdots) \mp F^{q} (-x, \cdots) \quad\mbox{for}\quad F\in\{H,E\}\,,
\\
F^{\qplmi}(x, \eta, t)& \equiv
F^{q}(x, \cdots) \pm  F^{q} (-x, \cdots)  \quad\mbox{for}\quad F\in\{\widetilde{H},\widetilde{E}\}\, ,
\end{align}
\end{subequations}
where $F^{\qpl}$  and  $F^{\qmi}$  refer to even and odd  charge parity, respectively.
A quark GPD $F^q$ can be decomposed in antiquark ($-1\le x\le-\eta$), meson-like ($-\eta\le x\le \eta$ ), and quark
($\eta \le x\le 1$) contributions, which one may write in analogy to PDF terminology (\ref{eq:Fq2pdf}) as
\begin{subequations}
\label{F^q-deomomposition}
\begin{eqnarray}
F^q(x,\cdots) = \mp F^{\overline{q}}(-x \ge \eta,\cdots) + F^q(|x|\le \eta ,\cdots)+  F^q(x \ge \eta,\cdots)\,,
\end{eqnarray}
where $-$ and $+$ applies for unpolarized quark GPDs $F\in\{H,E\}$  and polarized quark GPDs $F\in\{\widetilde{H},\widetilde{E}\}$,
respectively. Hence, charge even (odd) GPDs contain for $\eta\le x$ the sum (difference) of quark and aniquark GPDs,
\begin{eqnarray}
F^{\qplmi}(x\ge \eta,\cdots) =  F^q(x\ge \eta,\cdots) \pm F^{\overline{q}}(x\ge \eta,\cdots)\,.
\end{eqnarray}
We note that in PDF terminology a quark PDF  is decomposed into  valence and sea quark contributions, where the latter is equated with the antiquark ones.
In phenomenology it is common to adopt this PDF  terminology and we, therefore, write the charge even quantities in terms of valence quarks
and antiquarks:
\begin{eqnarray}
F^{q(+)}(x\ge \eta,\eta,\cdots)  &\!\!\! =\!\!\! & F^{q^{\rm val}}(x,\eta,\cdots) + 2 F^{\bar{q}}(x,\eta,\cdots)\,,
\\
F^{q(-)}(x\ge \eta,\eta,\cdots)  &\!\!\! =\!\!\! & F^{q^{\rm val}}(x,\eta,\cdots)\,,
\end{eqnarray}
where charge odd GPDs have only valence quark content.
\end{subequations}

Finally, a group theoretical SU$(n_f)$ decomposition of quark GPDs, e.g., for $n_f=3$ $(n_f=4)$
\begin{subequations}
\label{eq:SU(n)}
\begin{eqnarray}
F^{0} &
\!\!\!=\!\!\!&
F^{u}+F^{d}+F^{s}\, (+ F^{c})
\\
F^{3} &
\!\!\!=\!\!\!&
F^{u} -F^{d}
\\
F^{8} &
\!\!\!=\!\!\!&
F^{u} +F^{d} - 2 F^{s}
\\
(F^{15} &
\!\!\!=\!\!\!&
F^{u}
+F^{d}
+F^{s}
- 3 F^{c})
\,.
\end{eqnarray}
\end{subequations}
is  utilized to solve the mixing problem of charge even quark and gluon GPDs.
For flavor singlet $0^{(+)}$ quark and gluon GPDs we  utilize  a vector valued GPD, defined in (\ref{eq:gpdFvec}).
Note that the unpolarized $0^{(+)}$  GPD in the charge even sector is labeled by the superscript $\Sigma$.
Hard scattering amplitudes and evolution kernel may also contain a pure  singlet quark part. In our notation $\Sigma$ refers always to the net
contribution, containing both the group theoretical part $0^{(+)}$ and the pure  singlet quark piece, see the TFF decomposition for a neutral vector meson
TFF decomposition (\ref{eq:rhoTFFsing}). The sum of gluon and $\Sigma$ contribution is labeled by superscript S.

\subsection{Evolution kernels}
\label{app:def-evo}
The scale dependence of flavor non-singlet or charge odd DAs is governed by the well known
Efremov-Radyushkin-Brodsky-Lepage (ER-BL)  evolution equation which has the following form
\begin{equation}
\label{ER-BLequation}
\mu^2 \frac{d}{d\mu^2} \varphi^\text{A} (u,\mu^2) =
V\left(u, v | \alpha_s(\mu) \right) \stackrel{v}{\otimes}\varphi^\text{A}(v,\mu^2)\,, \quad  f\stackrel{v}{\otimes}g\equiv \int_0^1\!dv\,f g
\end{equation}
for  $\text{A}\in \{\NSpl,\qmi\}$.
In this case we have the restriction $0\le v \le 1$, and the support of the evolution kernel simplifies to
\begin{equation}
\label{V}
V(u, v|\alpha_s) = \theta(v-u) f(u,v|\alpha_s)
+   \theta(\bu-v) \overline{f}(u,v|\alpha_s)
+ \left\{{u\to \bu \atop v\to \bv}\right\}\,.
\end{equation}
Here, the $\overline{f}$-part stems from quark-antiquark mixing and appears at NLO \cite{Mueller:1998fv}.
Due to the appearance of $\overline{f}$,  the evolution kernels for symmetric or antisymmetric DAs differ in higher orders from each other, i.e.,
it depends on the signature. The LO approximation of the ER-BL kernel, known from 1980's,
(see \cite{Belitsky:1998gc} and references therein) is given in (\ref{V^{(0)}}).

For  flavor non-singlet or a charge odd GPD $F$ the
evolution equation can be obtained from the quoted one by replacing the $0\le v \le 1$ restriction  by the support
of the kernel \cite{Dittes:1988xz}:
$$
\theta(v-u) \rightarrow \Theta(u,v) = \theta\!\left(1-\frac{u}{v}\right)\theta\!\left(\frac{u}{v}\right) {\rm sign}(v)\,.
$$
In the variables we are using, i.e., $u=(\eta+x)/2\eta$ and $v=(\eta+y)/2\eta$, the evolution equation  reads
\begin{equation}
\mu^2 \frac{d}{d\mu^2} F^{\text{A}} (x, \eta, t,\mu^2) = \int_{-1}^1\!\frac{dy}{2\eta}\,
^\sigma V\!\left(\!\frac{\eta+x}{2 \eta}, \frac{\eta+y}{2 \eta} \Big| \alpha_s(\mu)\!\!\right)
F^{\text{A}}(y, \eta, t,\mu^2)
\label{eq:eveqNS-x}
\end{equation}
with the proper signature $\sigma(\text{A})$ for  $\text{A}\in \{\NS^{(\pm)},\qmi\}$.

In the flavor singlet channel the evolution of our vector valued GPD $\mbox{\boldmath $F$}$, given in \req{eq:gpdFvec}, reads
\begin{equation}
\mu^2 \frac{d}{d\mu^2} \mbox{\boldmath $F$} (x, \eta, t, \mu^2) = \int_{-1}^1\!\frac{dy}{2\eta}\,
\mbox{\boldmath $V$}\!\left(\!\frac{\eta+x}{2 \eta}, \frac{\eta+y}{2 \eta}; \eta \Big| \alpha_s(\mu)\! \right)\cdot
\mbox{\boldmath $F$}(y, \eta, t, \mu^2)\,.
\label{eq:eveqS}
\end{equation}
The quark entry of the GPD vector is the sum over all charge even quark GPDs (\ref{eq:Fqpm}) and it reads together with the evolution matrix as
\begin{subequations}
\label{bV}
\begin{equation}
\label{bV-matrix-1}
\left(\sum_q {F^\qpl } \atop {F^{\rm G} }\right)(\cdots)
\quad\mbox{and}\quad
\mbox{\boldmath $V$}(u, v;\eta| \alpha_s)
=
\left(\!\!\!
\begin{array}{cc}
\phantom{2\eta\,} {^{\Sigma\Sigma}}V   &   {^{\Sigma\g}}V/2\eta \\
        2\eta\, {^{\g\Sigma}}V        &   {^{\g\g} V}\phantom{/2\eta}
\end{array}
\!\!\!\right)\!\!
(u, v| \alpha_s)\,,
\end{equation}
respectively. Exploiting symmetry of the GPDs the entries can be represented as
\begin{eqnarray}
{^\text{AB} V}(u, v|\alpha_s)
=
\theta(v - u )
\, {^\text{AB}}v (u,v|\alpha_s) \pm
\left\{ {u \to \bar u \atop  v \to \bar v }\right\}
\mbox{\ for\ }
\left\{ {\text{A}=\text{B} \atop \text{A} \not= \text{B}}\right. ,
\label{bV-entries}
\end{eqnarray}
where the quark-quark channel consists of the charge even non-singlet, with definite signature, and the pure singlet part
\begin{equation}
{^{\Sigma\Sigma}}v (u,v|\alpha_s) = {^\sigma}v(u,v|\alpha_s) + {^\pS}v(u,v|\alpha_s) \quad
\sigma =\left\{
{
+\mbox{ for } F \in \{H,E\}
\atop
-\mbox{ for } F \in \{\widetilde{H},\widetilde{E}\}
}\right.  .
\end{equation}
\end{subequations}
With the GPD conventions (\ref{eq:defGPDV},\ref{eq:defGPDA}) the LO kernel reads as in \cite{Belitsky:1998gc}, shown for unpolarized
parton GPDs in (\ref{Vb^{(0)}}). Various other authors published these LO kernels, too. Be aware that the functional form can differ, which
only should affect the non-physical sector%
\footnote{This is certainly true if the evolution equations are used for symmetrized GPDs. If one  exploits symmetry to map all GPDs into the region $-\eta \le x \le 1$ \cite{Vinnikov:2006xw}, it is possibly necessary to symmetrize the GPDs afterwards.}. Those in (\ref{Vb^{(0)}}) are  the kernels from \cite{Cha80a}, while the improved ones provide for both kinds of moments (even and odd)
polynomials  \cite{Belitsky:1998vj}. These kernels were used as one ingredient to construct the NLO corrections for the entries in (\ref{bV-matrix-1}) \cite{Belitsky:1999hf}.

With these specifications, we have checked that the forward limit of the evolution equation (\ref{eq:eveqS}) yields for positive $x$ nothing but
the flavor singlet DGLAP equation,
\begin{equation}
\mu^2 \frac{d}{d\mu^2} \left( {\Sigma(x,\mu^2) \atop x g(x,\mu^2)} \right)   = \int_{x}^1\!\frac{dy}{y}\,
\left(\!\!\!
\begin{array}{cc}
\phantom{2\eta\,} {^{\Sigma\Sigma}}p\big(\frac{x}{y}|\alpha_s\big)   &  \frac{1}{y}\, {^{\Sigma\g}}p\big(\frac{x}{y}|\alpha_s\big)   \\
       x\, {^{\g\Sigma}}p\big(\frac{x}{y}|\alpha_s\big)       &   \frac{x}{y}\, {^{\g\g}}p\big(\frac{x}{y}|\alpha_s\big)
\end{array} \right)\cdot
\left({ \Sigma(y,\mu^2) \atop y\,g(y,\mu^2)}\right)\,,
\label{eq:eveqS-DGLAP}
\end{equation}
where the well-known expressions of LO kernels, see, e.g.~\cite{Furmanski:1980cm}, follows from (\ref{Vb^{(0)}}), showing that definitions and diagrammatical results as we use them here are consistent. The same holds true for the NLO corrections to the entries given in \cite{Belitsky:1999hf} and those for splitting kernels from Ref.~\cite{Furmanski:1980cm}.
Be aware that the flavor singlet evolution kernels for
$H/E$ and $\widetilde{H}/\widetilde{E}$ type GPDs are different.

Other forms of evolution equations are also known, e.g., with non-symmetrized quark GPDs (\ref{eq:defquarkGPDV},\ref{eq:defquarkGPDA}). However,  as long as one respects the proper
symmetry for gluon GPDs, only charge even quark GPDs have cross talk with gluonic GPDs, while the charge odd quark part decouples and evolves
autonomously.  Often the normalization remains confusing, if conventions are not entirely spelled out.  By  rescaling of quark and gluon GPDs
with constants $a$ and $b$, respectively, the  off diagonal entries  in the evolution kernel change, too,
\begin{subequations}
\label{bV-scaling}
\begin{equation}
\label{bV-matrix}
\left(\sum_q {F^\qpl } \atop {F^{\rm G} }\right) \rightarrow  \left( a\, \sum_q {F^\qpl } \atop b\, {F^{\rm G} }\right)\;\;
\mbox{implies:}\;\;
\left(\!\!\!
\begin{array}{cc}
\phantom{2\eta\,} {^{\Sigma\Sigma}}V   &   {^{\Sigma\g}}V/2\eta \\
        2\eta\, {^{\g\Sigma}}V        &   {^{\g\g} V}\phantom{/2\eta}
\end{array}
\!\!\!\right) \rightarrow
\left(\!\!\!
\begin{array}{cc}
\phantom{2\eta\,} {^{\Sigma\Sigma}}V   &  \frac{a}{b}\, {^{\Sigma\g}}V/2\eta \\
        2\eta\, \frac{b}{a}\, {^{\g\Sigma}}V        &   {^{\g\g} V}\phantom{/2\eta}
\end{array}
\!\!\!\right),
\end{equation}
and  also the  coefficient functions gets modified:
\begin{equation}
(T^\Sigma, T^\g) \rightarrow (T^\Sigma /a, T^\g /b)
\end{equation}
(analogous for singlet DAs, see e.g., Ref.~\cite{Kroll:2002nt}). Apart from checking factorization logarithms
in hard scattering amplitudes by means of evolution kernels, one may take the forward limit of kernels  or use
the energy-momentum sum rule for $H/E$ GPDs.
The case $a=1,b=1$ is taken  in \cite{Diehl:2007hd}, too, while
$a=1,b=1/2$ is quoted in  \cite{Belitsky:2005qn} and we also use it in conformal space. In all cases the charge even quark GPDs  $F^{\qpl} (x, \eta, t)$ are defined as in  \req{eq:Fqpm}.
\end{subequations}

\subsection{Anomalous dimensions}
\label{app:ad}

Local conformal  operators in the flavor singlet sector may be defined as in \cite{Belitsky:1998gc}
\begin{subequations}
\label{O_{nl}}
\begin{equation}
\label{O_{nl}-1}
{\cal O}_{nl}=
\left.\left(
{\hspace{5mm}
\frac{1}{2}( i \partial_+ )^l C^{\frac{3}{2}}_n \!
\left( \frac{\stackrel{\leftrightarrow}{\partial_+}}{\partial_+} \right)
{ {^\Sigma {\cal O}(\kappa_1,\kappa_2)}}
\atop
{( i \partial_+ )^{l-1} C^{\frac{5}{2}}_{n-1}\!
\left( \frac{\stackrel{\leftrightarrow}{\partial_+}}{\partial_+} \right)
{^\g{\cal O}(\kappa_1,\kappa_2)}}}
\right) \right|_{\kappa_1 = \kappa_2 = 0},
\end{equation}
where $\partial_+ = \partial_{\kappa_1} + \partial_{\kappa_2}$,
$\stackrel{\leftrightarrow}{\partial_+} = \partial_{\kappa_1} -
\partial_{\kappa_2}$  and  the bare non-local light ray operators read in light-cone gauge as follows
\begin{eqnarray}
{\cal O}(\kappa_1,\kappa_2) \equiv \left(
{
{ {^Q {\cal O}(\kappa_1,\kappa_2)}}
\atop
{^G{\cal O}(\kappa_1,\kappa_2)}}
\right) = \sum_q
\left(
{ \overline{q}(\kappa_2 n) \gamma^+ q (\kappa_1 n) - \overline{q}(\kappa_1 n) \gamma_+
\psi(\kappa_2 n)
\atop \frac{1}{N_f}
G^{+ \mu}_a (\kappa_2 n) g_{\mu\nu}  G^{\nu +}_a (\kappa_1 n)
}
\right),
\end{eqnarray}
\end{subequations}
where its entries appear also in the GPD definition (\ref{eq:defGPDV}).
An analogous definition holds true for operators with intrinsic odd parity ($\gamma^+ \to \gamma^+ \gamma_5$, $g_{\mu \nu} \to i \epsilon_{\mu\nu}^\perp$, symmetrized quark operator), see GPD definition (\ref{eq:defGPDA}).
The matrix valued anomalous dimensions are defined by the renormalization group equation
\begin{eqnarray}
\label{RGE}
\mu \frac{d}{d\mu} {\cal O}_{nl}(\mu)  =  -\frac{1}{2} \sum_{m=0 \atop n-m\; {\rm even}}^{n}  \mbox{\boldmath$\gamma$}^{\mbox{\tiny \cite{Belitsky:1998gc}}}_{nm}(\alpha_s(\mu)) {\cal O}_{ml}(\mu)\,,
\end{eqnarray}
and read explicitly
\begin{eqnarray}
\label{bgamma}
\mbox{\boldmath$\gamma$}^{\mbox{\tiny \cite{Belitsky:1998gc}}}_{nm}(\alpha_s)=\left(
\begin{array}{cc}
{^{QQ}}\gamma_{nm} & {^{Q G}}\gamma_{nm}\\
{^{QG}}\gamma_{nm} & {^{GG}}\gamma_{nm}
\end{array}\right)(\alpha_s)
\quad\mbox{with}\quad   {^{QQ}}\gamma_{nm} = {^\sigma}\gamma_{nm}^{\mbox{\tiny \cite{Belitsky:1998gc}}} + {^\pS}\gamma_{nm}^{\mbox{\tiny \cite{Belitsky:1998gc}}}\,.
\label{eq:andimmatrix}
\end{eqnarray}
Note that for flavor singlet operators with intrinsic even (odd) parity  the non-negative integers $n$ and $m$ are odd (even) and the signature of the flavor non-singlet anomalous dimensions is $\sigma = +1$ ($\sigma = -1$). To LO accuracy the anomalous dimensions are diagonal while at NLO the off-diagonal entries are evaluated to NLO accuracy in \cite{Belitsky:1998gc}. The explicit expressions for the complete NLO result in the minimal subtraction scheme are summarized in  \cite{Belitsky:1998uk}.

The conformal moments of GPDs or DAs are the (reduced) expectation values of conformal operators. However, some care is needed here with respect to
their overall normalization. Taking just Gegenbauer polynomials, entering in (\ref{O_{nl}}), it follows from the GPD definitions (\ref{eq:defGPDV},\ref{eq:defGPDA}),
\begin{subequations}
\label{<O_{nl}>}
\begin{eqnarray}
\frac{\langle s_2,p_2|^Q{\cal O}_{nn}|p_1,s_1\rangle_{(\mu^2)}}{(P^+)^{n+1}} &\!\!\! =\!\!\!&  \frac{1}{2} \sum_q \int_{-1}^1\!dx\, \eta^n C_{n}^{3/2}(x/\eta)\,
 \left[q(x, \eta, t,\mu^2) -  q(-x, \eta, t,\mu^2)\right],
 \\
\frac{\langle s_2,p_2|^G{\cal O}_{nn}|p_1,s_1\rangle_{(\mu^2)}}{(P^+)^{n+1}} &\!\!\! =\!\!\!& \frac{1}{4}  \int_{-1}^1\!dx\, \eta^{n-1} C_{n-1}^{5/2}(x/\eta)\,  G(x, \eta, t,\mu^2)\,,
\end{eqnarray}
\end{subequations}
which satisfy the renormalization group equation (\ref{RGE}). Note the factor $1/2$ in the quark singlet
entry (\ref{O_{nl}-1}) and the factor $4$ in the gluon GPD definitions (\ref{eq:defgluonGPDV},\ref{eq:defgluonGPDA}) are explicitly displayed here.
Consequently, one finds from the evolution equation (\ref{eq:eveqS}) for quark singlet GPDs
that the Gegenbauer moments of the evolution kernel (\ref{bV}) are given by the anomalous dimensions (\ref{bgamma})
\begin{eqnarray}
\label{eq:Vtogamma}
\int_0^1\!du\, C^{\nu(A)}_{n+3/2-\nu(A)}(2u-1)\, {^{A B} V}(u,v|\alpha_s) =- \frac{1}{2}\sum_{m=0}^n  {^{A B}\gamma}^{\mbox{\tiny \cite{Belitsky:1998gc}}}_{nm}(\alpha_s)\,  C^{\nu(B)}_{m+3/2-\nu(B)}(2v-1)\,,
\end{eqnarray}
where $\nu(\Sigma)=3/2$, $\nu(\g)=5/2$, $Q\equiv \Sigma$, and  $G\equiv \g$.

As said in Sec.~\ref{sec:prel-MB-TFF}, our (integral) conformal  GPD moments are defined in such a manner that in the forward limit they coincide with the common Mellin moments of PDFs. This is the convention of Ref.~\cite{Kumericki:2007sa}, where also the solution of the evolution equations is written down.
Here we repeat only that the transformation from the reduced matrix elements (\ref{<O_{nl}>}) to our conformal GPD moments (\ref{Fn}) is given by
\begin{equation}
 \mbox{\boldmath $N$}_{j}
=  \frac{\Gamma(3/2) \Gamma (j+1)}{2^{j}\Gamma\left(j+3/2\right)}
\left(
\begin{array}{cc}
1 & 0
\\
0 & \frac{6}{j}
\end{array}
\right)
\, , \quad\mbox{which implies}\quad
 \mbox{\boldmath $\gamma$}_{jk} =
 \mbox{\boldmath $N$}_{j}
 \mbox{\boldmath $\gamma$}_{jk}^{\mbox{\tiny {\cite{Belitsky:1998uk}}}}
 \mbox{\boldmath $N$}_{k}^{-1}
\label{eq:Utransf}
\end{equation}
or explicitly written as
\begin{eqnarray}
\mbox{\boldmath $\gamma$}_{jk}  =
\frac{2^k \Gamma(j+1) \Gamma(k+3/2)}{2^j \Gamma(k+1) \Gamma(j+3/2)}
 \left(
\begin{array}{cc}
{^{Q Q}\!\gamma}_{jk}^{\mbox{\tiny \cite{Belitsky:1998uk}}} & \frac{k}{6}
{^{Q G}\!\gamma}_{jk}^{\mbox{\tiny \cite{Belitsky:1998uk}}}
\\
\frac{6}{j} {^{GQ}\!\gamma}_{jk}^{\mbox{\tiny \cite{Belitsky:1998uk}}} &
\frac{k}{j} {^{GG}\!\gamma}_{jk}^{\mbox{\tiny \cite{Belitsky:1998uk}}}
\end{array}
\right) \, . \label{eq:Utrans2}
\end{eqnarray}
For the case of interest here, i.e., for vector valued singlet GPDs $\mbox{\boldmath $H/E$}$, assigned with signature $\sigma =+1$,
they are obtained from the analytic continuation of odd moments.
The diagonal entries
\begin{eqnarray}
\mbox{\boldmath$\gamma$}_{jj}(\alpha_s)=\left(
\begin{array}{cc}
{^{\Sigma\Sigma}}\gamma_{j} & {^{\Sigma \g}}\gamma_{j}\\
{^{\g \Sigma}}\gamma_{j} & {^{\g\g}}\gamma_{j}
\end{array}\right)(\alpha_s)
\quad\mbox{with}\quad   {^{\Sigma\Sigma}}\gamma_{j}(\alpha_s) = {^+}\gamma_{j}(\alpha_s) + {^\pS}\gamma_{j}(\alpha_s)
\label{eq:andimmatrix_jj}
\end{eqnarray}
coincide with the anomalous dimensions as used in unpolarized deeply inelastic scattering and are to LO given in \req{eq:gamma0} and \req{eq:andim}.

\section{Building blocks for separable contributions}
\label{app:ImRe}
\setcounter{equation}{0}

In this appendix we study some single variable building blocks,  which will appear in the perturbative expansion of (exclusive) hard scattering amplitudes. In particular, we consider  powers of logarithms and  dilog functions that are accompanied by negative powers of $u$ or $\bu$,
\begin{equation}
\label{f^p_i}
f^p_1(u|a,b) = -\frac{\ln^p\bu}{u^a \bu^b}\,,
\quad f^1_2(u|a,b) = \frac{{\rm Li}_2(u)}{u^a \bu^b}\,,  \quad \mbox{where}\quad a,b,p\in\{0,1,\cdots\}\,.
\end{equation}
Our notation is adopted from the standard convention ${\rm Li}_1(u) \equiv -\ln\bu$ and we may generalize these definitions
to functions that include higher order polylogarithms, which  appear  beyond the NLO approximation.
The building blocks possess $[1,\infty]$-cuts on the real axis in the complex $u$ plane and essentially behave at
infinity as $u^{-a-b} $, modified by some logarithmic corrections (except for $p=0$).
Moreover, they may possess poles at $u=0$, which we  remove below by subtraction.

\subsection{Relations among building blocks}

Numerous relations  among these building blocks (\ref{f^p_i}) exist, which can be exploited:
\begin{subequations}
\begin{itemize}
\item
The algebraic decomposition of the denominator in these functions implies for $a\ge1$ and $b\ge 1$ the recurrence relation
\begin{eqnarray}
\label{f^p-algebra}
f^p_i(u|a,b) = \sum_{m=1}^{a} f^p_i(u|m,b-1)  + f^p_i(u|0,b)\,,  \quad \mbox{for}\quad a\ge 1 \quad \mbox{and} \quad b \ge 1,
\end{eqnarray}
which allows us to reduce immediately these functions to $f^p_i(u|a=0,b)$ and by successive application to $f^p_i(u|a,b=0)$ ones.

\item
$f^p_1$ functions, i.e., powers of logarithms, are generated from $f_1(u|a,\beta) \equiv f^{p=0}_1(u|a,\beta)$ functions by means of
derivatives w.r.t.~the parameter $\beta$ which is now taken as a continuous variable,
\begin{equation}
\label{f^p-generate}
f^p_1(u|a,b)  = (-1)^p \frac{\partial^p}{\partial \beta^p}  f_1(u|a,\beta)\Big|_{\displaystyle \beta=b} \,
\quad\mbox{with}\quad
f_1(u|a,\beta) = u^{-a} \exp\{-\beta\ln\bu\} \,.
\end{equation}

\item Differentiation (integration) w.r.t.~$u$ can be used to reduce (increase) the power of logarithms in  $f_1$ functions and
to a simultaneously increase (decrease) of $b$,
\begin{eqnarray}
\label{f^p-dif}
\frac{u^{-a}\, \bu^{-b}}{p+1}\frac{d}{du} u^a \bu^b f^{p+1}_1(u|a,b) = -\,   f^{p}_1(u|a,b+1)\,.
\end{eqnarray}

\item Since $d{\rm Li}_{i+1}(u)/du = {\rm Li}_i(u)/u$, we employ for the  case  $p=1$ integration  (differentiation) w.r.t.~$u$  to increase (reduce) the index $i\ge 1$ of $ f^{p=1}_i$ functions and simultaneously decrease  (increase) the value of $a$:
\begin{eqnarray}
\label{f^p-i+1}
 f^1_{i+1}(u|a,b) =   u^{-a} \bu^{-b} \int_0^u\!dv\, v^{a} \bv^{b} f^1_{i}(v|a+1,b)\quad\mbox{for}\quad i\ge 1
\end{eqnarray}
and
\begin{eqnarray}
\label{f^p-i-1}
\frac{d}{du} u^{a} \bu^{b}  f^1_{i+1}(u|a,b) =    u^{a} \bu^{b} f^1_{i}(u|a+1,b)\quad\mbox{for}\quad i\ge 1\,.
\end{eqnarray}
\end{itemize}
\end{subequations}

Let us introduce a generating function for the building blocks (\ref{f^p_i}) in which a subtraction of $u=0$ poles is performed
in such a manner that the above quoted relations remain true,
\begin{equation}
\label{f^p_i-sub}
f_{1,+}(u|a,\beta) =  u^{-a} \left[\bu^{-\beta}  - \sum_{i=0}^{a-1}\frac{ \Gamma(i+\beta)}{i! \Gamma(\beta)} u^{i}\right]
= \frac{\Gamma(a+\beta)}{\Gamma(1+a)\Gamma(\beta)}\ {_2}F_1\!\left(\!{1,a+\beta\atop 1+a}\Big|u\!\right)\,.
\end{equation}
This function is nothing but a hypergeometric function  which for non integer $\beta$ possess only a $[1,\infty]$ cut.
Note that the algebraic relation (\ref{f^p-algebra}) holds true for this subtracted functions, too, which proofs an identity for a finite sum of hypergeometric functions.

\subsection{Values on the cut}

We can easily find the value of the  building blocks (\ref{f^p_i}) on the cut by means off the generating function (\ref{f^p_i-sub}).
On the real $u$ axis it reads for non integer $\beta$
\begin{eqnarray}
\label{f^p_i-sub-cont}
f_{1,+}(u\pm i\epsilon|a,\beta) =  u^{-a} \left[
\theta(\bu) \bu^{-\beta}  +  \theta(-\bu)\,\exp\{\mp i \beta \pi\}\, (-\bu)^{-\beta} -   \sum_{i=0}^{a-1}\frac{ \Gamma(i+\beta)}{ i! \Gamma(\beta)} u^{i}
\right],
\end{eqnarray}
where the cut arises for $u\ge 1$  and the sign of its imaginary part depends on whether the cut is approached from above ($+i\epsilon$) or below ($-i\epsilon$).

For the $b=0$ case with arbitrary non-negative integer $a$  we find by differentiation w.r.t.~$\beta$, see (\ref{f^p-generate}),  that powers of logarithms multiplied by $u^{-a}$ and subtracted $u=0$ poles have the values
\begin{subequations}
\label{f^p_1+(u|a,0)}
\begin{eqnarray*}
-f^p_{1,+}(u\pm i\epsilon|a,0) =
\theta(\bu) \frac{\ln^p\bu}{u^a}
+\theta(-\bu) \sum_{l=0}^p \left({p \atop l}\right) (\mp i\pi)^l\, \frac{\ln^{p-l}(-\bu)}{u^a}
+(-1)^p \sum_{i=1}^{a-1}\frac{ u^{i-a}}{i!} \frac{d^p}{d\beta^p} \frac{\Gamma(\beta+i)}{\Gamma(\beta)}\Big|_{\beta=0}\,.
\end{eqnarray*}
Consequently, the real and imaginary parts of these functions, introduced in Sec.~\ref{sec:NLO-blocks1} by the symbolic shorthand $\left[\ln^p(\bu \pm i\epsilon)/u^a\right]^{\rm sub}$, read
\begin{eqnarray}
\label{ref^p_1+(u|a,0)}
&&\!\!\!\!\!\!\!\!\!\!\!\! \re \left[\frac{\ln^p\bu}{u^a}\right]^{\rm sub}   =
\frac{\ln^p|\bu|}{u^a} +(-1)^p \sum_{i=1}^{a-1}\frac{ u^{i-a}}{i!} \frac{d^p}{d\beta^p} e^{\ln\frac{\Gamma(\beta+i)}{\Gamma(\beta)}}\Big|_{\beta=0}
+\theta(-\bu) \sum_{{m=2 \atop {\rm even}}}^{p}  \left({p \atop m}\right) (i \pi)^{m}\;  \frac{\ln^{p-m}|\bu|}{u^a}\,,
\nonumber\\
 &&\mbox{and}
 \\
\label{imf^p_1+(u|a,0)}
&&\!\!\!\!\!\!\!\!\!\!\!\!\im \left[\frac{\ln^p(\bu\mp i \epsilon)}{u^a}\right]^{\rm sub}   =
\mp \pi \theta(-\bu) \sum_{{m=1 \atop {\rm odd}}}^{p} \left({p \atop m}\right) (i \pi)^{m-1}\;  \frac{\ln^{p-m}|\bu|}{u^a}\,,
\end{eqnarray}
\end{subequations}
respectively,  where as indicated the sum over $m$ is restricted to even and odd $m$ values.  Note that apart from the additional
subtraction term the sums arise also from the well known formula  $$\ln^p(\bu \pm i\epsilon) = \left[\ln|\bu| \pm i \pi\theta(-\bu)\right]^p. $$

Employing the integration (\ref{f^p-i+1}), we can now successively  generate from the $p=1$ case of (\ref{f^p_1+(u|a,0)}) the value of  $f^1_{i+1,+}(u|a,0)$, i.e., subtracted polylog functions.
For the dilog we  quote the real part while for the imaginary parts of polylogs  a closed formula is quoted:
\begin{subequations}
\label{f^p_{2,+}(u|a,0)}
\begin{eqnarray}
\label{ref^p_{2,+}(u|a,0)}
\re \left[\frac{\Li(u)}{u^a}\right]^{\rm sub} &\!\!\! =\!\!\!  &
\frac{ \theta(\bu) \Li(u) + \theta(-\bu)\left[\zeta(2)-\ln u\ln|\bu| -\Li(\bu)\right]}{u^a}
-\sum_{i=0}^{a-2}\frac{ u^{i-a}}{(i+1)^2} \,,
\nonumber
\\
{\ }
\\
\label{imf^p_{2,+}(u|a,0)}
\im \left[\frac{{\rm Li}_{i+1}(u\pm i\epsilon)}{u^a}\right]^{\rm sub}  &\!\!\! =\!\!\!  &
\pm  \pi\theta(-\bu) \frac{\ln^{i}u}{i!\,  u^{a}} \quad\mbox{for}\quad i\in\{0,1,2,\cdots\}\,.
\end{eqnarray}
\end{subequations}

Utilizing  (\ref{f^p-dif}), we can  now obtain the $b>0$ cases.
First, we remind that the  generating functions for  non-negative integer $b$  are considered as generalized functions in the mathematical
sense, which are defined according to the $\pm i \epsilon$  prescription \cite{GelShi64}. Therefore, in some given space of
test functions $\tau(u)$,  the relation (\ref{f^p-dif}) reads for the $b=1$ case
\begin{eqnarray}
\label{f_{1,+}^p(u|1,a)}
\int_{-\infty}^\infty \!du\,  f_{1,+}^p(u\pm i\epsilon,a,1)   \tau(u)  =
\frac{1}{p+1} \int_{-\infty}^\infty \!du\,   u^a f_{1,+}^{p+1}(u\pm i\epsilon,a,0)  \frac{d\,u^{-a}\tau(u)}{du}\,.
\end{eqnarray}
Applying the differential operator to the generalized functions $u^a f_{1,+}^{p+1}(u\pm i\epsilon,a,0)$, in the following denoted for simplicity as
$\theta(\bu) f(u)+\theta(-\bu) f(u)$, yields a new generalized function.  We denote the differentiation in the region $u\ge 1$ as a $+$-prescription,
\begin{subequations}
\label{[[]]_+}
\begin{eqnarray}
\label{[[]]_+-1}
\big[\!\big[\theta(-\bu) f^\prime(u)\big]\!\big]_+  &\!\!\!\equiv\!\!\! &   \frac{d}{du}\left[ \theta(-\bu) f(u)\right]
=
\lim_{\epsilon\to 0} \left[\theta(-\bu)  f^\prime(u+\epsilon)  + \delta(\bu)   f(1+\epsilon)\right]\,,
\end{eqnarray}
where the (infinite) constant, concentrated in $u=1$, is regularized by $\epsilon$ and can be represented as
\begin{eqnarray}
\label{[[]]_+-2}
f(1+\epsilon) = - \int^{u_1}_1\!du\, f^\prime(u+\epsilon)  +c(u_1)\quad\mbox{with}\quad c(u_1) = f(u_1)\,.
\end{eqnarray}
\end{subequations}
Obviously, it cancels the non-integrable singularity  at $u=1$ in the integral that contains $f^\prime(u)$ and so
the limit $\epsilon \to 0$ can be interchanged with the integration.
Certainly, some care is needed in formal manipulations.  However, since  one set of generalized functions,
defined by the $\pm i\epsilon$ prescription, is expressed by another one,
a potentially ambiguous constant that is concentrated in the point $u=1$ is actually fixed.
Still we have the freedom to change the upper integration limit $u_1$ in (\ref{[[]]_+-2}) which alters then also the value of the finite constant
$c(u_1)=f(u_1)$.  Finally, if we include the region $u \le 1$,
the differentiation provides us $\theta(\bu) f^\prime(u) + \theta(-\bu) f^\prime(u)$  for $\lim_{\epsilon\to 0} \left[f(1-\epsilon)-f(1+\epsilon) \right] =0$,
which is nothing  but the principal value prescription for $f^\prime(u)$,
\begin{eqnarray}
\label{[]_+-2}
\frac{d}{du}\left[ \theta(\bu) f(u) + \theta(-\bu) f(u)\right] &\!\!\!\equiv\!\!\! &  {\cal P} f^\prime(u)\,.
\end{eqnarray}

Utilizing (\ref{f_{1,+}^p(u|1,a)}), we are now in the position to express, e.g., the  functions  $f^p_{1,+}(u \pm i\epsilon|a,b)$ by differentiation
of $f^{p+1}_{1,+}(u \pm i\epsilon|a,b-1)$ in terms of the $+$-prescription (\ref{[[]]_+}) and the principal value (\ref{[]_+-2}).
Differentiation of (\ref{f^p_1+(u|a,0)}) yields the real and imaginary parts for the $b=1$ and $a=0$ case:
\begin{subequations}
\label{f^p_{1,+}(u|0,1)}
\begin{eqnarray}
\label{ref^p_{1,+}(u|0,1)}
\re \frac{\ln^p(\bu)}{\bu}  &\!\!\! =\!\!\!  &
{\cal P}\frac{\ln^p|\bu|}{\bu}
+\sum_{{m=2 \atop {\rm even}}}^{p} (i \pi)^{m} \left({p \atop m}\right) \;
\bigg[\!\!\bigg[\frac{\theta(-\bu) \ln^{p-m}|\bu|}{\bu} \bigg]\!\!\bigg]_{+} - \frac{\re(i \pi)^{p+1}}{p+1} \delta(\bu) \,,\qquad
\\
\label{imf^p_{1,+}(u|0,1)}
\im \frac{\ln^p(\bu \mp i\epsilon)}{\bu \mp i\epsilon}    &\!\!\! =\!\!\!  &
\pm \pi \sum_{{m=1 \atop {\rm odd}}}^{p}  \left({p\atop m}\right)(i \pi)^{m-1}\; \bigg[\!\!\bigg[\frac{\theta(-\bu) \ln^{p-m}|\bu|}{\bu} \bigg]\!\!\bigg]_{+}
\pm\frac{\im(i \pi)^{p+1}}{p+1} \delta(\bu)\,,
\end{eqnarray}
where again the sum runs over even and odd $m$  values, respectively.  Besides  the
principal value ${\cal P} \log^p|\bu|/\bu$  additional terms appear in the real part (\ref{ref^p_{1,+}(u|0,1)}). They are the difference to the principal value of our original function, which was not defined  in terms of the modulus. Analogously, one may also derive the results for $b>1$, which yields then $+$-definitions that generate a truncated Taylor expansion of the order $b-1$.  The subtraction constant in this $+$-prescription $[\![f^\prime]\!]_+$ is fixed from the requirement (\ref{[[]]_+-2}). It reads as function of the integral limit $u_1$
\begin{eqnarray}
c_{f^\prime}(u_1)  =  \frac{\ln^{p+1}\!(u_1-1)}{p+1}  \quad \mbox{for} \quad  f^\prime= \frac{\ln^{p}(u-1)}{\bu}\,.
\end{eqnarray}
\end{subequations}

We can now employ the integral relation (\ref{f^p-i+1}), to find from the $p=1$ case of (\ref{f^p_{1,+}(u|0,1)}) the real and imaginary values of  $\Li(u\pm i\epsilon)/(\bu\mp i\epsilon)$. Alternatively, we can use the identity (\ref{eq:Li2-identity}) to evaluate both parts
\begin{subequations}
\label{f^2_{2,+}(u|0,1)}
\begin{eqnarray}
\label{f^2_{2,+}(u|0,1)-re}
\re \frac{\Li(u)}{\bu} &\!\!\! =\!\!\!  & {\cal P} \frac{\re\,\Li(u)}{\bu}\,,
\qquad
\\
\im  \frac{\Li(u \pm i\epsilon)}{\bu \mp i\epsilon}   &\!\!\! =\!\!\!  & \mp \theta(-\bu)\frac{\pi  \ln u}{-\bu} \pm \frac{\pi^3}{6} \delta(\bu)\,,
\end{eqnarray}
where the imaginary part arises from $\ln\bu$ and the real part of the dilogarithm  can be written as
\begin{eqnarray}
\re\,\Li(u) \equiv \theta(\bu) \Li(u) + \theta(-\bu) \left[\zeta(2)-\ln u \ln|\bu|-\Li(\bu) \right].
\end{eqnarray}
\end{subequations}
We also encounter the $b=2$ case in the flavor non-singlet channel. It can be treated in an analogous manner, most easily  by  considering the subtracted function (\ref{eq:uv-func-5}),
$$
\frac{\Li(u)-\zeta(2)-\bu\ln\bu +\bu}{\bu^2}\,,
$$
and the known results (\ref{f^p_{1,+}(u|0,1)}) for the subtraction terms ($b=1$, $a=0$, $p \in\{0,1\}$),
\begin{subequations}
\label{f^2_{2,+}(u|0,2)}
\begin{eqnarray}
\label{f^2_{2,+}(u|0,2)-re}
\re \frac{\Li(u )-\zeta(2)}{(\bu)^2}  &\!\!\! =\!\!\!  &
{\cal P} \frac{\re\, \Li(u)-\zeta(2)}{\bu^2}  + \frac{\pi^2}{2}\delta(\bu) \,,
\\
\im  \frac{\Li(u \pm i\epsilon)-\zeta(2)}{(\bu \mp i\epsilon)^2}   &\!\!\! =\!\!\!  &
\pm  \theta(-\bu)\frac{\pi  [\ln u + \bu]}{\bu^2}  \pm \pi \bigg[\!\!\bigg[ \frac{\theta(-\bu)}{-\bu} \bigg]\!\!\bigg]_{+}  \mp  \pi \delta(\bu)\,,
\end{eqnarray}
\end{subequations}
where  one subtraction is still used to  remove the second order pole. Note that in convolution integrals the pole at $u=1$,
appearing in the real parts (\ref{f^2_{2,+}(u|0,1)-re}) and (\ref{f^2_{2,+}(u|0,2)-re}), is treated as Cauchy's principal value integral.

\renewcommand{\arraystretch}{2}
\begin{table}
{\footnotesize
\begin{center}
\begin{tabular}{|l|l|l|l|}
\multicolumn{4}{c}{} \\ \hline
$\displaystyle \frac{f(u+i\epsilon)}{(u-i\epsilon)^{a} (\bu-i\epsilon)^{b}} $
& $\displaystyle  \re\frac{f(u+i\epsilon)}{(u-i\epsilon)^{a} (\bu-i\epsilon)^{b}}
$
& $\displaystyle  \frac{1}{\pi} \, \im\frac{f(u+i\epsilon)}{(u-i\epsilon)^{a} (\bu-i\epsilon)^{b}} $ &
$\displaystyle  \frac{x}{2\xi\,\pi} \, \im\frac{f(u+i\epsilon)}{(u-i\epsilon)^{a} (\bu-i\epsilon)^{b}} $
\\[-0.0cm] \hline\hline
$a\in \{0,1,2\}, b=0$  & &  &\\[-0.1cm] \hline
$\displaystyle \frac{\ln(\bu-i \eps)}{(u-i \eps)^a} $ &
$\displaystyle {\cal P} \frac{\ln|\bu|}{u^a} $&
$ \displaystyle - \frac{\theta(-\bu)}{u^a} -  \delta_{a,2}\, \delta(u) $ &
$\displaystyle -\frac{(2r)^{a-1}}{(1+r)^{a}} +  \delta_{a,2}\, \sigma\, \delta(1-r)$
\\[0.0cm]
$\displaystyle \frac{\ln^2(\bu - i \eps)}{u^a}$ &
$\displaystyle \frac{\ln^2|\bu|}{u^a} - \frac{\pi^2}{u^a} \, \theta(-\bu)$ &
$\displaystyle -  \frac{\theta(-\bu)\, 2\ln|\bu|}{u^a} $ &
$\displaystyle  -  \frac{2(2r)^{a-1}\ln\frac{1-r}{2r}}{(1+r)^{a}}$
\\[0.0cm]
$\displaystyle \frac{\Li(u + i \eps)}{(u- i \eps)^a} $ &
$\displaystyle {\cal P} \frac{\re\,\Li(u)}{u^a} $&
$ \displaystyle \frac{\theta(-\bu)\, \ln u}{u^a}  + \delta_{a,2}\, \delta(u)$ &
$\displaystyle  \frac{(2r)^{a-1}\ln\frac{1+r}{2r}}{(1+r)^{a}} - \delta_{a,2}\, \sigma\, \delta(1-r)$
\\[-0.0cm]\hline\hline
$a=0, b=1$  & & & \\[-0.1cm] \hline
$ \displaystyle \frac{1}{\bu-i \eps} $ &
$ \displaystyle
{\cal P} \frac{1}{\bu} $ &
$ \delta(\bu) $ &
$\delta(1-r)$
\\[0.0cm]
$\displaystyle \frac{\ln(\bu-i \eps)}{\bu-i \eps} $ &
$\displaystyle {\cal P} \frac{\ln|\bu|}{\bu} + \frac{\pi^2}{2} \delta(\bu)$ &
$ \displaystyle  \bigg[\!\!\bigg[ \frac{\theta(-\bu)}{-\bu} \bigg]\!\!\bigg]_{+}$ &
$ \left\{ \frac{1}{1-r} \right\}_{+} $
\\[0.0cm]
$\displaystyle \frac{\ln^2(\bu-i \eps)}{\bu-i \eps} $ &
$\displaystyle {\cal P}\frac{\ln^2 |\bu|}{\bu} + \pi^2 \bigg[\!\!\bigg[ \frac{\theta(-\bu)}{-\bu} \bigg]\!\!\bigg]_{+}\!\!\!  $ &
$ \displaystyle \bigg[\!\!\bigg[ \frac{\theta(-\bu)\,2 \ln|\bu|}{- \bu} \bigg]\!\!\bigg]_{+}\!\!\! - \frac{\pi^2}{3} \delta(\bu) $ &
$\displaystyle \left\{ \frac{2\ln\frac{1-r}{2r}}{1-r} \right\}_{+}$
\\[0.0cm]
$\displaystyle \frac{\Li(u+i \eps)}{\bu-i \eps} $ &
$\displaystyle {\cal P}\frac{\re\,\Li(u)}{\bu}  $ &
$ \displaystyle  \frac{\theta(-\bu)\,\ln u}{\bu}  + \frac{\pi^2}{6} \delta(\bu) $ &
$\displaystyle - \frac{\ln\frac{1+r}{2r}}{1-r}  +\frac{\pi^2}{6}\delta(1-r)$
\\[0.1cm]\hline\hline
$a=0, b=2$  & & & \\[-0.1cm] \hline
$\displaystyle \frac{\Li(u+i \eps)-\zeta(2)}{(\bu-i \eps)^2} $ &
$\displaystyle {\cal P}\frac{\re\,\Li(u)-\zeta(2)}{\bu^2}  $ &
$ \displaystyle  \frac{\theta(-\bu)[\ln u+\bu]}{\bu^2}  $ &
$\displaystyle \frac{2r \ln\frac{1+r}{2r} - 1+r}{(1-r)^2} $
\\
$$ &
$ \displaystyle\phantom{\theta(-\bu)} + \frac{\pi^2}{2}\delta(\bu) $ &
$ \displaystyle\phantom{\theta(-\bu)} + \bigg[\!\!\bigg[ \frac{\theta(-\bu)}{-\bu} \bigg]\!\!\bigg]_{+} - \delta(\bu) $ &
$ \displaystyle\phantom{\theta(-\bu)}   +\left\{ \frac{1}{1-r} \right\}_{+}  - \delta(1-r)  $
\\[0.1cm]
\hline
\end{tabular}
\end{center}
\caption{\small Equivalence of selected general functions, where the explicit expressions for real and imaginary parts arise
from (\ref{f^p_1+(u|a,0)},\ref{f^p_{2,+}(u|a,0)},\ref{f^p_{1,+}(u|0,1)},\ref{f^2_{2,+}(u|0,1)},\ref{f^2_{2,+}(u|0,2)}).
The $[\![\cdots]\!]_+$-prescriptions is defined in  (\ref{[[]]_+}).  In the forth column we give the imaginary part in the notation
of the main body of this paper for $0 \le r= \xi/x \le 1$, see for example (\ref{eqs:DRtffFqC}), and the $\sigma$ signature factor (\ref{eq:sigma4F}), where the
$\{\cdots\}_+$-prescriptions are defined in (\ref{{}_+}) and they are explicitely given in (\ref{f+}). Note that the principal value is only needed to
treat remaining $1/u$ or $1/\bu$ singularities and that it can be dropped for $a\in \{0,1\}$ of the $b=0$ case.
\label{tab:ReImf} }
}
\end{table}

In our NLO case, considered in Sec.~\ref{sec:NLO-blocks1}, the non-negative integer $p$ of $f^p_{1}(u|a,b) $ is limited to $p \le 2$ and  we can restrict ourselves
to the  cases $a \le 2$ for $b=0$ and $b \le 1$ for $a=0$. The same choices we need for  $f^1_2(u|a,b)$ and in addition the case $b = 2$ for $a=0$.
To be very explicit, we finally collect the results for the cases of interest in the common nomenclature in Tab.~\ref{tab:ReImf} for the original building blocks (\ref{f^p_i}). Here, we  take the $-i\epsilon$ prescription according to the variable $u=\frac{\xi+x-i\epsilon}{2(\xi-i\epsilon)}$, see Sec.~\ref{sec:prel-DR}.  Consequently, we have $\bu=\frac{\xi-x-i\epsilon}{2(\xi-i\epsilon)}$ and set
$$
\frac{\ln^p(\bu-i\epsilon)}{(u-i\epsilon)^a (\bu-i\epsilon)^b}\,, \quad  \mbox{however}\,, \quad \frac{\Li(u+i\epsilon)}{(u-i\epsilon)^a (\bu-i\epsilon)^b}\,.
$$
In passing from the considered functions, having only a $[1,\infty]$ cuts, to the building blocks (\ref{f^p_i}) we also include poles at $u=0$,  which appear in the $a=2$ case:
$$
\frac{\ln(\bu- i\epsilon)}{(u- i\epsilon)^2}
= \frac{\ln\bu +u}{(u- i\epsilon)^2} - {\cal P} \frac{1}{u}  - i \pi \delta(u) \,, \quad
\frac{\Li(u + i\epsilon)}{(u-i\epsilon)^2}
=  \frac{\Li(u+ i\epsilon) -u}{(u-i\epsilon)^2} + {\cal P} \frac{1}{u}  + i \pi \delta(u).
$$
We also include in the Tab.~\ref{tab:ReImf} the expressions for the imaginary part in terms of the variable $r=\xi/x$ as introduced in Sec.~\ref{sec:prel-DR}, i.e., $u= (1+r)/2r$, and used in the presentation of NLO corrections in Sec.~\ref{sec:NLO}. For convenience we use in the main body the $+$-prescriptions
\begin{subequations}
\label{{}_+}
\begin{eqnarray}
\left\{\frac{\theta(r)}{1-r} \right\}_+ &\!\!\!=\!\!\!& \frac{1}{2r} \bigg[\!\!\bigg[ \frac{\theta(-\bu)}{- \bu} \bigg]\!\!\bigg]_{+}
\qquad \mbox{with} \quad  c_f = \ln\frac{1-\xi}{2\xi}\,,
\\
 \left\{\frac{\theta(r) \ln\frac{1-r}{2r}}{1-r} \right\}_+ &\!\!\!=\!\!\!& \frac{1}{2r}\bigg[\!\!\bigg[ \frac{\ln|\bu|\,\theta(-\bu)}{- \bu} \bigg]\!\!\bigg]_{+} -\zeta(2) \delta(\bu)
\qquad \mbox{with} \quad  c_f = \frac{1}{2}\ln^2\frac{1-\xi}{2\xi}  -\zeta(2)\,,
\qquad\quad
\end{eqnarray}
\end{subequations}
in which the additional $\zeta(2)$ term is absorbed in the subtraction constant $c_f$ and they are fixed in such a manner that GPD convolution
integrals (\ref{f+}) take a simple form. Note that the support restriction $ r\le 1$ of $\{\cdots\}_+$ is not explicitly indicated.

\section{The non-separable function $L(u,v)/(u-v)$}
\label{app:H0}
\setcounter{equation}{0}

As explained in the main text, in momentum fraction representation the non-separable contributions are proportional to $1/(u-v)^n$, which can be expressed
in terms of the function
\begin{eqnarray}
\frac{L(u,v)}{u-v}\quad\mbox{with}\quad L(u,v) &\!\!\! = \!\!\! &
\Li(\bv) -\Li(\bu)   +\ln \bu \ln v -\ln u \ln \bu\,,
\nonumber\\
&\!\!\! = \!\!\! &  \Li(\bv)  +\Li(u)   +\ln \bu \ln v   -\zeta(2)\,.
\label{eq:Lovumv-elementar}
\end{eqnarray}

\subsection{Representations and holomorphic properties}
\label{app:L}
To introduce an integral representation of the function \req{eq:Lovumv-elementar}, we may follow \cite{Mueller:1998qs} and expand this function as
\begin{equation}
\frac{L(u,v)}{u-v} =  \sum_{m=0}^\infty
\,
 \sum_{n=0}^\infty
\left[\frac{1}{(n+m+1)^2}- \frac{\ln \bu}{n+m+1}
\right]
\bu^m \, \bv^n
\, .
\end{equation}
To convert now this sum into an integral, we employ the ``kernel'' method. Expressing the coefficients
in the expansion by the convenient integrals
$\int_0^1 dz \, z^{m+n}$ and $\int_0^1 dz \,z^{m+n} \ln z$ and
noticing that
$\sum_{m=0}^\infty (\bu z)^m=1/(1-\bu z)$ one gets rid of the sums
and we obtain
\begin{eqnarray}
\frac{L(u,v)}{u-v}=
 -\int_0^1 \! dz\, \frac{\ln (\bu z)}{(1-\bu z) (1-\bv z)}
\, .
\label{eq:Ltoint}
\end{eqnarray}
Finally, to convert this integral into a double dispersion integral, we plug in the representation
\begin{equation}
\frac{\ln \bu z}{1-\bu z} = \int_0^1 dy \frac{-1}{y+z-y z} \frac{1}{1-u y}
 = \int_0^1 dy \frac{-1}{1-\bz \by} \frac{1}{1-u y}
\label{eq:lnuzint}
\end{equation}
into the integral (\ref{eq:Ltoint}) and find the desired representation
\begin{eqnarray}
\label{L-conv}
\frac{L(u,v)}{u-v}=
 \int_0^1\!  dy \int_0^1 \! dz\,
 \frac{1}{1-u y}\, \frac{1}{1-\bz \by}\, \frac{1}{1-\bv z}\,,
\label{eq:Ltodoubleint}
\end{eqnarray}
where we use the shorthand notation $\by\equiv 1-y$ and $\bz\equiv 1-z$. In this representation the $u\leftrightarrow \bv $ symmetry is manifest.

We add that from the symmetric integral representation (\ref{L-conv}) one can easily write
down a symmetric sum representation. The expansion of the integral kernel in  (\ref{L-conv}) in powers of $\bz \by$ yields simple integrals that allow to express the expansion coefficients of the  sum  representation in  terms  of hypergeometric ${_3}F_2$ functions
\begin{equation}
\frac{L(u,v)}{u-v} =  \sum_{m=0}^\infty
\,
 \sum_{n=0}^\infty {_3}F_2\!\left({1,1,1 \atop m+2, n+2}\Big|1\! \right) \frac{u^m\,  \bv^n}{(m+1)(n+1)}\,.
\end{equation}
In the class of ${_3}F_2$ functions with unit argument this function can be represented in various manner, e.g.,
$$
{_3}F_2\!\left({1,1,1 \atop m+2, n+2}\Big|1\! \right)= \frac{\Gamma(m+2)\Gamma(n+2)}{(2+m+n)\Gamma(2+m+n)} \,
{_3}F_2\!\left({m+1,n+1,m+n+1 \atop m+n+2, m+n+2}\Big|1\! \right)\,,
$$
which follows from Thomae's identity.
However, it is not given as a ratio of $\Gamma$ functions rather it can be expanded in terms of subtracted harmonic sums.

Next the unsubtracted double dispersion relation (\ref{eq:Ltodoubleint}) can be derived in the common manner where one may start from
\begin{eqnarray}
\frac{L(u,v)}{u-v}= \int_{-\infty}^\infty\! du^\prime\! \int_{-\infty}^\infty\!  dv^\prime\,
\frac{1}{u^\prime-u}
\left(\frac{1}{\pi^2} \im_{u^\prime}  \im_{v^\prime} \frac{L(u^\prime,v^\prime)}{u^\prime-v^\prime}\right)
\frac{1}{v^\prime-v}\,.
\label{eq:L-DR}
\end{eqnarray}
Obviously, (\ref{eq:Ltodoubleint}) tells us that the function $L(u,v)/(u-v)$ contains in the complex $u$ and $v$ planes cuts along on the real axes  $u \ge 1$ and $v \le 0$, which we may directly evaluate from the $\ln \bu \ln v$ term of the representation (\ref{eq:Lovumv-elementar}),
$$
\frac{1}{\pi^2} \im_{u}  \left(\im_{v} \frac{L(u,v)}{u-v} \right)=\frac{1}{\pi^2} \im_{u}  \im_{v} \frac{\ln \bu \ln v}{u-v} = \frac{\theta(u-1) \theta(-v)}{u-v}\,,
$$
where we used $u\pm i\epsilon$ (or $\bu \mp i\epsilon$) and $v\pm i\epsilon$  prescriptions.
Plugging this into the dispersion relation (\ref{eq:L-DR}) and mapping the integral regions $0 \le u^\prime \le \infty $ and $-\infty  \le v^\prime \le 0$ to
$0\le y \le 1$ and $0\le z \le 1$ by the SL$(2,\mathbb{R})$ transformations
$$ u^\prime =\frac{1}{y} \quad\mbox{and}\quad v^\prime =-\frac{1-z}{z},$$
respectively, yields the integral (\ref{eq:Ltodoubleint}), where the imaginary part, i.e., $1/(u^\prime-v^\prime)$, translates into the integral kernel $1/(1-\by \bz)$.

The double dispersion relation \req{eq:Ltodoubleint}, used here to represent $L(u,v)/(u-v)$, appears in the first place
as a mathematical construct. We are interested on its physical value on the branch cut $u\ge 1$ (or positive $x\ge \xi$) for $0\le v\le 1$,
which arises from the $\xi-i \epsilon$ prescription. Hence, the following term in the double DR (\ref{eq:Ltodoubleint}) has to be decorated with
a $-i\epsilon$ prescription
$$
\frac{1}{1- u y} \Rightarrow \frac{1}{1-\frac{\xi-i\epsilon +x}{2(\xi-i\epsilon)} y}  \Rightarrow \frac{1}{1-u y - i \epsilon^\prime}\,,
$$
where $\epsilon^\prime = \epsilon(2-y)/2\xi$ is a positive quantity in the integration region. Consequently, the imaginary part of the integrand is $i\pi \delta(1-u y)$
and we obtain
\begin{eqnarray}
\frac{1}{\pi} \im \frac{L(u,v)}{u-v}= \theta(u-1)
  \int_0^1 \! dz\, \frac{1}{1-\bu z }\, \frac{1}{1-\bv z}
  =\frac{\theta(u-1) }{u - v} \ln\frac{u}{v}\,.
\label{eq:Ltodoubleint-Im}
\end{eqnarray}
According to this finding, the physical value of the $L(u,v)$ function is given by the prescription
\begin{eqnarray}
L(u,v) &\!\!\! = \!\!\!& \Li(\bv)  +\Li(u+i\epsilon)   +\ln(\bu-i\epsilon) \ln\bv   -\zeta(2) \,,
\nonumber\\
&\!\!\! = \!\!\!&   \Li(\bv)  -\Li(\bu+i\epsilon)   + \ln(\bu-i\epsilon)\ln v -\ln u \ln (\bu-i\epsilon)\,.
\label{eq:L-phys}
\end{eqnarray}

We add that the same exercise can be repeated for the function $L(\bu,\bv)/(\bu-\bv)$, which now possess an imaginary part for
negative $u\le 0$ or for negative $x \le -\xi$. The result for the physical value of this function arises simply from (\ref{eq:L-phys})
by the replacement $u\to \bu$ and $v\to \bv$. For the function $H_0(u,v)=L(u,v)-L(\bu,\bv)$, we have then
\begin{eqnarray}
H_0(u,v)=
\Li (u+i \eps)
-\Li (\bu+i \eps)
-\Li (v)
+\Li (\bv)
-\ln (u-i \eps) \ln \bv
+\ln (\bu-i \eps) \ln v
\, ,
\label{eq:fullH0}
\qquad
\end{eqnarray}
and thus the imaginary part of
\begin{eqnarray}
\frac{1}{\pi} \im \frac{H_0(u,v)}{u-v} =
 \frac{\theta(u-1)}{u-v} \ln\frac{u}{v}
- \frac{\theta(-u)}{u-v} \ln\frac{\bu}{\bv}
=
\frac{\theta(-\bu)}{u-v} \ln\frac{u}{v}
+\frac{\theta(-u)}{\bu-\bv} \ln\frac{\bu}{\bv}
\, .
\label{eq:ImH0}
\end{eqnarray}
is symmetric under $u\to \bu, v\to \bv$ exchange. Note that
the non-separable terms in the diagrammatical results%
\footnote{The $H$ and $R$ functions from \cite{Ivanov:2004zv} read
$H(z,y)=L(u,v)-L(\bu,\bv)$ and $R(z,y)= v \bv \partial \left(L(u,v)-L(\bu,\bv)\right)/\partial v$ with $z=\bv$ and $y=-\bu$,
while in \cite{Melic:1998qr} the functions $H$ and $R$ have quite different definitions.}
can be expressed also in terms of the antisymmetric function (\ref{eq:fullH0})
which has now $[-\infty,0]$ and $[1,\infty]$ cuts  on both sides of the real axes.
If we allow for an analytic extension in $v$, the function can be also understood as symmetric under $u\leftrightarrow v$ exchange. This is explicitly implemented in the DR representation,
\begin{eqnarray}
\frac{H_0(u,v)}{u-v} =
 \int_0^1\!dy\!\int_0^1\!dz \frac{1}{1-\bz \by}
\left(\!\frac{1}{1-u y - i \epsilon}\frac{1}{1-\bv z- i \epsilon}
+\frac{1}{1-\bu z- i \epsilon}\frac{1}{1-v y- i \epsilon}
\!\right).
\label{eq:H0todoubleint}
\end{eqnarray}

\subsection{Diagrammatical origin}
\label{app:diag}

The origin of non-separable terms in the hard scattering amplitude can be traced back to
the appearance of scalar three-point Feynman integrals that occur
in several Feynman diagrams contributing to
$\gamma^\ast_L  q\to (q \bar q) q $ and $\gamma^\ast_L  g \to (q \bar{q}) g$
subprocesses.

The scalar three-point integral
\begin{equation}
I_3(p^2,k^2,2p k)=
 \int \frac{d^{4} l}{(2 \pi)^4}
  \frac{1}{[l^2 + i \eps][(l-p)^2 + i \eps][(l-k)^2 + i \eps]}
\label{eq:scalar3point}
\end{equation}
with $p^2\ne0$, $k^2\ne0$, $2p\cdot k \ne 0 \ne p^2+k^2$,
i.e., $(p-k)^2\ne0$ is a (UV and IR) finite integral.
We give here the most simple form derived in \cite{Duplancic:2002dh}:
\begin{subequations}
\begin{eqnarray}
I_3 & =&
               \frac{i}{(4 \pi)^2} \,
               \frac{1}{\nu_3 (x_1-x_2)} \,
 \left\{ 2 \mathrm{Li}_2\left(\frac{1}{x_2}\right)
       - 2 \mathrm{Li}_2\left(\frac{1}{x_1}\right)
\right.
\nonumber
\\[0.3cm]
& &
 \left.
+\mathrm{ln}(x_1 x_2 + i \eps \, \mathrm{sign} \, \nu_3)
   \left[
  \mathrm{ln}\frac{1-x_1}{-x_1}
  -\mathrm{ln}\frac{1-x_2}{-x_2}
   \right]
\right\} \, , \qquad
\label{eq:barI3}
\end{eqnarray}
where $x_{1,2}$ are solutions of the equation
\begin{equation}
x \nu_1 + (1-x) \nu_2 - x (1-x) \nu_3 = 0 \, ,
\label{eq:x12eqn}
\end{equation}
$D$ is its discriminant, while
\begin{equation}
\{ \nu_1, \nu_2, \nu_3 \}  = {\cal P}(p^2,k^2, (p-k)^2) \, ,
\label{eq:nu123}
\end{equation}
is a  permutation chosen  such that $x_{1,2} \not \in [0,1]$
-- in practice, $\nu_3$ should have the smallest absolute value or opposite sign.
Note,
$\nu_3 (x_1-x_2)=\sqrt{D}=\nu_1^2 + \nu_2^2 + \nu_3^2 - 2 \nu_1 \nu_2 - 2 \nu_1 \nu_3
   -2 \nu_2 \nu_3$ is invariant under $\nu_{1,2,3}$ permutations,
and $x_1 x_2=\nu_2/\nu_3$.
\end{subequations}
Alternatively, one can express \req{eq:x12eqn} as
\begin{subequations}
\begin{equation}
 (q_2+x q_3)^2=0 \, ,
\label{eq:x12eqnNew}
\end{equation}
where $q_1+q_2+q_3=0$ and
\begin{equation}
\{ q_1, q_2, q_3 \}  = {\cal P}(p,-k, -(p-k)) \, ,
\label{eq:q123}
\end{equation}
while $q_i^2=\nu_i$.
\end{subequations}

In our process of interest one encounters
$
\{ \nu_1, \nu_2, \nu_3 \}
={\cal P}(-Q^2,-u v Q^2, -\bu\bv Q^2 )
$
and one can take $\nu_3=-\bu\bv Q^2$
or $\nu_3=-u v Q^2$.
For $u \in \mathbb{R}$ while $0<v<1$,
the result takes the form
\begin{eqnarray}
I_3&=&\frac{i}{(4\pi)^2} \frac{1}{u-v}
\big[
\mbox{Li}_2(v) - \mbox{Li}_2(1-v)
-\mbox{Li}_2(u + i \eps ) - \mbox{Li}_2(1-u + i \eps)
\nonumber \\ & &
+ \mbox{ln}(u - i \eps) \, \mbox{ln}(1-v)
- \mbox{ln}(1-u - i \eps) \, \mbox{ln}(v)
\big] \, ,
\nonumber
\end{eqnarray}
i.e.,
\begin{equation}
I_3 = -\frac{i}{(4\pi)^2} \frac{H_0(u,v)}{u-v}
\, ,
\end{equation}
with $H(u,v)$ given by \req{eq:fullH0} (i.e., the expression is in agreement with the $\xi-i\epsilon$ prescription).


\end{document}